\documentclass[twocolumn,twocolappendix,trackchanges]{aastex631}
\usepackage{amsmath,bm}
\usepackage{subfigure}
\usepackage{xcolor}
\usepackage{appendix}

\begin{document}

\title{EUV polarimetric diagnostics of the solar corona: the Hanle effect of Ne {\sc viii} 770 \AA}

\author[0000-0001-6104-8938]{Raveena Khan}
\affiliation{Indian Institute of Astrophysics, Bengaluru 560034, India}
\affiliation{High Altitude Observatory, National Center for Atmospheric Research, Boulder, CO 80307, USA}
\affiliation{University of Calcutta, College Street, Kolkata 700073, India}

\author[0000-0001-9831-2640]{Sarah E. Gibson}
\affiliation{High Altitude Observatory, National Center for Atmospheric Research, Boulder, CO 80307, USA}

\author[0000-0001-6990-513X]{Roberto Casini}
\affiliation{High Altitude Observatory, National Center for Atmospheric Research, Boulder, CO 80307, USA}

\author[0000-0002-0465-8032]{K. Nagaraju}
\affiliation{Indian Institute of Astrophysics, Bengaluru 560034, India}

\begin{abstract}
Magnetic fields are the primary driver of the plasma thermodynamics in the upper solar atmosphere, especially in the corona. However, magnetic field measurements in the solar corona are sporadic, thereby limiting us from the complete understanding of physical processes occurring in the coronal plasma. In this paper, we explore the diagnostic potential of a coronal emission line in the extreme-ultraviolet (EUV), i.e., Ne {\sc viii} 770 \AA\ to probe the coronal magnetic fields. 
We utilize 3D 'Magneto-hydrodynamic Algorithm outside a Sphere' (MAS) models as input to the FORWARD code to model polarization in Ne {\sc viii} line produced due to resonance scattering, and interpret its modification due to collisions and the magnetic fields through the Hanle effect. The polarization maps are synthesized both on the disk as well as off-the-limb. The variation of this polarization signal through the different phases of solar cycle 24 and the beginning phase of solar cycle 25 is studied in order to understand the magnetic diagnostic properties of this line owing to different physical conditions in the solar atmosphere. The detectability of the linear polarization signatures of the Hanle effect significantly improves with increasing solar activity, consistently with the increase in the magnetic field strength and the intensity of the mean solar brightness at these wavelengths. We finally discuss the signal-to-noise ratio (SNR) requirements by considering realistic instrument designs. 
\end{abstract}


\section{Introduction} \label{sec:intro}
The magnetic field of the solar corona is a key ingredient to understand the fundamental plasma processes in the upper solar atmosphere, such as hydro-magnetic instabilities, acceleration of energetic particles, and the energization of the million-degree solar corona. Manifestations of these magnetic fields are the structures of inhomogeneous and dynamic hot plasma observed in the corona. However, the formation and evolution of these structures are poorly understood, mainly due to the lack of routine measurements of their prime driver: the coronal magnetic field.

To tackle these problems, there have been numerical modeling efforts that attempt to predict the coronal magnetic structure through  extrapolation from photospheric fields \citep[and references therein] {Mackay2012LRSP....9....6M}. Observations of ubiquitous waves in the corona (`coronal seismology') have been used to provide synoptic measurements of the plane-of-sky (POS) component of the coronal field \citep{YangGlobal2020Sci...369..694Y}. The few existing measurements of the circular polarization profiles of the Fe~{\sc xiii} line at 10747\,\AA\ due to the longitudinal Zeeman effect have inferred coronal field strengths between approximately 10 to 30 gauss above active regions \citep{Lin2000ApJ...541L..83L}. There has been a single spectro-polarimetric observation of the linear polarization of O {\sc vi} 1032\,\AA, which was performed by rotating the SUMER (Solar Ultraviolet Measurements of Emitted Radiation) spectrometer, exploiting the instrument polarization, and a field strength of about 3 gauss was derived above a coronal hole \citep{RaouafiOVIdetect1999A&A...345..999R,RaouafiOVIBmeasure2002A&A...396.1019R,RaouafiOVImeasurepol2002A&A...390..691R}. Such an impromptu and unoptimized measurement of resonance line polarization demonstrates that ultraviolet (UV) spectro-polarimeters onboard space telescopes are capable of providing critical magnetic field measurements in the corona.

The application of the Hanle effect in the far ultraviolet (FUV) and extreme ultraviolet (EUV) spectral ranges is one of the potential methods to diagnose the coronal magnetic fields. The Hanle effect \citep{Hanle1924ZPhy...30...93H} refers to the modification of the polarization degree and rotation of the plane of polarization of the scattered radiation in the presence of an external magnetic field. Extensive theoretical studies have been performed in the FUV coronal lines, i.e., O {\sc vi} 1032\,\AA\ \citep{SahalOVI1986A&A...168..284S, Trujillo2017SSRv..210..183T, Zhao2019ApJ...883...55Z} and Ly-$\alpha$ 1215\,\AA\ \citep{bommier1982SoPh...78..157B, Fineschi1991SPIE.1343..376F,  Hebbur2021ApJ...920..140H}. Recently, \cite{Khan2022SoPh..297...96K} have reported several electric-dipole (E1) transition lines in the UV with different sensitivity regimes to the unsaturated Hanle effect. These lines
can be exploited as potential diagnostics of the coronal magnetic field vector \citep{Fineschi1993SPIE.1742..423F,Fineschi1995sowi.confR..68F}, possibly enabling the technique of differential Hanle effect \citep{House1982SoPh...80...53H, Stenflo1998A&A...329..319S, Bueno2012ApJ...746L...9T} to make the coronal diagnostic more robust to model dependencies.

A fairly recent work by \cite{Raouafi2016FrASS...3...20R} used a 3D magnetohydrodynamic (MHD) model to derive the Hanle regime polarization signals in the solar corona (off-limb) in the FUV (H {\sc i} Lyman-$\alpha$) and the IR (He {\sc i} 10830\,\AA) lines. \cite{Zhao2019ApJ...883...55Z,Zhao2021ApJ...912..141Z} similarly examined the linear polarization of Lyman-$\alpha$ in the presence of both the Hanle effect and symmetry breaking induced by Doppler dimming for 3D coronal MHD models, and \cite{Hebbur2021ApJ...920..140H} used 3D coronal MHD models to investigate the linear polarization signals in the FUV (H {\sc i} Ly-$\alpha$ at 1215\,\AA) and the EUV (He {\sc ii} Ly-$\alpha$ at 304\,\AA) lines within 0.5 R$_{\odot}$ above the Sun's visible limb. All these diagnostics require a coronagraph instrument to detect the off-limb coronal signal and its polarization unhindered by the disk radiation. To our knowledge, MHD models have so far not been used to estimate the polarization signals of EUV coronal emission lines observed directly on the disk of the Sun. In the present work, we have utilized MHD simulation data cubes as inputs into the FORWARD code and have synthesized polarization maps of Ne {\sc viii} 770\,\AA\ both on-disk and off-limb. 

The FORWARD SolarSoft IDL package \citep{Gibson2016FrASS...3....8G} is an extensive toolset consisting of various analytic magnetostatic equilibrium solutions, in the form of physical models and datacubes, the predictions of which can be compared with the observations. FORWARD also computes synthetic observables, such as the Stokes polarization maps produced by scattering processes and the Hanle effect. By forward modeling the EUV coronal emission line of Ne {\sc viii} at 770\,\AA\, we illustrate its potential for probing the weak magnetic field of the solar corona through its sensitivity to the unsaturated Hanle effect. We have simulated the polarization signals for different phases of Solar Cycles 24 (SC24) and 25 (SC25). In Section \ref{sec:pol_ne}, we describe the scattering processes inducing polarization and the assumptions involved in the synthesis of the polarization maps of Ne {\sc viii} 770 \AA. The 3D MHD model used in FORWARD is described in Section \ref{sec:model}. In Section \ref{sec:results}, we discuss and illustrate the effects of the phases of the solar cycle, and the impact of collisions on the linear polarization signals of Ne {\sc viii} 770\,\AA. We also discuss the requirements on the signal-to-noise ratio (SNR) during the various phases of the solar cycle.

\section{Polarization of Ne {\sc viii}} \label{sec:pol_ne}

Polarization in the corona is mostly produced by the scattering of the anisotropic radiation from the underlying atmospheric layers of the Sun. In the case of atomic transitions, this anisotropic radiation induces population imbalance and quantum coherence among the atomic sublevels (mostly in the excited state), which cause the scattered radiation  to be predominantly linearly polarized, even in the absence of magnetic fields. This linear polarization tends to be larger when the scattered direction is orthogonal to the direction of the incident radiation, as in the case of scattering by coronal structures off the limb (cf.~Eqs~(\ref{eq:qoi_90}) and (\ref{eq:qoi_90_max}), and Figure~\ref{fig:rt_geo}). It tends instead to zero when the line-of-sight (LOS) approaches disk center (the case of forward scattering), unless other symmetry breaking mechanisms (such as an inclined magnetic field, or inhomogeneities of the solar radiation on the disk) introduce a preferential direction of linear polarization on the POS.

In this section, we describe the sensitivity of Ne {\sc viii} 770 \AA\ resonance line to the unsaturated Hanle effect in the weakly magnetized coronal plasma, and state the assumptions used in calculating the line's polarized emissivity (Stokes $I,Q,U$). When defining linear polarization, the reference axis for positive Stokes $Q$ is taken as the radial direction from disk center through the scattering point. Of course, such a direction is ill-defined at exact disk center.

\subsection{Magnetic sensitivity of scattering polarization in the unsaturated Hanle regime} 
The linear polarization of emission lines formed by scattering in a magnetized plasma are most sensitive to the Hanle effect in the domain
\begin{equation}
    0.1\lesssim g_{u} \omega_{B}\tau \lesssim 10
        \label{eq:HE_cond2}
\end{equation}
\cite[e.g.,][]{Bommier1978A&A....69...57B,Bommier1981A&A...100..231B},
where $g_{u}$ is the Land\'e factor of the upper atomic level of the transition, $\omega_B$ is the Larmor frequency, and $\tau$ is the lifetime of the excited level. The condition $g_{u} \omega_B \tau = 1$ determines the so-called \textit{critical} Hanle-effect field strength. The resonance line Ne {\sc viii} 770 \AA\ corresponds to the E1 transition between the two lowest atomic terms of the ion, $^{2}$S$_{1/2}$ and $^{2}$P$_{3/2}$. For this line, the critical Hanle field ($B_{\rm H}$) is 49 gauss, and its \emph{polarizability} coefficient is $W_{2}=0.5$ (see, e.g., \citealt{Stenflo1994ASSL..189.....S} for its analytic expression). 

\begin{figure}[htbp]      
   {\includegraphics[width=0.53\textwidth,trim={1.6cm 15cm 0cm 8cm},clip]{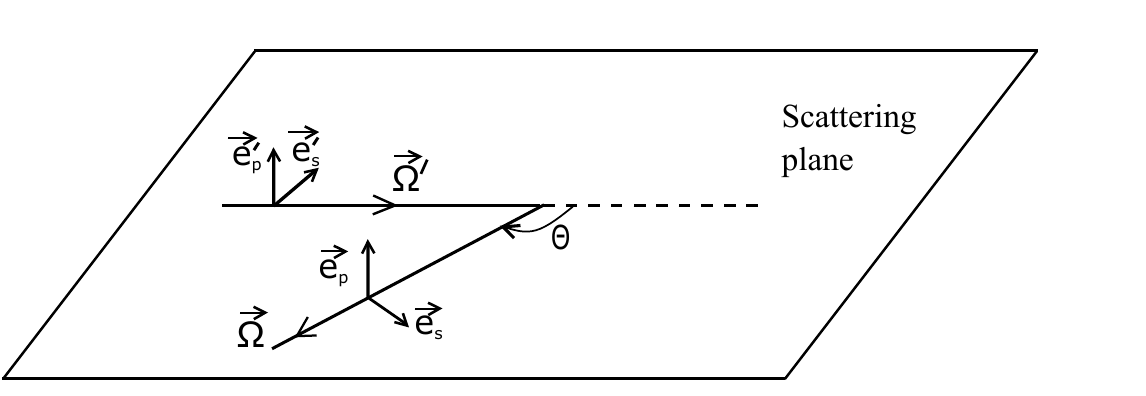}}
    \caption{Geometry of a simple scattering event.  The incident beam of unpolarized radiation propagating in $\Vec{\Omega'}$ direction gets scattered in the direction $\Vec{\Omega}$. The unit vectors $\Vec{e_{p'}}$ and $\Vec{e_{p}}$ perpendicular to the scattering plane ($\Vec{\Omega'},\Vec{\Omega}$) denote the reference direction of positive Stokes $Q$ for the incident and the scattered beam, respectively. (Adapted from Fig. 10.1 of \citealt{Landi2004ASSL..307.....L})}
    \label{fig:rt_geo}
\end{figure}

The polarizability $W_{2}$ determines the maximum linear polarization by scattering attainable for a given transition. In fact, following \cite{Landi2004ASSL..307.....L}, in the absence of a magnetic field, the linear polarization produced in the scattering process depicted in Fig.~\ref{fig:rt_geo} is given by 

\begin{equation} \label{eq:qoi_90}
\frac{Q}{I} = \frac{3W_{2} \sin^{2}{\Theta}}{4-W_{2}+3W_{2}\cos^{2}{\Theta}}\;,\qquad
\frac{U}{I} = 0\;,
\end{equation} 
and for $\Theta=90^{\circ}$, we get

\begin{equation} \label{eq:qoi_90_max}
   \biggl(\frac{Q}{I}\biggl)_{\rm max} = \frac{3W_{2}}{4-W_{2}}.
\end{equation} 
Therefore, the maximum scattering polarization for Ne {\sc viii} 770 \AA\ (in the absence of magnetic fields) is $(Q/I)_{\rm max} \approx 43$\%. In this case, $Q/I$ is positive  because the reference direction adopted for the definition of Eqs.~(\ref{eq:qoi_90}) lies along the unit vector $\Vec{e_{p}}$ of Fig.~\ref{fig:rt_geo}.

Because of the typical low levels of coronal polarization, spectro-polarimetric observations are notoriously photon starved, and therefore it is critical to identify coronal diagnostics with significant photon fluxes. Among the Hanle sensitive lines of the EUV listed by \cite{Khan2022SoPh..297...96K}, Ne {\sc viii} 770 \AA\ appears to be the brightest one. It has a line formation temperature of about 800,000~K, so it originates between the transition region (TR) and the lower solar corona \citep{Fludra2021A&A...656A..38F}. Thanks to its significant polarizability, and its sensitivity to the Hanle effect over expected coronal field strengths, the linear polarization of Ne {\sc viii} 770 \AA\ can be exploited as a quantitative diagnostic of both coronal magnetic field strength and field orientation in the POS (sometimes, also called field azimuth).

\subsection{Model assumptions for the $1/2{-}3/2$ transition} \label{sec:assum}
In the weak field approximation, the equations used to calculate the line emissivity for the Stokes parameters are the same as those used in Section 2.4 of \cite{Zhao2021ApJ...912..141Z}. We have adopted the equations for a two-level atomic model having a lower unpolarized level\footnote{Unpolarized level refers to a non-degenerate level with negligible population imbalances and quantum coherences between its magnetic sublevels.} with $J_{l}=1/2$ and an upper level with $J_{u}=3/2$. This is a good approximation for the resonance line of Ne {\sc viii} at 770 \AA. We also assume that the two-level atom is anisotropically illuminated by unpolarized and cylindrically symmetric (around the local solar vertical to the scattering ion) radiation field coming from the TR. The geometry underlying the problem of radiation scattering in a magnetized plasma is shown in Figure 1 of \cite{Casini2002ApJ...568.1056C}. 

By considering an optically thin coronal plasma in the EUV, we have integrated the emission coefficients (i.e., equation (1) of \citealt{Zhao2021ApJ...912..141Z}) along the LOS to obtain the emergent signals for the Stokes parameters as 
\begin{equation} \label{eq:stokes}
    I_{i}(\bm{\Omega}) = \int_{LOS} \epsilon_{i}(\bm{\Omega},s)\; ds ~,
\end{equation}
where $i=0,1,2$ refer to Stokes $I,Q,U$, respectively; \textit{s} is the coordinate along the LOS; and $\bm{\Omega}$ is the propagation direction of the line emission. We have considered different values for the mean TR brightness that drives the radiative pumping of the Ne {\sc viii} 770 \AA\ line, and which is responsible for the scattered radiation both on the disk and off the limb of the Sun. Evidently, this TR radiation must be added as a background intensity term to the LOS-integrated Stokes $I$ signal given by Eq.~(\ref{eq:stokes}), when observing on the disk. 

We have also taken into account the contribution of collisional excitation to the Ne {\sc viii} 770 \AA\ transition, which is responsible for a significant
depolarization of the radiation emitted in the lower atmosphere, where the electron density is larger (see Figure \ref{fig:plots_chrovary} and caption therein).

There are additional effects, including non-radial Doppler dimming and temperature anisotropies, which could potentially impact scattering polarization \citep{Raouafi2002A&A...386..721R,Zhao2021ApJ...912..141Z}. However, for the sake of simplicity and focus of this investigation, we have chosen to defer their detailed study to future work.  

\begin{figure*}[htbp]
    \subfigure[]{\includegraphics[width=0.3\textwidth,trim={1.1cm 2.5cm 0cm 2.9cm},clip]{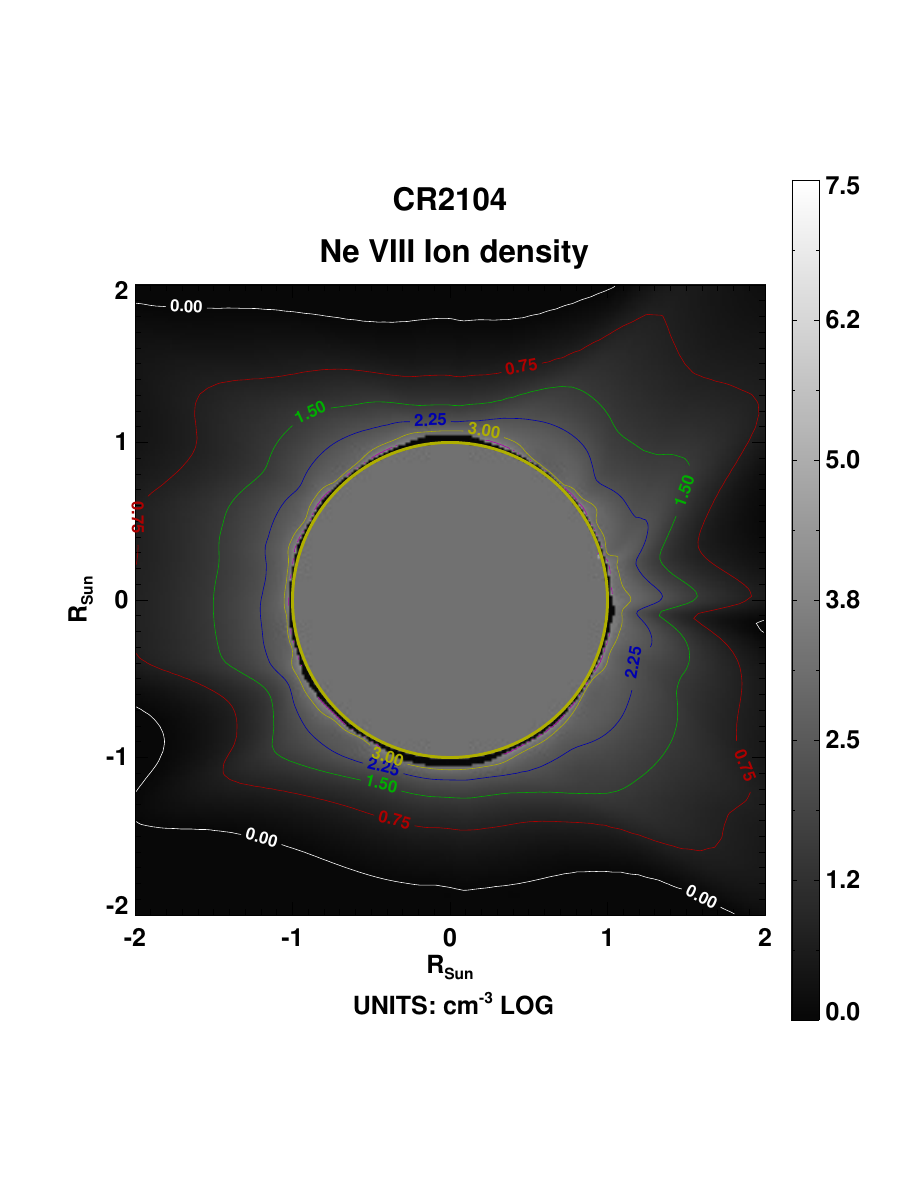}}
    %\hspace{0.4cm}
    \subfigure[]{\includegraphics[width=0.3\textwidth,trim={1.1cm 2.5cm 0cm 2.9cm},clip]{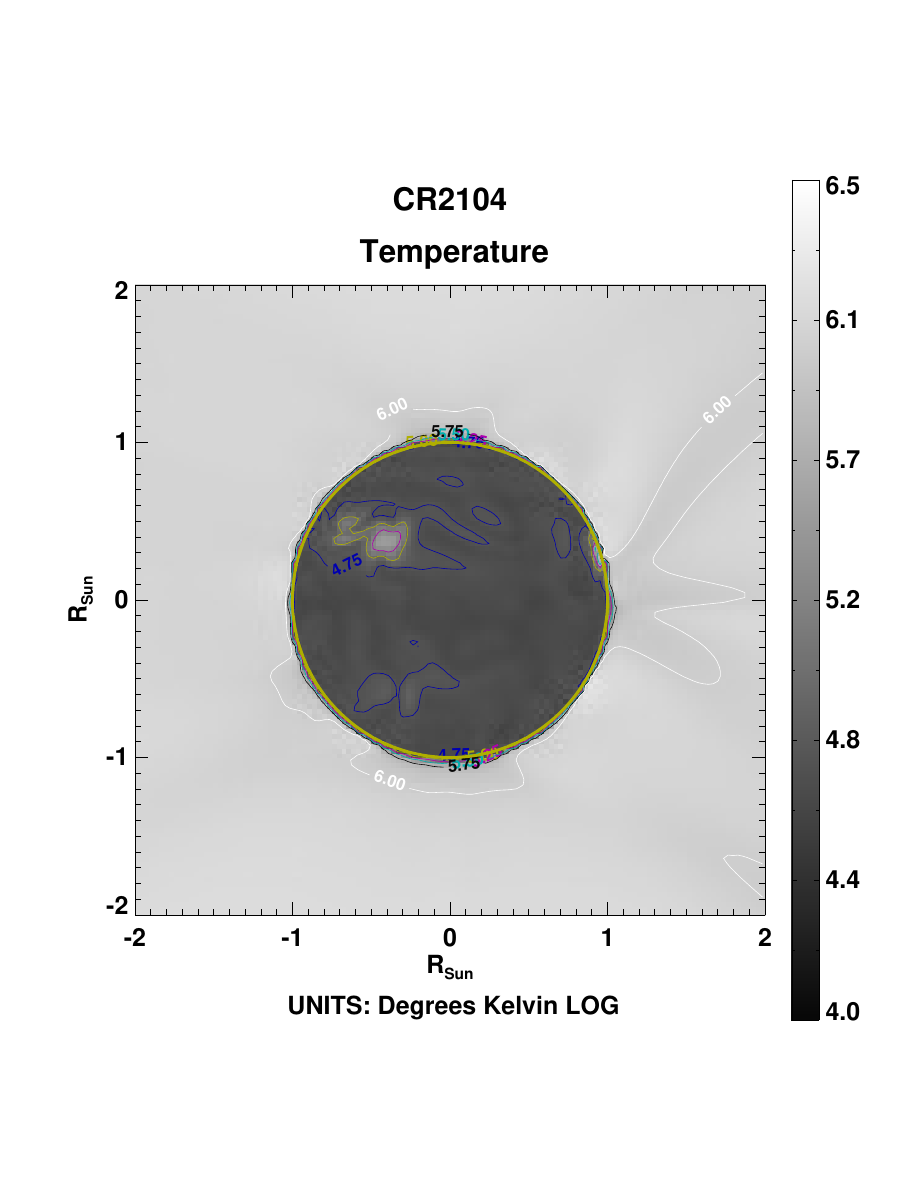}}
     %\hspace{0.4cm}
    \subfigure[]{\includegraphics[width=0.3\textwidth,trim={1.1cm 2.5cm 0cm 2.9cm},clip]{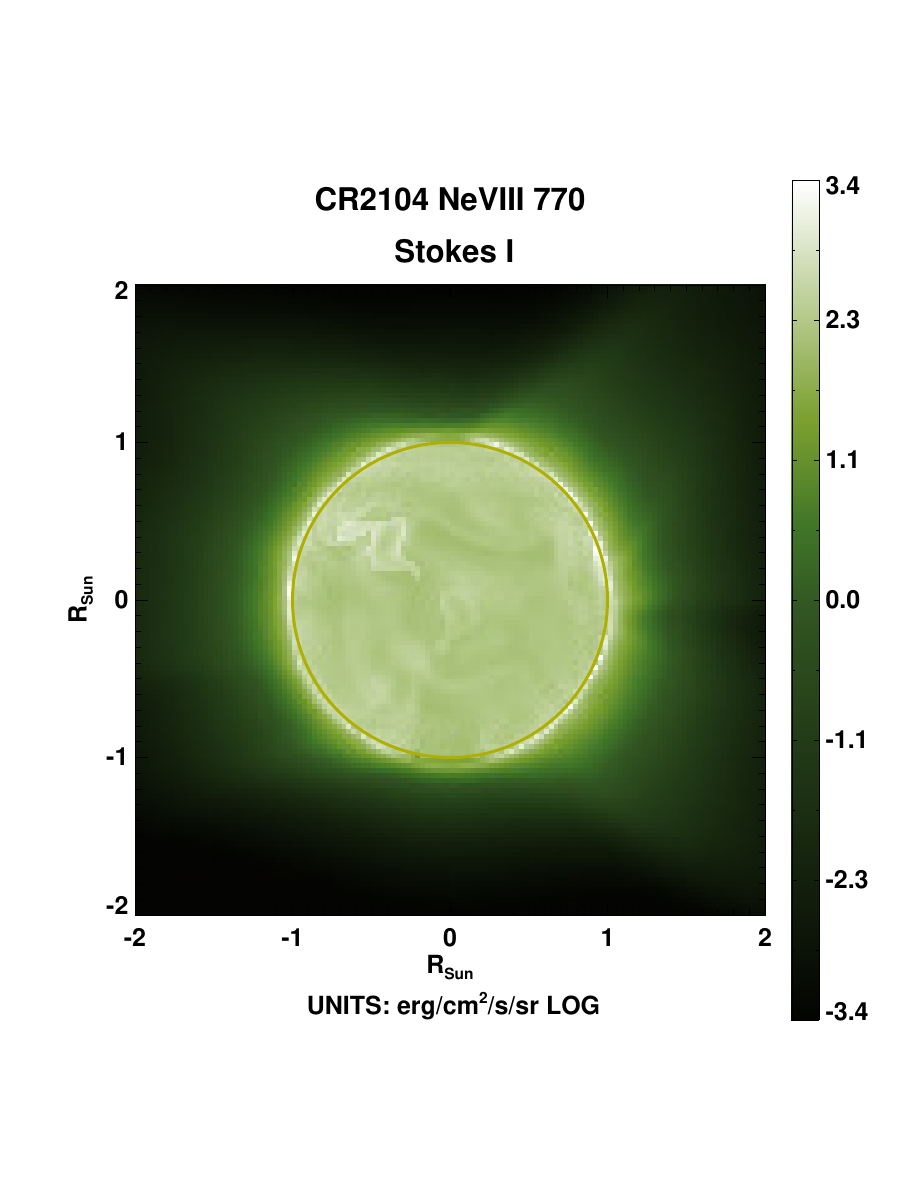}}
    \subfigure[]{\includegraphics[width=0.3\textwidth,trim={1.1cm 2.5cm 0cm 2.9cm},clip]{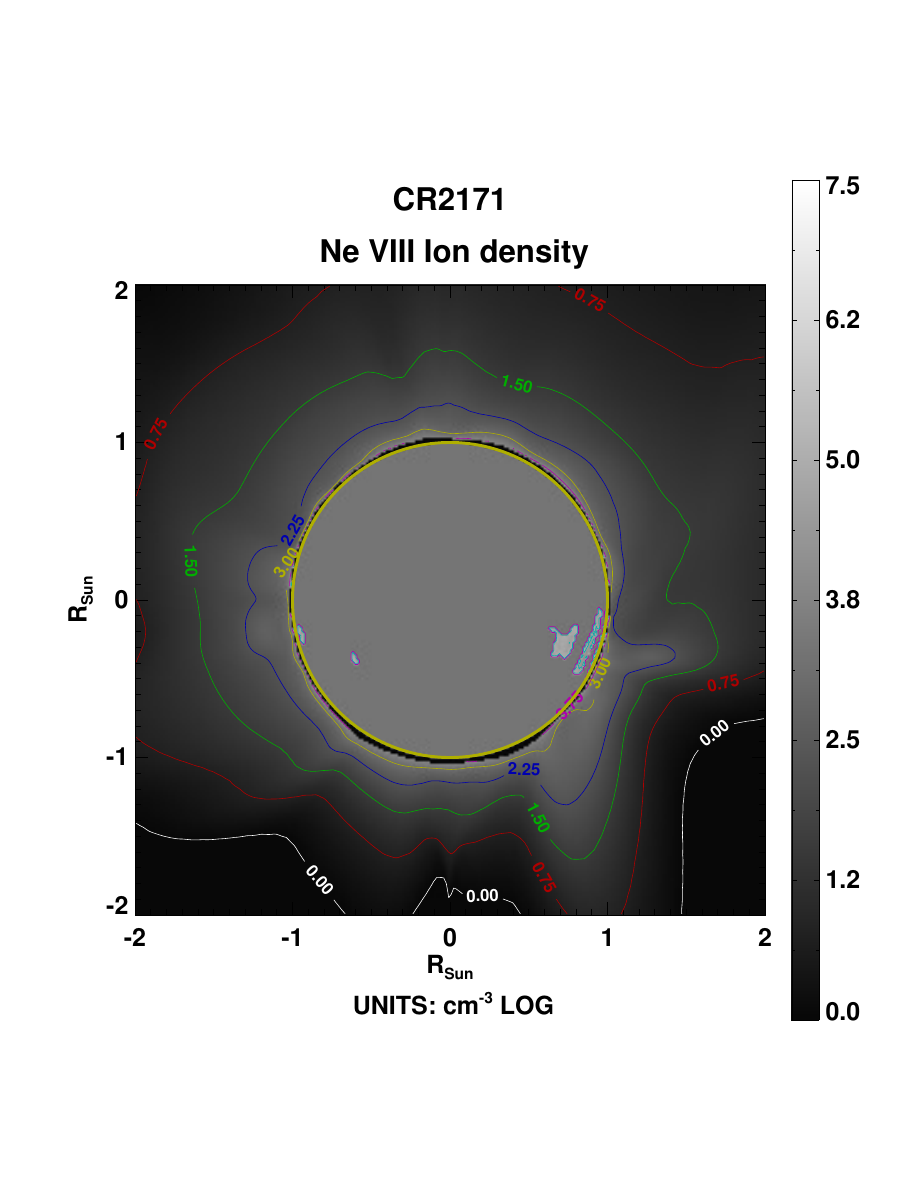}}
    %\hspace{0.4cm}
    \subfigure[]{\includegraphics[width=0.3\textwidth,trim={1.1cm 2.5cm 0cm 2.9cm},clip]{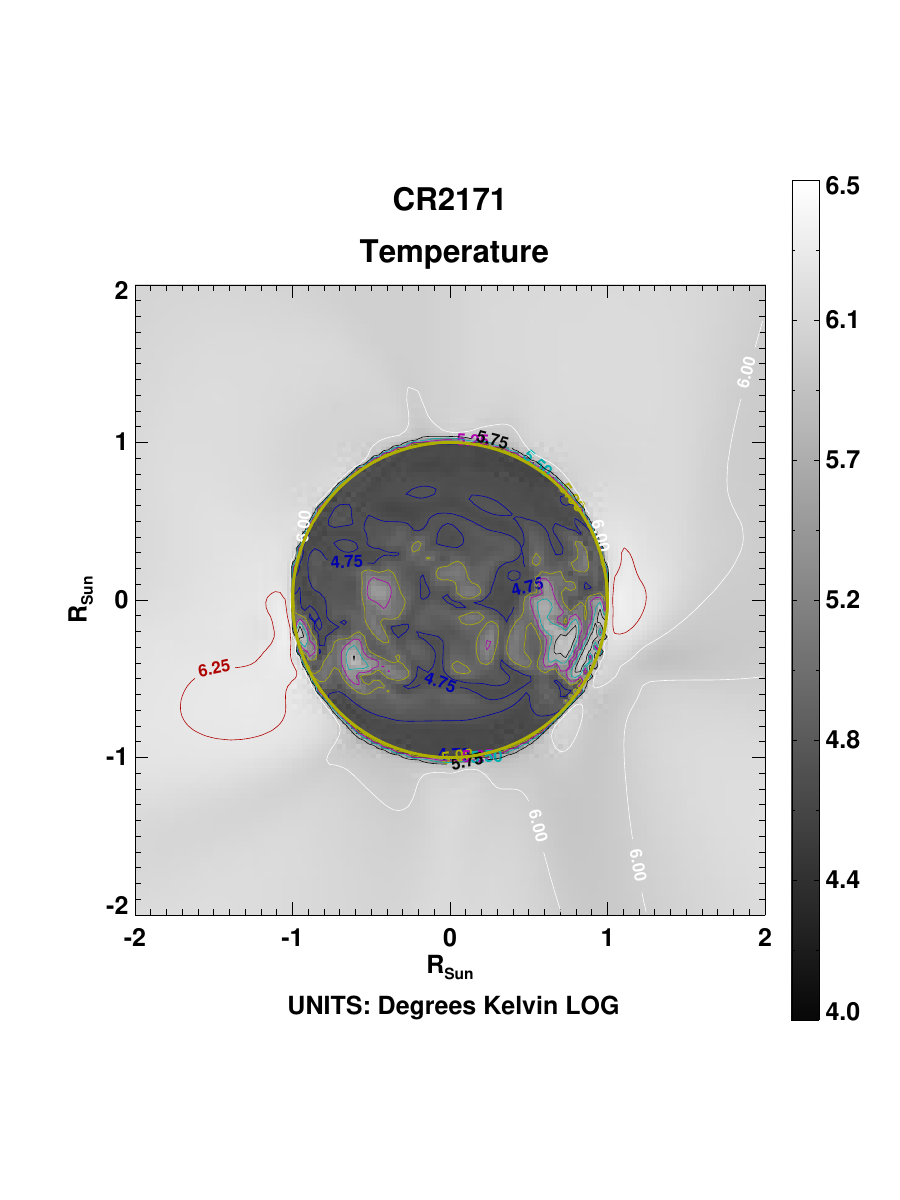}}
     %\hspace{0.4cm}
    \subfigure[]{\includegraphics[width=0.3\textwidth,trim={1.1cm 2.5cm 0cm 2.9cm},clip]{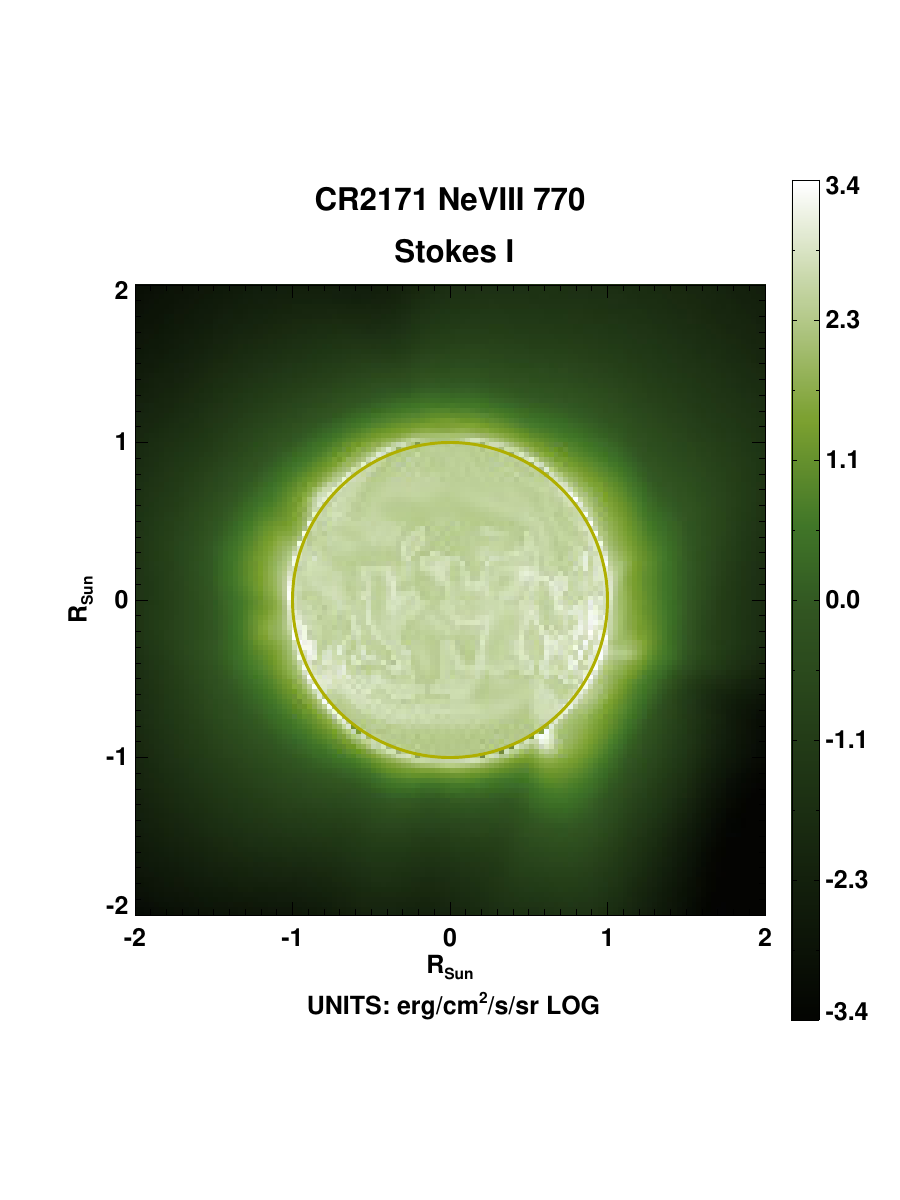}}
    \subfigure[]{\includegraphics[width=0.3\textwidth,trim={1.1cm 2.5cm 0cm 2.9cm},clip]{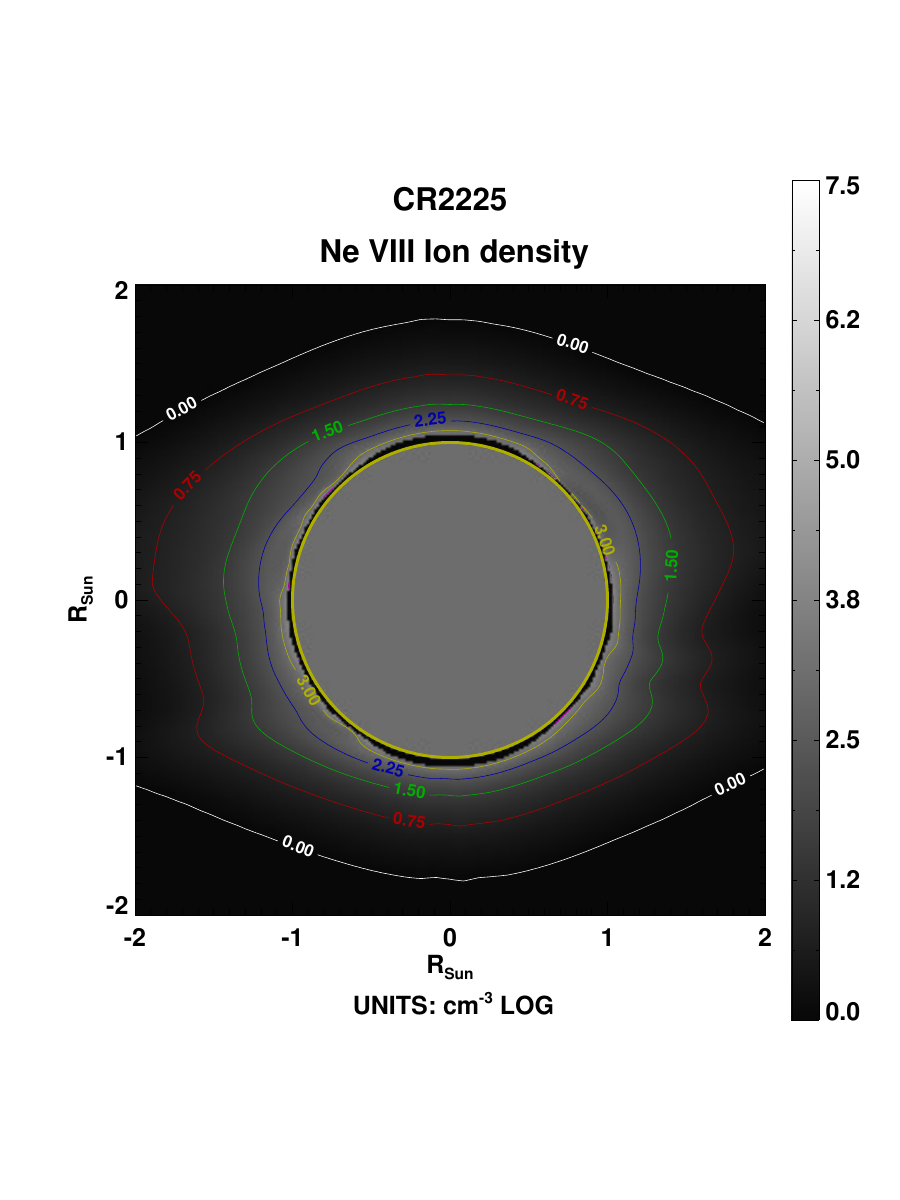}}
    \hspace{0.7cm}
    \subfigure[]{\includegraphics[width=0.3\textwidth,trim={1.1cm 2.5cm 0cm 2.9cm},clip]{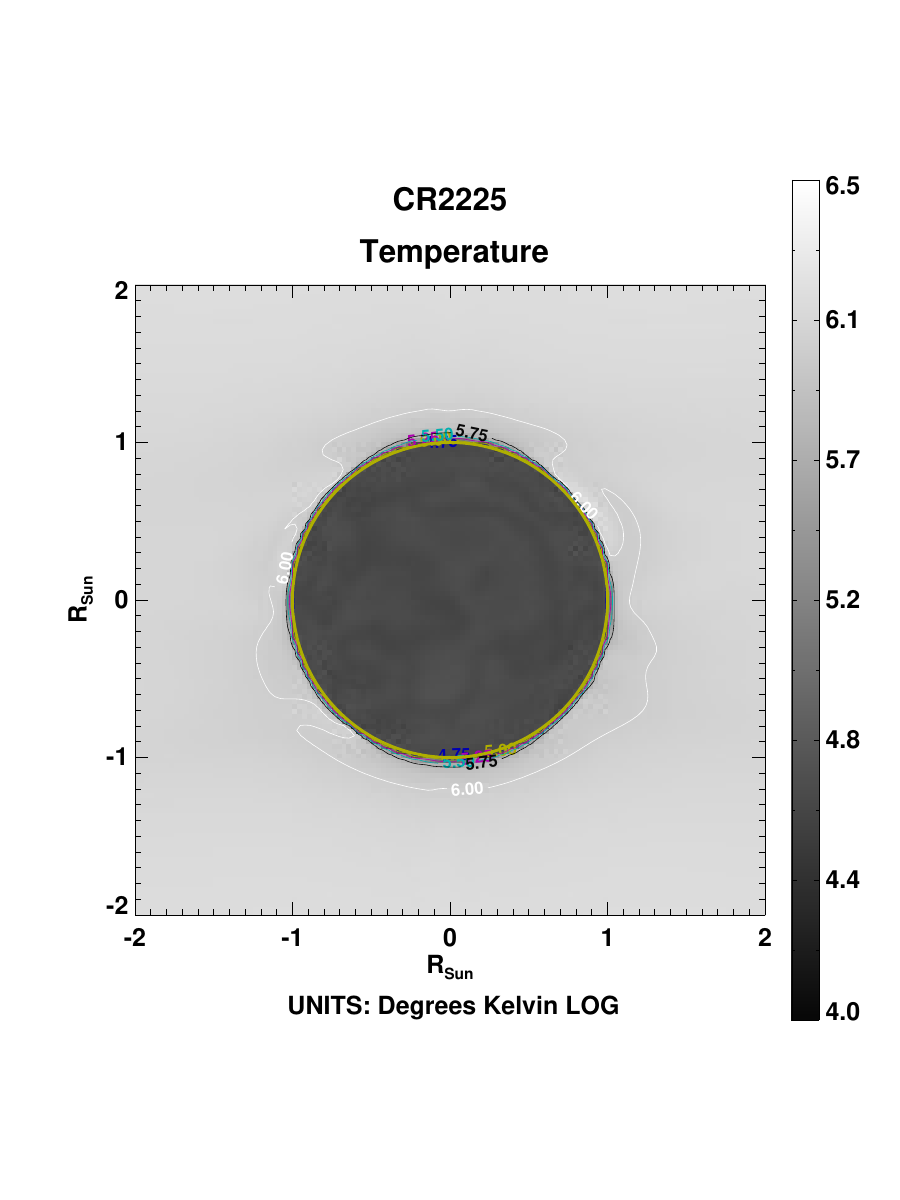}}
    \hspace{0.7cm}
    \subfigure[]{\includegraphics[width=0.3\textwidth,trim={1.1cm 2.5cm 0cm 2.9cm},clip]{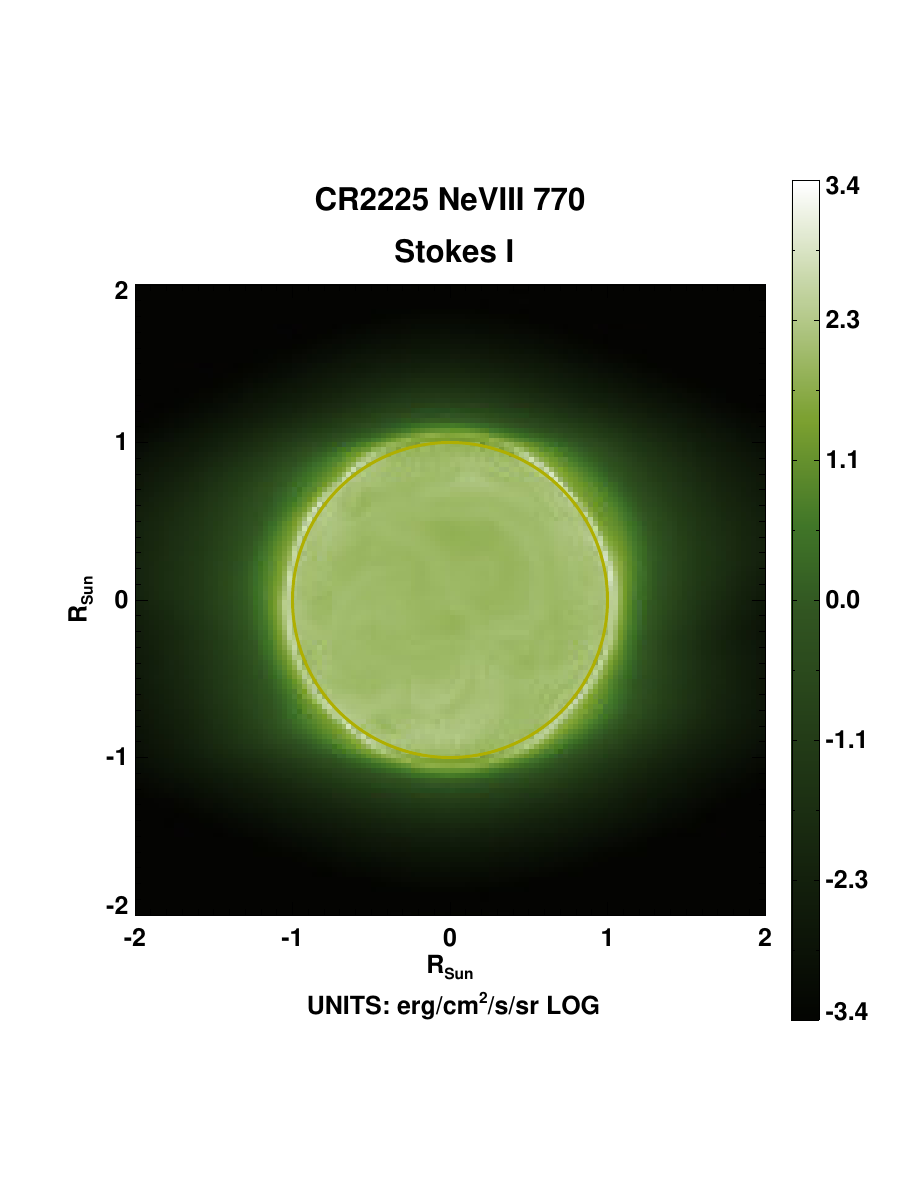}}
    \caption{\textit{Top row:} (a) derived Ne {\sc viii} ion density map, (b) PSIMAS model temperature map, and (c) synthesized Stokes $I$ (LOS integrated) map during the rising phase (CR2104) of SC24. \textit{Middle row} and \textit{bottom row} illustrate the same maps, but during the maximum phase (CR2171) and the minimum phase (CR2225), respectively. Note that MAS coronal ion densities are not shown if simulation temperature is lower than 500,000 K at the plotted height (1.01 $R_{\odot}$ on the disk), and they appear grey (the color of the ion density corresponding to the assumed TR brightness below them; see Section~\ref{sec:bkg_variation}). The yellow circle of radius 1 $R_{\odot}$ demarcates the circumference of the simulated solar surface. Contours of a particular color in a given map represent iso-curves of the depicted physical quantity shown in logarithmic scale. Note that Stokes $I$ includes the contribution of collisional excitation to the scattered radiation.}
    \label{fig:models_stki}
\end{figure*}

\section{PSIMAS model} \label{sec:model}
The PSI (Predictive Science Inc.; refer to \url{https://www.predsci.com/portal/home.php}) have developed three-dimensional MAS (Magnetohydrodynamic Algorithm outside a Sphere) 
numerical simulations of the solar corona and the inner heliosphere. It provides us with realistic models of EUV emission which help us visualize the occurrence of the critical Hanle effect on polarization both on the disk of the Sun and at the limb. The MAS simulations used in this work are based on a 3D magnetohydrodynamic model which incorporates improved energy transport mechanisms such as coronal heating, radiative losses, Alfv\'en wave acceleration and parallel thermal conduction \citep{Mikic1999PhPl....6.2217M,Lionello2001ApJ...546..542L}. We have obtained MAS simulations for a time period of 11 years starting from the rising phase of the SC24 (i.e., from 2010) until the rising phase of the SC25 (i.e., until 2020). The selected eleven Carrington rotation (CR) simulations are: CR2104, CR2118, CR2131, CR2144, CR2158, CR2171, CR2185, CR2198, CR2211, CR2225, and CR2238, which coincide with mid-December of each year from 2010 to 2020. We have then synthesized the LOS integrated spectro-polarimetric signals of Ne {\sc viii} 770 \AA\ and studied the variation in the signals during the different phases of the solar cycles. For the purpose of this paper, we consider CR maps describing three phases of the solar cycle, i.e., the rising, the maximum, and the end phase of SC24. The rest of the MAS model maps and the corresponding Stokes maps are provided as supplementary material.

\begin{figure*}[htbp]
    \subfigure[]{\includegraphics[width=0.22\textwidth,trim={1.1cm 2.5cm 0cm 2.9cm},clip]{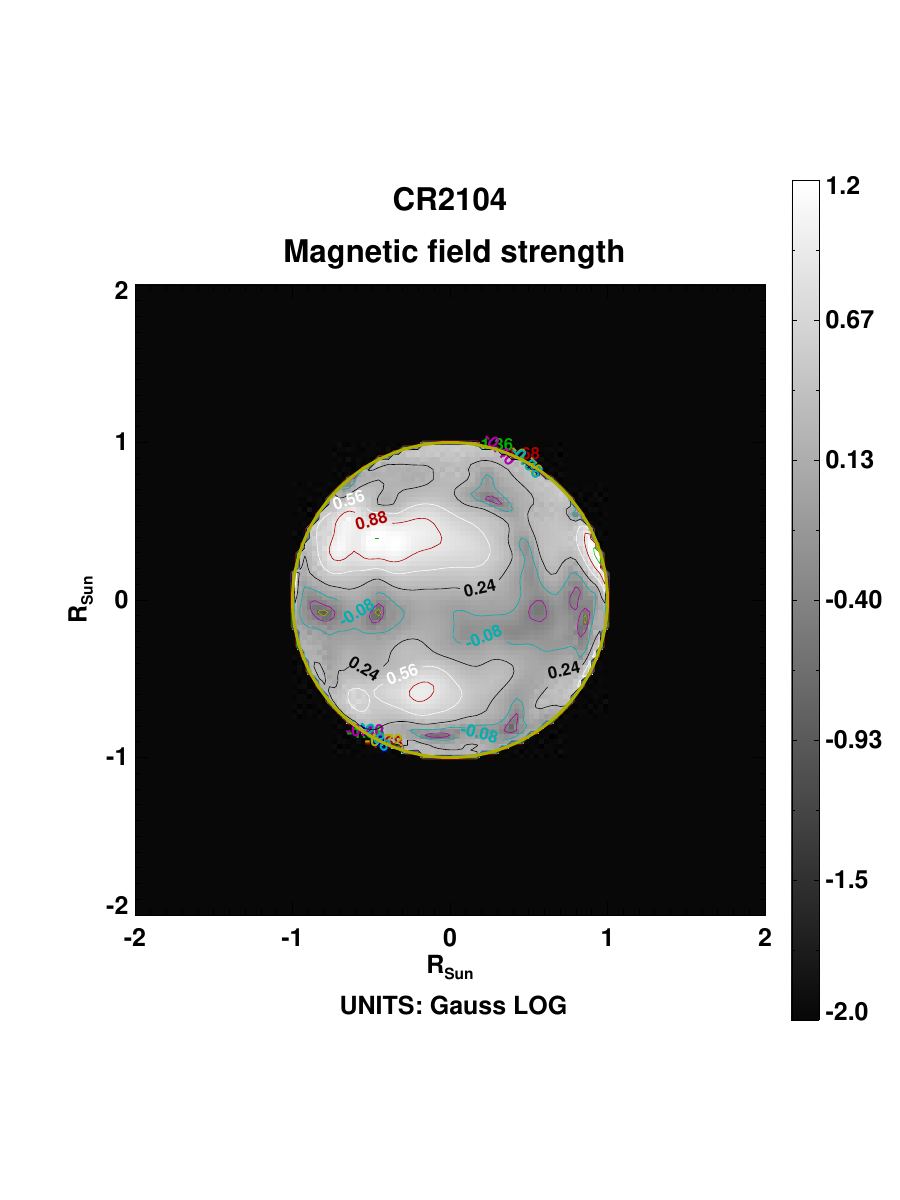}}
    %\hspace{0.8cm}
    \subfigure[]{\includegraphics[width=0.22\textwidth,trim={1.1cm 2.5cm 0cm 2.9cm},clip]{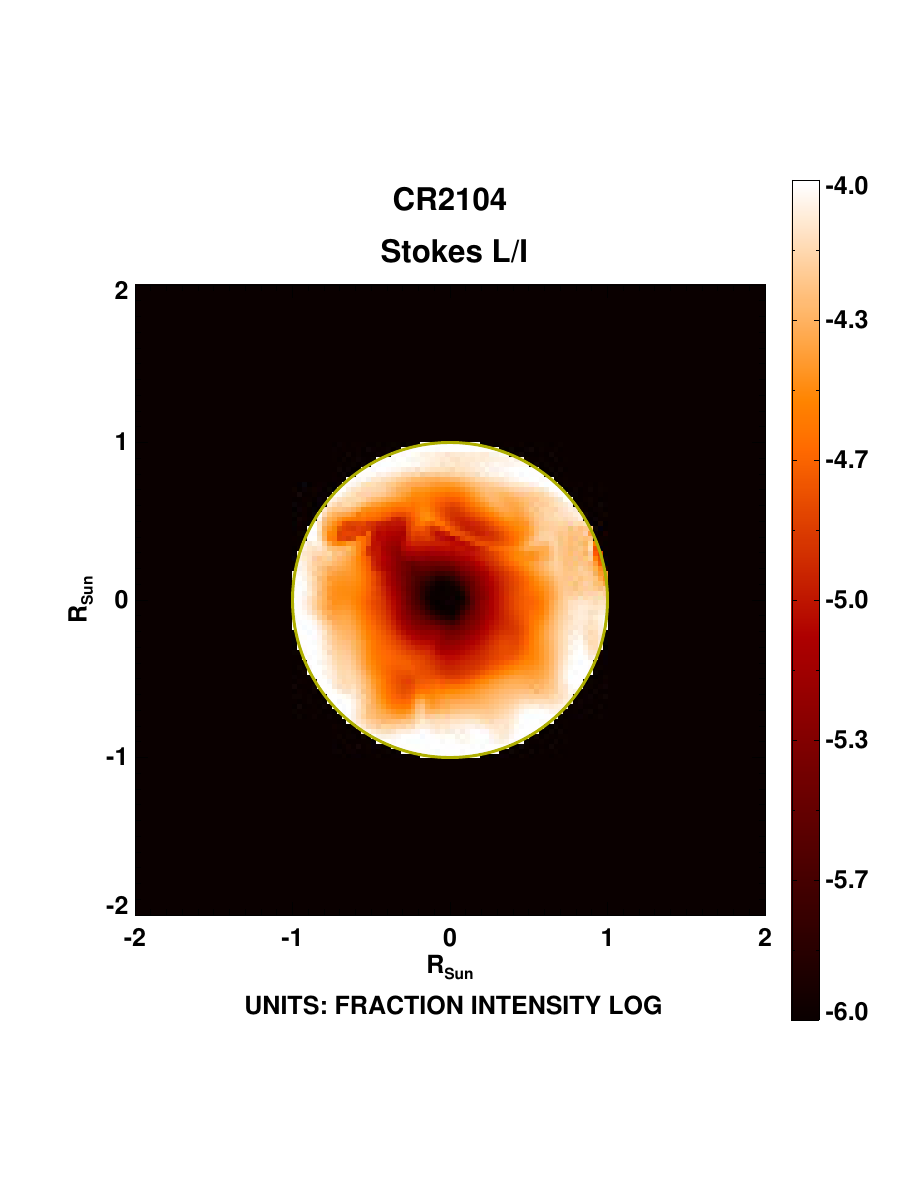}}
    \subfigure[]{\includegraphics[width=0.22\textwidth,trim={1.1cm 2.7cm 0cm 2.9cm},clip]{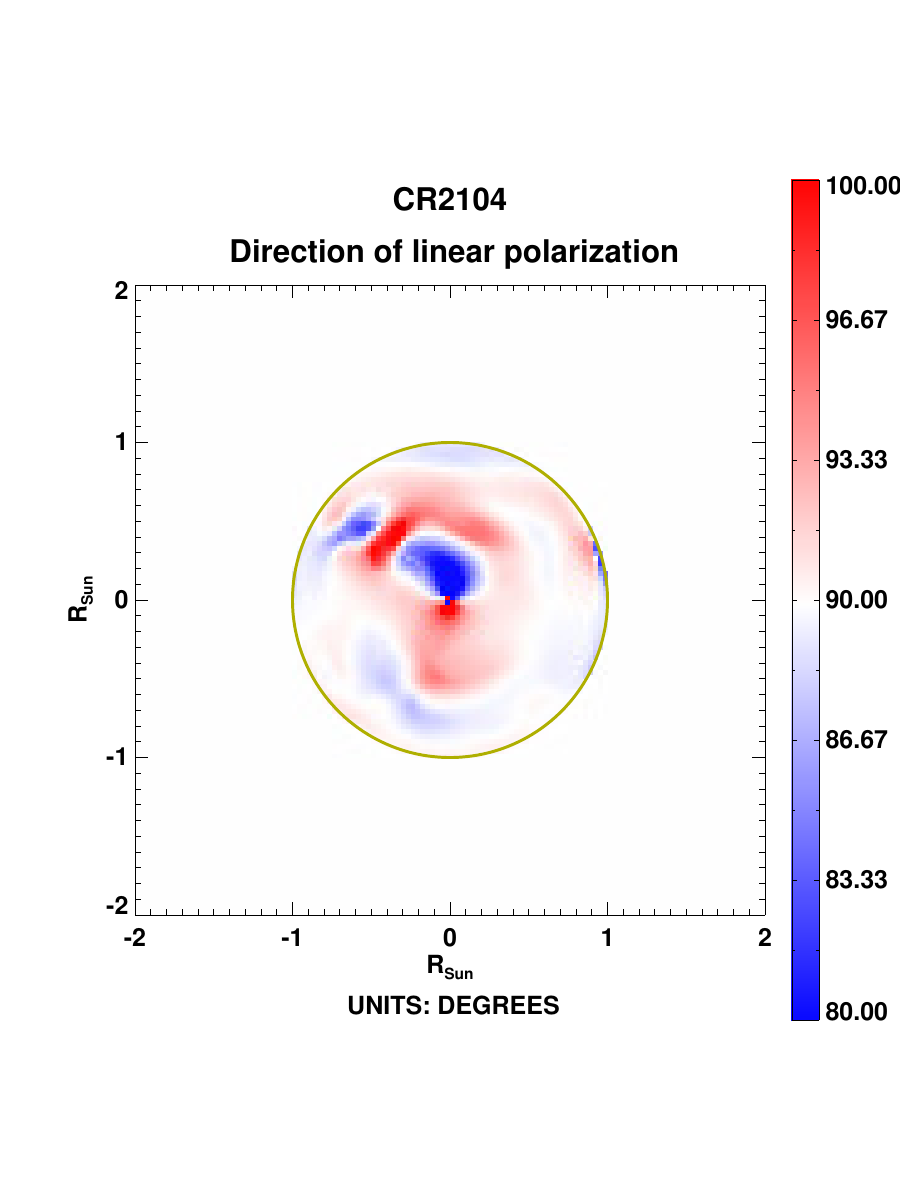}}
    %\hspace{0.8cm}
    \subfigure[]{\includegraphics[width=0.235\textwidth,trim={0.3cm 1.2cm 0cm 0.5cm},clip]{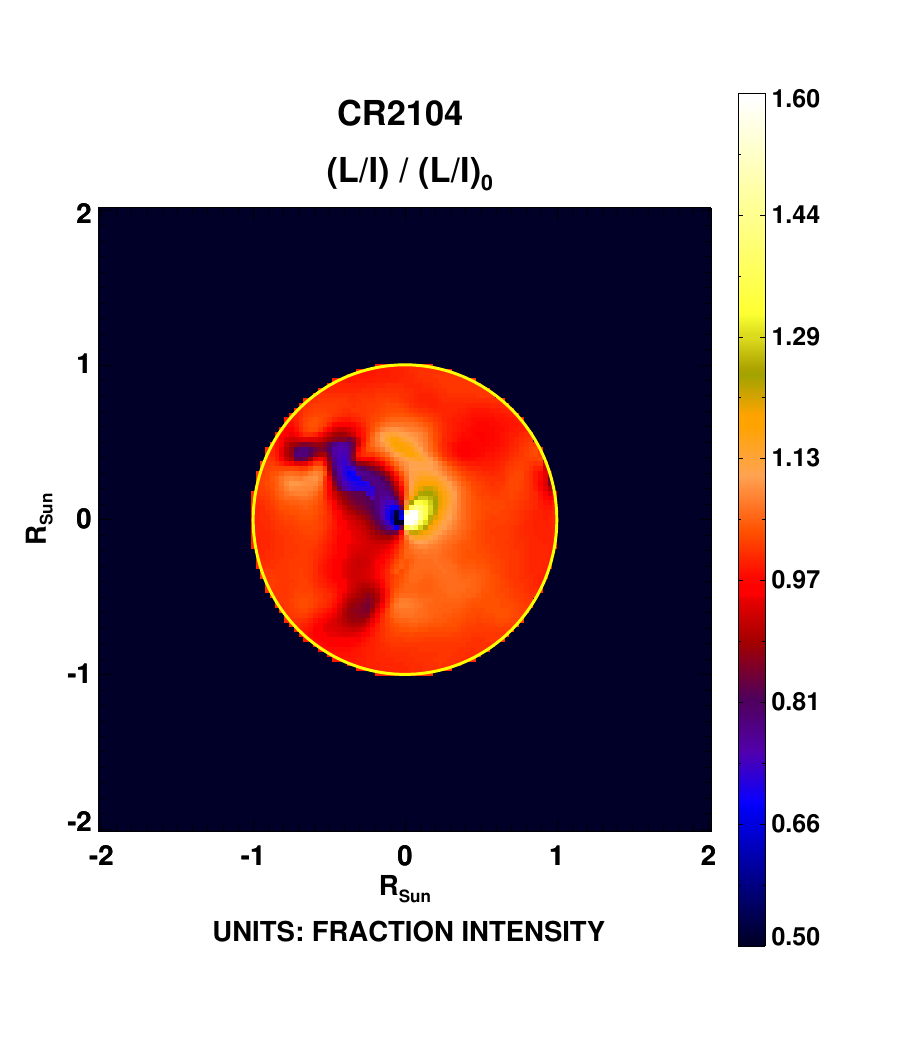}}
    %\hspace{0.8cm}
    \subfigure[]{\includegraphics[width=0.22\textwidth,trim={1.1cm 2.5cm 0cm 2.9cm},clip]{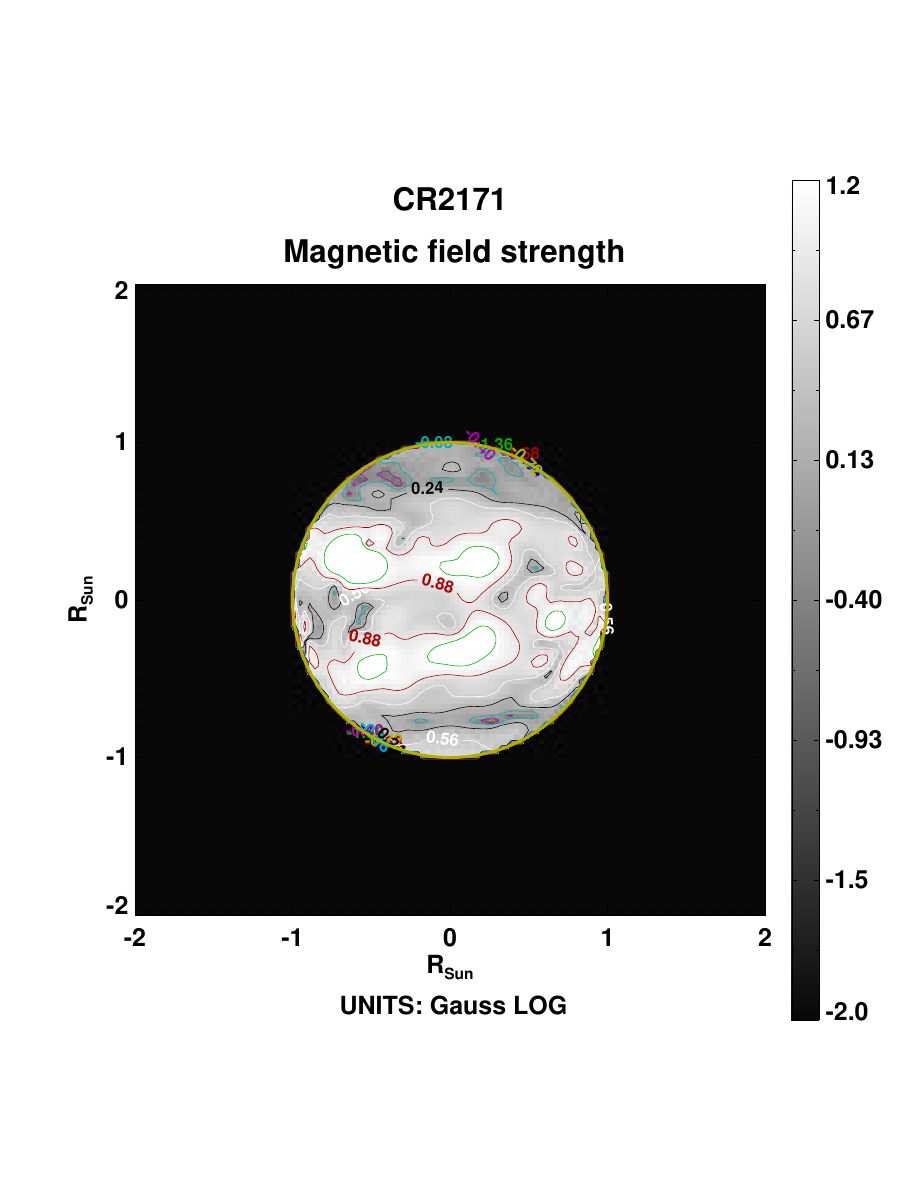}}
    %\hspace{0.8cm}
    \subfigure[]{\includegraphics[width=0.22\textwidth,trim={1.1cm 2.5cm 0cm 2.9cm},clip]{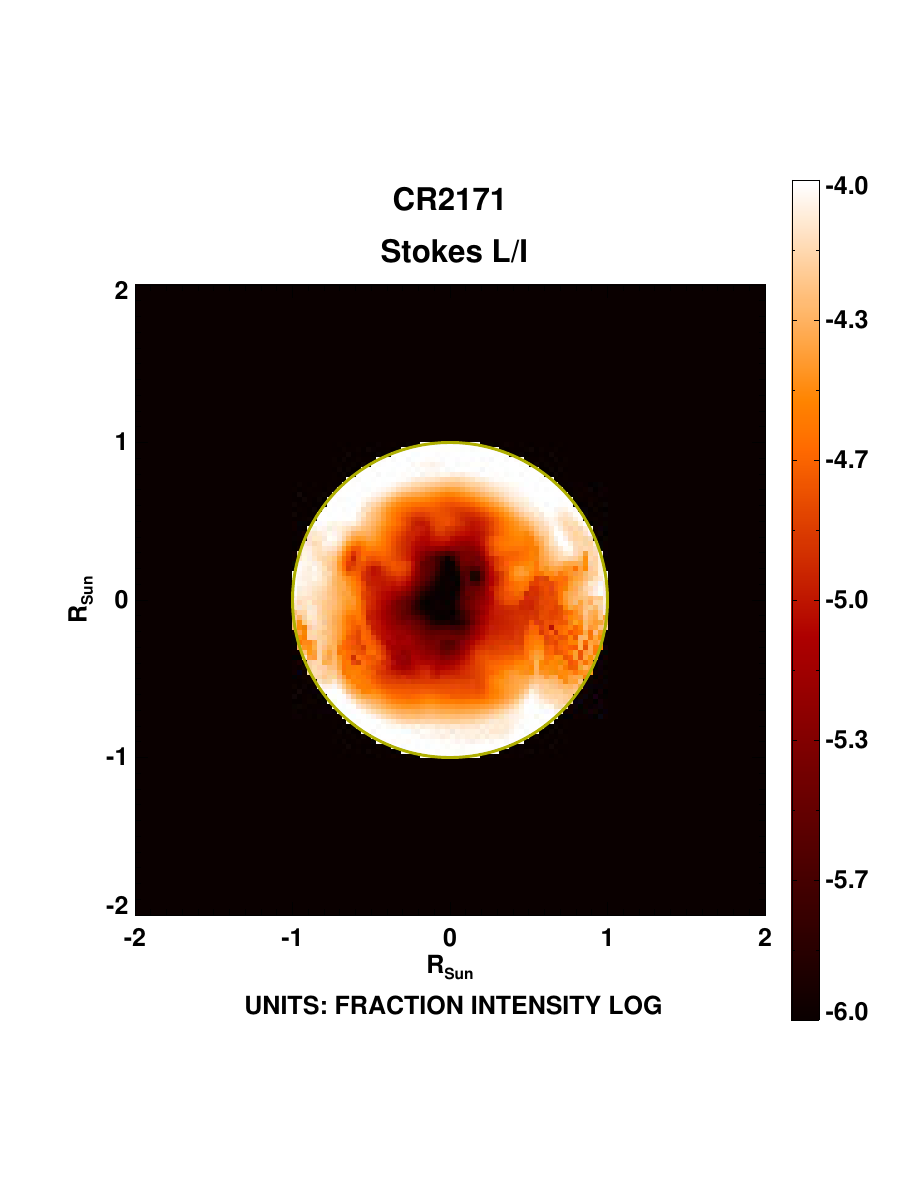}}
    \subfigure[]{\includegraphics[width=0.22\textwidth,trim={1.1cm 2.7cm 0cm 2.9cm},clip]{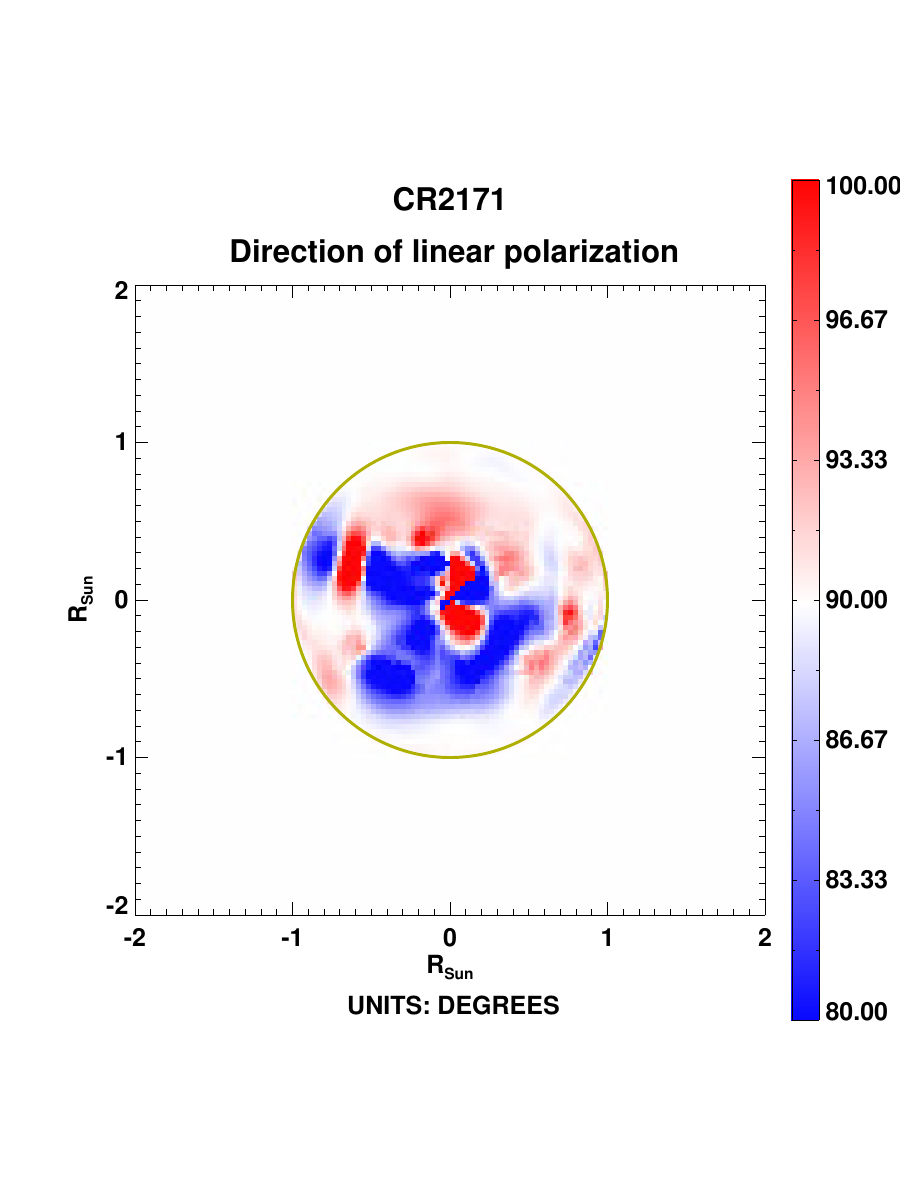}}
    %\hspace{0.8cm}
    \subfigure[]{\includegraphics[width=0.235\textwidth,trim={0.3cm 1.2cm 0cm 0.5cm},clip]{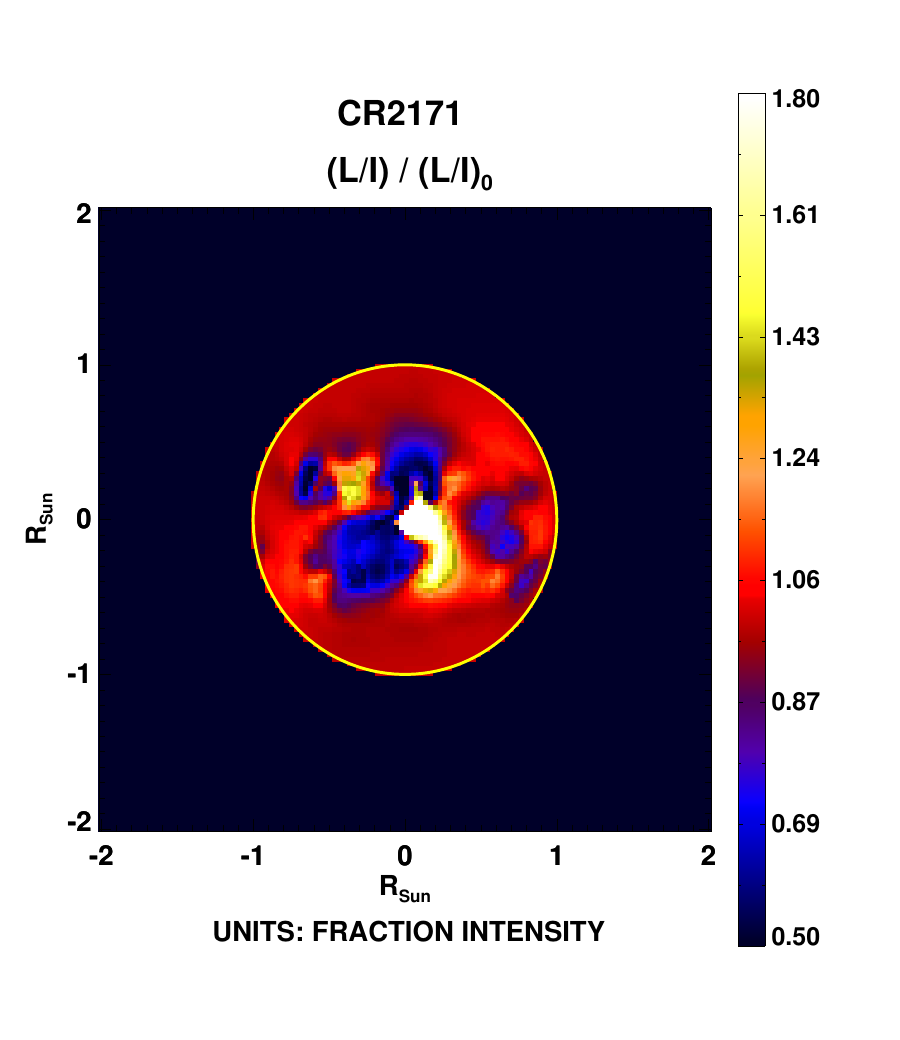}}
    %\hspace{0.8cm}
    \subfigure[]{\includegraphics[width=0.22\textwidth,trim={1.1cm 2.5cm 0cm 2.9cm},clip]{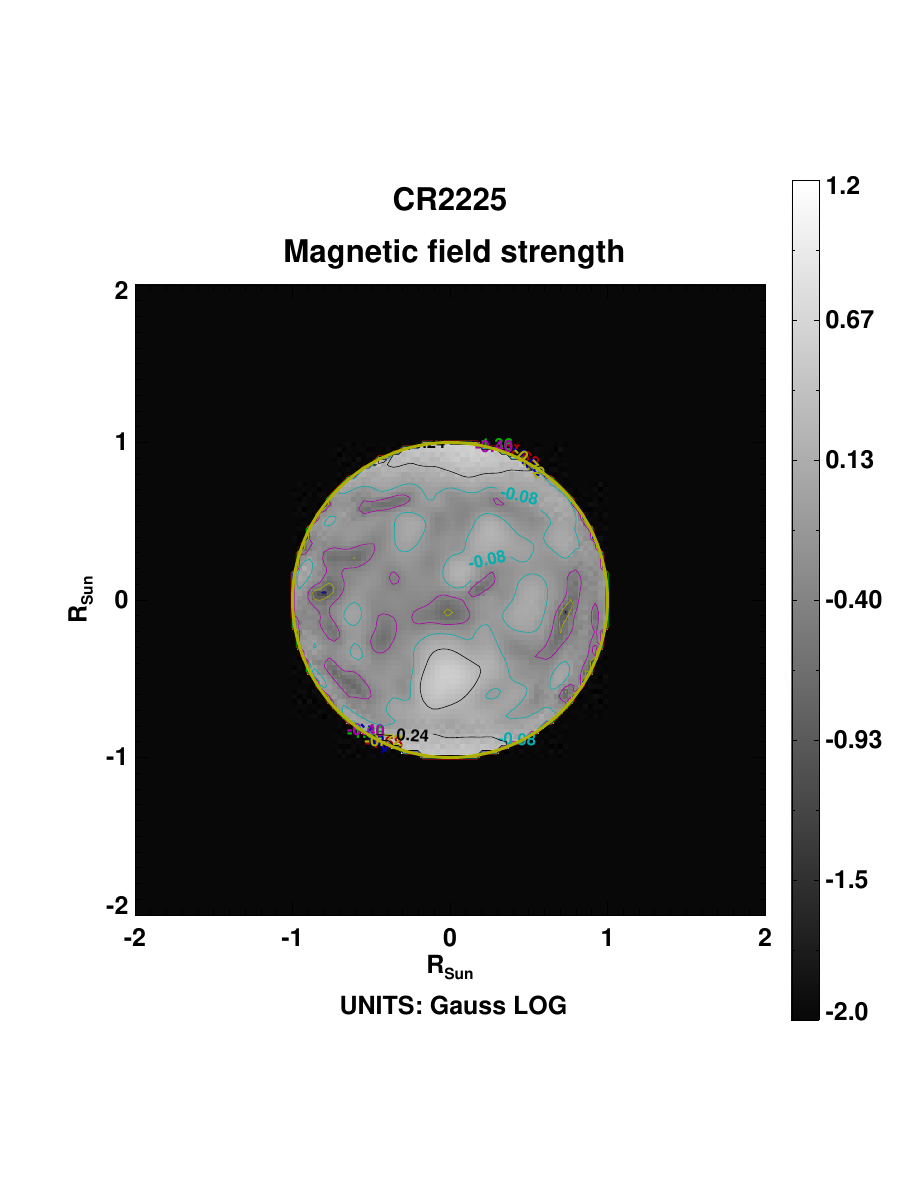}}
    \hspace{0.4cm}
    \subfigure[]{\includegraphics[width=0.22\textwidth,trim={1.1cm 2.5cm 0cm 2.9cm},clip]{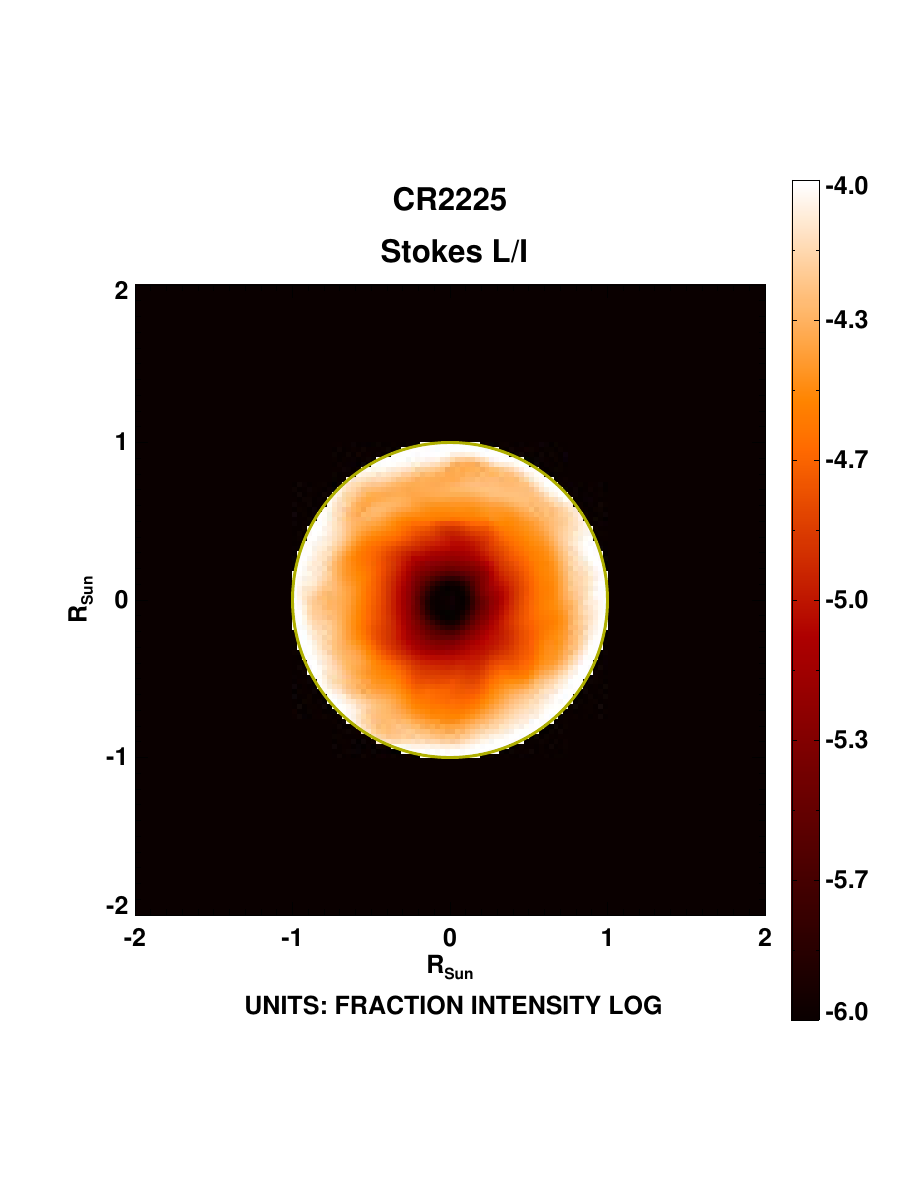}}
    \hspace{0.5cm}
    \subfigure[]{\includegraphics[width=0.22\textwidth,trim={1.1cm 2.7cm 0cm 2.9cm},clip]{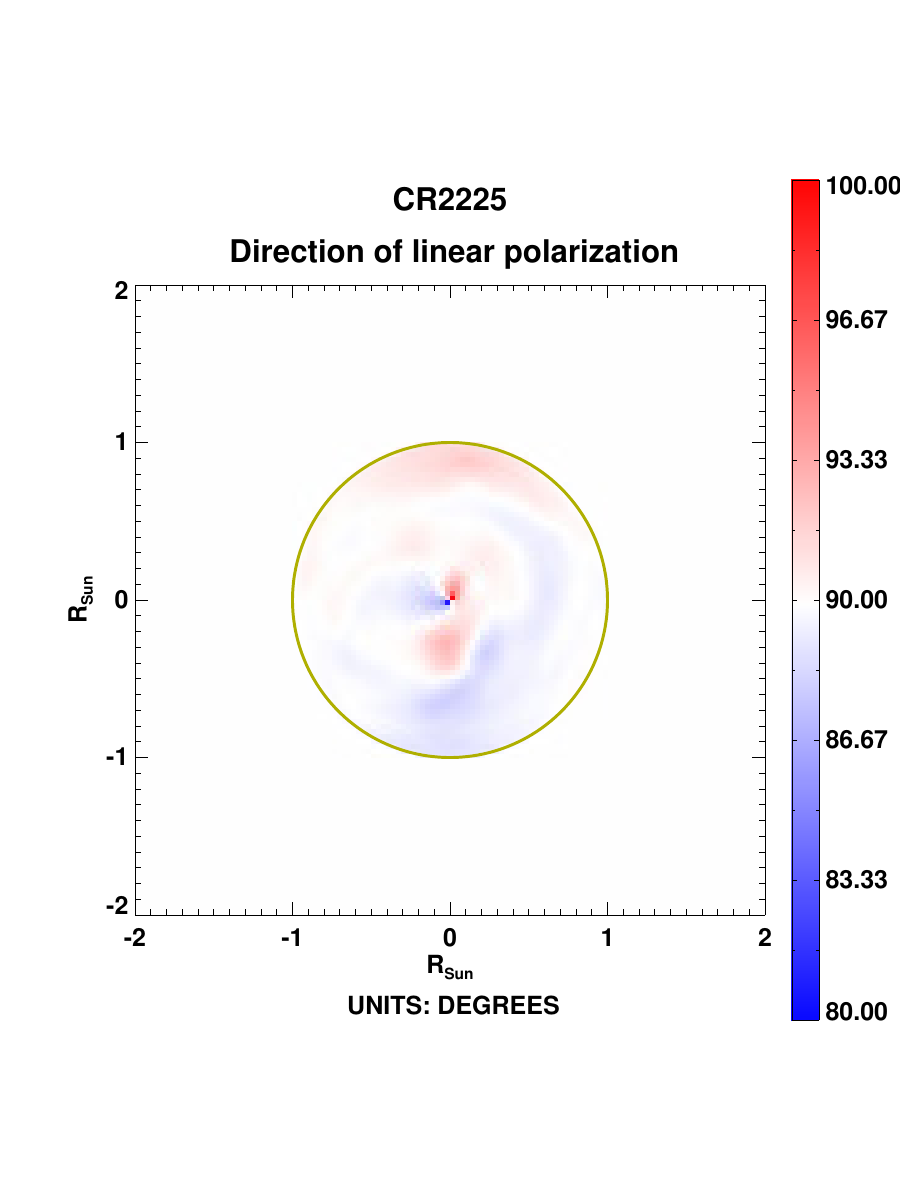}}
    \hspace{0.4cm}
    \subfigure[]{\includegraphics[width=0.235\textwidth,trim={0.3cm 1.2cm 0cm 0.5cm},clip]{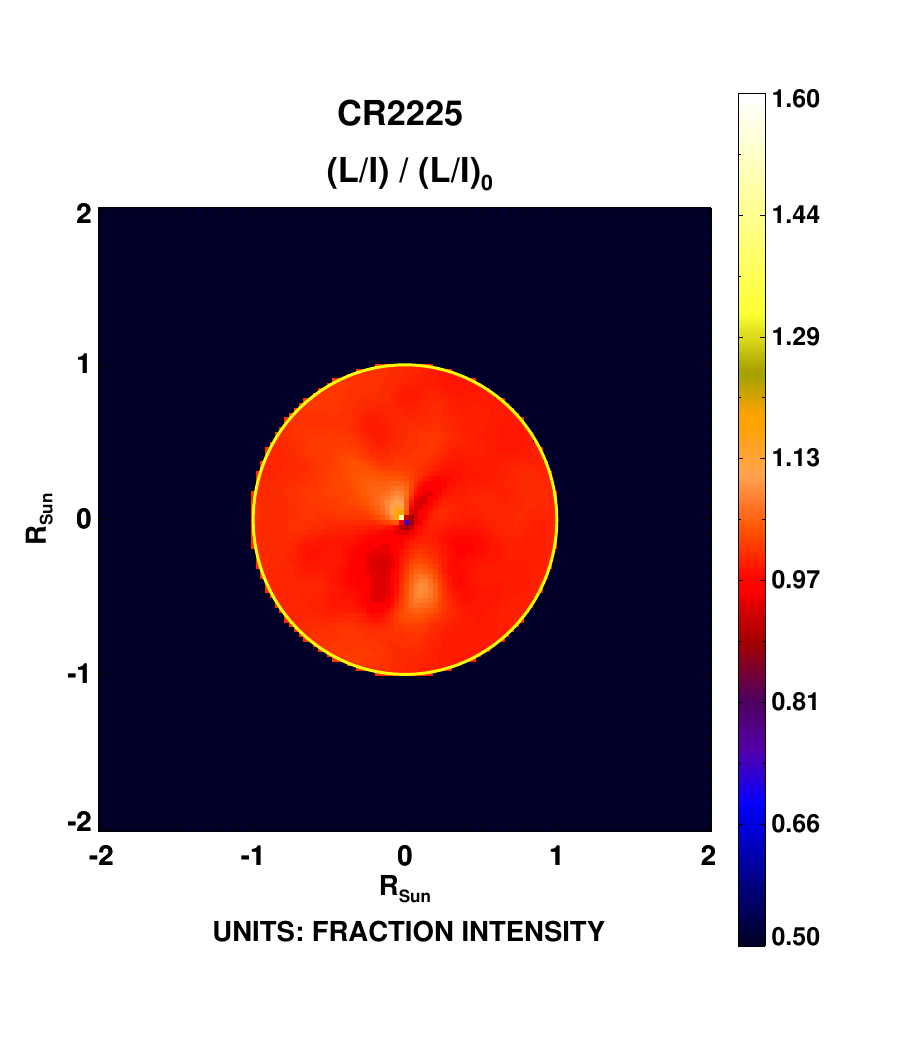}}
    \caption{\textit{Top row:} Rising phase (CR2104) of SC24: (a) PSIMAS model map of magnetic field, (b) LOS-integrated Stokes $L/I$ in the presence of magnetic fields, (c) LOS-integrated linear polarization azimuth (relative to the radial direction through the point of calculation), and (d) synthesized ratio between LOS integrated $L/I$ in presence and $(L/I)_{0}$ in absence of model magnetic fields. Only on-disk information within 1 $R_{\odot}$ is shown here. \textit{Middle row} and \textit{bottom row} illustrate the same maps, but during the maximum phase (CR2171) and the minimum phase (CR2225), respectively. Contours of a particular color in a given map represent iso-curves of the depicted physical quantity shown in logarithmic scale. Note that collisional excitation has been included here.}
    \label{fig:bmag_stokesloi_az_ondisk}
\end{figure*}
\begin{figure*}[htbp]
    \subfigure[]{\includegraphics[width=0.22\textwidth,trim={1.1cm 2.5cm 0cm 2.9cm},clip]{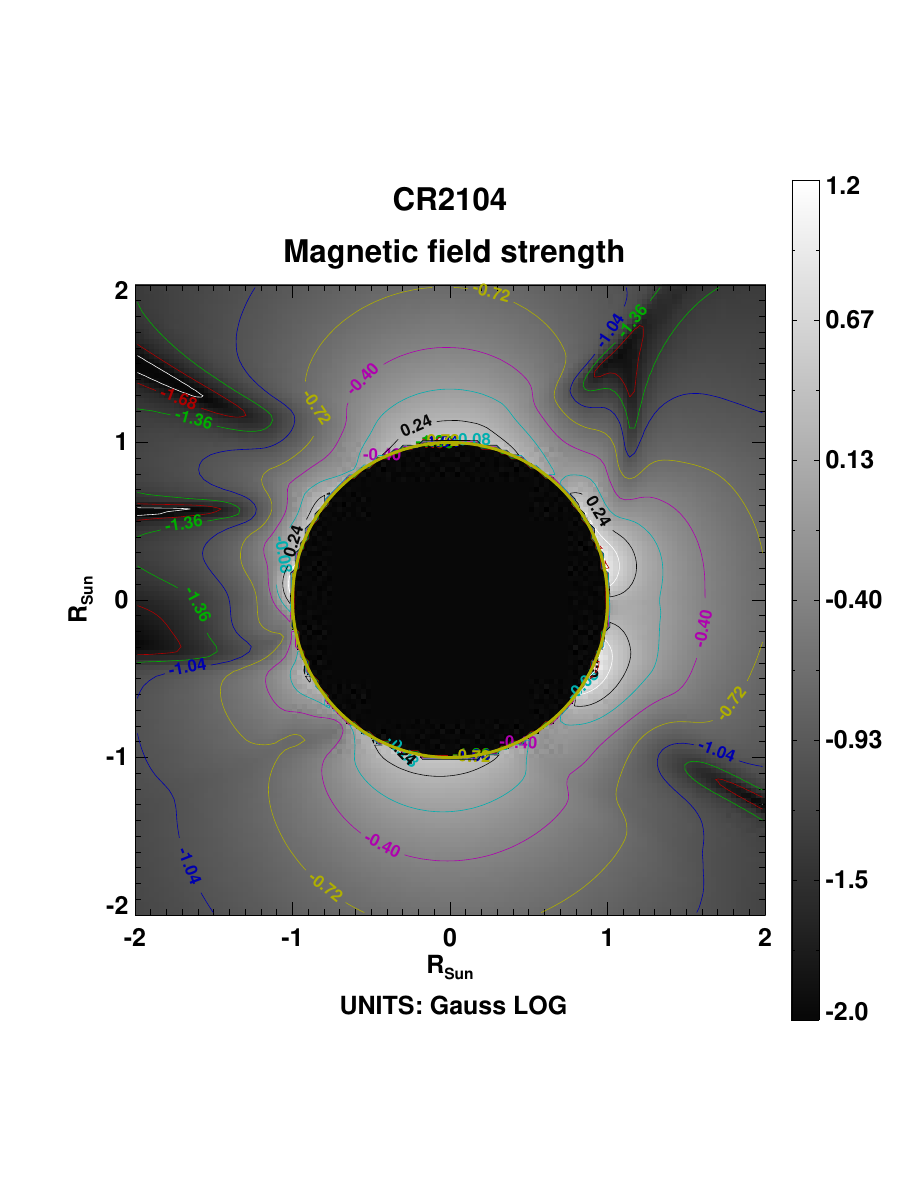}}
    %\hspace{0.8cm}
    \subfigure[]{\includegraphics[width=0.22\textwidth,trim={1.1cm 2.5cm 0cm 2.9cm},clip]{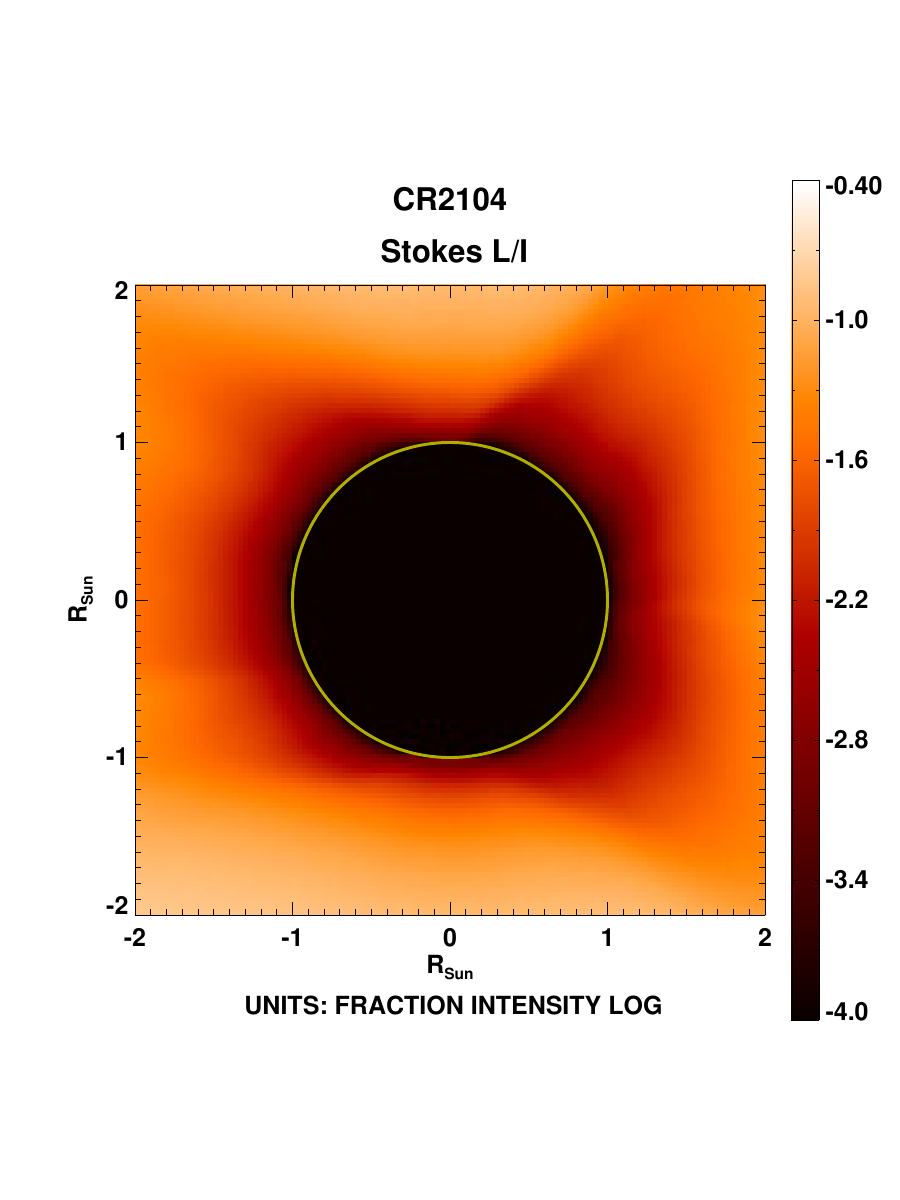}}
    \subfigure[]{\includegraphics[width=0.22\textwidth,trim={1.1cm 2.7cm 0cm 2.9cm},clip]{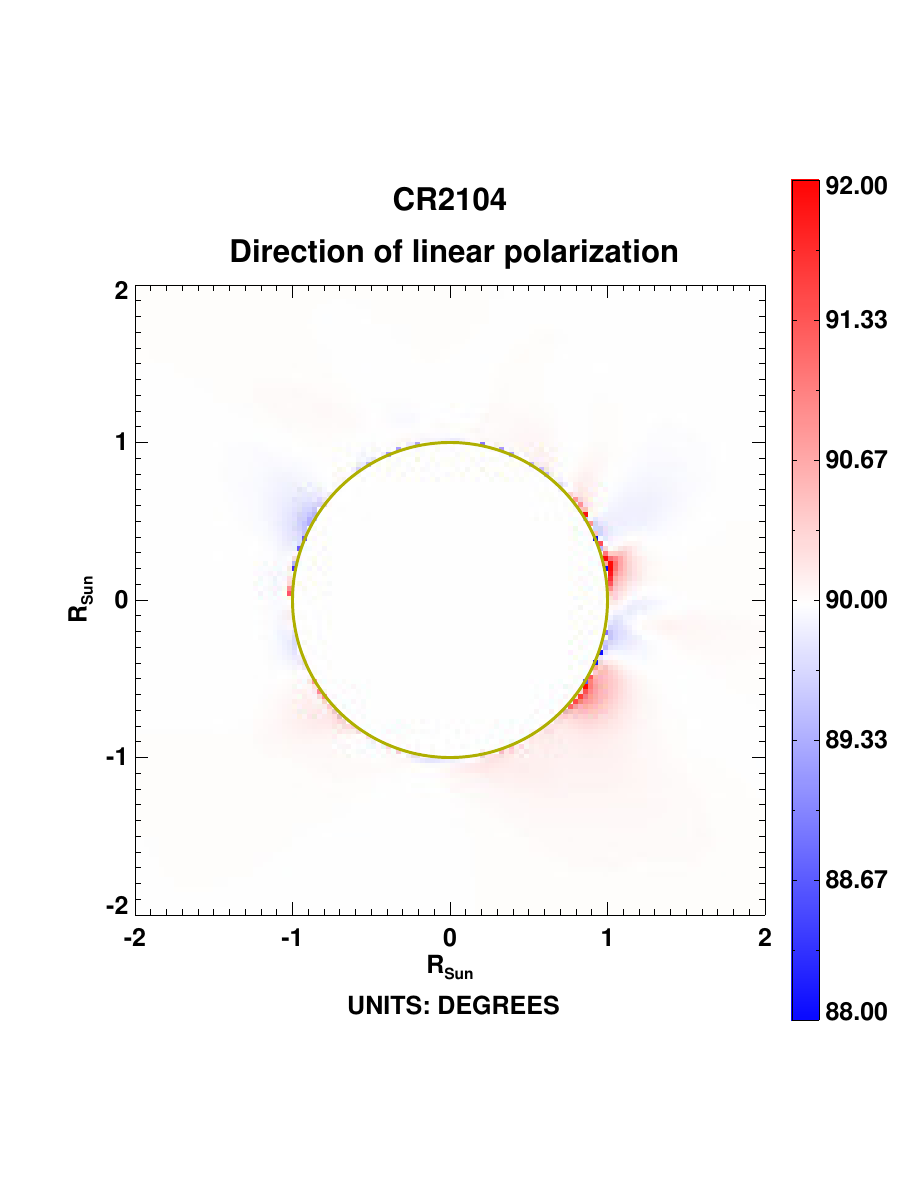}}
    %\hspace{0.8cm}
    \subfigure[]{\includegraphics[width=0.235\textwidth,trim={0.3cm 1.2cm 0cm 0.5cm},clip]{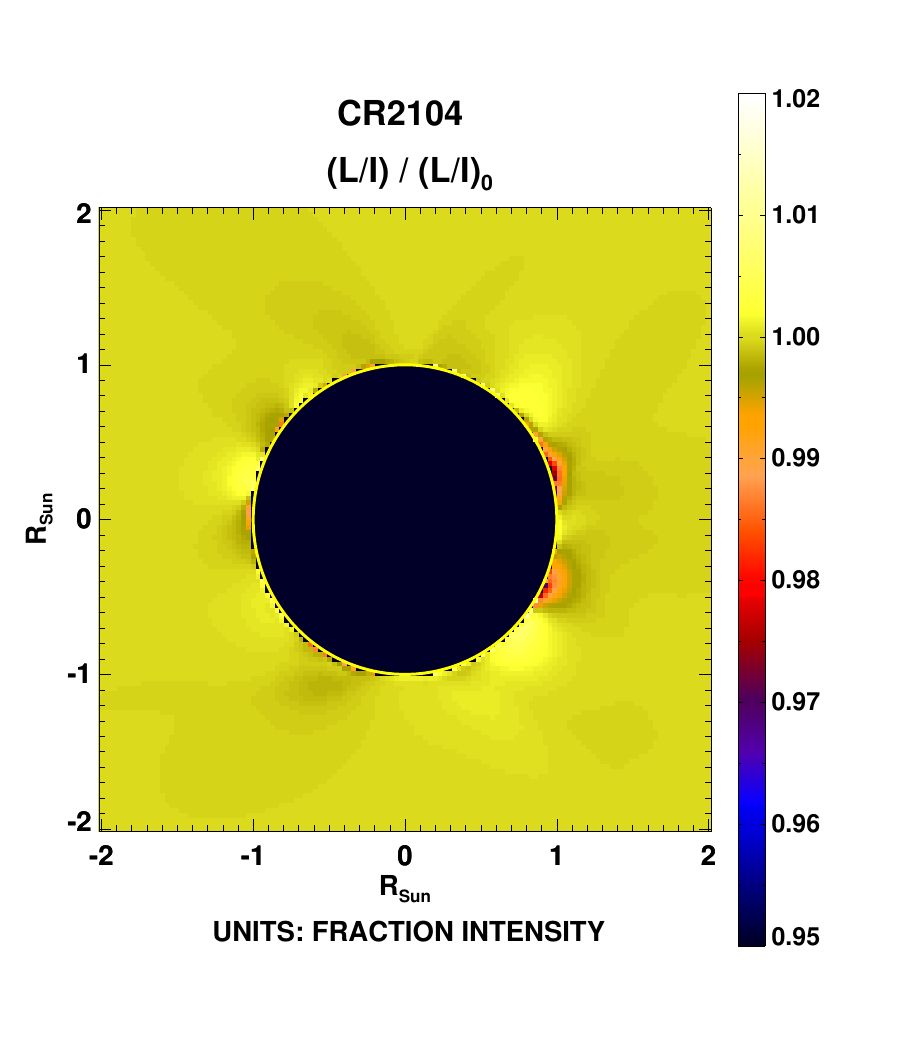}}
    %\hspace{0.8cm}
    \subfigure[]{\includegraphics[width=0.22\textwidth,trim={1.1cm 2.5cm 0cm 2.9cm},clip]{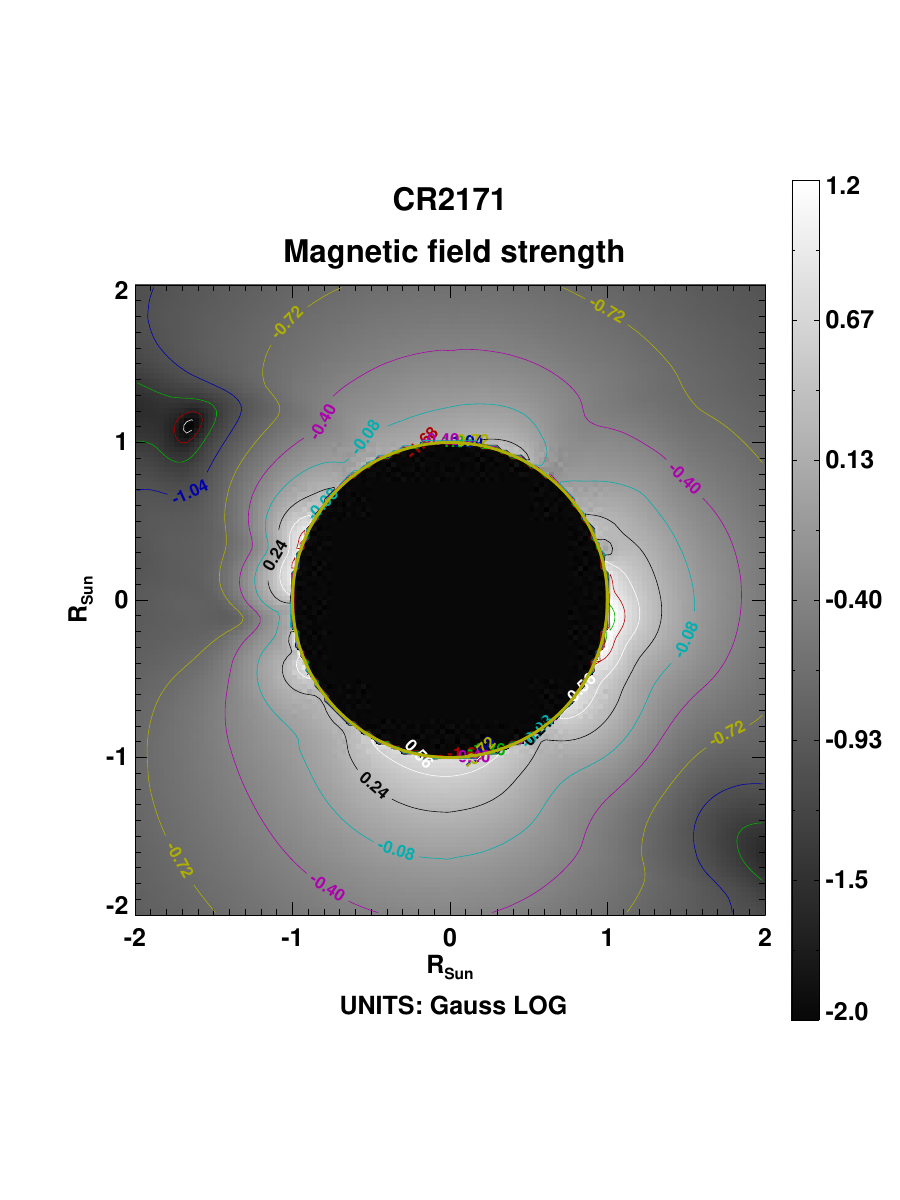}}
    %\hspace{0.8cm}
    \subfigure[]{\includegraphics[width=0.22\textwidth,trim={1.1cm 2.5cm 0cm 2.9cm},clip]{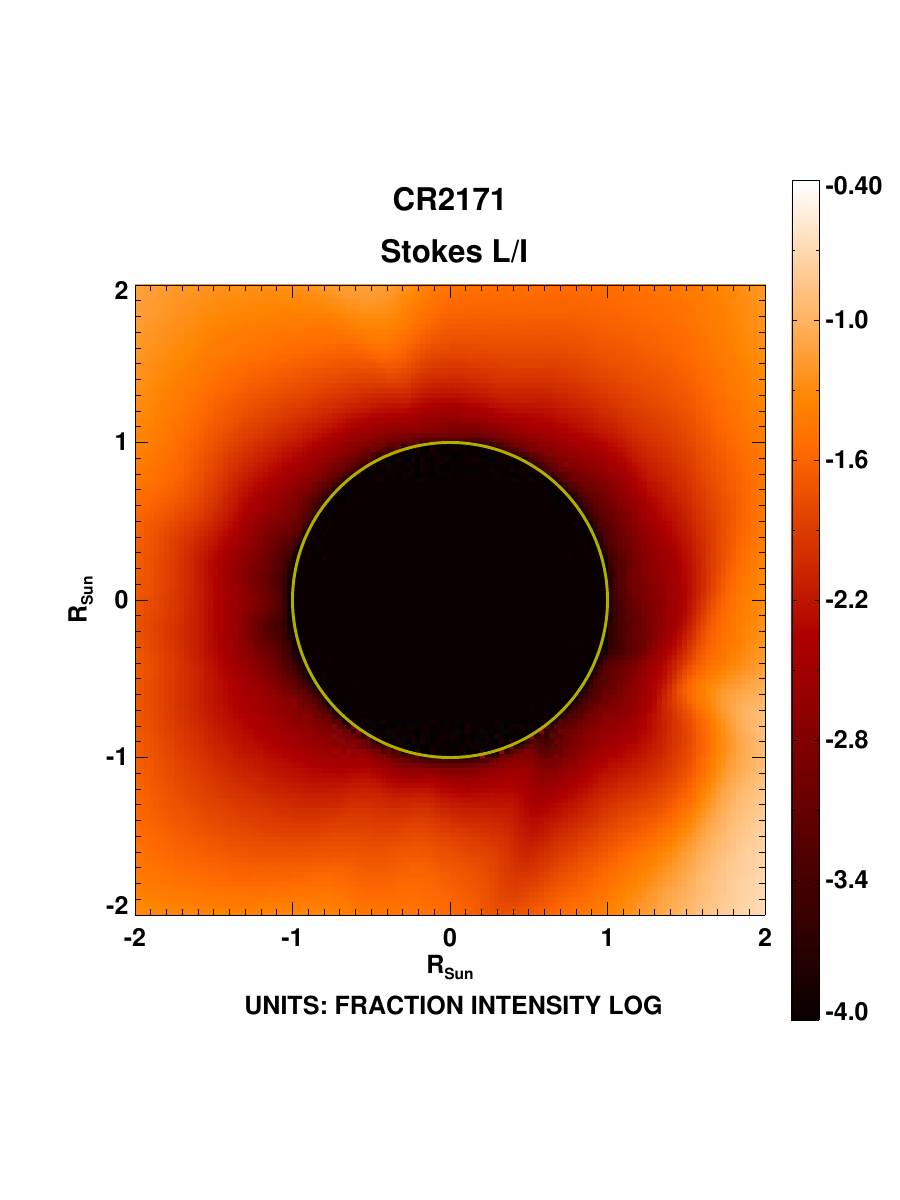}}
    \subfigure[]{\includegraphics[width=0.22\textwidth,trim={1.1cm 2.7cm 0cm 2.9cm},clip]{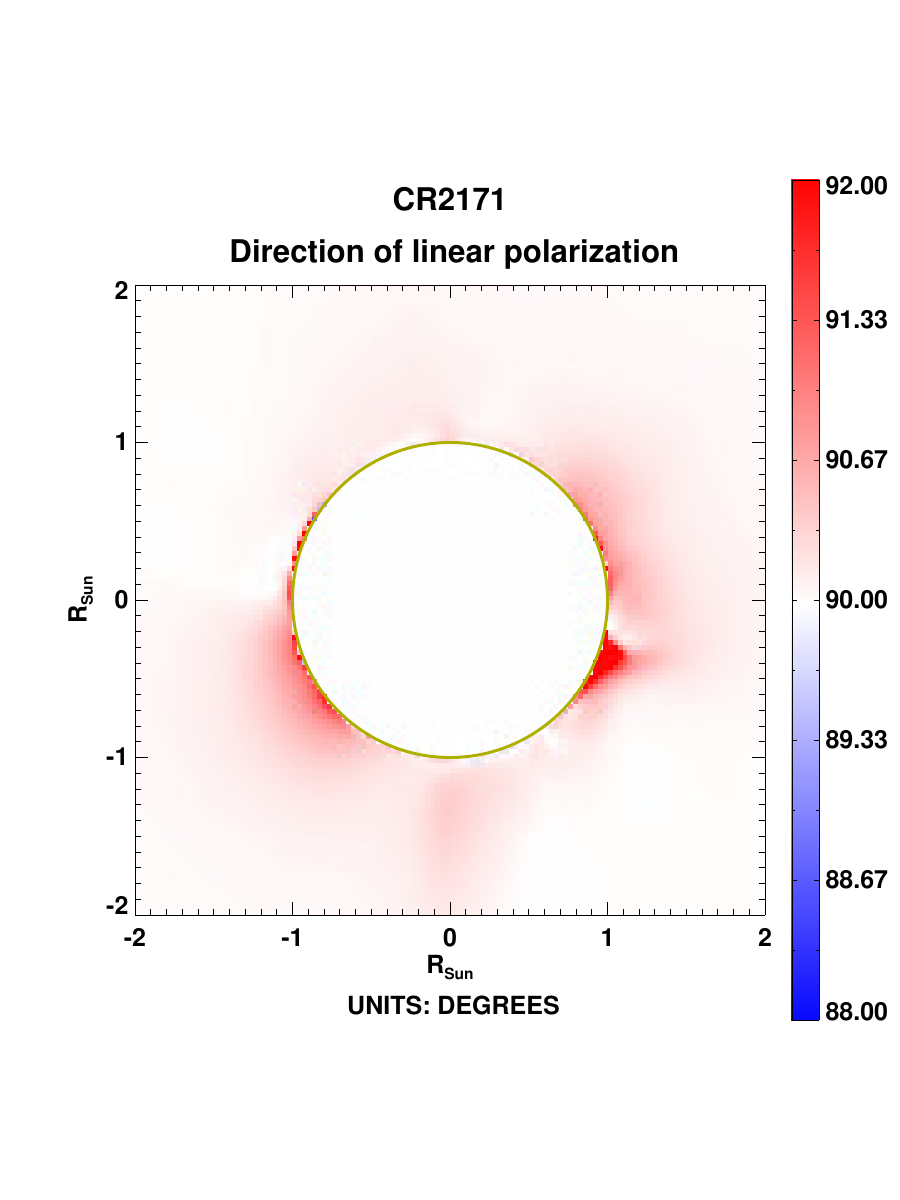}}
    %\hspace{0.8cm}
    \subfigure[]{\includegraphics[width=0.235\textwidth,trim={0.3cm 1.2cm 0cm 0.5cm},clip]{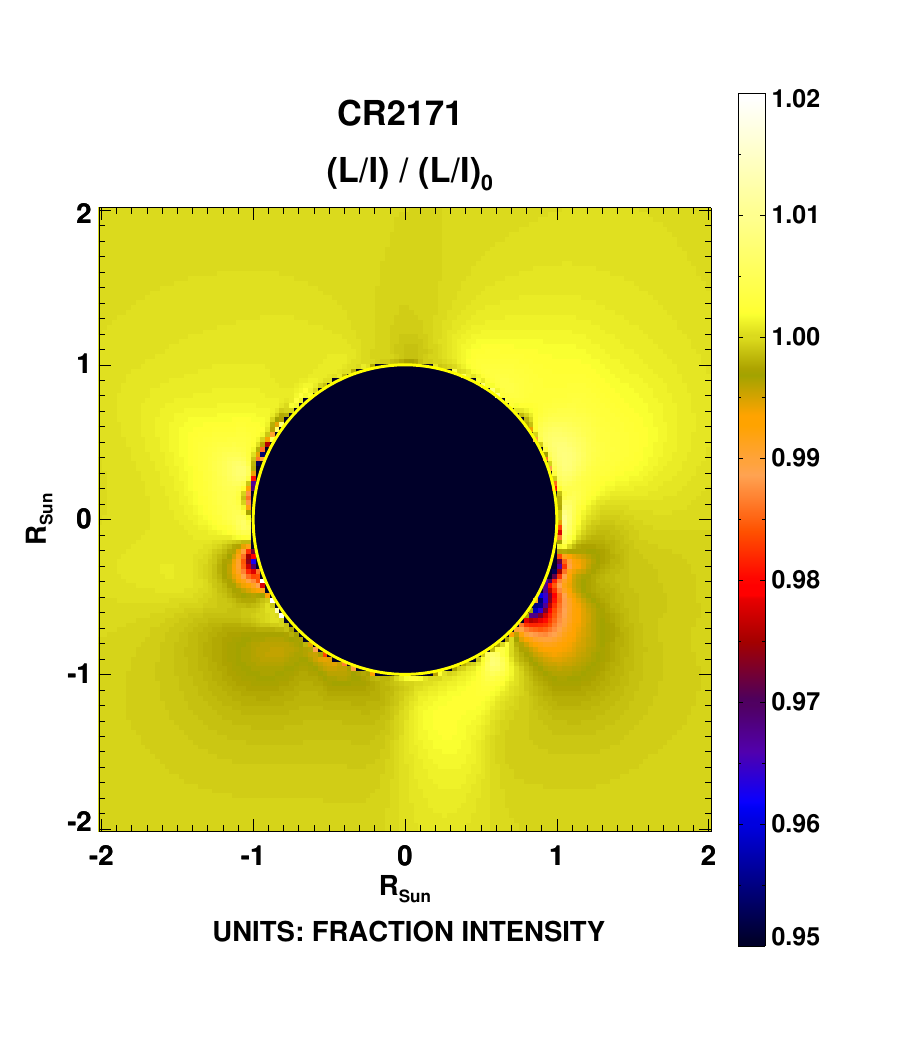}}
    %\hspace{0.8cm}
    \subfigure[]{\includegraphics[width=0.22\textwidth,trim={1.1cm 2.5cm 0cm 2.9cm},clip]{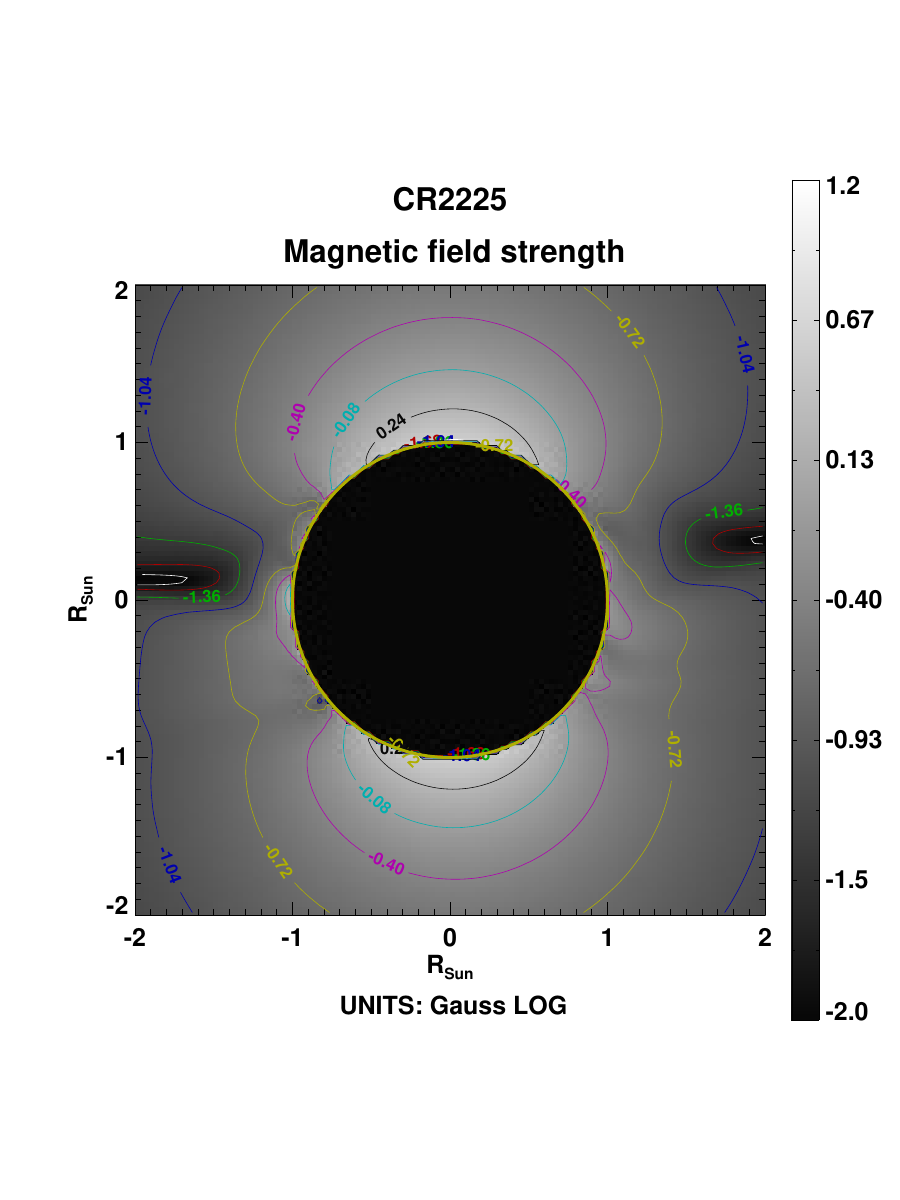}}
    \hspace{0.4cm}
    \subfigure[]{\includegraphics[width=0.22\textwidth,trim={1.1cm 2.5cm 0cm 2.9cm},clip]{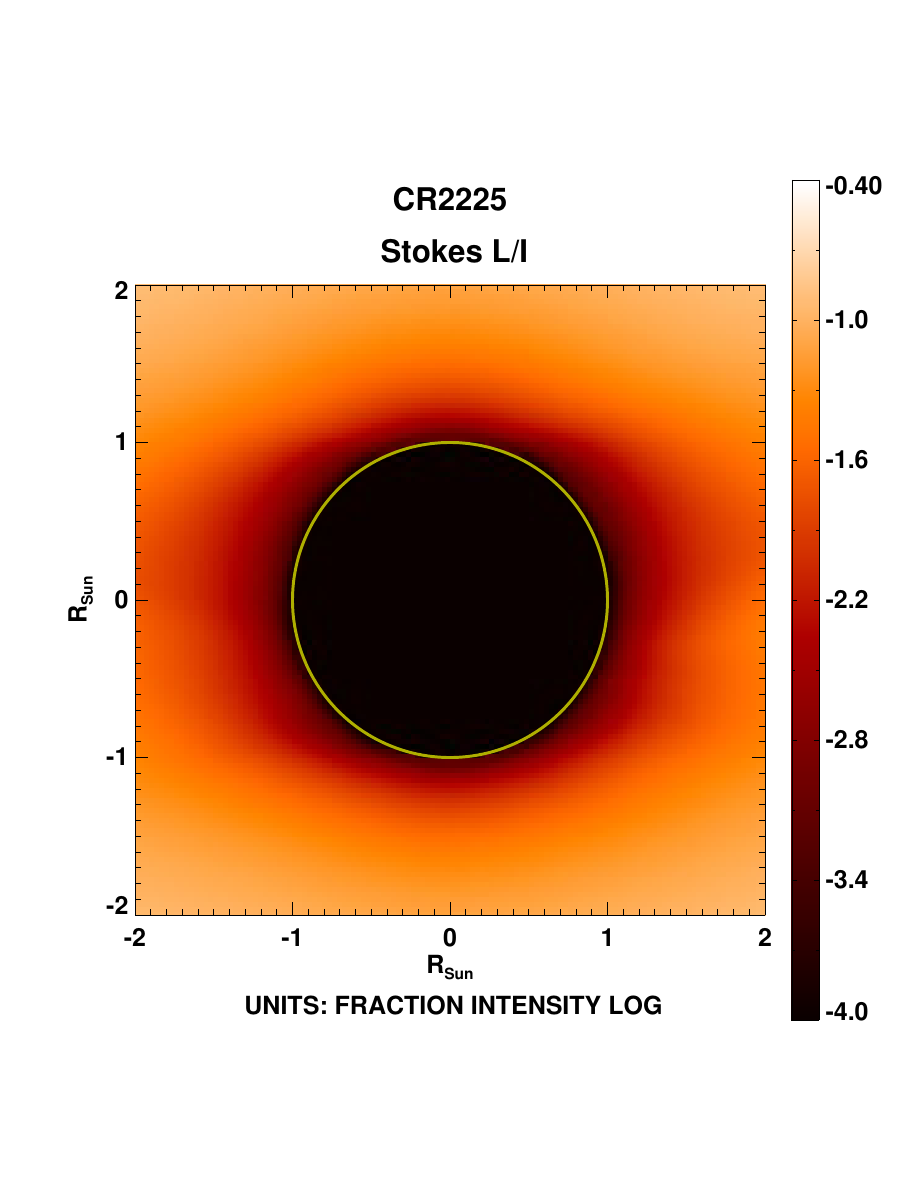}}
    \hspace{0.5cm}
    \subfigure[]{\includegraphics[width=0.22\textwidth,trim={1.1cm 2.7cm 0cm 2.9cm},clip]{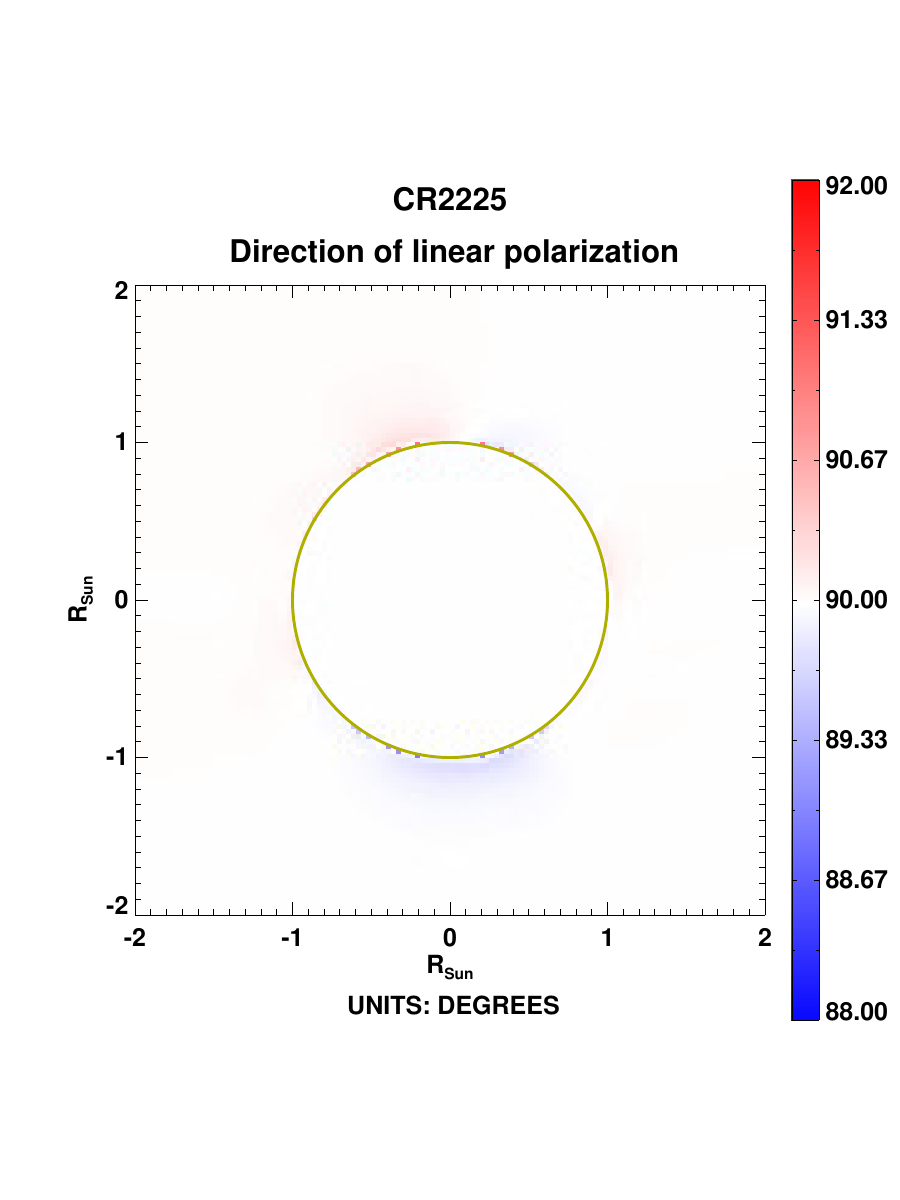}}
    \hspace{0.4cm}
    \subfigure[]{\includegraphics[width=0.235\textwidth,trim={0.3cm 1.2cm 0cm 0.5cm},clip]{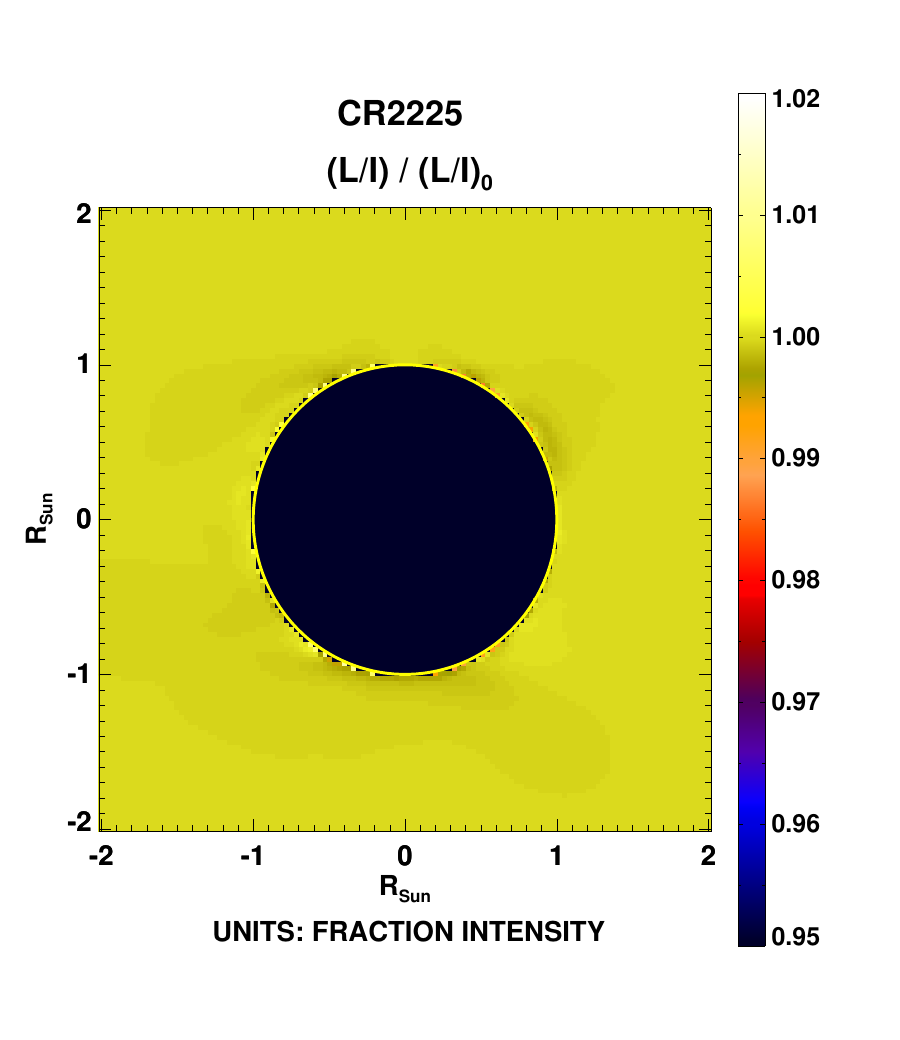}}
    \caption{Similar maps as shown in Figure \ref{fig:bmag_stokesloi_az_ondisk}, with the solar disk masked in order to emphasize the off-limb scales.}
    \label{fig:bmag_stokesloi_az_offdisk}
\end{figure*}

The 3D MAS models also provide us information of physical quantities such as magnetic field, density, and temperature of the coronal plasma. Note that as we do not want to introduce Doppler dimming effects into this analysis, we have replaced MAS simulated velocities with zero values. The second column of Figure \ref{fig:models_stki} and the first column of Figures \ref{fig:bmag_stokesloi_az_ondisk} and \ref{fig:bmag_stokesloi_az_offdisk} represent the temperature and magnetic field maps of the MAS model, respectively, on the spherical "solar surface" lower boundary (circumscribed by the yellow circle) and continuing at the limb in the $X=0$ plane out to $r=2.0~R_{\odot}$. On the disk, all the model maps are shown for a shell at 1.01~$R_{\odot}$ from where the LOS integrations of the observables (intensity and linear polarization maps) are initiated. 

The first column of Figure \ref{fig:models_stki}
represents the ion density maps of 
Ne {\sc viii}, a derived quantity which has been calculated using the electron density and temperature from the MAS simulations, assuming element abundances and ionization fractions of Ne {\sc viii} from the latest version 10.0 of the CHIANTI database \citep{Dere1997A&AS..125..149D, DelZanna2021ApJ...909...38D}. On the disk, the ion densities are shown for the same shell at 1.01~$R_{\odot}$ as the temperature plots in the second column, however, because we are interested in the coronal plasma, we only show MAS coronal ion densities for points possessing simulation temperatures greater than 500,000~K (the coronal base for the PSIMAS model; see \citealt{Lionello2001ApJ...546..542L}).
Points with temperatures below this threshold are also excluded from the line-of-sight integrals of intensity (column three of Figure \ref{fig:models_stki}) and linear polarization (columns two to four of Figures \ref{fig:bmag_stokesloi_az_ondisk} and \ref{fig:bmag_stokesloi_az_offdisk}). Note that the MAS model assumes a single fluid, i.e., ion temperature equals to electron temperature.

The top, middle and bottom rows of Figures \ref{fig:models_stki}, \ref{fig:bmag_stokesloi_az_ondisk} and \ref{fig:bmag_stokesloi_az_offdisk} depict the aforementioned maps at the rising, the maximum, and the end phase of the SC24, respectively. 
\begin{figure*}[ht!]
\centering
    \includegraphics[width=0.3\textwidth,trim={0.42cm 0.3cm 0cm 0.4cm},clip]{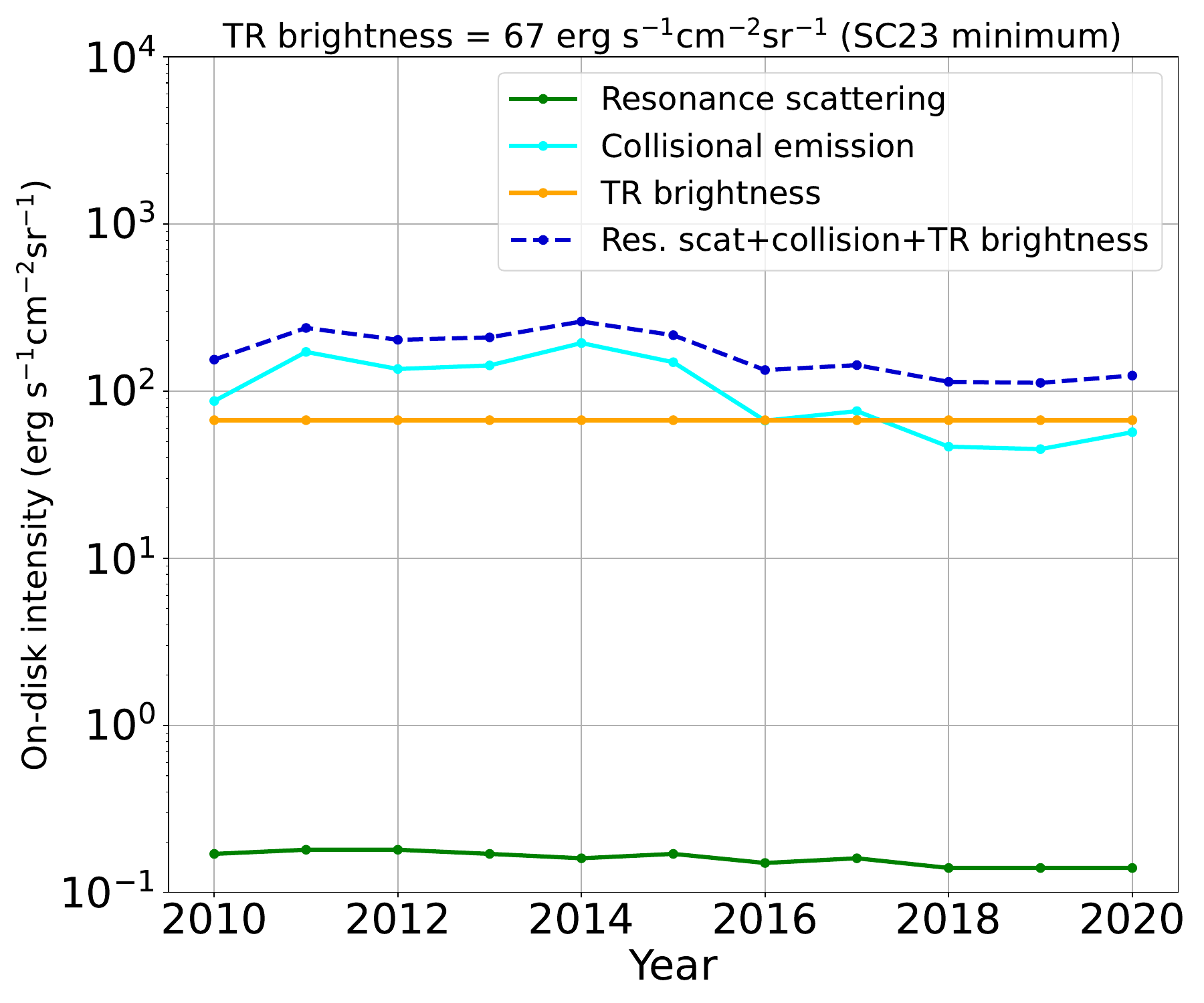}
    \includegraphics[width=0.3\textwidth,trim={0.42cm 0.3cm 0cm 0.4cm},clip]{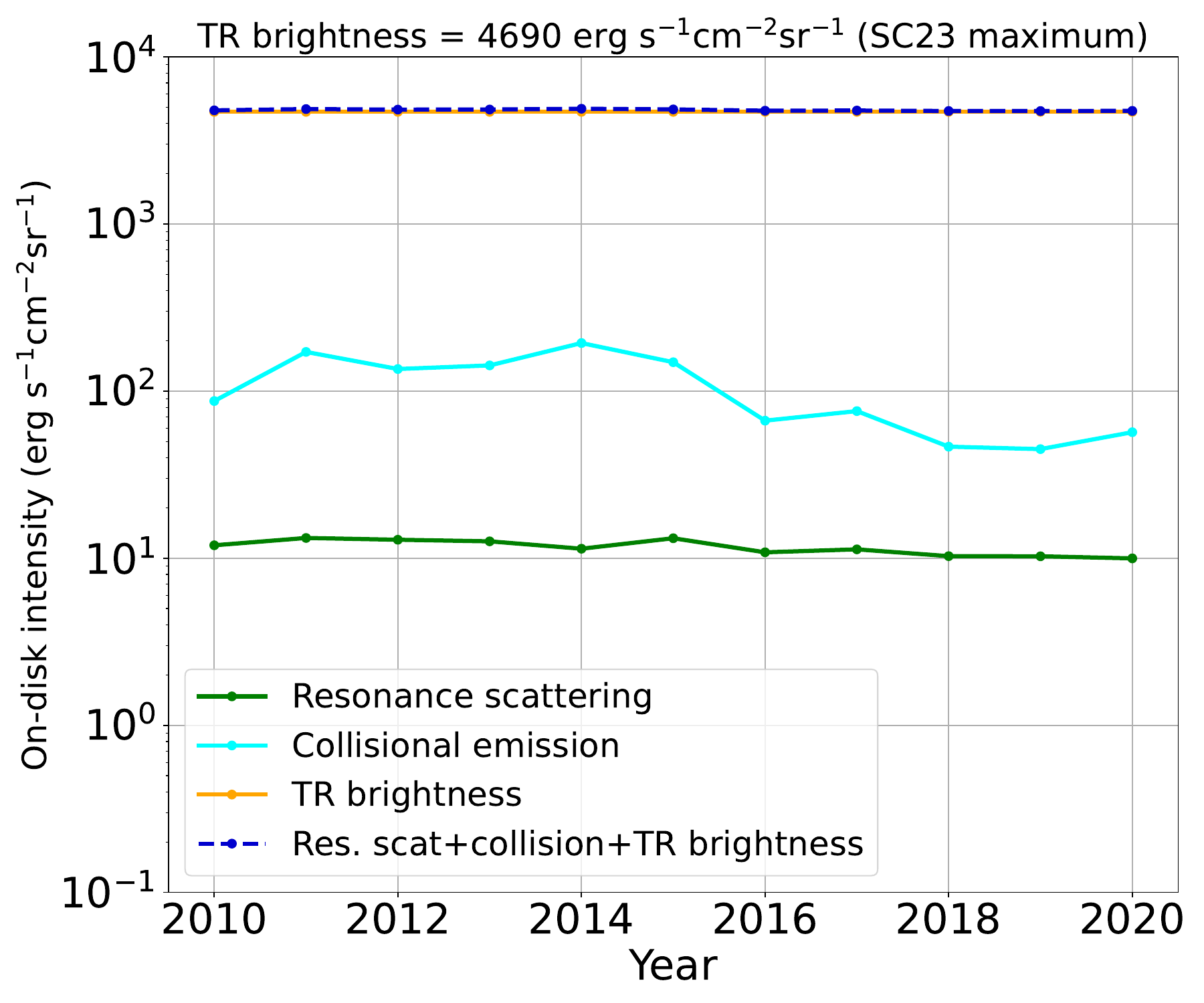}
    \includegraphics[width=0.3\textwidth,trim={0.42cm 0.3cm 0cm 0.4cm},clip]{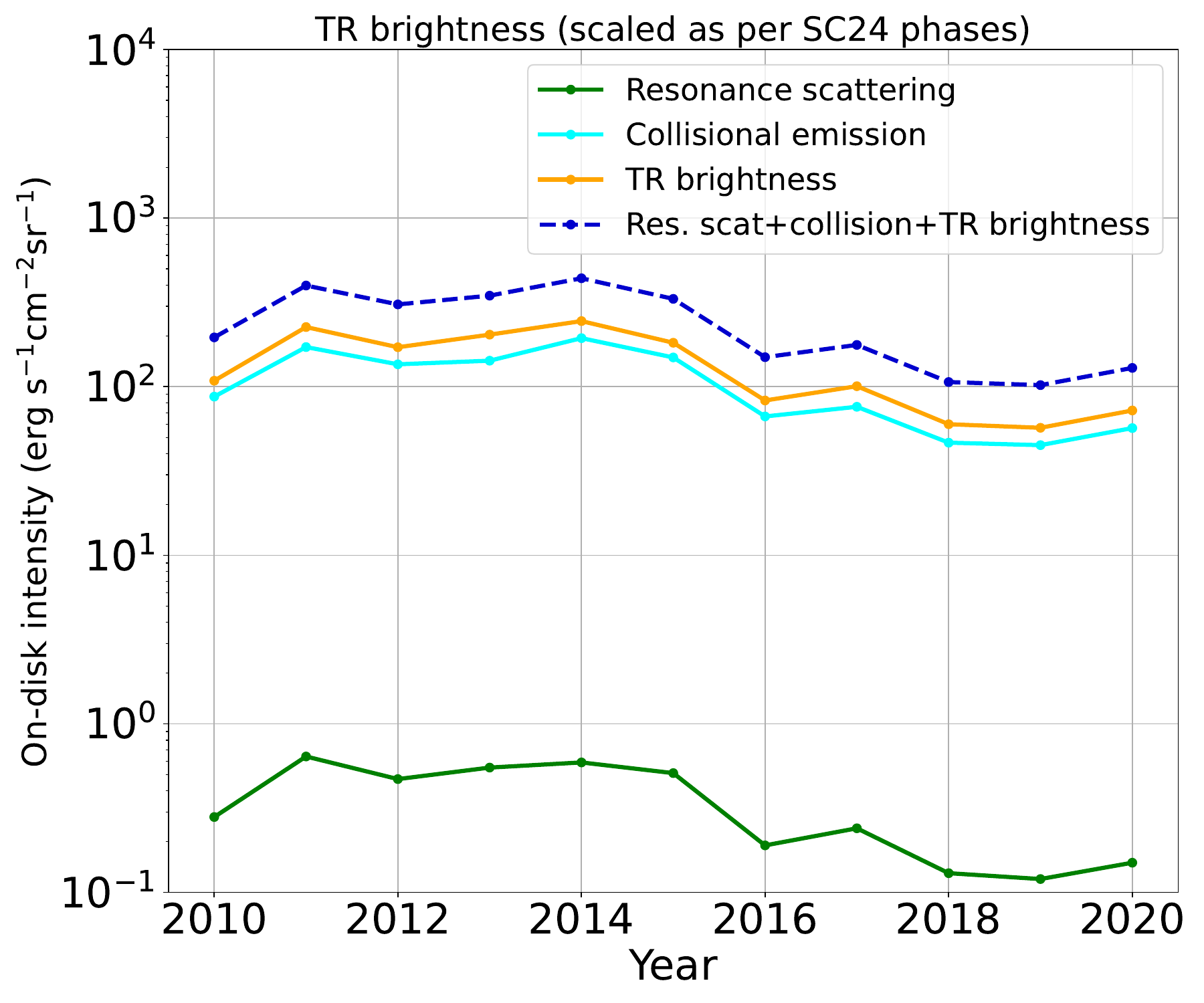}
    \includegraphics[width=0.3\textwidth,trim={0.42cm 0.3cm 0cm 0.4cm},clip]{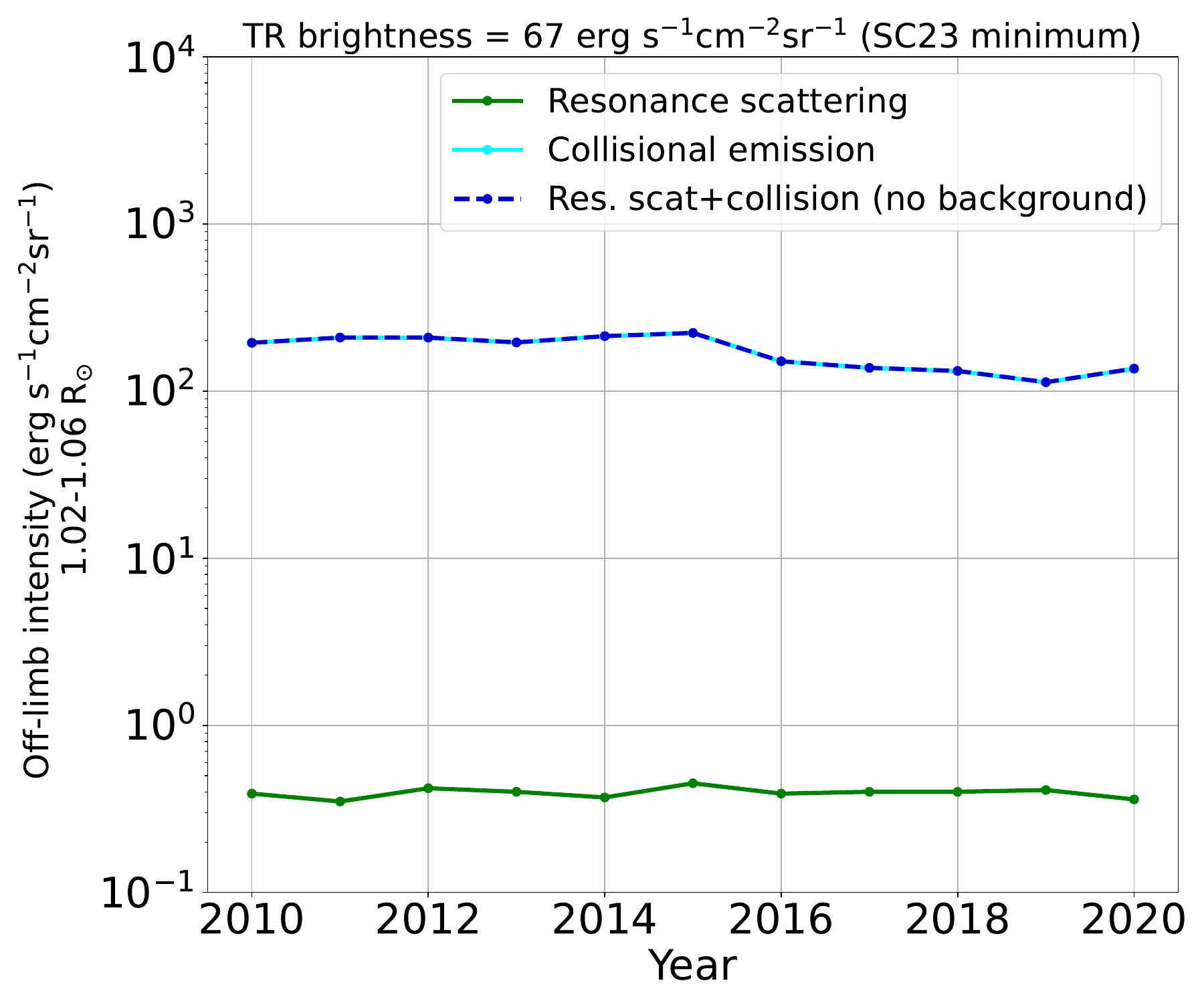}
    \includegraphics[width=0.3\textwidth,trim={0.42cm 0.3cm 0cm 0.4cm},clip]{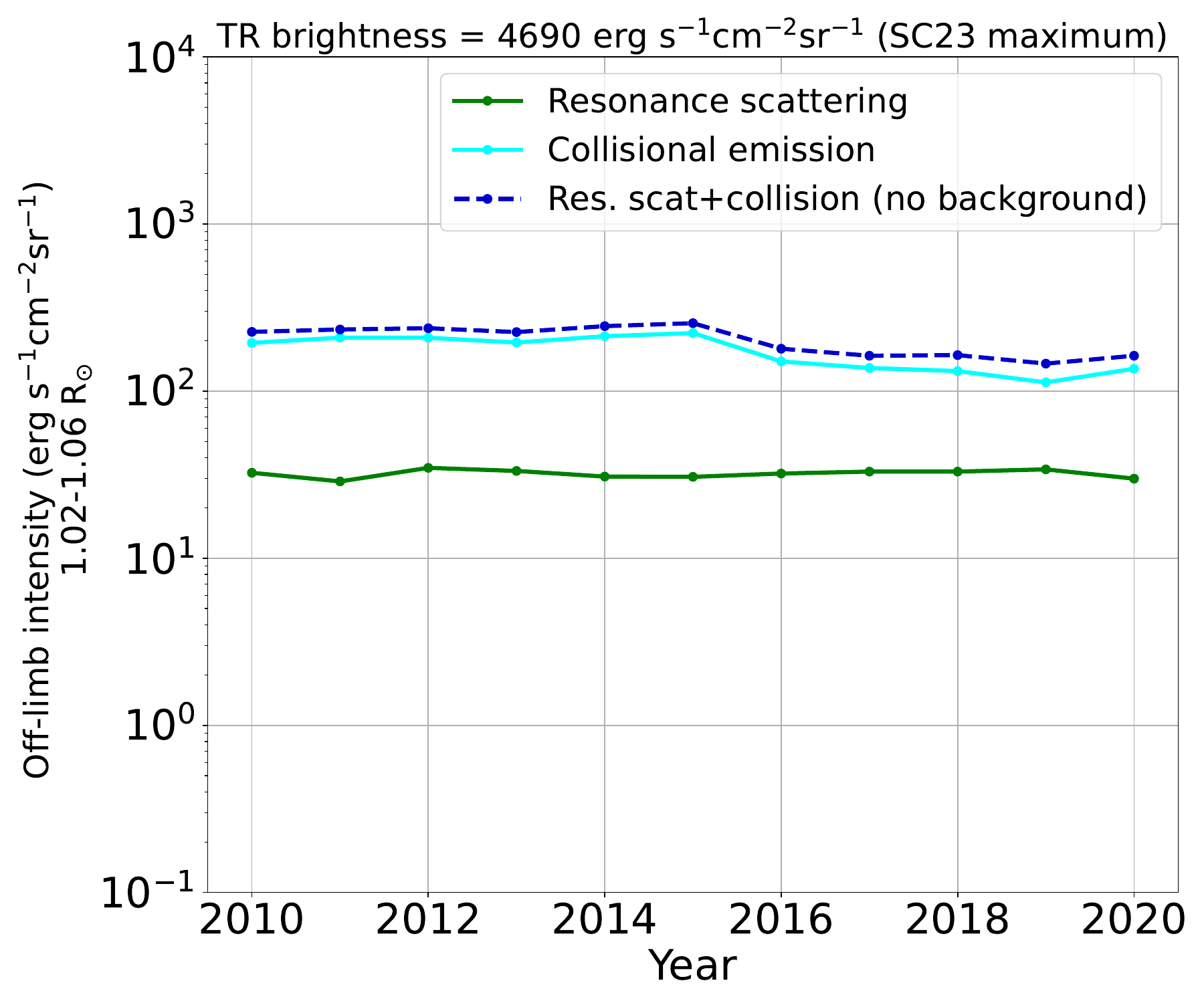}
    \includegraphics[width=0.3\textwidth,trim={0.42cm 0.3cm 0cm 0.4cm},clip]{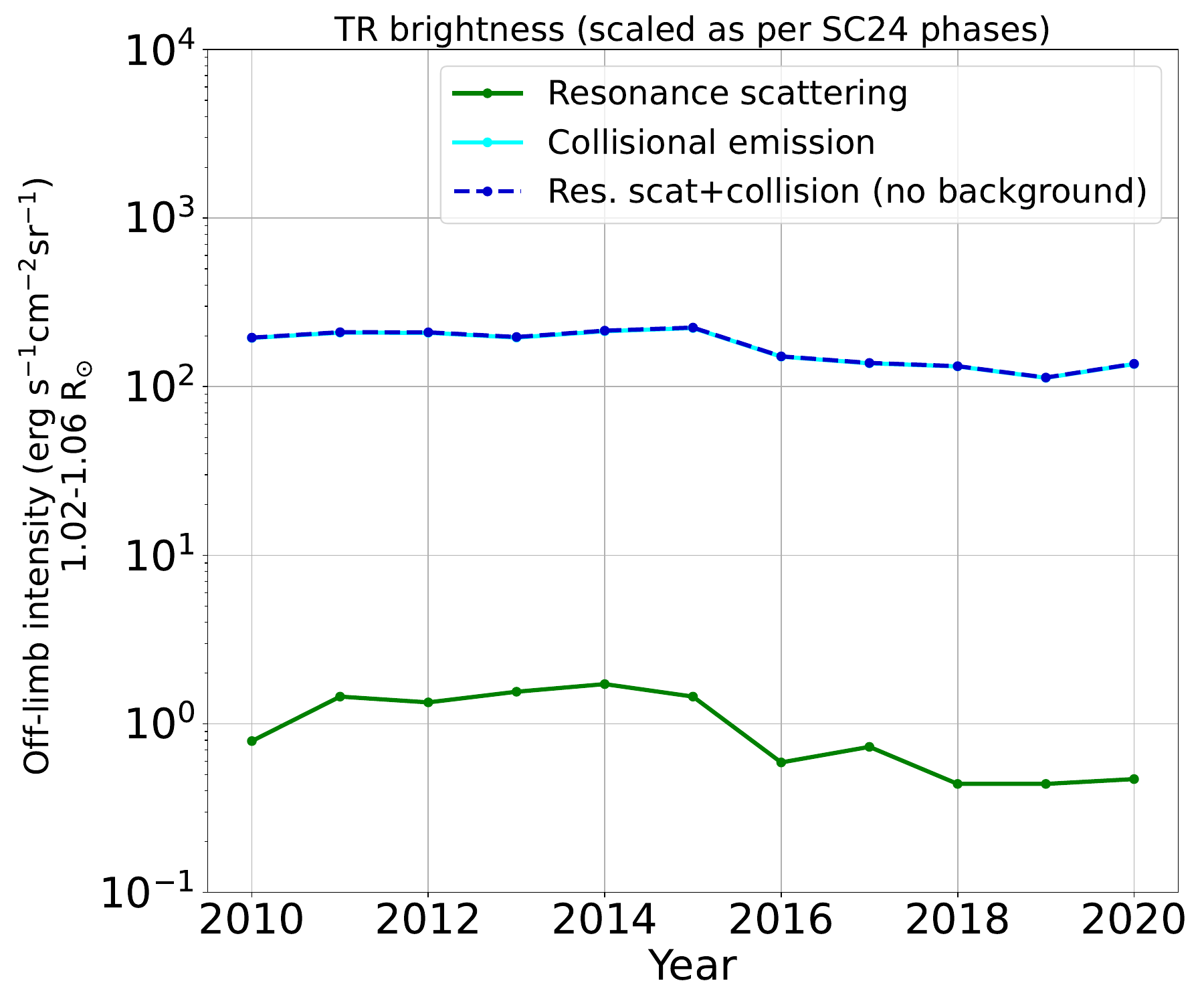}
    \caption{\textit{Top row} and \textit{bottom row} depict, respectively, on-disk Stokes $I$ (median value) and off-limb Stokes $I$ (median value between 1.02 and 1.06 $R_{\odot}$) as a function of SC24 period. \textit{First}, \textit{second} and \textit{third} columns describe, respectively, the three cases of TR brightness,
    i.e., observed during SC23 minimum, SC23 maximum, and linearly scaled values as per the MAS simulations for SC24 and SC25. The Y-axis represents the intensity in logarithmic scale, while the X-axis covers the phases of SC24 and the beginning phase of SC25. Colored lines separate out the contributions from resonance scattering (green), collisional excitation (cyan),  TR brightness (orange, on-disk), and the total of all sources of emissions (blue dashed). Note how the observed radiation on the disk is completely dominated by the TR brightness during peaks of solar activity (second column). Instead, the collisional component to the line radiation is always the dominant contribution to the brightness of the off-limb corona, so the visibility of its polarized component via resonance scattering is enhanced near solar maximum.}
    \label{fig:plots_chrovary}
\end{figure*}

\section{Results} \label{sec:results}
FORWARD can model UV polarization from the unsaturated Hanle effect in two-level atom transitions \citep{Landi1988A&A...192..374L},
such as H {\sc i} 1215 \AA\ and O {\sc vi} 1032 \AA\ \citep{Zhao2019ApJ...883...55Z}, also taking into account the effects of symmetry breaking processes due to non-radial Doppler dimming and anisotropic distributions of the microscopic velocity \citep{Zhao2021ApJ...912..141Z}. For the current work, we have extended the FORWARD capabilities to facilitate analyses of other two-level atom transitions in the EUV, e.g., Ne {\sc viii} 770 \AA\, and in particular to enable modeling signals on the disk. We have simulated the linear polarization maps of the 770 \AA\ resonance line, considering the presence of both magnetic fields and collisional excitation \citep{Susino2018A&A...617A..21S}. 

Figures \ref{fig:models_stki}, \ref{fig:bmag_stokesloi_az_ondisk} and \ref{fig:bmag_stokesloi_az_offdisk} step through examples of the rising
(\textit{top row:} model CR2104), maximum
(\textit{middle row:} model CR2171)  
and minimum
(\textit{bottom row:} model CR2225)
phases of SC24. 
The present section elucidates the changes in the following observables across the various phases of SC24:
\begin{itemize}
    \item the spectrally integrated coronal brightness, Stokes $I$ (in erg\,s$^{-1}$\,cm$^{-2}$\,sr$^{-1}$) shown in Figure \ref{fig:models_stki} (third column);

    \item the degree of linear polarization defined as  $L/I=\sqrt{Q^{2}+U^{2}}/I$ shown in Figures \ref{fig:bmag_stokesloi_az_ondisk} and \ref{fig:bmag_stokesloi_az_offdisk} (second column);

    \item the rotation angle of the plane of polarization, or polarization azimuth   
    $\beta$, shown in Figures \ref{fig:bmag_stokesloi_az_ondisk} and \ref{fig:bmag_stokesloi_az_offdisk} (third column);\footnote{Operationally, the 
    polarization azimuth corresponds to the angle by which the polarimeter should be rotated in order to attain the condition $Q'=L$ in the new reference, and it is always contained between $\pm 90^\circ$; this is a direct consequence of the properties of transformation under rotation of the Stokes parameters $Q$ and $U$ \cite[e.g.,][Ch.\ 1]{Landi2004ASSL..307.....L}}
    its definition is given by Eqs.~(1.8) of \cite{Landi2004ASSL..307.....L}, and in FORWARD it is directly attained by using the two-argument form of the arctan function, $\beta=\frac{1}{2}\arctan(U,Q)$, with the additional folding of the negative branch $[-90^\circ,0^\circ)$ to the positive branch via addition of $180^\circ$; this also ensures that the tangent to the limb in our maps corresponds to $90^\circ$, in order to reflect our choice of the reference direction for $Q>0$ which is everywhere aligned with the radius vector from the disk center to the observed point.
    
    \item the ratio between linear polarization in a magnetized vs non-magnetized plasma given in Figures \ref{fig:bmag_stokesloi_az_ondisk} and \ref{fig:bmag_stokesloi_az_offdisk} (fourth column).
\end{itemize}

\subsection{Effects of TR brightness variations} 
\label{sec:bkg_variation}
We discuss here the impact of the pumping radiation change on the emitted intensity at Ne {\sc viii} 770 \AA\ due to physical processes such as resonance scattering. As pointed out earlier in Section \ref{sec:pol_ne}, this radiation comes from heights below the resonantly scattered coronal emission, and for Ne {\sc viii} 770 \AA\ these heights correspond to the high TR and the lower corona \citep{Fludra2021A&A...656A..38F}.

In the first column of Figure \ref{fig:plots_chrovary} (orange curve), we have assumed a TR brightness
at Ne {\sc viii} 770 \AA\ of 67 erg~s$^{-1}$cm$^{-2}$sr$^{-1}$, based on \cite{Curdt2001AIPC..598...45C} (SC23 minimum phase). In the second column of Figure \ref{fig:plots_chrovary}, we have chosen 4690 erg~s$^{-1}$cm$^{-2}$sr$^{-1}$ from \cite{Sarro2011A&A...528A..62S}, which corresponds to the average TR brightness observed during the maximum phase of SC23. In the third column, we have linearly interpolated the 
TR brightness according to the median value of the full-disk collisional emission obtained from the MAS simulations for each of the phases of SC24 and also the beginning of SC25. We now explain the justification for the scaling of this interpolation. 

\cite{Carrasco2021RNAAS...5..181C} reported that the beginning of SC25 was similar to that of SC24,
which was in turn weaker than SC23. Hence, we chose
57 erg~s$^{-1}$cm$^{-2}$sr$^{-1}$ (taken as 15\% lower than the above mentioned SC23 value of 67 erg~s$^{-1}$cm$^{-2}$sr$^{-1}$ as justified by
\citealt{Didkovsky2010ASPC..428...73D}) as the TR brightness
for both SC24 and SC25 minimum, and applied it to
the year 2019. We further used linear interpolation, relating to the median value of collisional emission on the disk, to derive the TR brightness for the rest of the years from 2010 to 2020. We then analyzed how the variation of this radiation
with solar cycle affects the signal on the disk (shown in the top row of Figure \ref{fig:plots_chrovary}), and off the limb between a height of 1.02 R$_{\odot}$ and 1.06 R$_{\odot}$ (shown in the bottom row of Figure \ref{fig:plots_chrovary}). In addition, we have calculated the LOS-integrated intensity above the disk considering only resonance scattering (green points) and only collisional emission (cyan points). 

\begin{figure}[htbp] 
 \centering
    \includegraphics[width=0.23\textwidth,trim={1cm 2.5cm 0cm 2.5cm},clip]{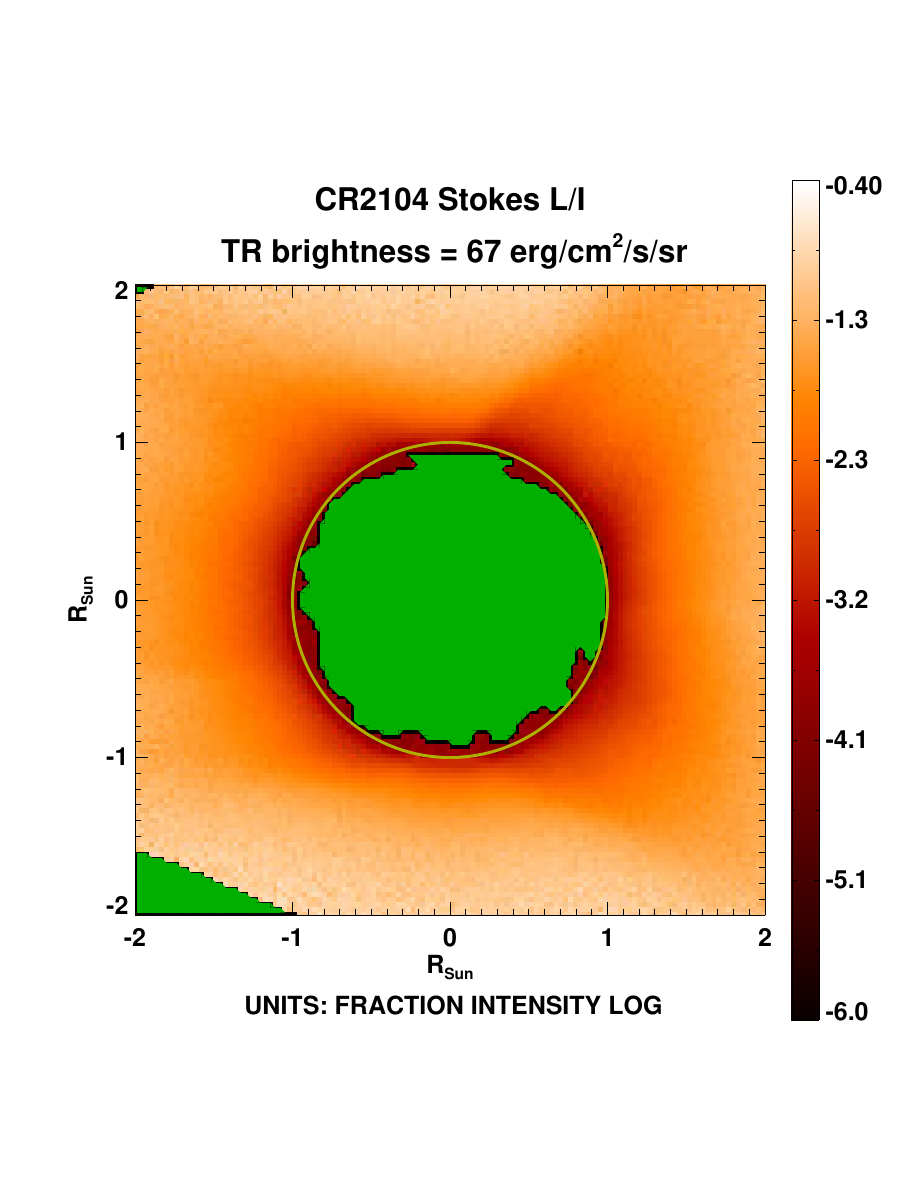}
    \includegraphics[width=0.23\textwidth,trim={1cm 2.5cm 0cm 2.5cm},clip]{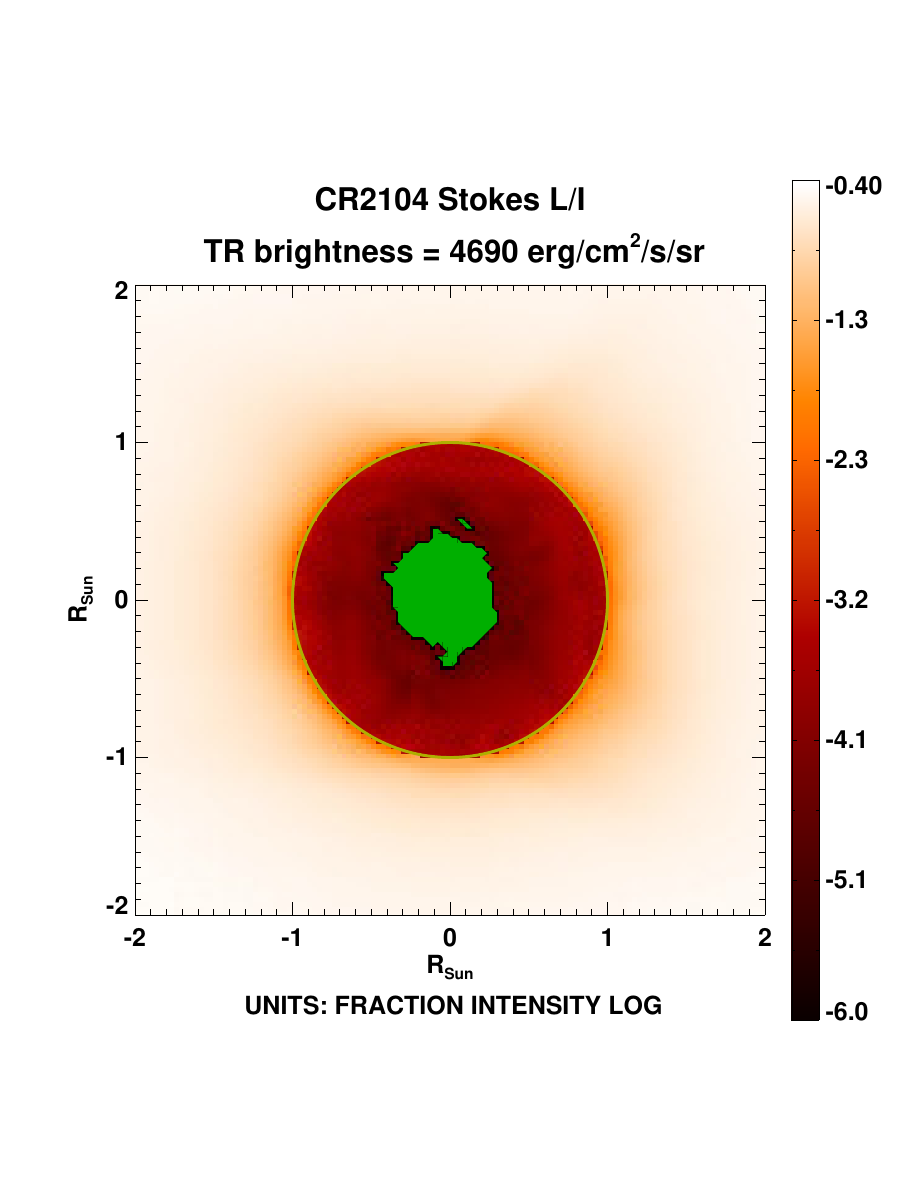}
    \includegraphics[width=0.23\textwidth,trim={1cm 2.5cm 0cm 2.5cm},clip]{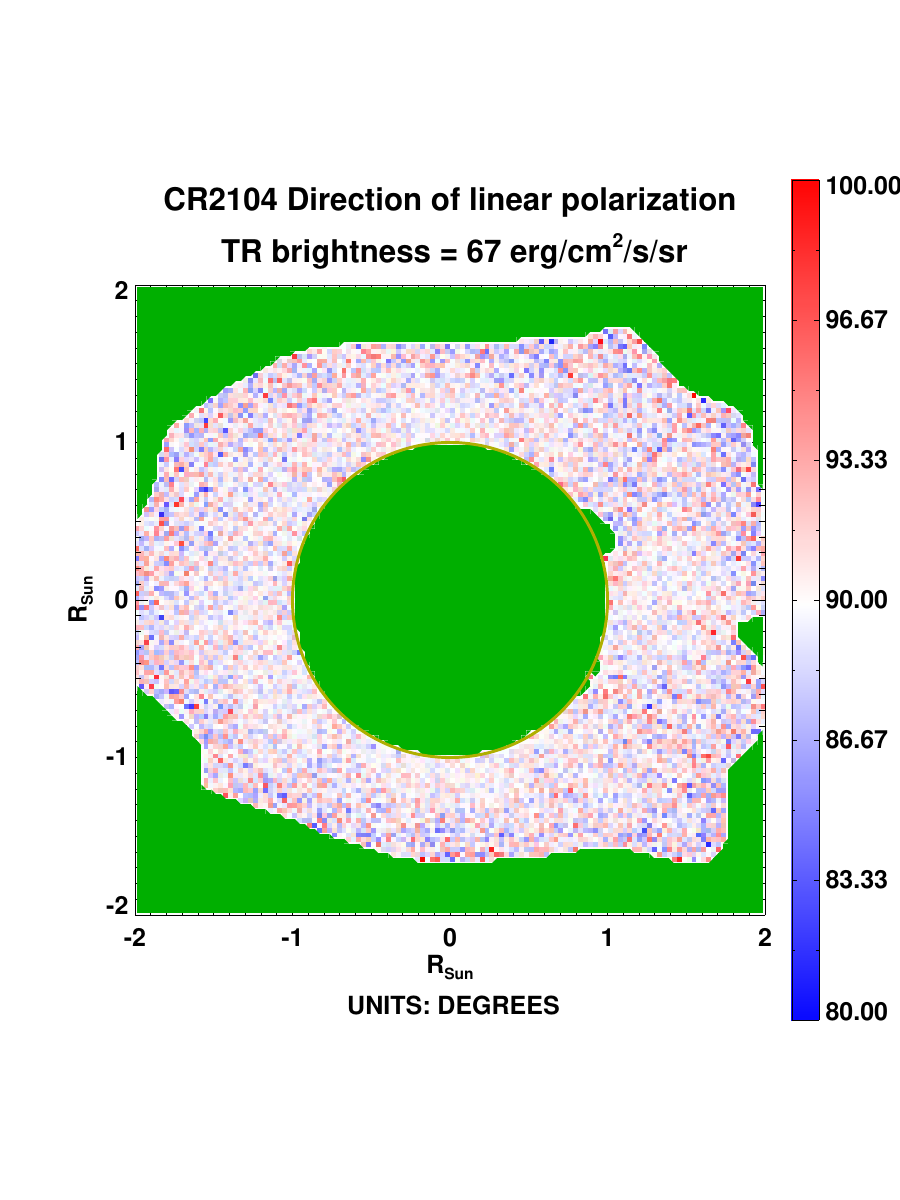}
    \includegraphics[width=0.23\textwidth,trim={1cm 2.5cm 0cm 2.5cm},clip]{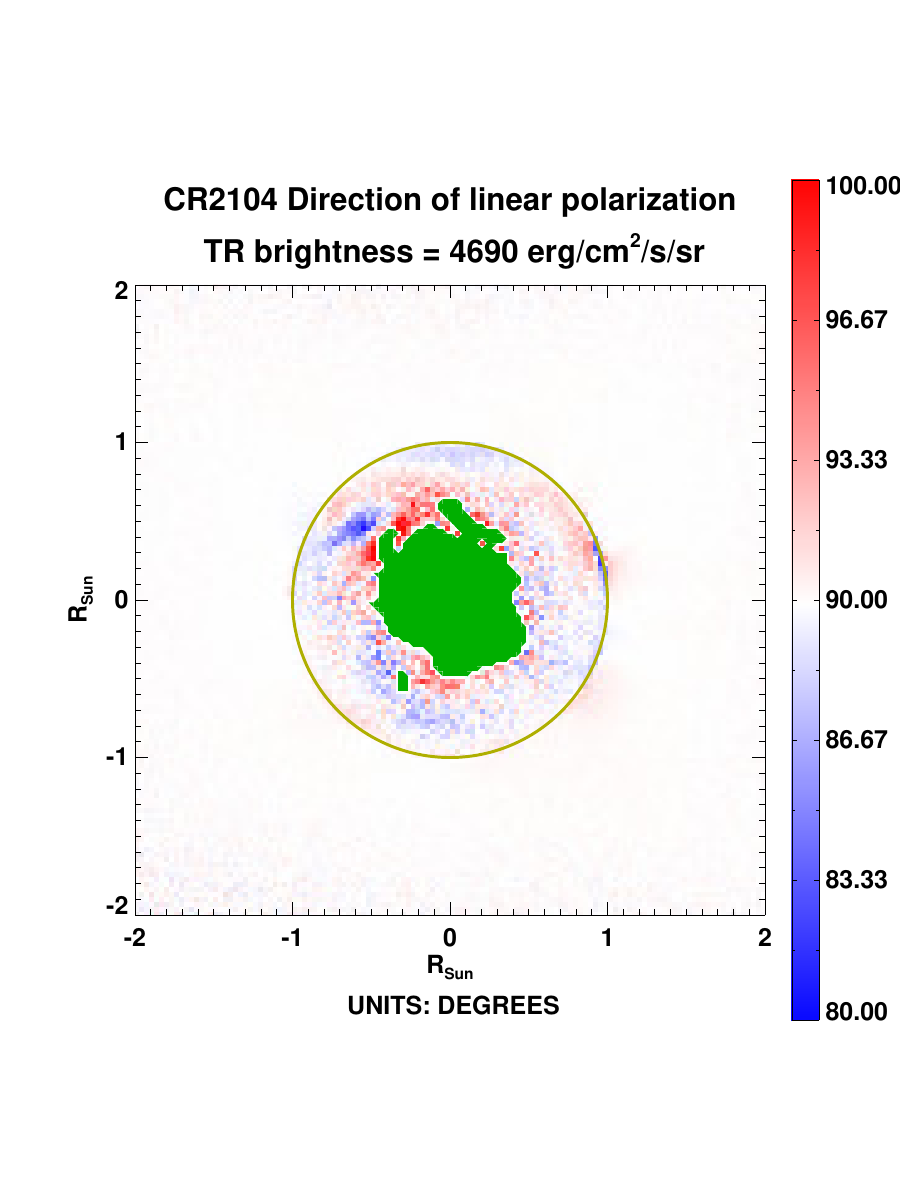}
    \caption{\textit{Left column} and \textit{right column} correspond to polarization maps considering a TR brightness 
    of 67 $\rm erg\,cm^{-2}\,s^{-1}\,sr^{-1}$ (observed during SC23 minima) and 4690 $\rm erg\,cm^{-2}\,s^{-1}\,sr^{-1}$ (observed during SC23 maxima), respectively, for a telescope of aperture 50~cm, 30 arcsecond pixel size, instrument throughput of 0.5\% and total integration time of 50 minutes. Green masked regions in azimuth maps indicate the areas where error on azimuth measurement is greater than $\pm 3^{\circ}$, while in $L/I$ maps they indicate areas where SNR on $L/I$ is less than 5$\sigma$.}
    \label{fig:stokesloi_az_chrovary}
\end{figure}

The effect of the mean TR brightness on the LOS-integrated linear polarization parameters (Stokes $L/I$ and polarization azimuth), and the resulting SNR, are demonstrated by the synthetic polarization maps generated with the two extreme brightness values from the SC23 as shown in Figure \ref{fig:stokesloi_az_chrovary}. A detailed derivation of the equations for calculating the SNR on the linear Stokes parameters can be found in \autoref{sec:appendix_noise}. For this study, we have considered a telescope of aperture 50~cm with a throughput of 0.5\% (from Raveena et al. 2024, in preparation) and a 30 arcsecond pixel. In order to calculate the maximum feasible time of integration on the disk of the Sun, let us consider to be observing a parcel of coronal plasma moving along the equator. Based on a synodic rotation rate at the equator of 26.24 days, this results in approximately 8 arcsecond displacement at disk center over 50 minutes for a feature at the solar surface. Assuming a (highly) conservative estimate of two solar radii for the highest coronal feature to be contributing to the LOS integral, the maximum smearing would be 16 arcseconds at disk center, roughly half a pixel, and hence tolerable.

\begin{figure}[ht!] 
 \centering
    \includegraphics[width=0.23\textwidth,trim={1cm 2.5cm 0cm 2.5cm},clip]{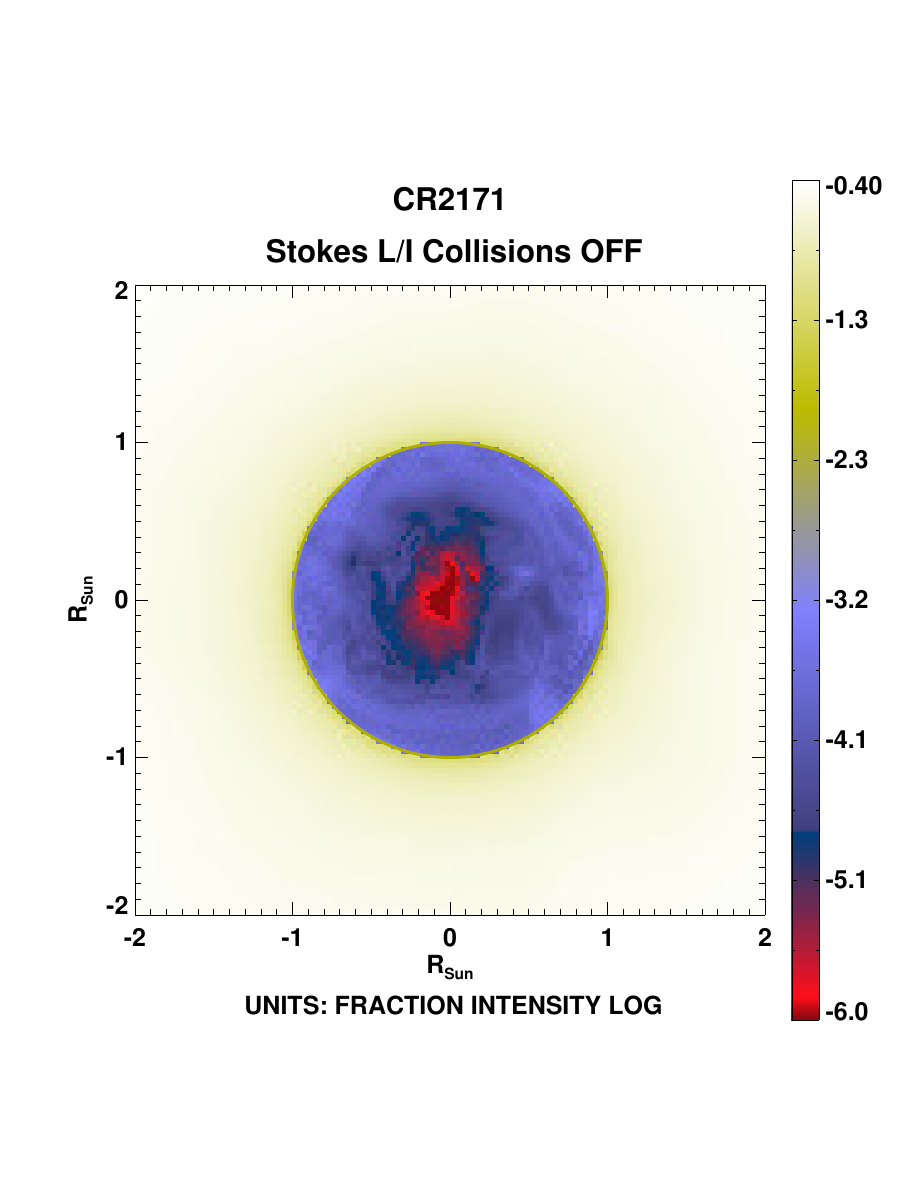}
    \includegraphics[width=0.23\textwidth,trim={1cm 2.5cm 0cm 2.5cm},clip]{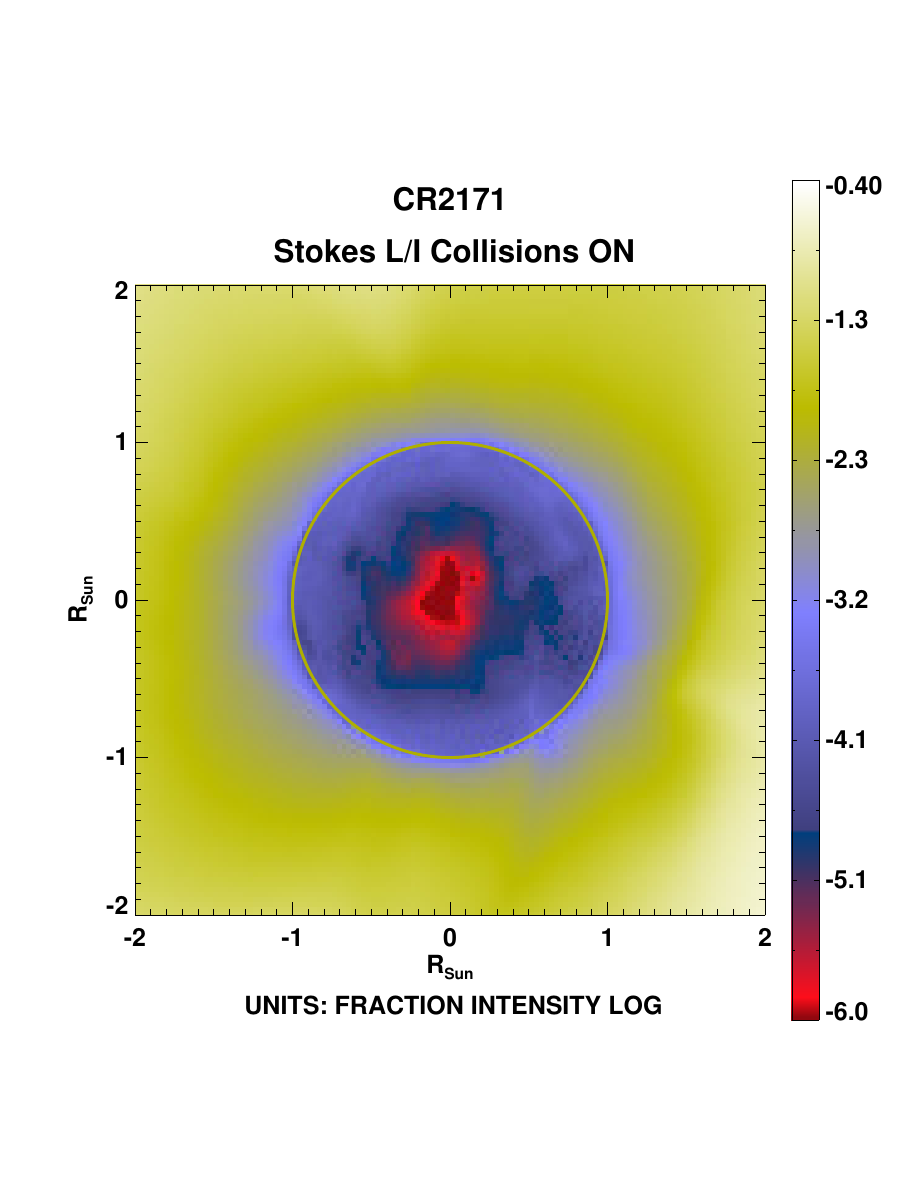}
    \includegraphics[width=0.23\textwidth,trim={1cm 2.5cm 0cm 2.5cm},clip]{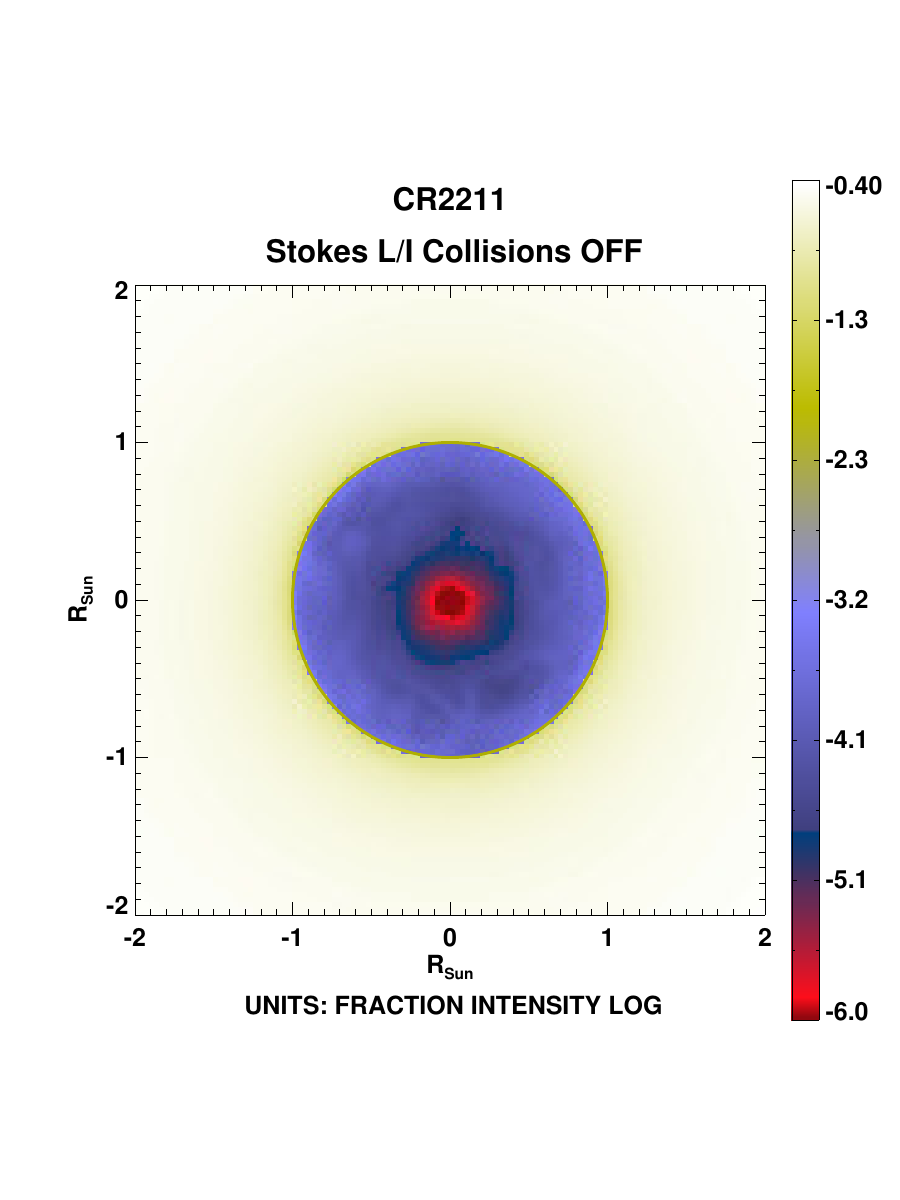}
    \includegraphics[width=0.23\textwidth,trim={1cm 2.5cm 0cm 2.5cm},clip]{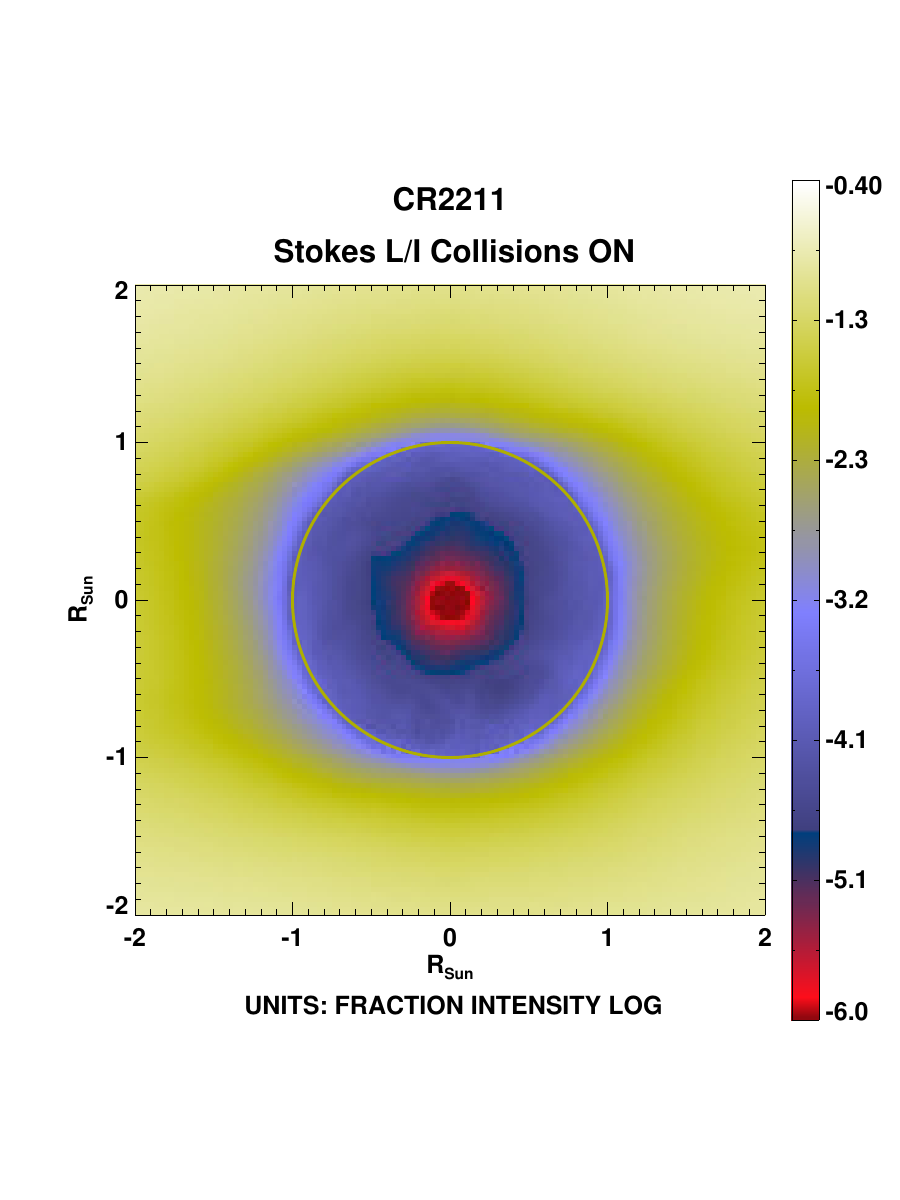}
    \caption{\textit{Top row} and \textit{bottom row} corresponds to the solar maximum and minimum phase of SC24, respectively. \textit{Left column} shows linear polarization fraction ($L/I$) maps in absence of collisions, while \textit{right column} shows $L/I$ when collisions are switched on. SNR requirements are not considered here. Note a different color table is used for ($L/I$) than was shown in earlier figure, to enable visualization of limb and disk simultaneously.}
    \label{fig:stokesloi_coll}
\end{figure}

\begin{figure*}[htbp] 
    \subfigure[]{\includegraphics[width=0.23\textwidth,trim={1.1cm 2.5cm 0cm 2.9cm},clip]{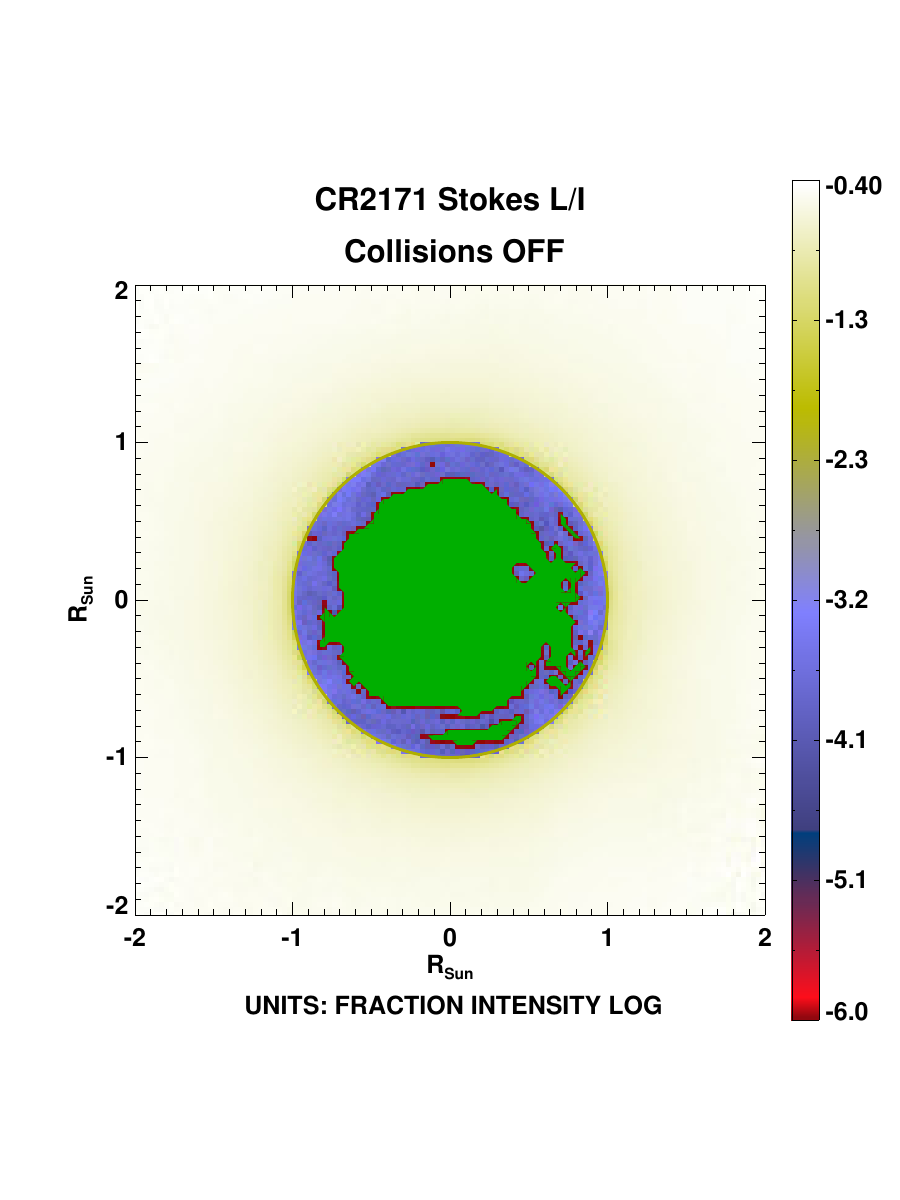}}
    \subfigure[]{\includegraphics[width=0.23\textwidth,trim={1.1cm 2.5cm 0cm 2.9cm},clip]{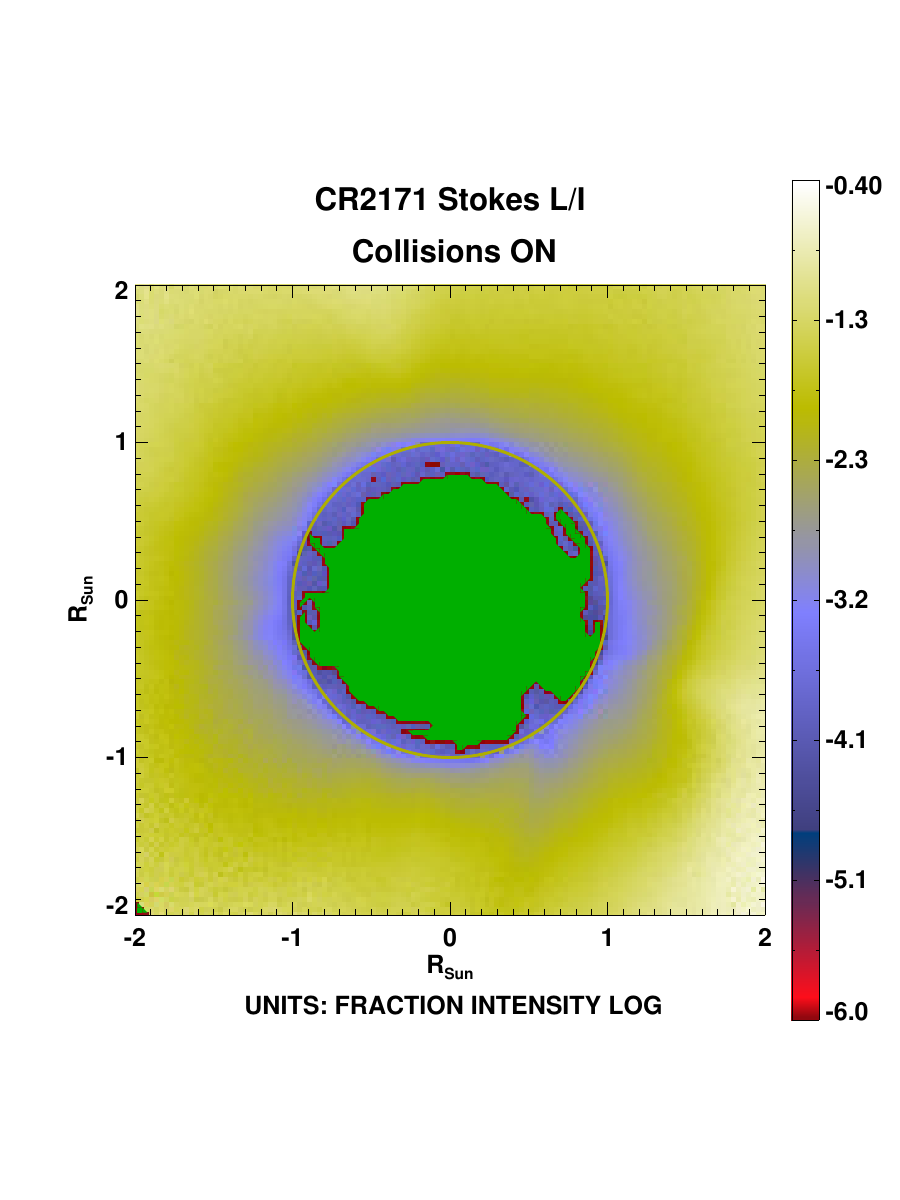}}
    \subfigure[]{\includegraphics[width=0.23\textwidth,trim={1.1cm 2.5cm 0cm 2.9cm},clip]{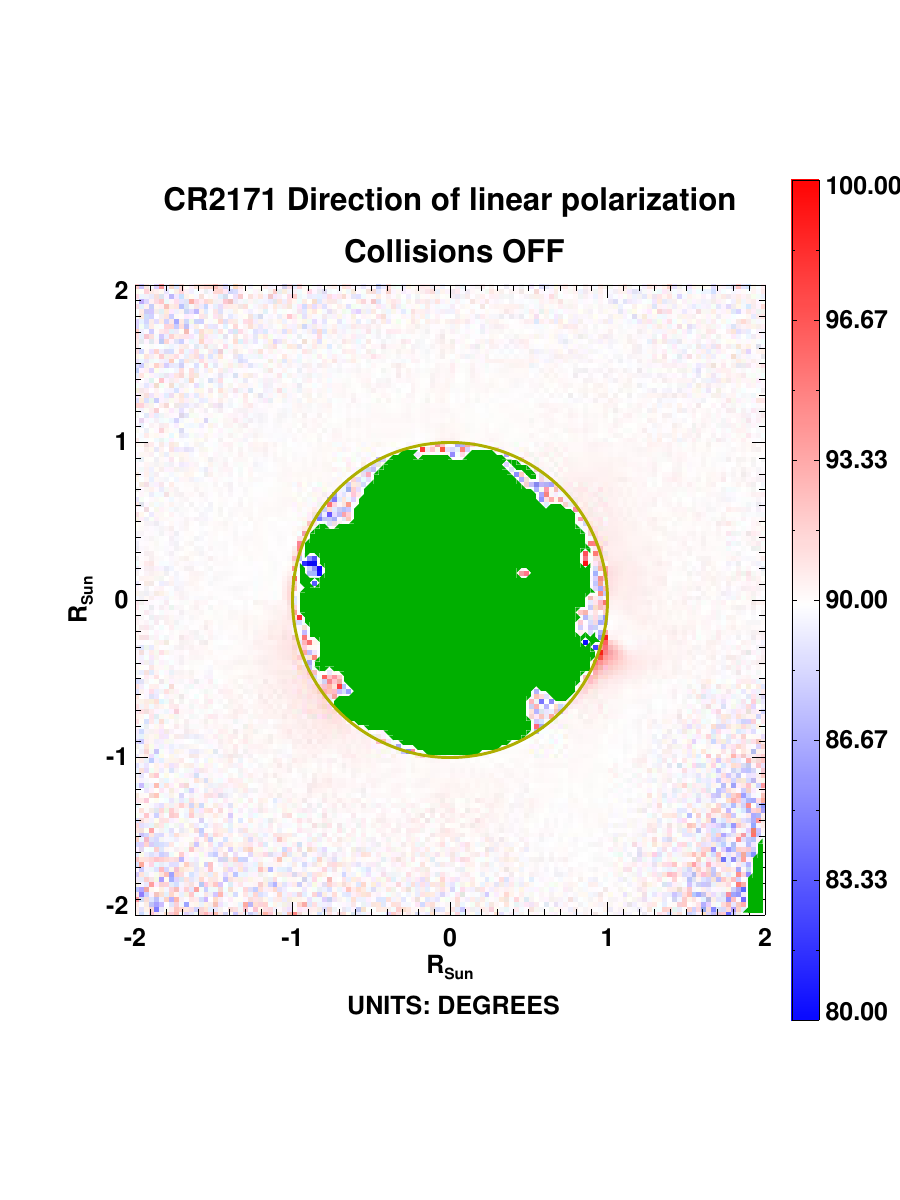}}
    \subfigure[]{\includegraphics[width=0.23\textwidth,trim={1.1cm 2.5cm 0cm 2.9cm},clip]{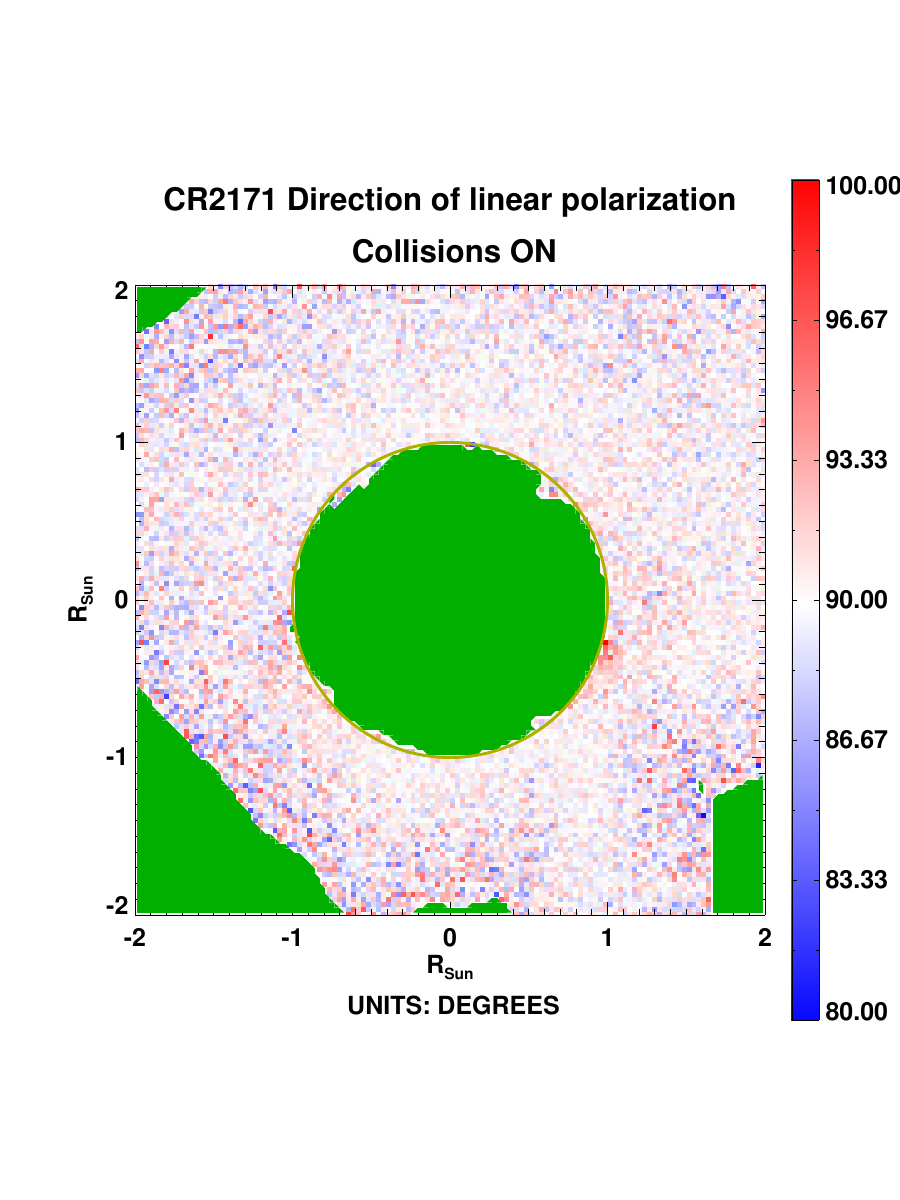}} 
    \subfigure[]{\includegraphics[width=0.23\textwidth,trim={1.1cm 2.5cm 0cm 2.9cm},clip]{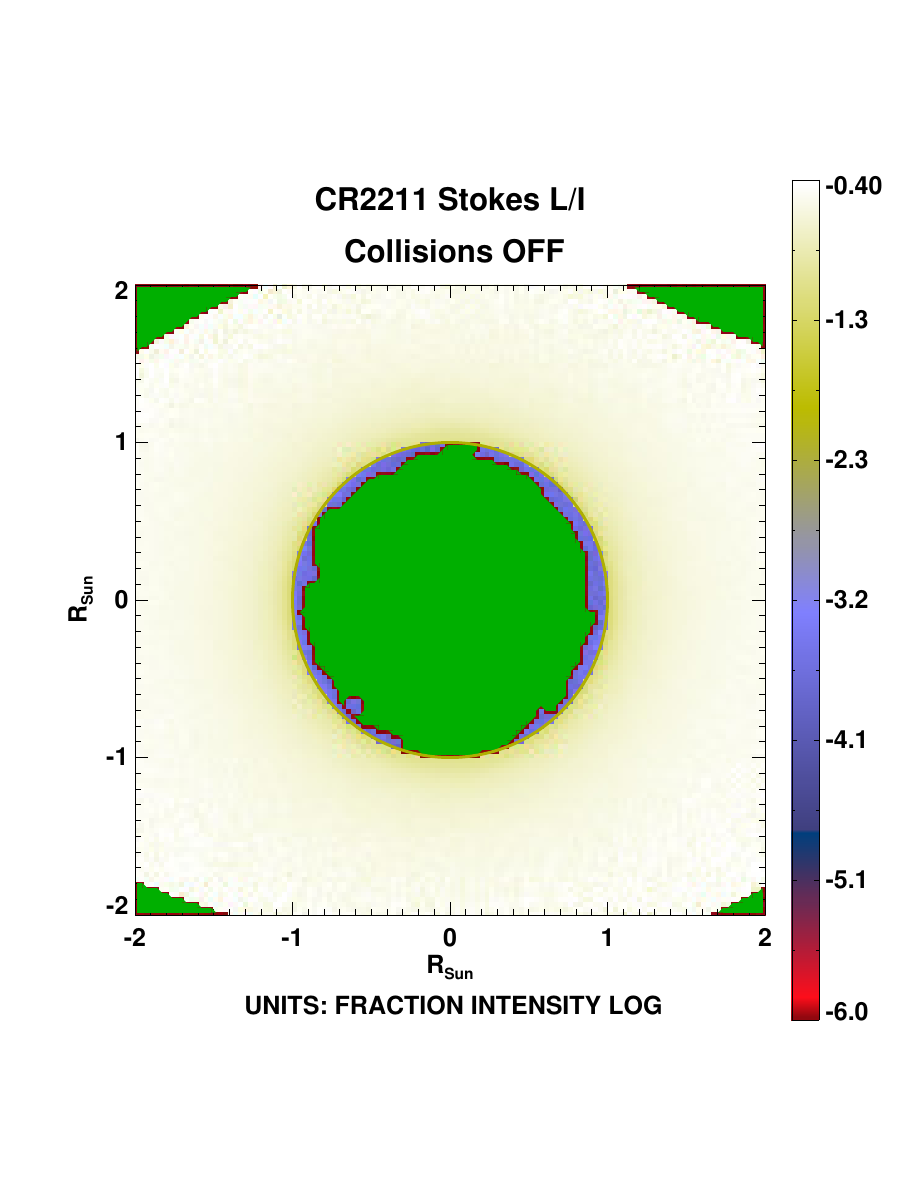}}
    \hspace{0.1cm}
    \subfigure[]{\includegraphics[width=0.23\textwidth,trim={1.1cm 2.5cm 0cm 2.9cm},clip]{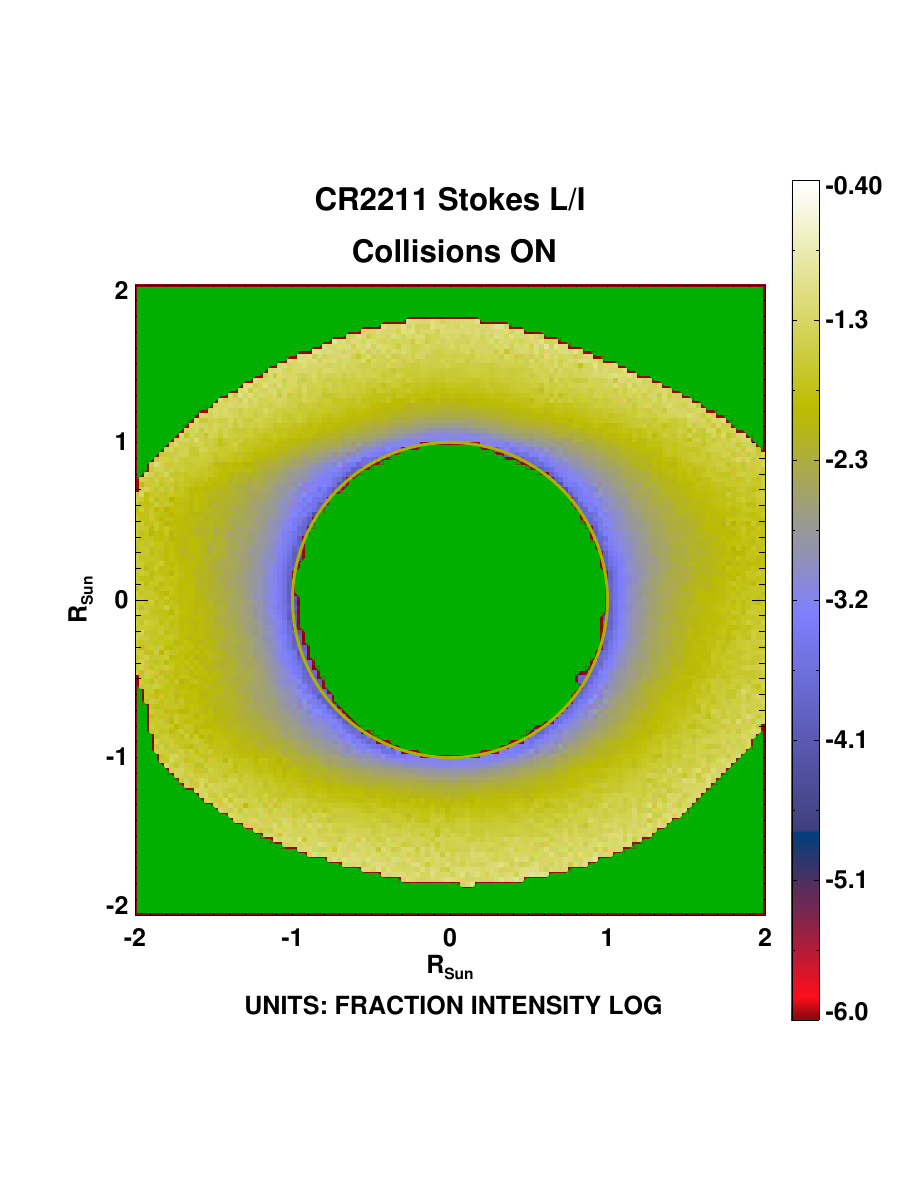}}
    \hspace{0.1cm}
    \subfigure[]{\includegraphics[width=0.23\textwidth,trim={1.1cm 2.5cm 0cm 2.9cm},clip]{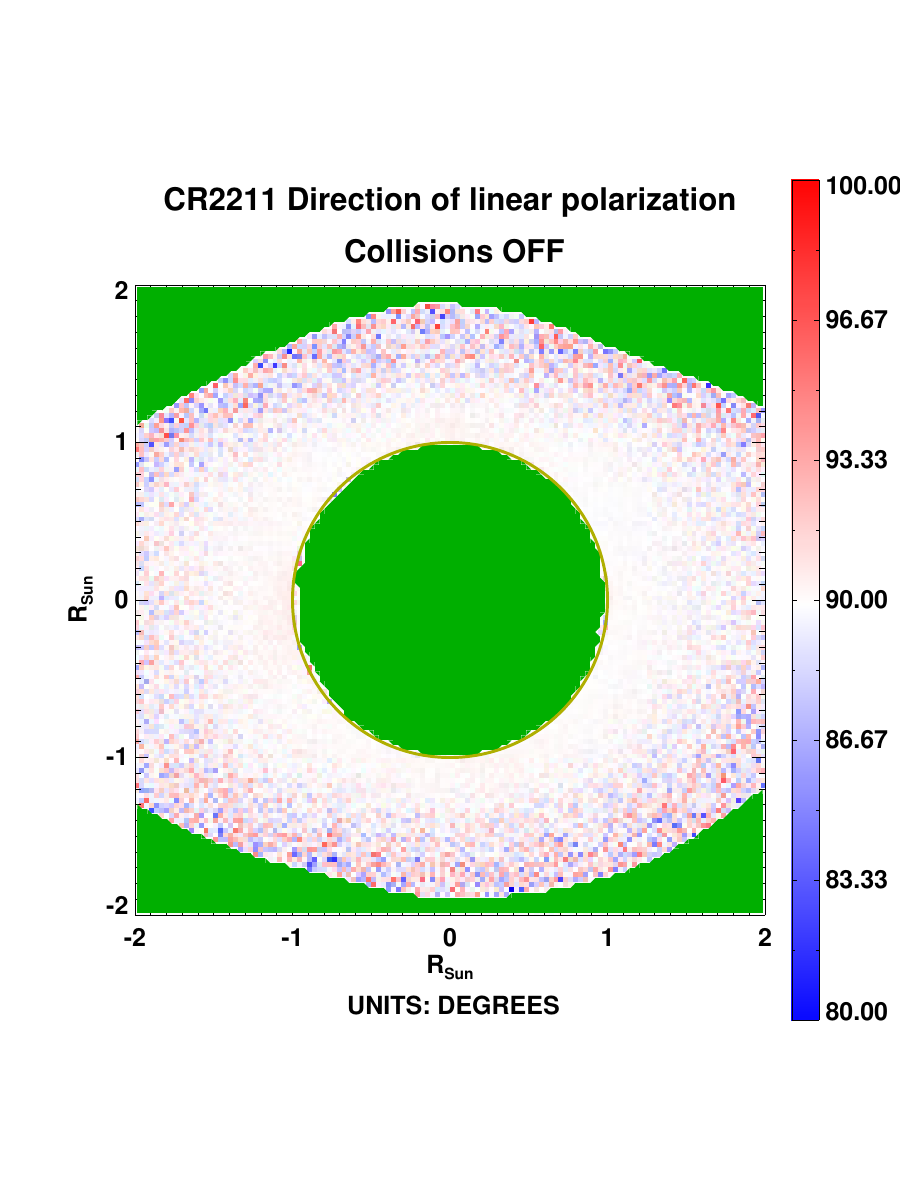}}
    \subfigure[]{\includegraphics[width=0.23\textwidth,trim={1.1cm 2.5cm 0cm 2.9cm},clip]{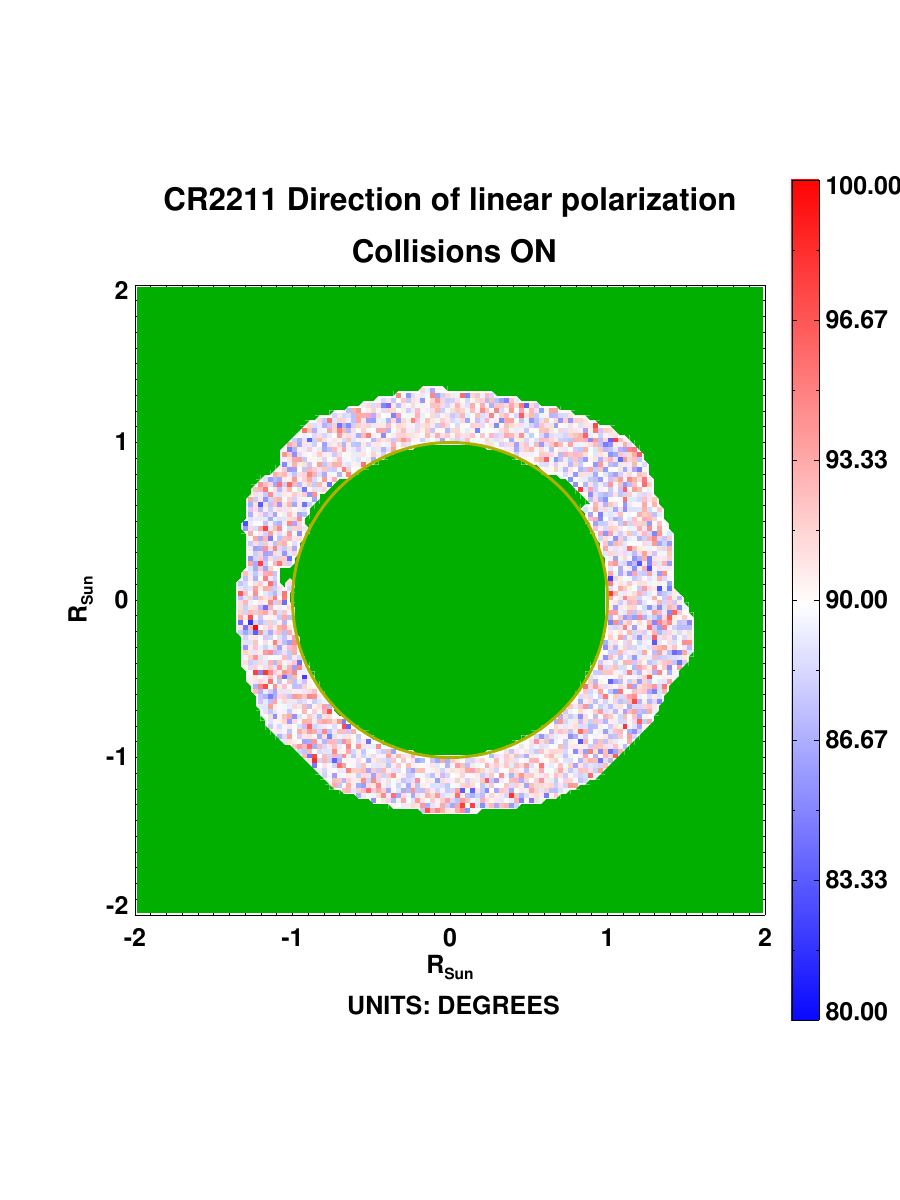}}
    \caption{\textit{Top row} and \textit{bottom row} corresponds to the solar maximum and minimum phase of SC24, respectively. \textit{First column} and \textit{third column} show the maps without collisions, while \textit{second} and \textit{fourth} columns show the maps with collisions, for a telescope of aperture 50~cm, 30 arcsecond pixel size, instrument throughput of 0.5\% and total integration time of 50 minutes. Green masked regions in azimuth maps indicate the areas where error on azimuth measurement is greater than $\pm 3^{\circ}$, while in $L/I$ maps they indicate areas where SNR on $L/I$ is less than 5$\sigma$.}
    \label{fig:stokesloi_az_snr}
\end{figure*}

By comparing between the right and left columns of Figure \ref{fig:stokesloi_az_chrovary}, we observe the drastic increase in the Stokes $L/I$ signal, especially off the limb, and also notice the improved SNR (reduced green masked regions on the disk) in both Stokes $L/I$ and azimuth maps. Due to such variability in the scattered radiation, it is not a good approximation to choose a single TR brightness value 
while forward modeling the Hanle regime signals for all phases of a particular solar cycle. However, in the absence of adequate observations at Ne {\sc viii} 770 \AA\ to be able to include routinely observed TR brightness values throughout SC24 and SC25, we have scaled this radiation
as a function of SC24 phases (shown in the third column of Figure \ref{fig:plots_chrovary}), for use in our analyses of the Hanle polarization signal at Ne {\sc viii} 770 \AA\, as described above and applied in the results shown in Figures \ref{fig:models_stki}-\ref{fig:bmag_stokesloi_az_offdisk} and \ref{fig:stokesloi_coll}-\ref{fig:stokesloi_az_snr}. 

\subsection{Variation with phases of the Solar Cycle} \label{sec:var24}
Figures \ref{fig:models_stki}(c), (f) and (i) depict simulated Stokes~$I$ maps for different phases across the entire SC24, including all of the contributary terms discussed in Section \ref{sec:bkg_variation} and shown in Figure \ref{fig:plots_chrovary}. As SC24 progresses, the total photon energy emitted (Stokes~$I$) at Ne {\sc viii} 770 \AA\ keeps increasing until SC24 reaches its maximum (Figure \ref{fig:models_stki}(f)), after which Stokes $I$ decreases again until the end of SC24 (Figure \ref{fig:models_stki}(i)). 

Similarly, the effect of the different phases of SC24 is observed on the degree of linear polarization. Figures \ref{fig:bmag_stokesloi_az_ondisk} and \ref{fig:bmag_stokesloi_az_offdisk} show the on-disk and the off-limb maps, respectively. Panels marked (b), (f) and (j) describe the fractional $L/I$ maps, while those marked (d), (h) and (l) describe the ratio between $L/I$ in the presence of magnetic fields and $(L/I)_{0}$ in the absence of magnetic field.

As we proceed towards the maximum phase of SC24 (i.e., columnwise from Figure \ref{fig:bmag_stokesloi_az_ondisk}(l) to (d) to (h)), it is noticed that the overall contrast in fractional $L/I$ on the disk increases with respect to the zero field case, owing to the increase in magnetic field strength and different field geometries along the LOS. In addition, there is an increased complexity of structures as opposed to the faint cylindrical structures visible in (l) and (d) (and in the equivalent magnetic structures in the first column of Figure \ref{fig:bmag_stokesloi_az_ondisk}). 

The increase of the polarization contrast $(L/I)$/$(L/I)_{0}$ on the disk is easily explained, since the amplitude of scattering polarization goes to zero as one approaches the condition of pure forward scattering at disk center (under our simplifying hypothesis that the radiation field remains axially symmetric around the local vertical through the scattering point). Thus, the increased linear polarization is simply due to the presence of plasma structures harboring inclined magnetic fields generating polarization on top of a nearly zero background of scattering polarization.

In contrast, as the magnetic field strength intensifies towards the maximum phase of SC24, a notable decrease of polarization amplitude (i.e., the ratio $(L/I)$/$(L/I)_{0}$ becoming less than 1) is observed off the limb. This phenomenon is particularly evident in right-angle scattering, as the Hanle effect causes a reduction of the linear polarization induced by the radiation anisotropy, which reaches its maximum for $90^\circ$ scattering.

The 3rd column of Fig.~\ref{fig:bmag_stokesloi_az_ondisk} shows the polarization azimuth across the disk, whereas Fig.~\ref{fig:bmag_stokesloi_az_offdisk} shows the same quantity within the off-limb corona. The white regions on these maps indicate areas where the linear polarization is tangent to the nearest limb, which is consistent with the case of radiation scattering from E1 transitions in the absence of magnetic fields (\citealt{Zhao2019ApJ...883...55Z}). Conversely, in the presence of magnetic fields, the blue (red) regions show where the polarization direction tilts clockwise (counterclockwise) relative to the reference direction. The CR2225 simulation during the solar minimum (panel (k) of Figures \ref{fig:bmag_stokesloi_az_ondisk} and \ref{fig:bmag_stokesloi_az_offdisk}) shows that the linear polarization of the scattered Ne {\sc viii} radiation remains along the tangent to the limb for most of the regions in the model, due to relatively weak magnetic fields. However, with the increase in magnitude of the magnetic field towards the rising phase (panel (c) of Figures \ref{fig:bmag_stokesloi_az_ondisk} and \ref{fig:bmag_stokesloi_az_offdisk}) and the maximum phase (panel (g) of Figures \ref{fig:bmag_stokesloi_az_ondisk} and \ref{fig:bmag_stokesloi_az_offdisk}) of the SC24, these red and blue regions become more pronounced, indicating greater deviations of the linear polarization vector from the reference axis.


\subsection{Effect of collisional excitation} \label{sec:coll}
The electron collisional rates have been calculated using the analytical formula given in \cite{Susino2018A&A...617A..21S}, i.e.
\begin{equation} \label{eq:coll_rates}
    C^{e}_{ij}(T) = 2.73 \times 10^{-15} \frac{f_{ij}g}{E_{ij}T^{1/2}} e^{\frac{-Eij}{k_{B}T}},
\end{equation}
where, $C^{e}_{ij}$ is in cgs units, $f_{ij}$ is the absorption oscillator strength for the considered transition, $g$ is the average electron-impact Gaunt factor, $E_{ij}$ is the transition energy, $k_{B}$ is the Boltzmann constant and $T$ is the electron temperature. The above formula can be further simplified in terms of the collision strength. Hence, we can write
\begin{equation} \label{eq:coll_rates_sim}
    C^{e}_{ij}(T) = \frac{8.63 \times 10^{-6}}{\omega_{i} T^{1/2}} e^{\frac{-Eij}{k_{B}T}} \Omega_{ij},
\end{equation}
where, $\omega_{i}$ is the statistical weight of level $i$, and $\Omega_{ij}$ is the collision strength for the specified ion, its transition $i \rightarrow j$ and temperature. The collision strengths for Ne {\sc viii} line at 770 \AA\ are obtained from the CHIANTI 10.0 version.

To visualize the effect of the collisional contribution to the emitted radiation, we consider the fractional linear polarization maps considering a model CR2171 from the solar maximum and a model CR2211 from the solar minimum of the SC24. Figure \ref{fig:stokesloi_coll} shows the $L/I$ signal for Ne {\sc viii} in the absence (left column) and presence (right column) of collisions. As can be seen from the right column of Figure \ref{fig:stokesloi_coll}, more areas of the Stokes $L/I$ map darken in the presence of collisions. In general, collisions do not affect the linear polarization $L$ due to their isotropic nature, but they increase the unpolarized intensity. Therefore, with the increase in collisions, the overall emitted intensity $I$ -- which is the sum of both polarized and unpolarized intensities -- increases, thereby reducing the $L/I$ signal.  

The left column maps of Figure \ref{fig:stokesloi_coll} illustrate a gradual change in the orientation of the LOS with respect to the local solar vertical, as we move from the disk center to the limb, which results in variation of Stokes $L/I$ with maximum being at the limb of the model while negligible $L/I$ around the disk center. In the right column maps of Figure \ref{fig:stokesloi_coll}, a relatively similar trend of increase in $L/I$ is observed from disk center to the limb except that collisions result in lower value of $L/I$ at each point as compared to the respective points in the left column maps. In addition, the coronal helmet streamer belt is clearly seen around the equator, with the  characteristic dipolar structure typical of the solar minimum, and which manifests itself also through a similarly shaped depression of the $L/I$ signal.

\subsection{Signal-to-noise ratio in Ne {\sc viii}}
\label{sec:snr}
In the previous sections, we have discussed the sensitivity of Ne {\sc viii} 770 \AA\ to the Hanle effect across the entire SC24 and the beginning of SC25. Here we discuss the change in the required SNR during the various phases of the solar cycle for the detection of those polarization signatures. The instrument parameters considered here have been discussed in Section \ref{sec:bkg_variation}. For coronal modeling on the disk of the Sun, apart from the resonance line scattering intensity, we have accounted for an additional background intensity as explained in Section \ref{sec:assum}.  
We have then computed the SNR on the quantities Stokes~$L/I$ and azimuth ($Az$). Considering photon-noise limited polarization measurements, we have masked the regions on the Stokes~$L/I$ maps (shown as green shaded areas in the first and second columns of Figure \ref{fig:stokesloi_az_snr}) which have SNR of $\leq 5\sigma$. Similarly, the green masked areas on the azimuth maps (third and fourth column of Figure \ref{fig:stokesloi_az_snr}) represent the points which have an error of greater than $\pm 3^{\circ}$ in determining the direction of the linear polarization vector.

Considering the first and second columns of Figure \ref{fig:stokesloi_az_snr}, we observe how the area covered by the green shaded regions, where the SNR on Stokes $L/I$ is below 5$\sigma$, increases in the presence of collisions. At the same time, there is a significant reduction in the off-limb polarization signal, adding significant complexity to the interpretation of the observed L/I in terms of magnetic sensitivity (as was also seen in Figure~\ref{fig:stokesloi_coll}). In addition, we observe that the noise on $L/I$ improves during the solar maximum phase (top row of Figure \ref{fig:stokesloi_az_snr}) as compared to the solar minimum phase (bottom row of Figure \ref{fig:stokesloi_az_snr}) due to the overall increase 
of the TR brightness (as was also seen in Figure~\ref{fig:stokesloi_az_chrovary}). This implies that underestimating the pumping radiation at the spectral line considered can easily lead to under-calculating the SNR on the polarization signal.

Azimuth is independent of the collisional contribution to total intensity and this makes its interpretation in terms of magnetic sensitivity much more straightforward. However, collisions still affect the measurement error in determining the azimuth. Comparing between the third and fourth column maps in the top row of Figure \ref{fig:stokesloi_az_snr}, we observe a trend of increase in the green shaded areas when collisions are turned on. Moreover, we note that there is lesser noise in the azimuth measurement during the maximum phase of the solar cycle (as seen in the third and fourth column maps in top row of Figure \ref{fig:stokesloi_az_snr}), while the error in azimuth measurement both on the disk and off the limb increases during the solar minimum (third and fourth column maps in bottom row of Figure \ref{fig:stokesloi_az_snr}) phase as the TR brightness reduces.

\section{Discussion and Conclusion} \label{sec:conc}
In this study, we have explored the possibility of using the EUV emission line Ne {\sc viii} 770 \AA ~as a potential diagnostic for the coronal magnetic fields -- both for the off-limb corona and on the disk. We have utilized the 3D thermodynamic models developed by the Predictive Science Inc.\ to simulate linear polarization maps, and study the change in the Hanle-regime polarization signals during the different phases of the solar cycle. 

Further, we have performed a critical study of the effect of variation in the TR brightness on the scattered radiation at Ne {\sc viii} 770 \AA. We have also examined the effect of collisions on the linear polarization and azimuth signals, and calculated the required SNR on these quantities. Although collisions have a significant impact on the Hanle polarization signals at all solar phases, we showed how the higher 
TR brightness characterizing the solar maximum helps achieve a larger SNR values in both Stokes $L/I$ and azimuth. In particular, the azimuth values are not affected by the presence of collisional excitation, although the noise on their measurement is, because of the impact of its contribution to the intensity on the detectability of polarized signals. 

The present study has also shown that Ne {\sc viii} 770 \AA\ has the potential to be used as an overall coronal diagnostic without the need of a coronagraphic instrument. This EUV line is a useful complement to other coronal field diagnostics in the FUV, such as O {\sc vi} 1032 \AA\ and H {\sc i} 1216 \AA, which share comparable critical Hanle fields of about 35 gauss and 53 gauss, respectively. 

EUV coronal diagnostics can open a new era of non-coronagraphic measurements of the coronal magnetic field, and will complement other off-limb spectropolarimetric measurements in the visible and infrared wavelengths, such as those obtained with the Upgraded Coronal Multi-Channel Polarimeter (UCoMP) and the Daniel K. Inouye Solar Telescope (DKIST).

\section{Acknowledgement}
The work is supported by the National Center for Atmospheric Research, a major facility sponsored by the National Science Foundation under Cooperative Agreement No. 1852977. CHIANTI is a collaborative project involving George Mason University, the University of Michigan (USA), University of Cambridge (UK) and NASA Goddard Space Flight Center (USA). This research has made use of the High Performance Computing (HPC) resources
(\url{https://www.iiap.res.in/?q=facilities/computing/nova}) made available by the Computer Center of the Indian Institute of Astrophysics, Bangalore. We acknowledge Cooper Downs for insights on the thermodynamic models developed by Predictive Science Inc. We acknowledge Terry Kucera for implementation of codes for calculating ion density.

\begin{appendix}
\newcommand{\snr}{\mbox{SNR}}

\section{Noise estimation for linear-polarization degree and
azimuth measurements} \label{sec:appendix_noise}

We consider a telescope with throughput $\tau$, equipped with a polarimeter that employs a modulation scheme consisting of $n$ independent signal measurements (modulation states), and characterized by  
Stokes efficiencies $\epsilon_i$, where $i=0,1,2,3$ for $S_i=I,Q,U,V$. We follow \cite{delToro2000ApOpt..39.1637D} for the definition of the Stokes efficiencies of a polarization modulation scheme, in terms of which the polarimetric errors on the Stokes parameters at the entrance of the polarimeter are given by 
\begin{equation} \label{eq:noise}
\sigma_i=\frac{1}{\sqrt{n}}\frac{\sigma_s}{\epsilon_i}\;,\qquad
i=0,1,2,3\;,
\end{equation}
where $\sigma_s$ is the noise associated with 
the measurement of the signal by the detector at a given modulation state $s=1,\ldots,n$.
It is important to remark that the definition of Stokes efficiencies implied by Eq.~(\ref{eq:noise}) assumes that the polarimeter has unit transmissivity, so the actual throughput of the polarimeter, including any optical losses and the quantum efficiency of the detector, must be taken into account by the throughput $\tau$ of the instrument (see Sect.~\ref{sec:snr}). In addition, Eq.~(\ref{eq:noise}) assumes that $\sigma_s$ is constant over the modulation cycle, which generally is a good approximation for a temporal modulation scheme (e.g., a rotating retarder), under the condition of weakly polarized signals commonly encountered in observations of the solar corona.

We now consider the total linear-polarization signal
\begin{displaymath}
L=\sqrt{Q^2+U^2}\;,
\end{displaymath}
and the corresponding polarimetric error $\sigma_L$, given by 
\begin{displaymath}
\sigma_L^2=\biggl(\frac{Q}{L}\biggr)^2\,\sigma_Q^2
+\biggl(\frac{U}{L}\biggr)^2\,\sigma_U^2\;.
\end{displaymath}
For a balanced modulation scheme,
$\epsilon_L\equiv\epsilon_{Q,U}$,\footnote{In the ideal case of a balanced and optimally efficient
linear-polarization scheme,
$\epsilon_I=1$ and $\epsilon_L=1/\sqrt{2}$.} 
thus Eq.~(\ref{eq:noise}) gives
\begin{equation} \label{eq:sigL}
\sigma_L\equiv\sigma_{Q,U}=\frac{1}{\sqrt{n}}\,\frac{\sigma_s}{\epsilon_L}\;.
\end{equation}
It is often preferable to work instead with the linear-polarization
\emph{degree} $P=L/I$, for which error propagation gives
\begin{displaymath}
\sigma_P^2
=\biggl(\frac{L}{I^2}\biggr)^2\sigma_I^2+\frac{\sigma_L^2}{I^2}
=P^2\,\frac{\sigma_I^2}{I^2}+\frac{\sigma_L^2}{I^2}\;.
\end{displaymath}
Using again Eq.~(\ref{eq:noise}), we can rewrite Eq.~(\ref{eq:sigL}) as
\begin{equation} \label{eq:sigma_L}
\sigma_L=\frac{1}{\sqrt{n}}\,\frac{\sigma_s}{\epsilon_I}\,
	\frac{\epsilon_I}{\epsilon_L}
=\sigma_I\,\frac{\epsilon_I}{\epsilon_L}\;,
\end{equation}
which we substitute in the expression of $\sigma_P^2$ above to get
\begin{equation} \label{eq:sigma_P}
\sigma_P
=\frac{\sigma_I}{I}\sqrt{P^2+\frac{\epsilon_I^2}{\epsilon_L^2}}\;.
\end{equation}
Hence, the signal-to-noise ratio $\snr=I/\sigma_I$ necessary to achieve  
a sensitivity $\sigma_P$ on the measurement of $P$ is
\begin{equation} \label{eq:SNR_P}
(\snr)_P=\frac{1}{\sigma_P}\sqrt{P^2+\frac{\epsilon_I^2}{\epsilon_L^2}}\;.
\end{equation}

Similarly, for the linear polarization \emph{azimuth} $\beta$, we adopt the definition
\begin{equation}
\beta=\frac{1}{2}\arctan\frac{U}{Q}+\beta_0\;,
\end{equation}
where $\beta_0$ is an appropriate offset necessary to identify the direction of polarization for different sign combinations of $Q$ and $U$ \cite[cf.][Eqs.~(1.8), and the definition of the azimuth we gave in Section \ref{sec:results}]{Landi2004ASSL..307.....L}. We thus have
\begin{eqnarray*}
\sigma_\beta^2
&=&\frac{1}{4}\biggl(\frac{U}{L^2}\biggr)^2\sigma_Q^2+
\frac{1}{4}\biggl(\frac{Q}{L^2}\biggr)^2\sigma_U^2 \\
&=&\frac{1}{4}\,\frac{Q^2+U^2}{L^4}\,\sigma_L^2=\frac{1}{4L^2}\,\sigma_L^2 \\
&=&\frac{1}{4P^2}\,\frac{\sigma_L^2}{I^2}\;,
\end{eqnarray*}
and using again Eq.~(\ref{eq:sigma_L}),
\begin{equation} \label{eq:sigma_beta}
\sigma_\beta=\frac{1}{2}\,\frac{\epsilon_I}{\epsilon_L}\,
\frac{1}{P}\,\frac{\sigma_I}{I}\;.
\end{equation}
Hence, the signal-to-noise ratio necessary to achieve  
a sensitivity $\sigma_\beta$ on the measurement of $\beta$ is
\begin{equation} \label{eq:SNR_beta}
(\snr)_\beta=
\frac{1}{2}\,\frac{\epsilon_I}{\epsilon_L}\,
\frac{1}{P}\,\frac{1}{\sigma_\beta}\;.
\end{equation}

\subsection{Presence of a background signal.}
The previous analysis assumed that the observed intensity signal 
$I$ is produced exclusively by physical processes responsible for the line emission, possibly including collisional excitation
and de-excitation. 
In the presence of ambient and instrumental stray light, or a physical 
continuum that does not pertain to the resonance scattering
process (e.g., a Planckian continuum from the disk, or Thomson scattered
radiation off the limb), the target 
signal $I$ is only a contribution to the total observed signal, i.e.,
\begin{equation} \label{eq:Itot}
I_{\rm obs}=I+I_{\rm bg}\;,
\end{equation}
where we indicated with $I_{\rm bg}$ the sum of all contributions to the
observed intensity signal other than resonance line scattering. We call
these contributions ``background'' because it is possible in theory
(e.g., through the spectral analysis of the incoming radiation) to separate
those contributions from the observed signal in order to isolate the target line
signal. Thus, the 
 intensity signal $I$ needed to compute the observables of the previous 
analysis and their noise (we note that $\beta$ is independent of $I$, but $\sigma_\beta$ is not)
must be obtained via background subtraction, i.e.,
\begin{equation}
I=I_{\rm obs}-I_{\rm bg}\;.
\end{equation}
If the intensity signals $I_{\rm obs}$ and $I_{\rm bg}$ are expressed in
photon counts, then the variance on $I$ (assuming the independent 
measurement of the background) is given by
\begin{eqnarray} \label{eq:sigma+bkgd}
\sigma^2_I
&=&\sigma^2_{\rm obs}+\sigma^2_{\rm bg} \nonumber \\
&\approx &I_{\rm obs}+I_{\rm bg}=I+2I_{\rm bg}\;,
\end{eqnarray}
where for the approximation in the second line we assumed that all 
intensity measurements are photon-noise limited, so the photon count 
$N$ satisfies the Poisson-statistics condition $\sigma_N\approx\sqrt N$,
and in the last equivalence we used again Eq.~(\ref{eq:Itot}).
Equation~(\ref{eq:sigma+bkgd}) can finally be rewritten as
\begin{equation} \label{eq:usethis}
\sigma^2_I
=\sigma^2_{\rm obs}\biggl(1+\frac{\sigma^2_{\rm bg}}{\sigma^2_{\rm obs}}\biggr)
\approx \sigma^2_{\rm obs}\left(1+\frac{I_{\rm bg}}{I_{\rm obs}}\right)\;.
\end{equation}

It is important to remark that the two 
\emph{measured} signals in Eq.~(\ref{eq:Itot}) are $I_{\rm obs}$ 
and $I_{\rm bg}$, and for these it is typically desirable (or required) 
to achieve photon-noise limited acquisition, so that the 
Poisson-statistics relation for the signal's noise implied by Eq.~(\ref{eq:sigma+bkgd}) 
is valid. On the other hand, the noise on the \emph{derived} signal $I$ does 
not satisfy the same relation, as shown by
Eq.~(\ref{eq:sigma+bkgd}), even if the coronal photons that contribute to the $I$ signal obviously still obey the Poisson statistics.

The expression (\ref{eq:usethis}) for $\sigma_I$ must be used in
Eqs.~(\ref{eq:sigma_P}) and (\ref{eq:sigma_beta}) in order 
to calculate the proper target SNR in the presence of a background, 
for a given noise target on $P$ and $\beta$, respectively. 
After some tedious but straightforward algebra, we find
\begin{eqnarray} \label{eq:manipulation}
\frac{\sigma_I}{I}
&=&\left(\frac{\sigma_{\rm obs}}{I_{\rm obs}}\right)\frac{I_{\rm obs}}{I}
	\sqrt{1+\frac{I_{\rm bg}}{I_{\rm obs}}} \nonumber \\
&\equiv&(\snr)^{-1}\,
	\sqrt{\frac{I_{\rm obs}^2}{I^2}
	\left(1+\frac{I_{\rm bg}}{I_{\rm obs}}\right)} \nonumber \\
&=&(\snr)^{-1}\,\sqrt{
	\left(1+\frac{I_{\rm bg}}{I}\right)
	\left(1+\frac{2I_{\rm bg}}{I}\right) }\;,
\end{eqnarray}
showing that, in the presence of a background, the required SNR of
Eqs.~(\ref{eq:SNR_P}) and (\ref{eq:SNR_beta}) are augmented
by the factor
\begin{equation} \label{eq:corrector}
k_{\rm bg}=\sqrt{
	\left(1+\frac{I_{\rm bg}}{I}\right)
	\left(1+\frac{2I_{\rm bg}}{I}\right) }\;.
\end{equation}
We note that this correction factor is dimensionless and does not depend on the
specific intensity units adopted. In particular, it can also be
calculated using energy units instead of photon counts.

In conclusion, \emph{in the presence of a background signal}, the
required SNR for a given sensitivity target on the quantitites $P$ and
$\beta$ becomes, respectively,
\begin{equation} \label{eq:SNR_P+bkgd}
(\snr)_P=\frac{k_{\rm bg}}{\sigma_P}\sqrt{P^2+\frac{\epsilon_I^2}{\epsilon_L^2}}\;,
\end{equation}
\begin{equation} \label{eq:SNR_beta+bkgd}
(\snr)_\beta=
\frac{1}{2}\,\frac{\epsilon_I}{\epsilon_L}\,
\frac{1}{P}\,\frac{k_{\rm bg}}{\sigma_\beta}\;.
\end{equation}

The effect of a background signal on the noise of a polarimetric observable has also been considered by \cite{Penn2004SoPh..222...61P}. In order to clarify a seeming discrepancy between our expression (\ref{eq:corrector}) for $k_{\rm bg}$ and the result derived by those authors (cf.\ the expression of $f_{\rm bg}$ given on page 64 of their paper), we consider again Eq.~(\ref{eq:sigma+bkgd}) 
to observe that
\begin{equation} \label{eq:clarify}
\sigma_I=\sqrt{I+2I_{\rm bg}}
=(\sigma_I)_{\rm no.bg}\,\sqrt{1+\frac{2I_{\rm bg}}{I}}\;,
\end{equation}
where we defined $(\sigma_I)_{\rm no.bg}\equiv\sqrt I$, with $I$ expressed in photon counts. Upon 
substitution of this expression into 
Eqs.~(\ref{eq:sigma_P}) and (\ref{eq:sigma_beta}), the errors 
$\sigma_P$ and $\sigma_\beta$, \emph{for a prescribed set of 
observing conditions}, are indeed augmented by $f_{\rm bg}^{-1}=(1+2I_{\rm bg}/I)^{1/2}$, using the notation of \cite{Penn2004SoPh..222...61P}, but this cannot constrain in any useful way the required $\snr$ of the observation, which is the purpose of Eqs.~(\ref{eq:SNR_P}) and (\ref{eq:SNR_beta}). The different form of our factor $k_{\rm bg}$ follows precisely from this purpose of determining the $\snr$ of the \emph{observed} signal $I_{\rm obs}$ necessary to achieve a target error on  the \emph{derived} signal $I$, as implied by Eqs.~(\ref{eq:SNR_P+bkgd}) and (\ref{eq:SNR_beta+bkgd}). In doing so, $k_{\rm bg}$ takes the 
form (\ref{eq:corrector}) as a result of the algebraic manipulation
shown in Eq.~(\ref{eq:manipulation}).

\section{Supplementary materials}
\label{sec:appendix_figs}
\begin{figure}[htbp]
\centering
    \includegraphics[width=0.126\textwidth,trim={1.1cm 2.5cm 0cm 2.9cm},clip]{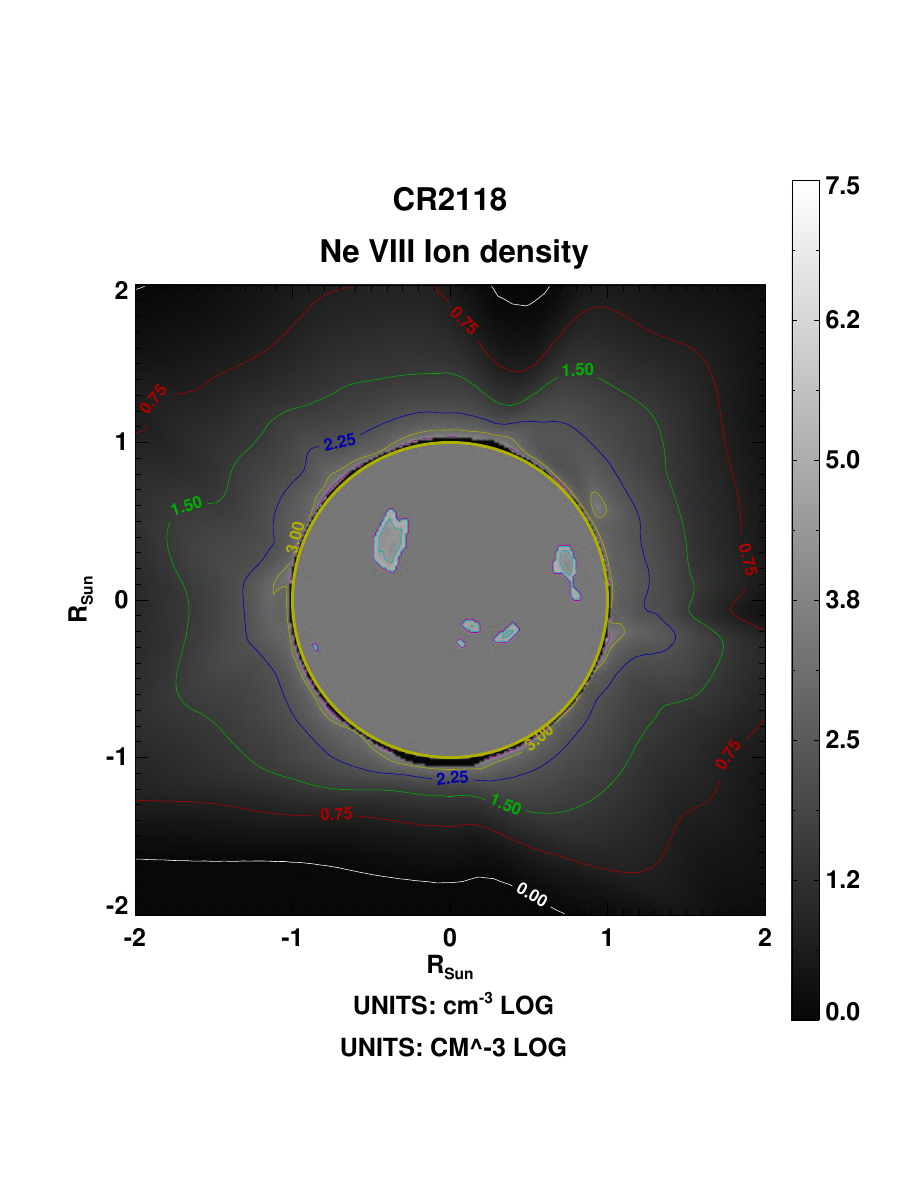}
    \includegraphics[width=0.126\textwidth,trim={1.1cm 2.5cm 0cm 2.9cm},clip]{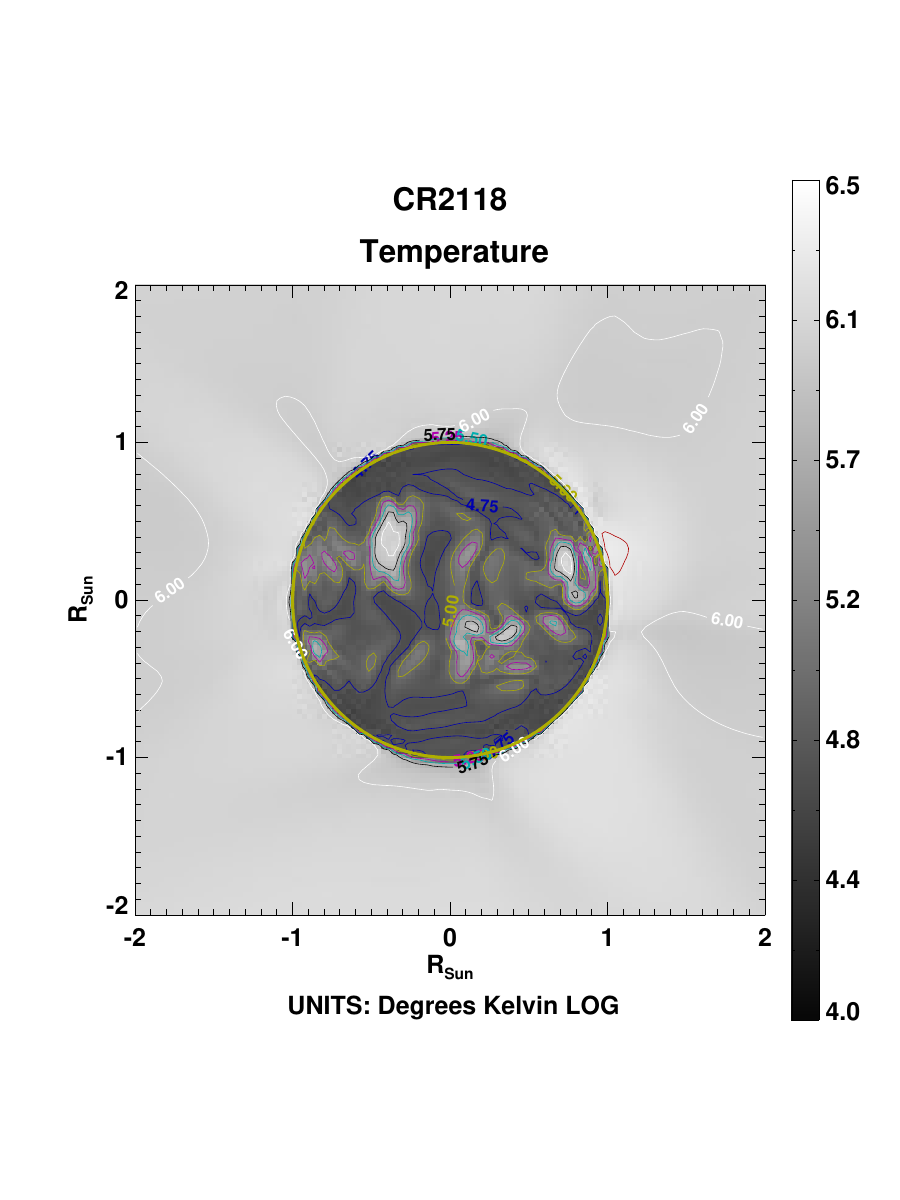}
    \includegraphics[width=0.126\textwidth,trim={1.1cm 2.5cm 0cm 2.9cm},clip]{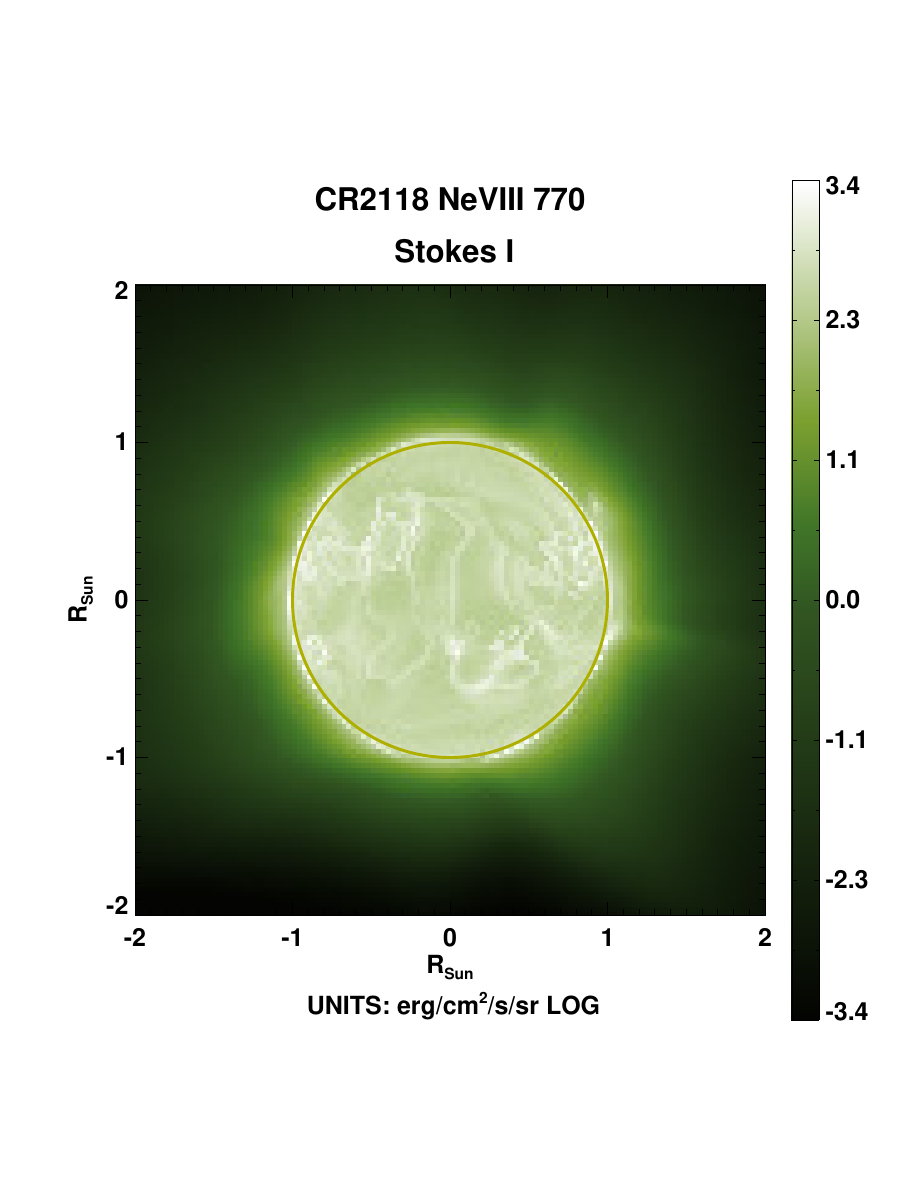}
    \includegraphics[width=0.126\textwidth,trim={1.1cm 2.5cm 0cm 2.9cm},clip]{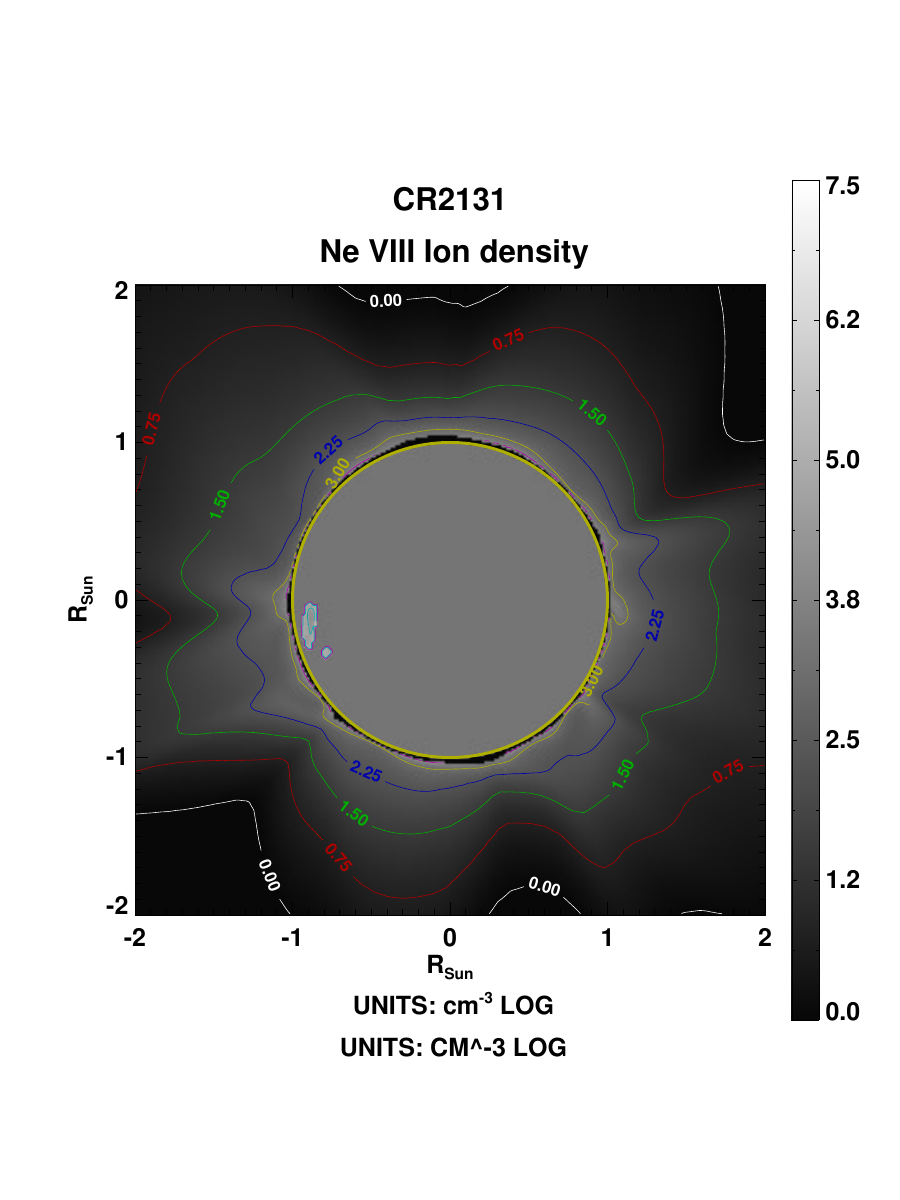}
    \includegraphics[width=0.126\textwidth,trim={1.1cm 2.5cm 0cm 2.9cm},clip]{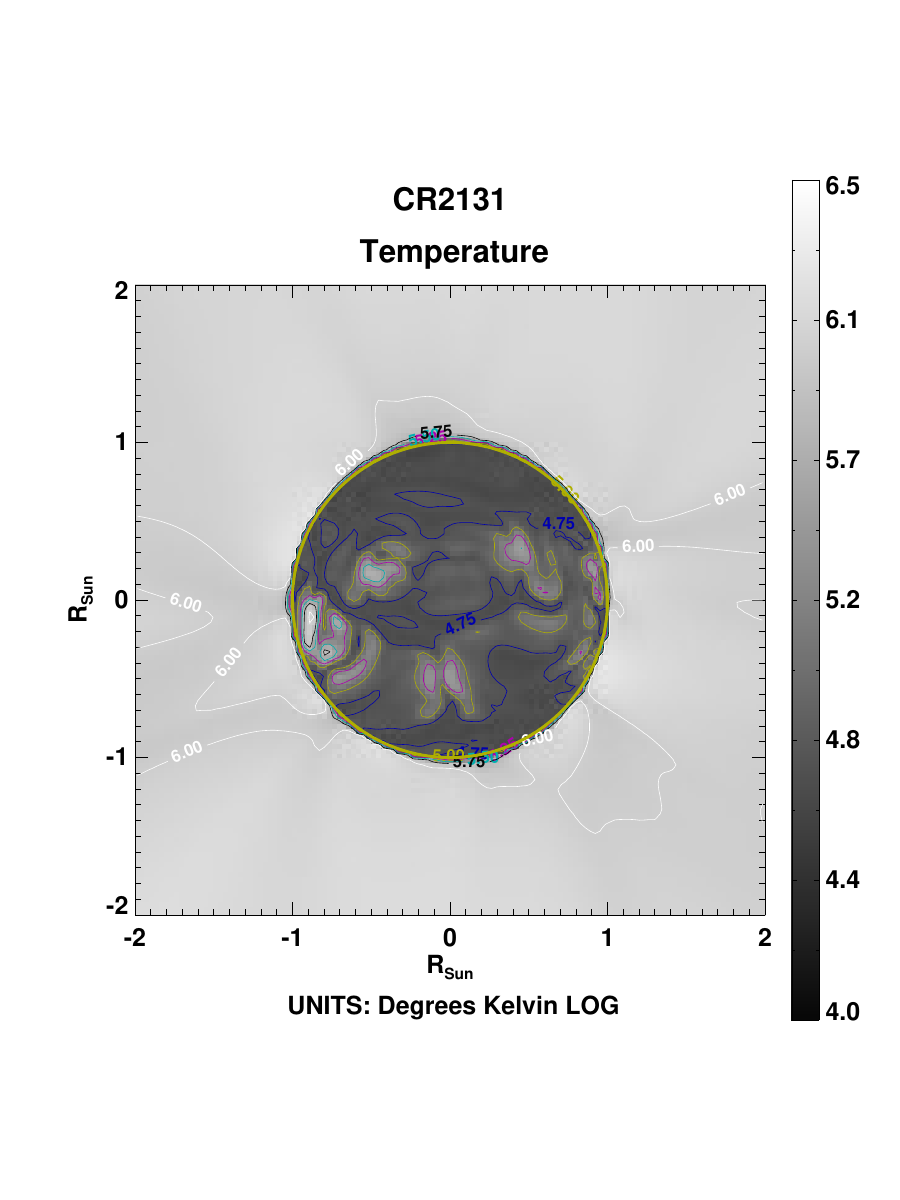}
    \includegraphics[width=0.125\textwidth,trim={1.1cm 2.5cm 0cm 2.9cm},clip]{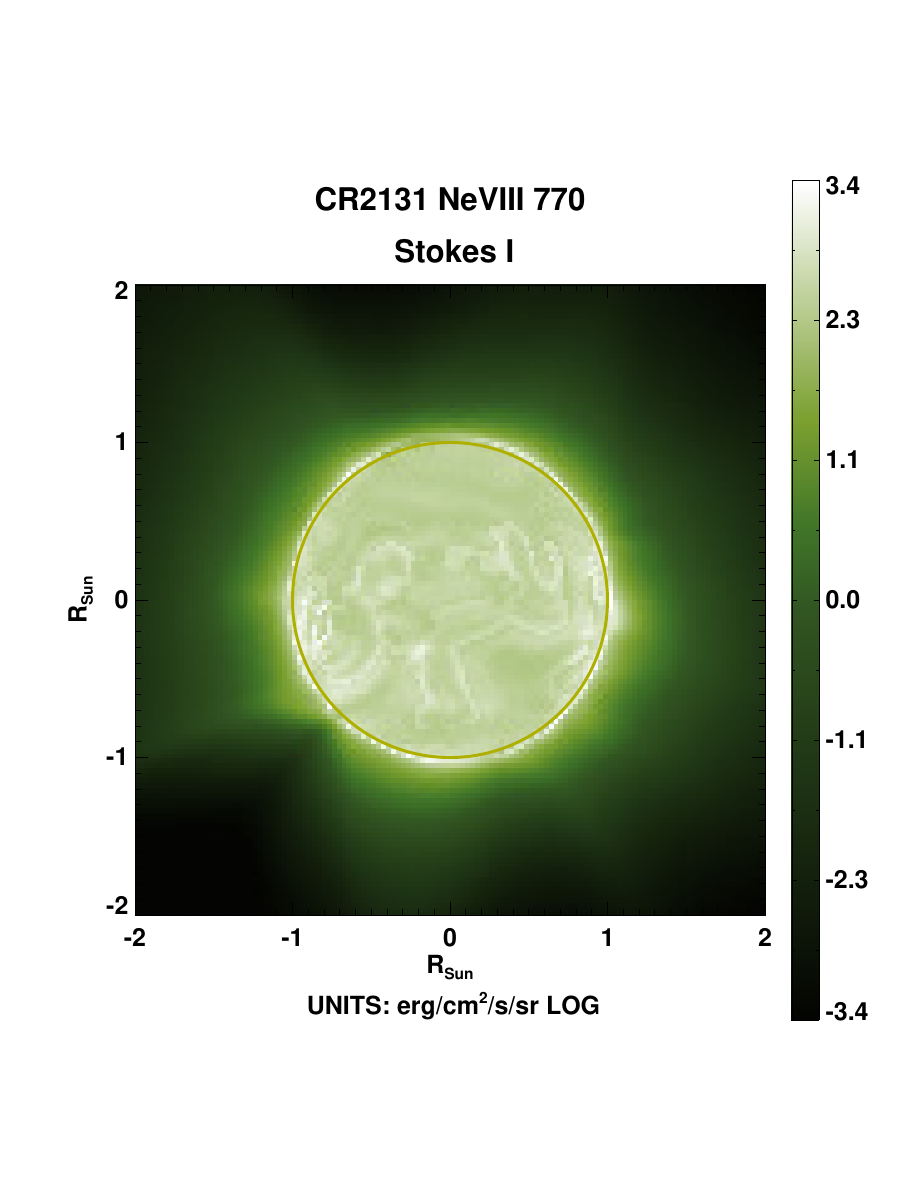}
    \includegraphics[width=0.125\textwidth,trim={1.1cm 2.5cm 0cm 2.9cm},clip]{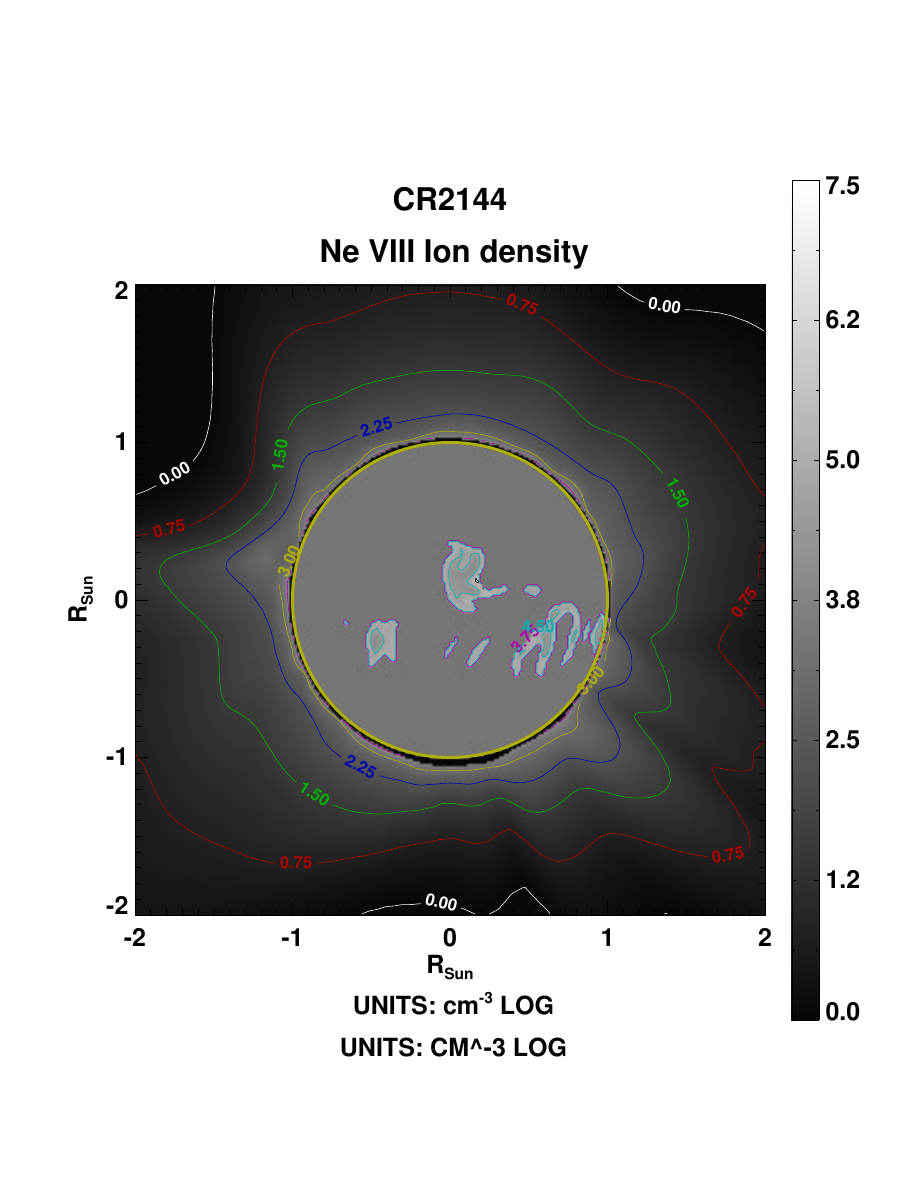}
    \includegraphics[width=0.125\textwidth,trim={1.1cm 2.5cm 0cm 2.9cm},clip]{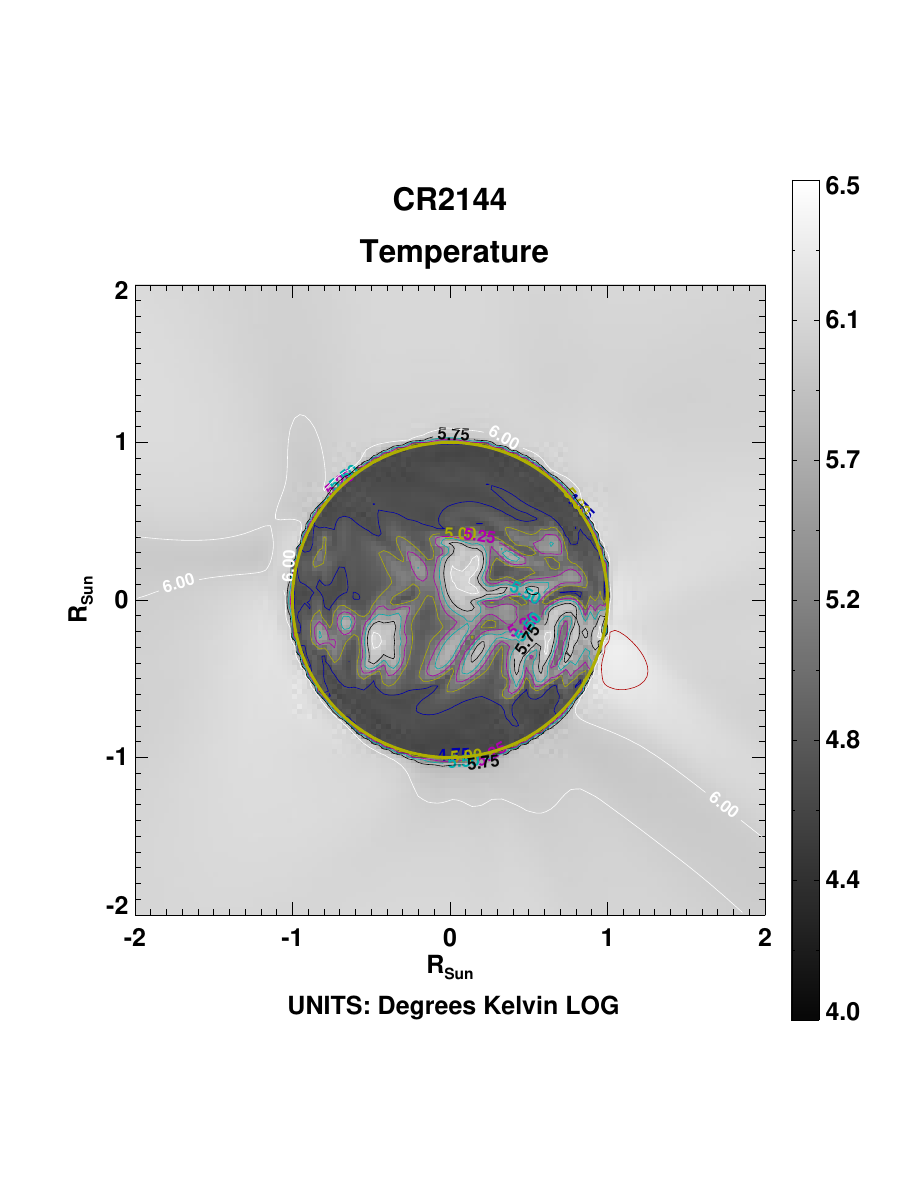}
    \includegraphics[width=0.125\textwidth,trim={1.1cm 2.5cm 0cm 2.9cm},clip]{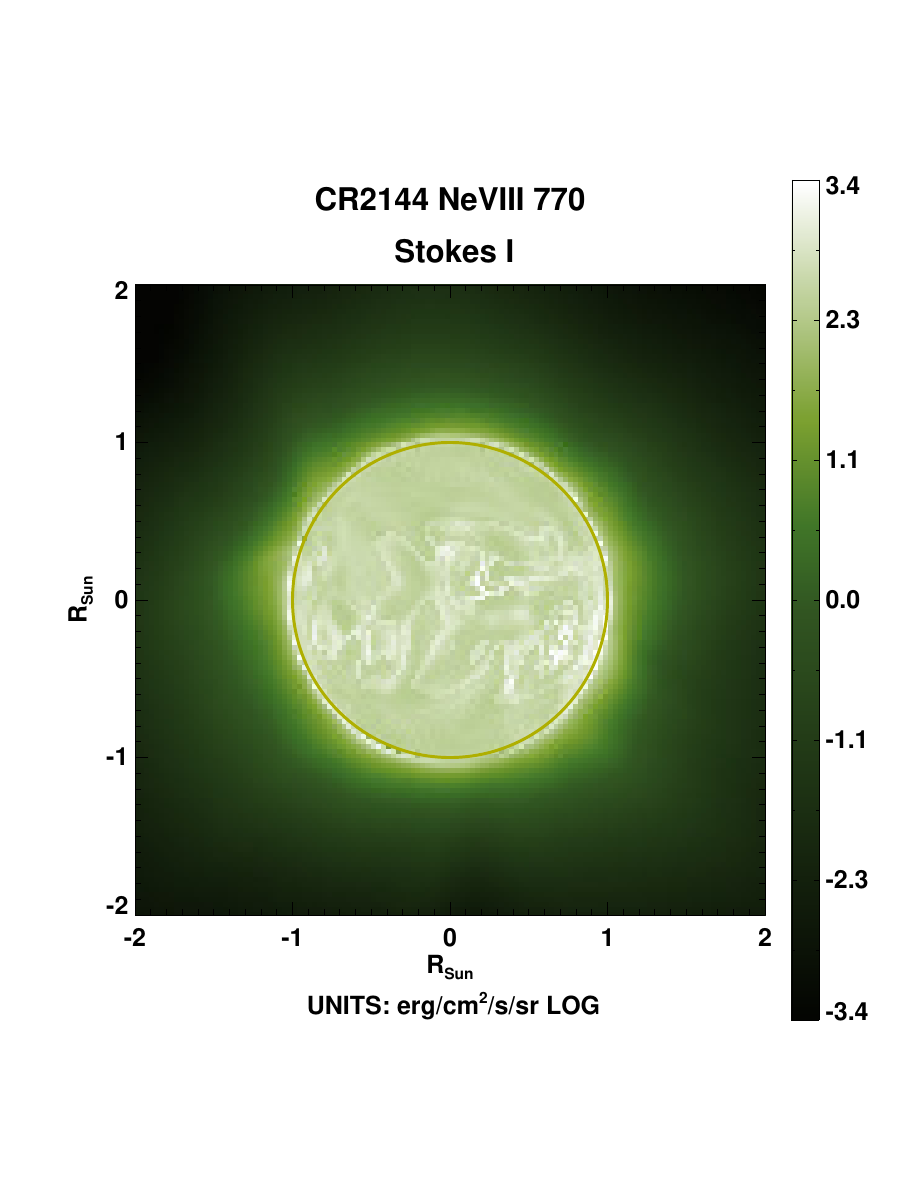}
    \includegraphics[width=0.125\textwidth,trim={1.1cm 2.5cm 0cm 2.9cm},clip]{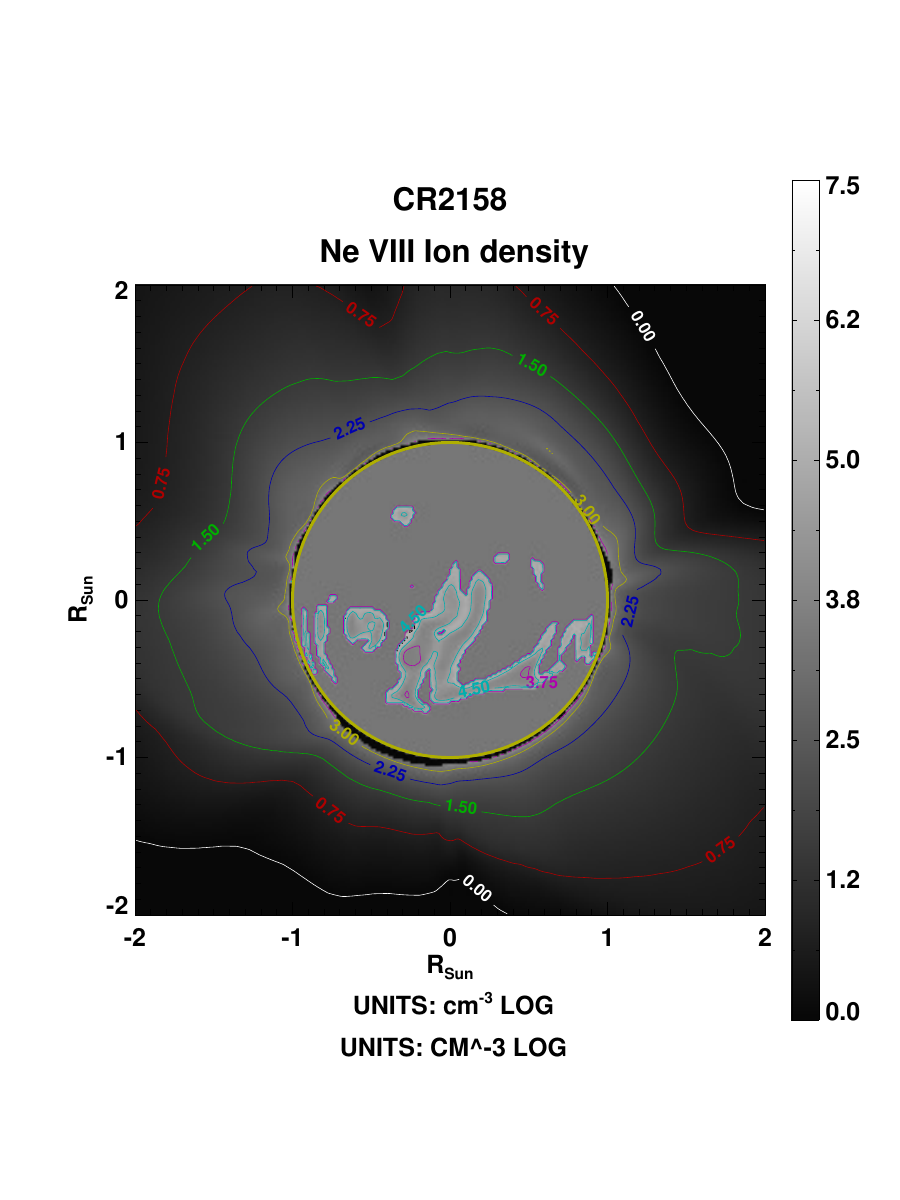}
    \includegraphics[width=0.125\textwidth,trim={1.1cm 2.5cm 0cm 2.9cm},clip]{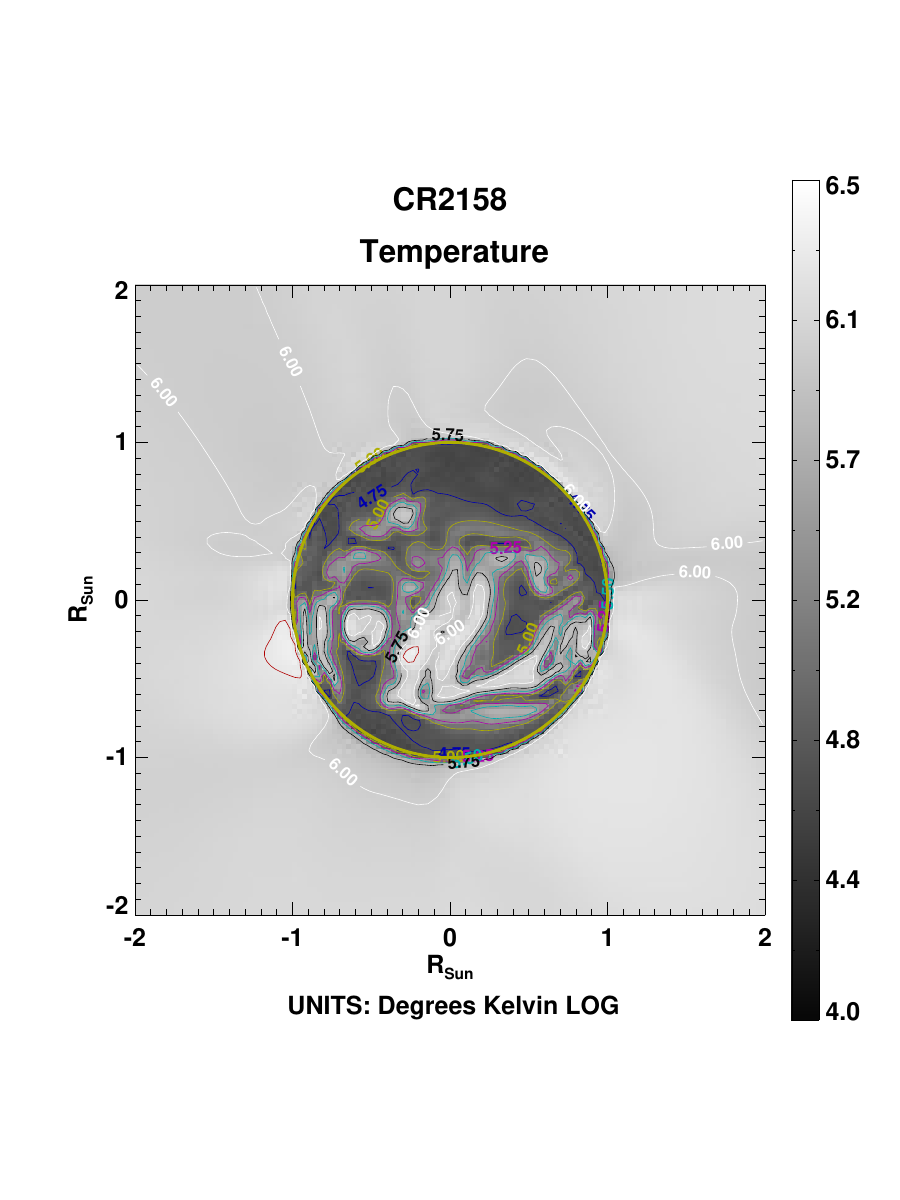}
    \includegraphics[width=0.125\textwidth,trim={1.1cm 2.5cm 0cm 2.9cm},clip]{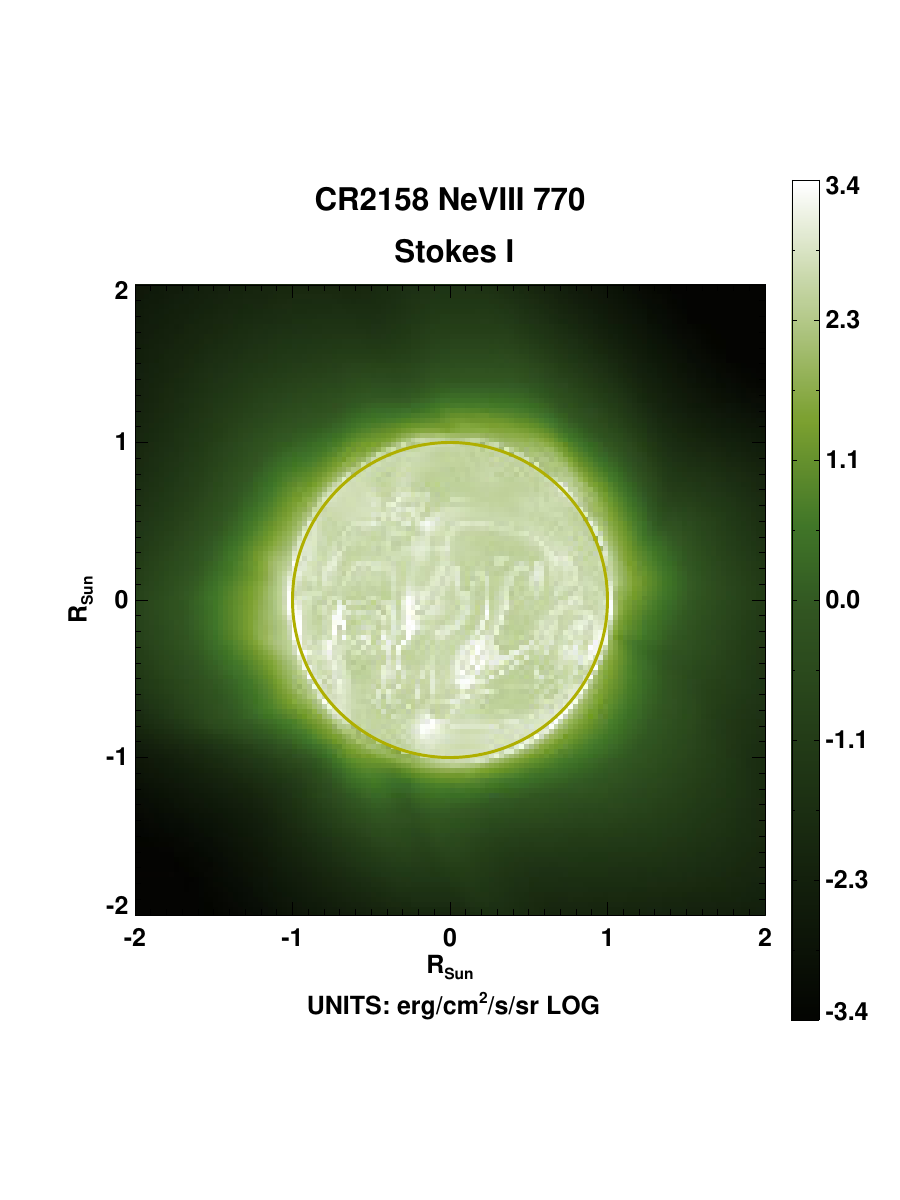}
    \includegraphics[width=0.125\textwidth,trim={1.1cm 2.5cm 0cm 2.9cm},clip]{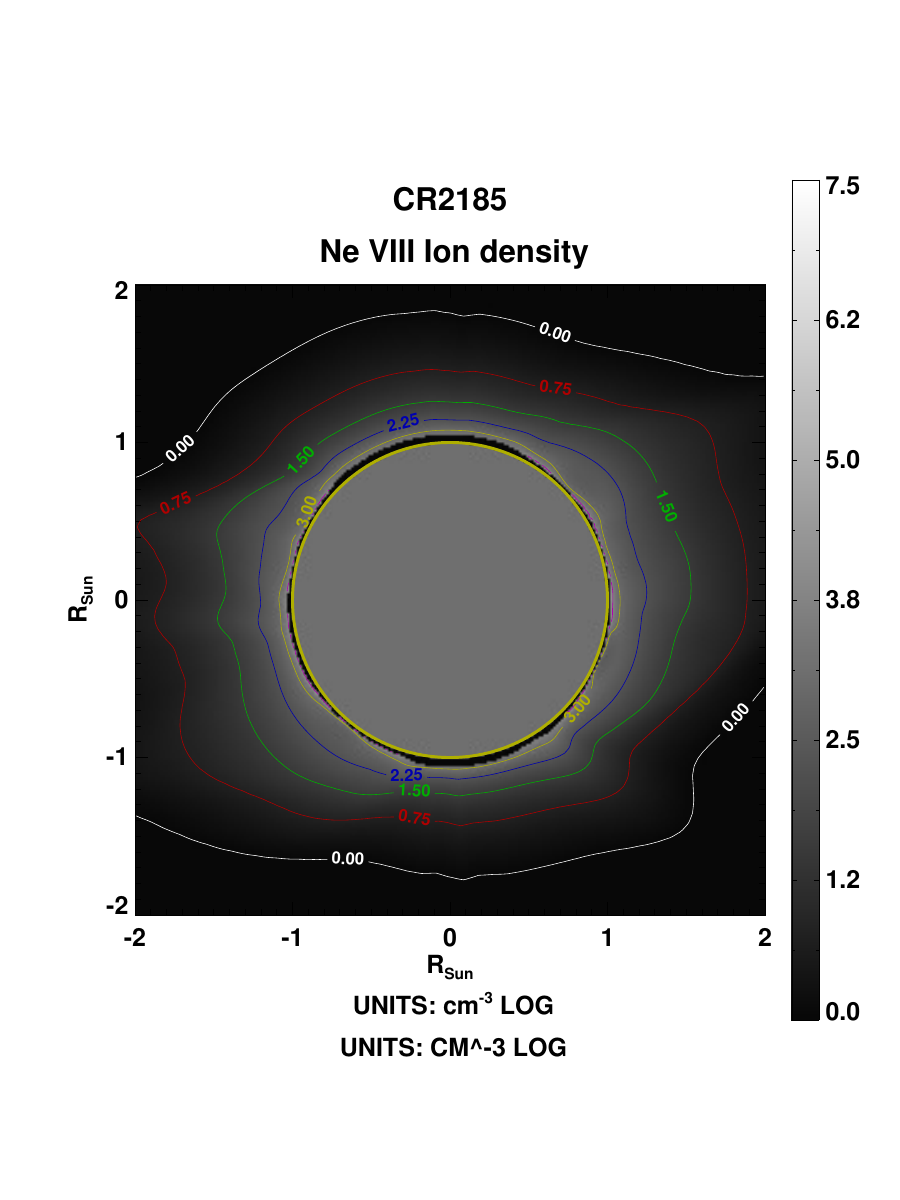}
    \includegraphics[width=0.125\textwidth,trim={1.1cm 2.5cm 0cm 2.9cm},clip]{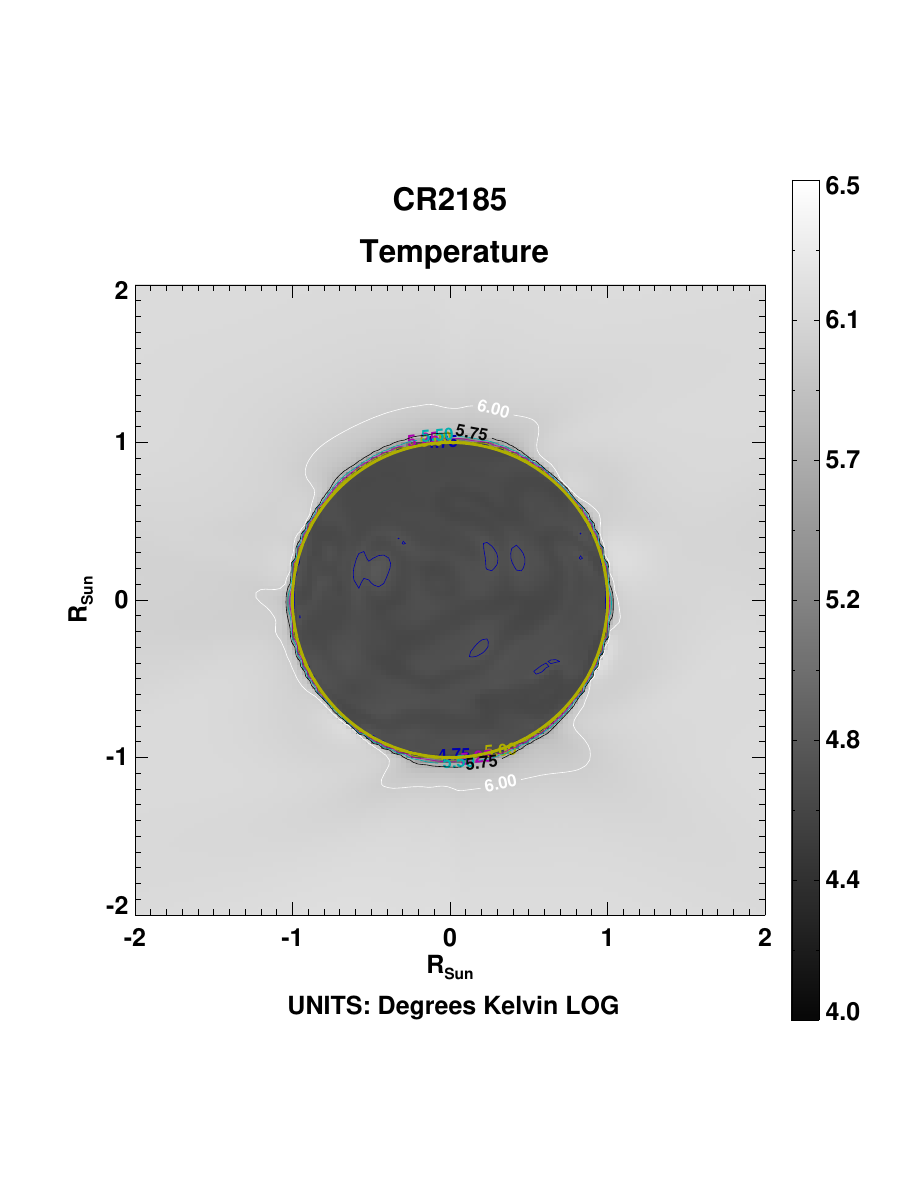}
    \includegraphics[width=0.125\textwidth,trim={1.1cm 2.5cm 0cm 2.9cm},clip]{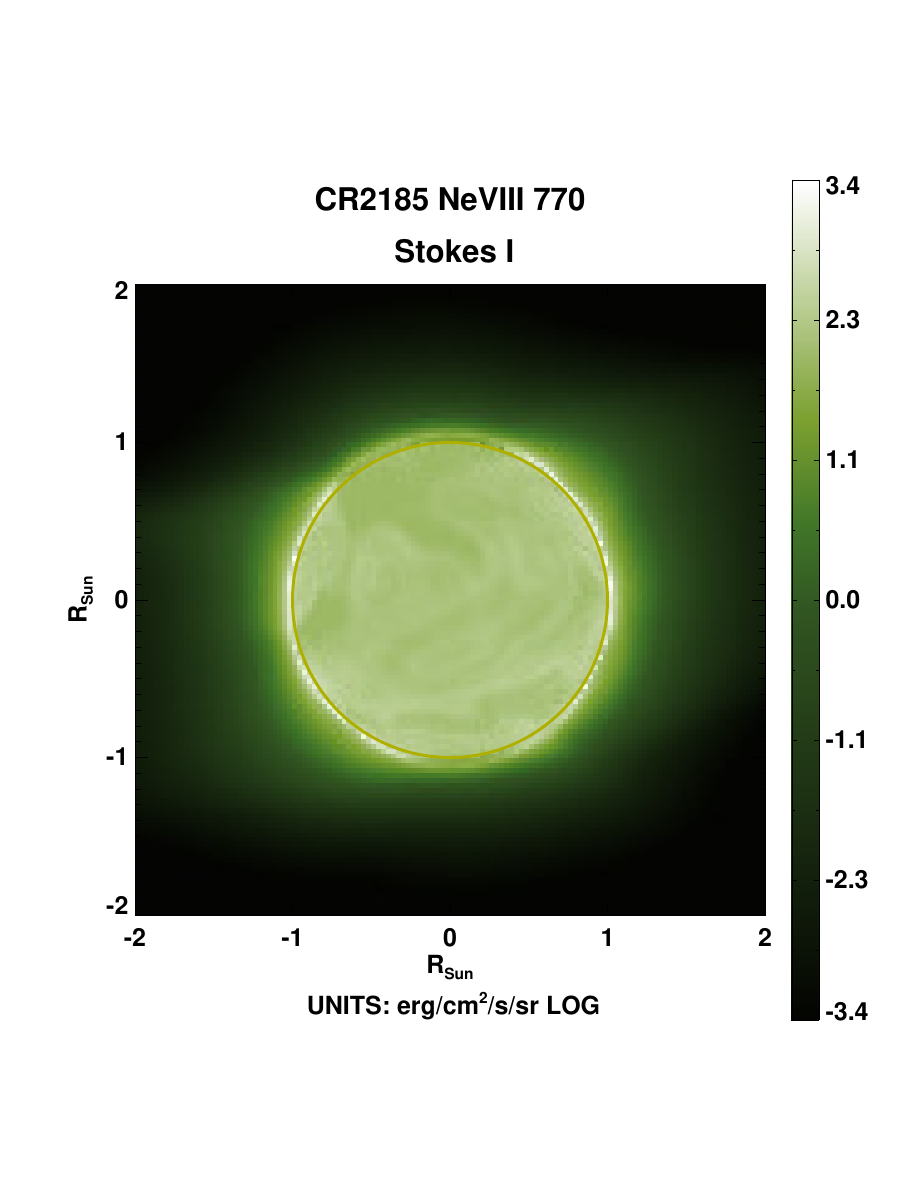}
    \includegraphics[width=0.125\textwidth,trim={1.1cm 2.5cm 0cm 2.9cm},clip]{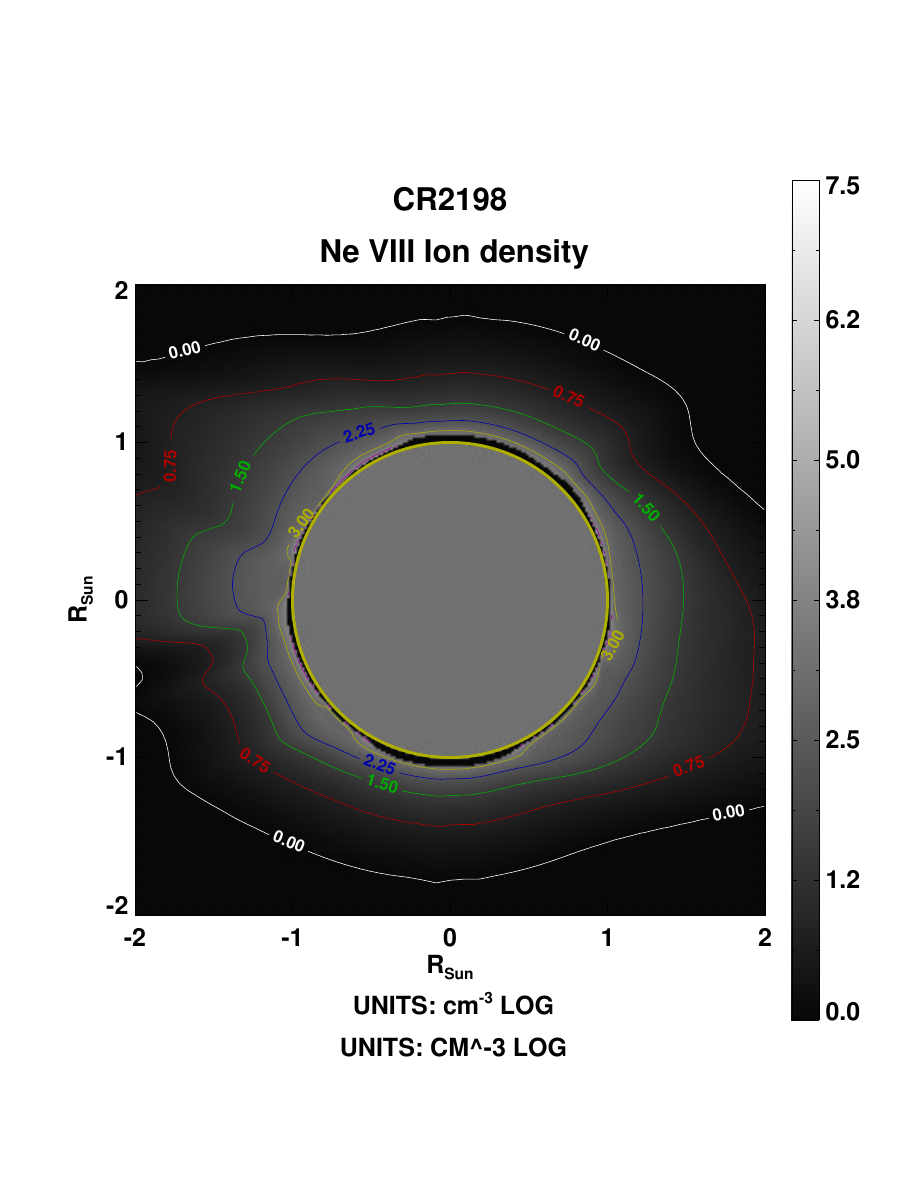}
    \includegraphics[width=0.125\textwidth,trim={1.1cm 2.5cm 0cm 2.9cm},clip]{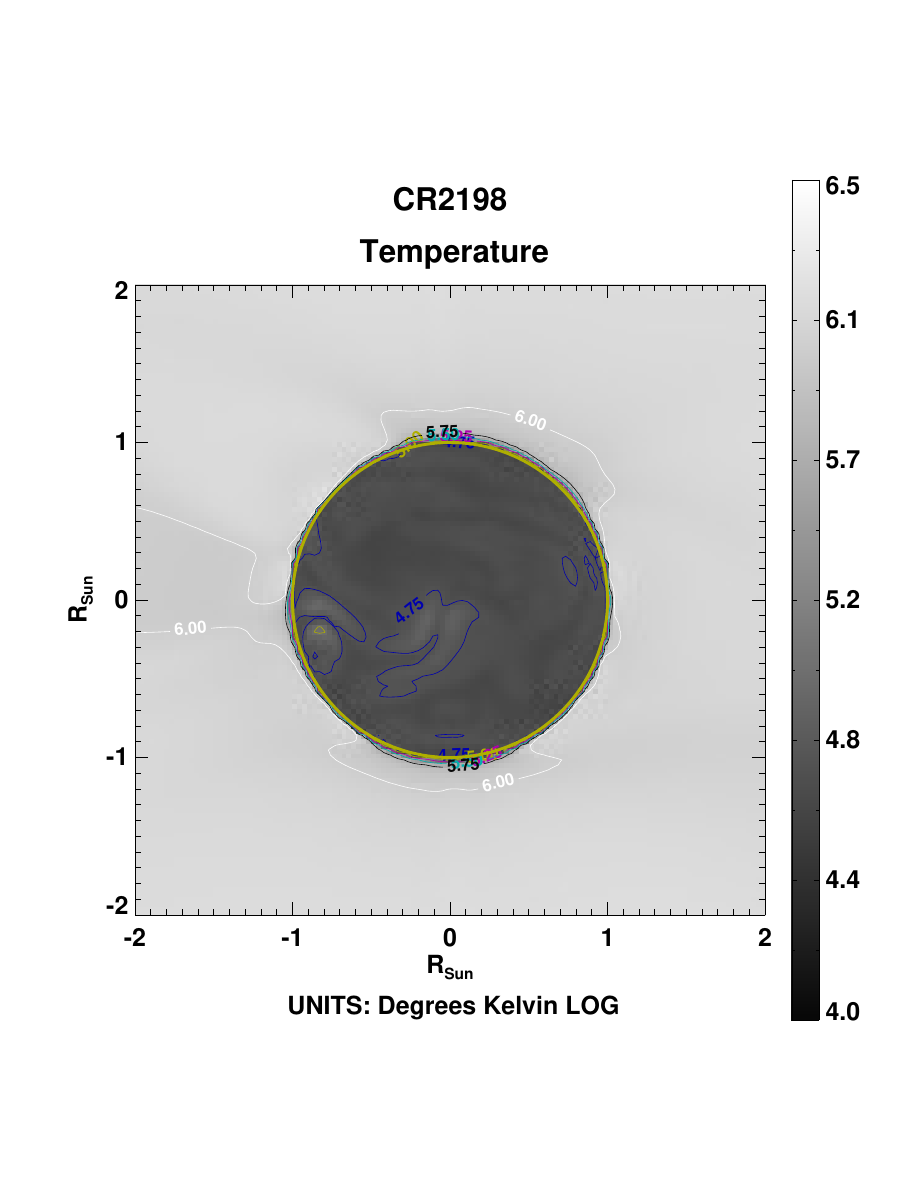}
    \includegraphics[width=0.125\textwidth,trim={1.1cm 2.5cm 0cm 2.9cm},clip]{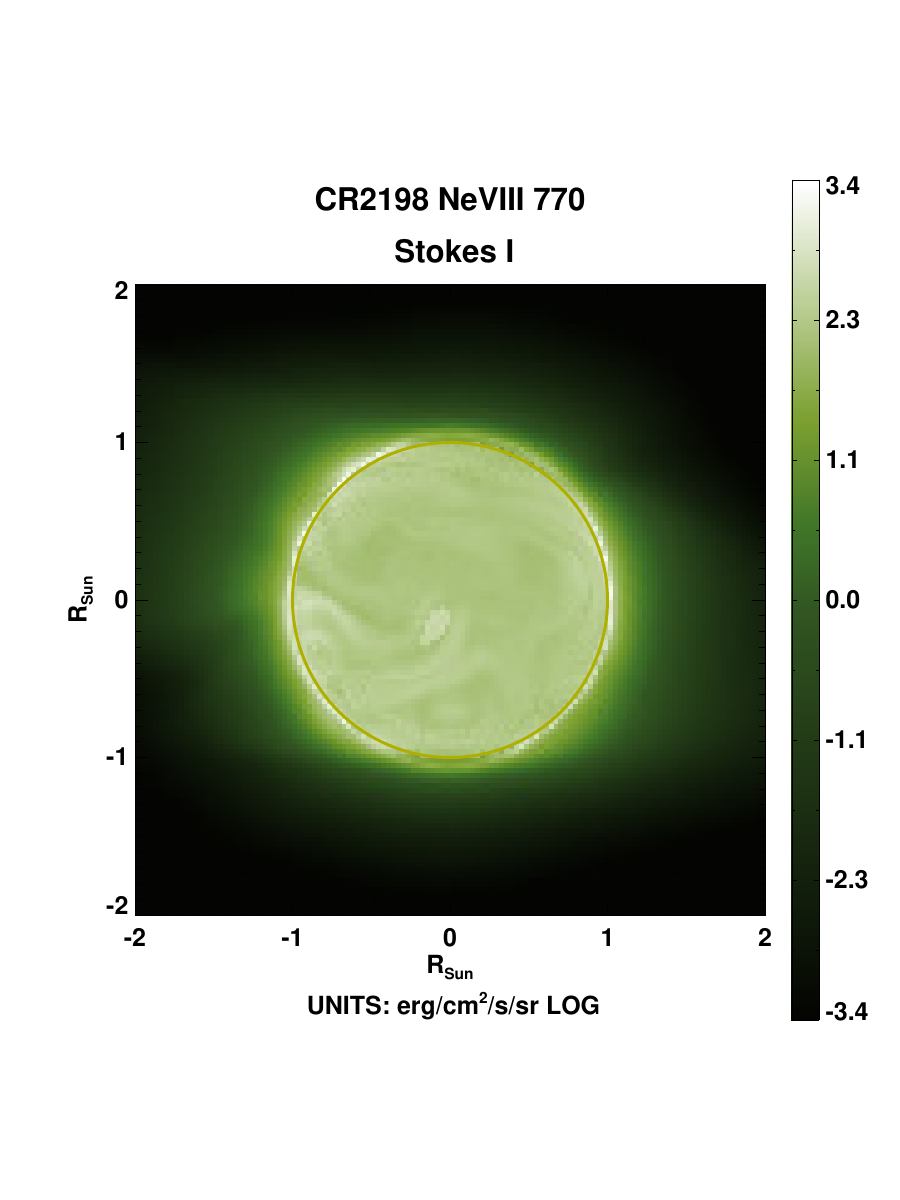}
    \includegraphics[width=0.125\textwidth,trim={1.1cm 2.5cm 0cm 2.9cm},clip]{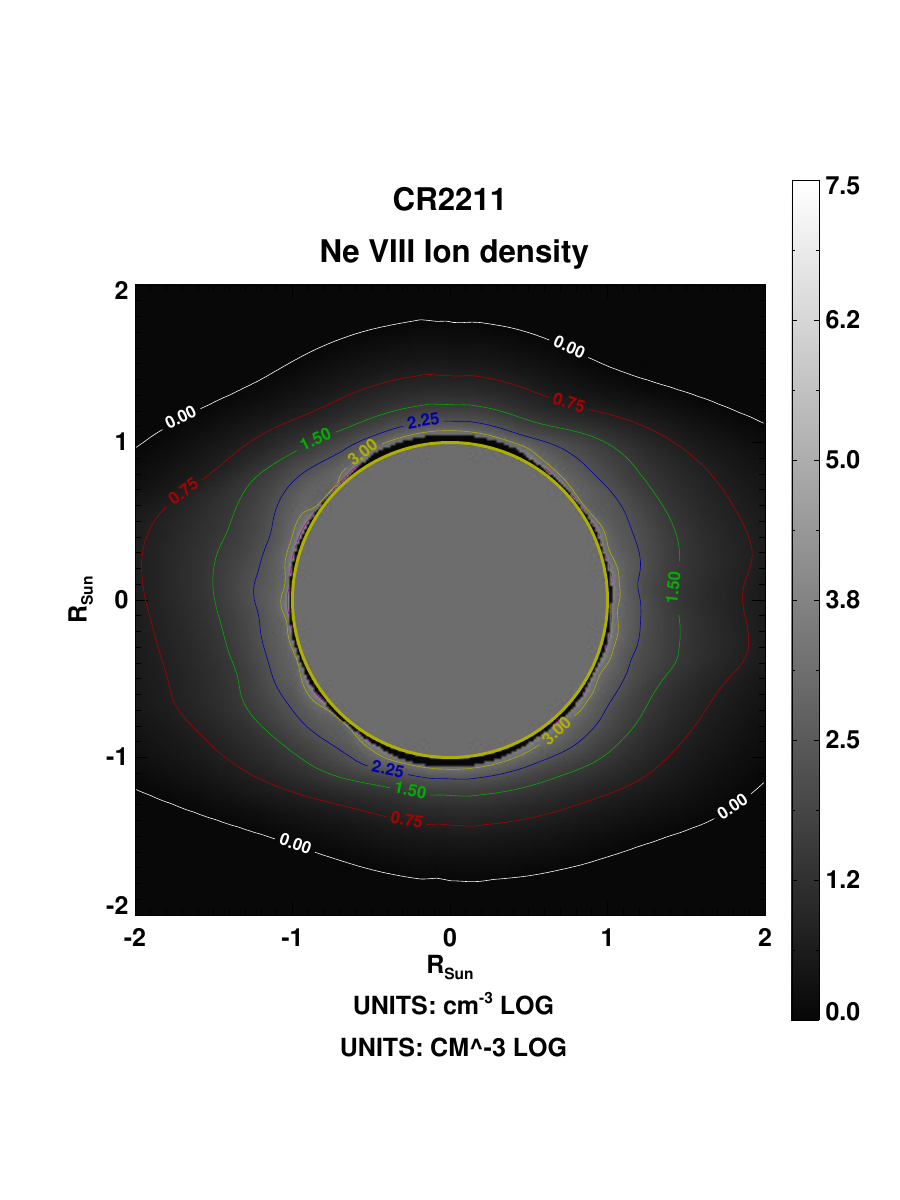}
    \includegraphics[width=0.125\textwidth,trim={1.1cm 2.5cm 0cm 2.9cm},clip]{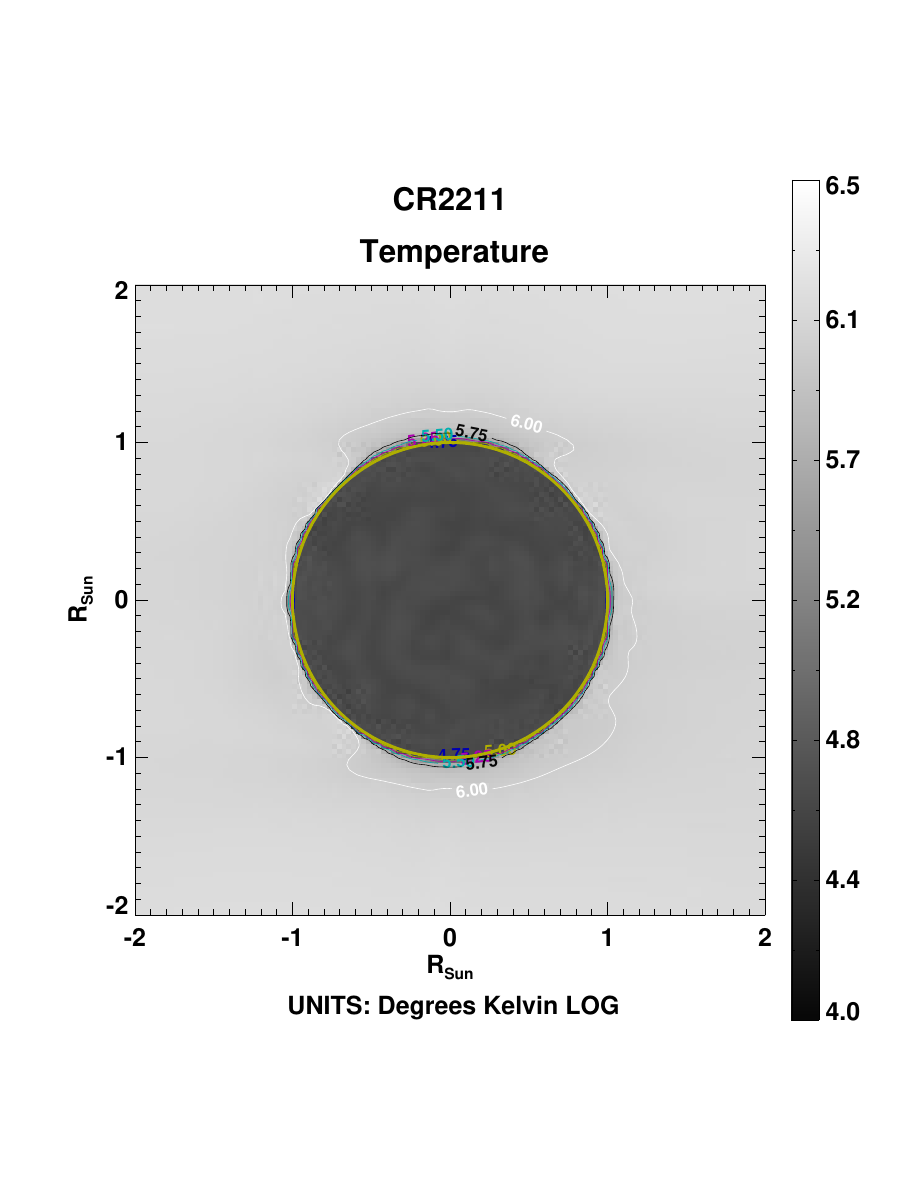}
    \includegraphics[width=0.125\textwidth,trim={1.1cm 2.5cm 0cm 2.9cm},clip]{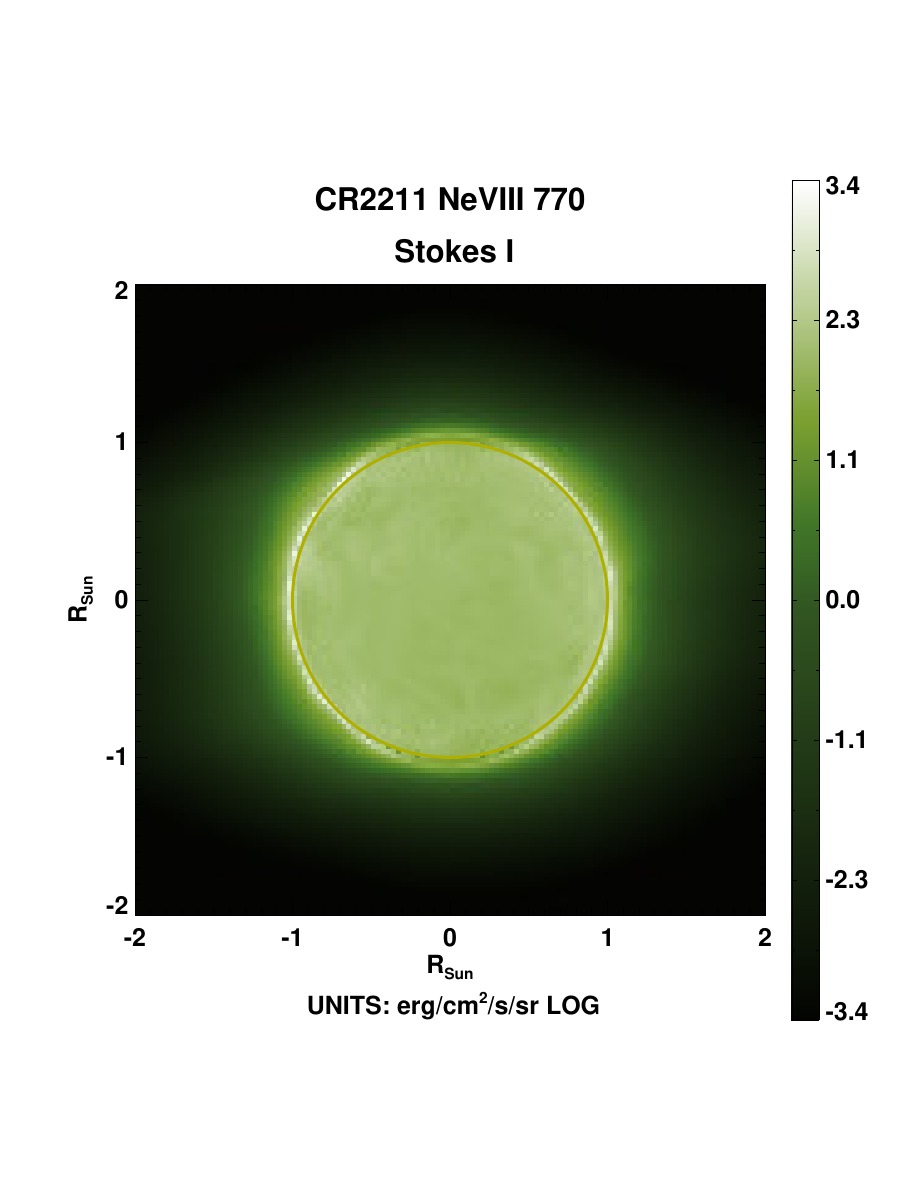}
    \includegraphics[width=0.125\textwidth,trim={1.1cm 2.5cm 0cm 2.9cm},clip]{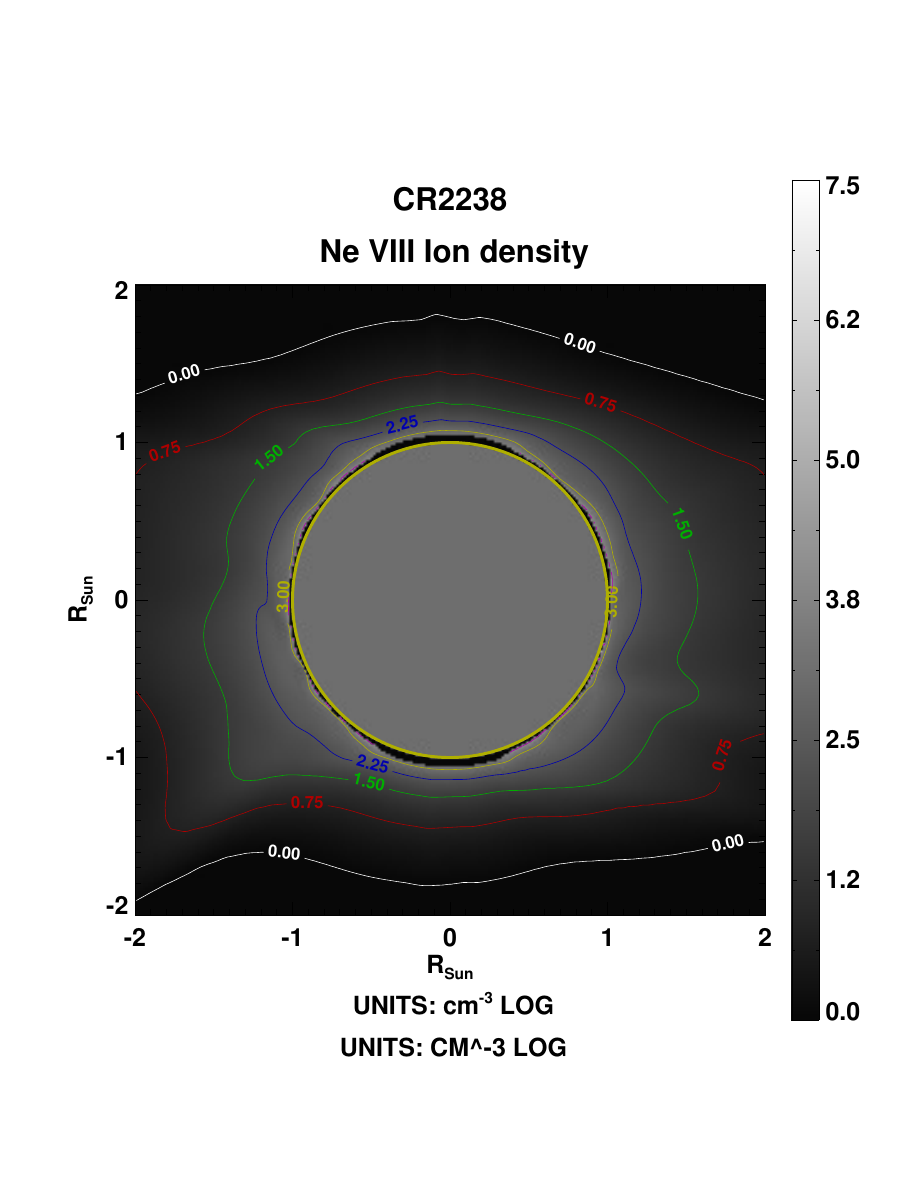}
    \includegraphics[width=0.125\textwidth,trim={1.1cm 2.5cm 0cm 2.9cm},clip]{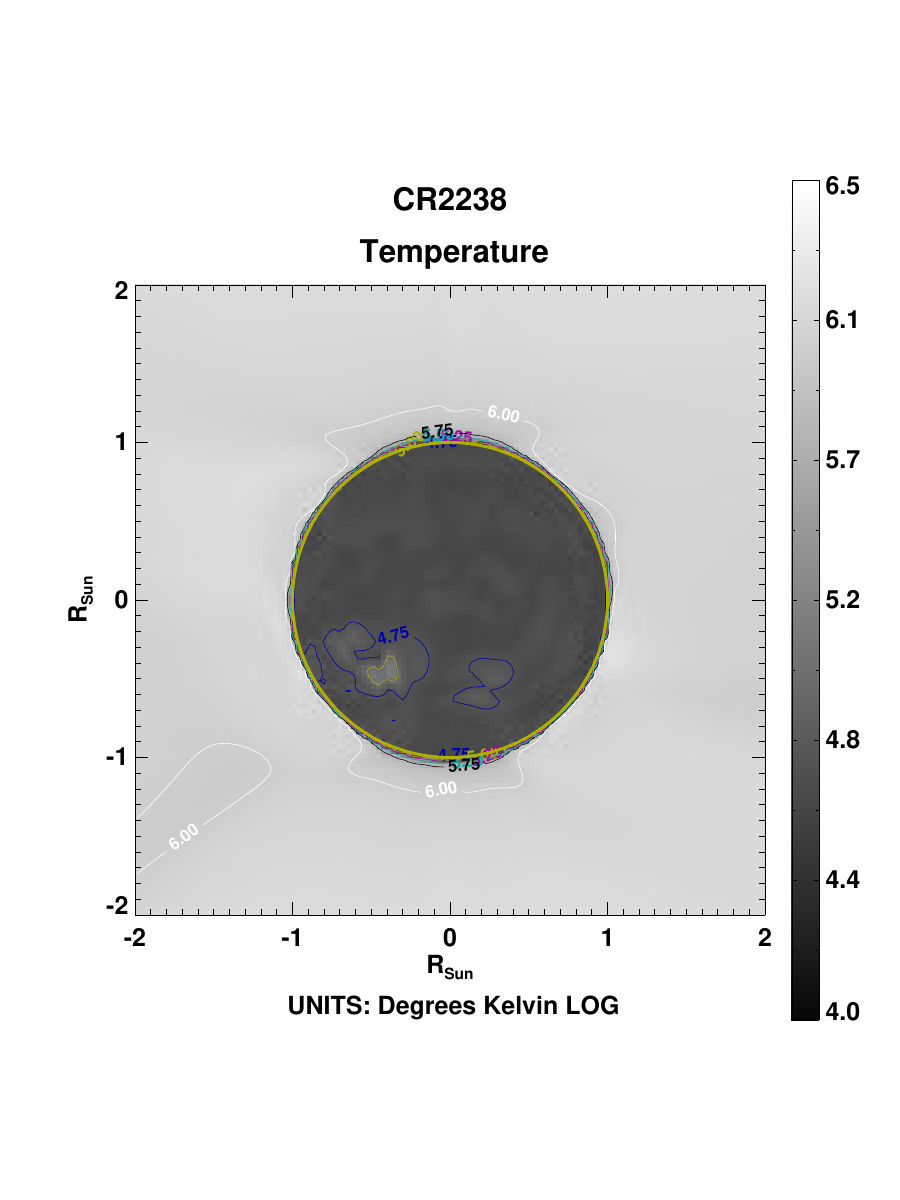}
    \includegraphics[width=0.125\textwidth,trim={1.1cm 2.5cm 0cm 2.9cm},clip]{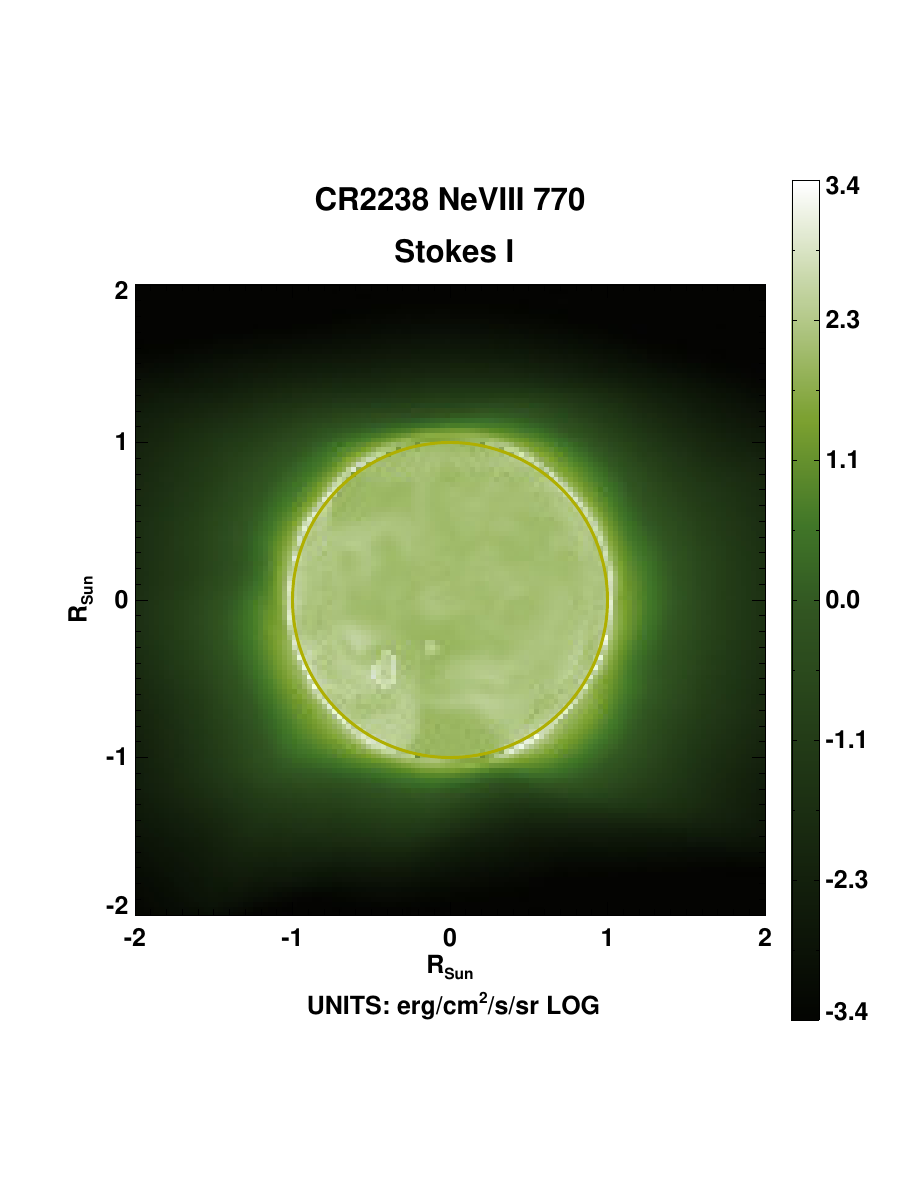}
    \caption{Same as Figure \ref{fig:models_stki}, but for rest of the CR simulations.}
    \label{fig:supp_models_stki}
\end{figure}

\begin{figure}[htbp]
\centering
    \includegraphics[width=0.11\textwidth,trim={1.1cm 2.5cm 0cm 2.9cm},clip]{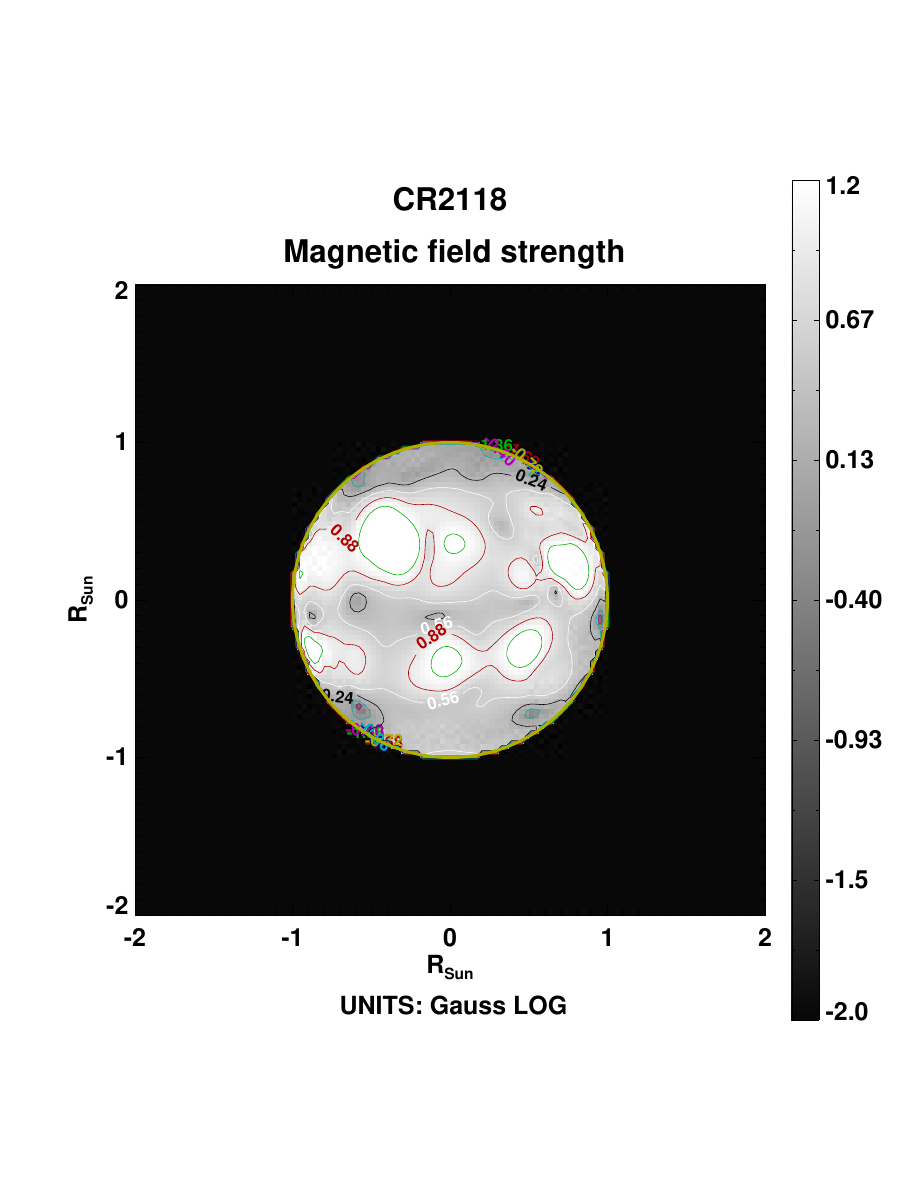}
    \includegraphics[width=0.11\textwidth,trim={1.1cm 2.5cm 0cm 2.9cm},clip]{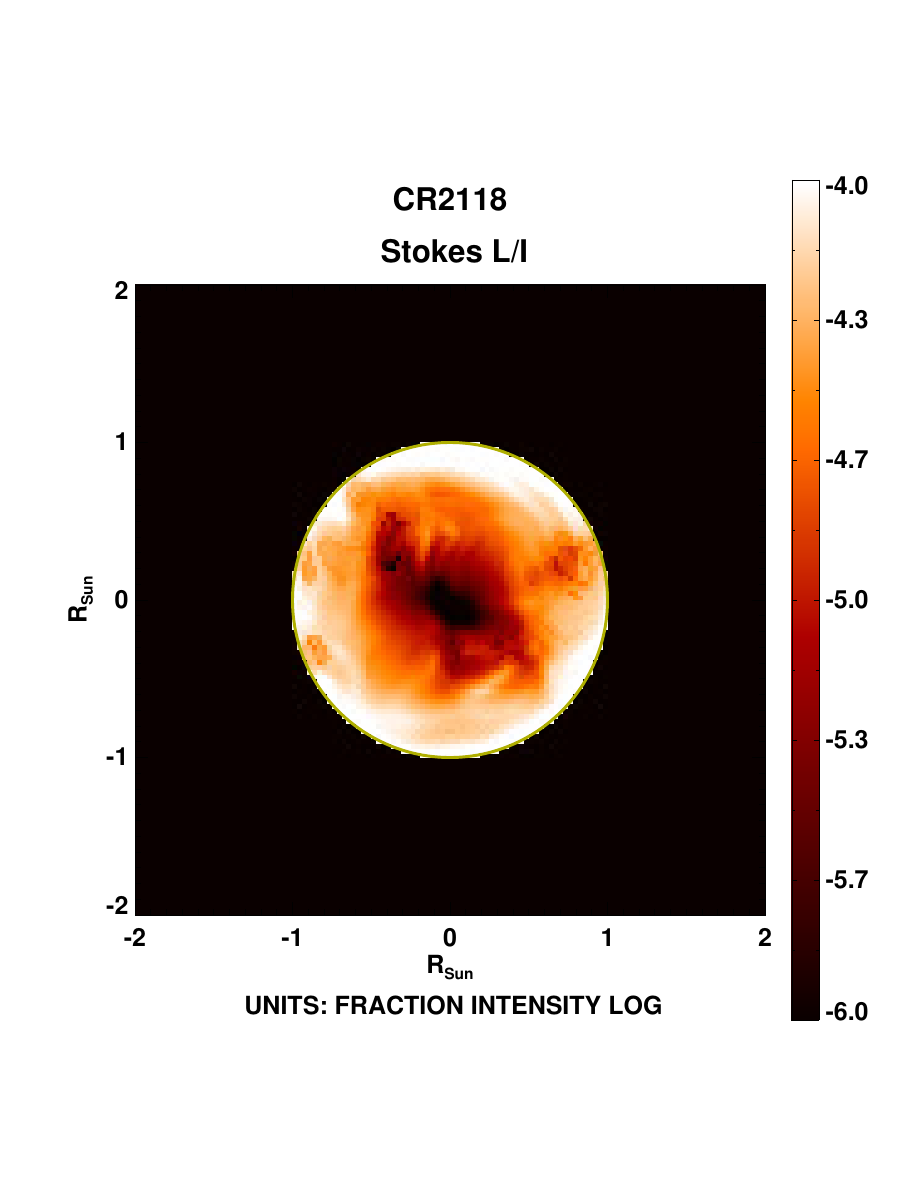}
    \includegraphics[width=0.11\textwidth,trim={1.1cm 2.7cm 0cm 2.9cm},clip]{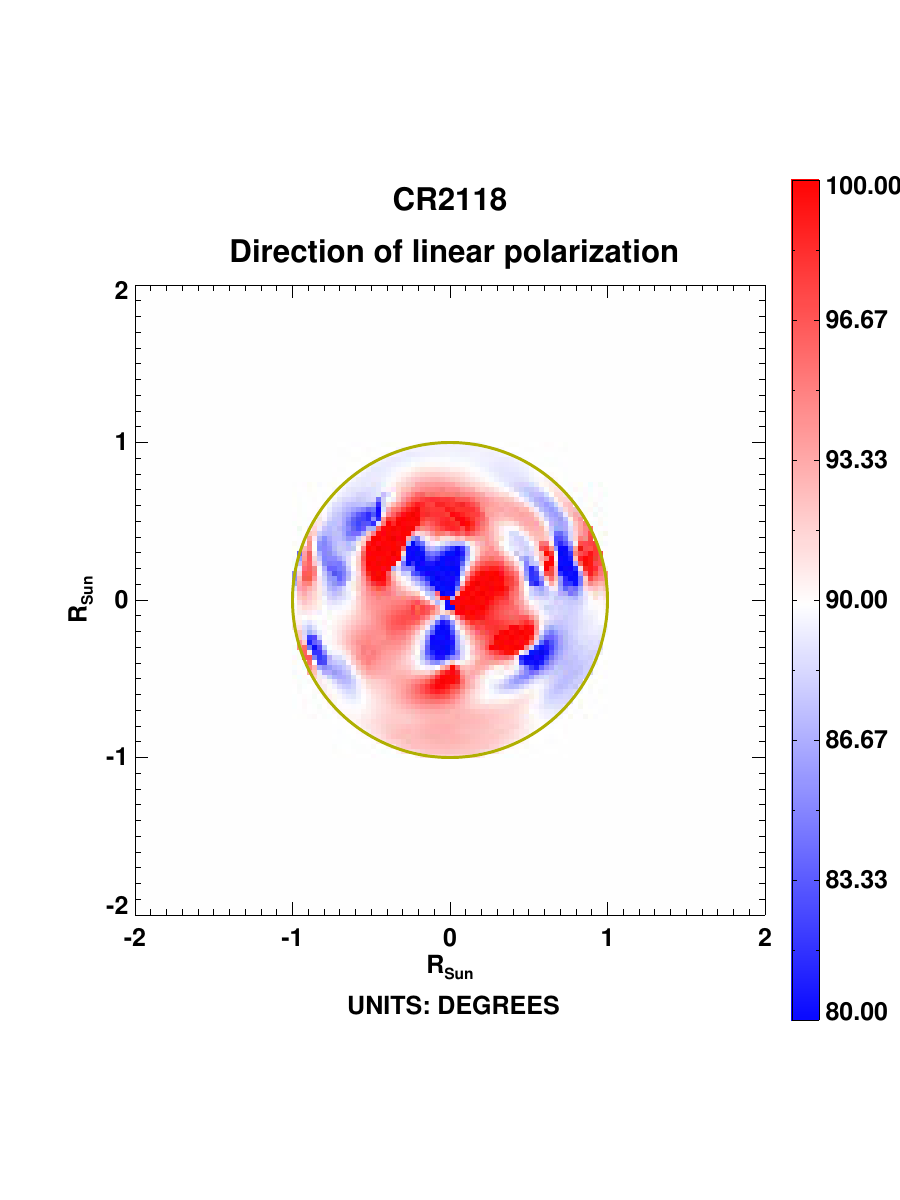}
    \includegraphics[width=0.115\textwidth,trim={0.3cm 1.2cm 0cm 0.5cm},clip]{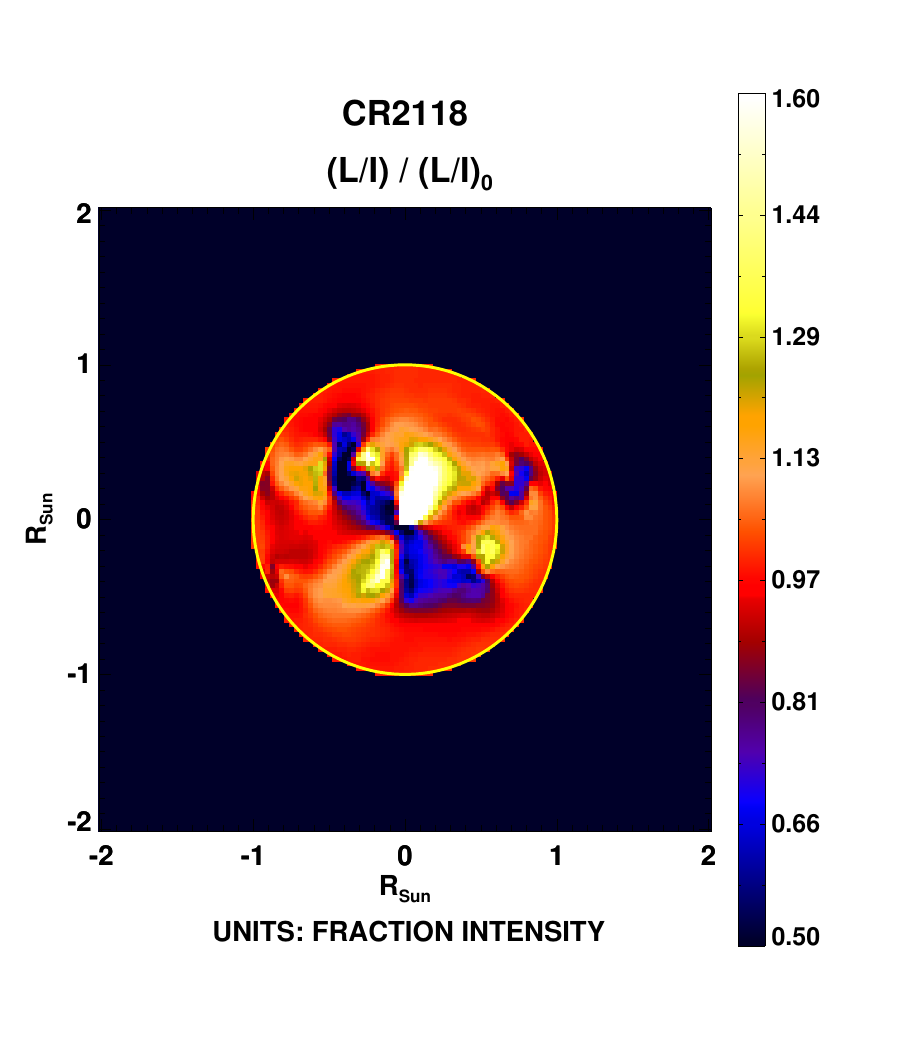}
    \includegraphics[width=0.11\textwidth,trim={1.1cm 2.5cm 0cm 2.9cm},clip]{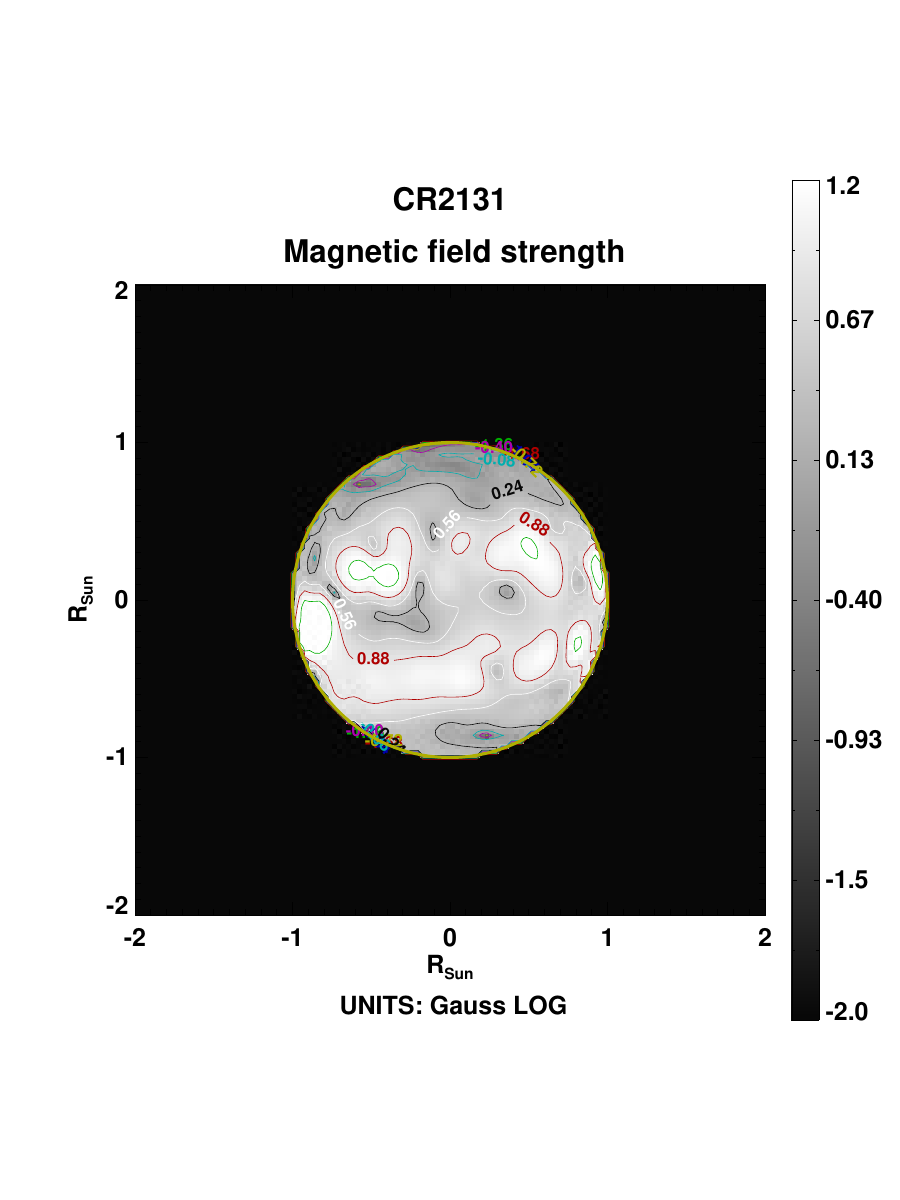}
    \includegraphics[width=0.11\textwidth,trim={1.1cm 2.5cm 0cm 2.9cm},clip]{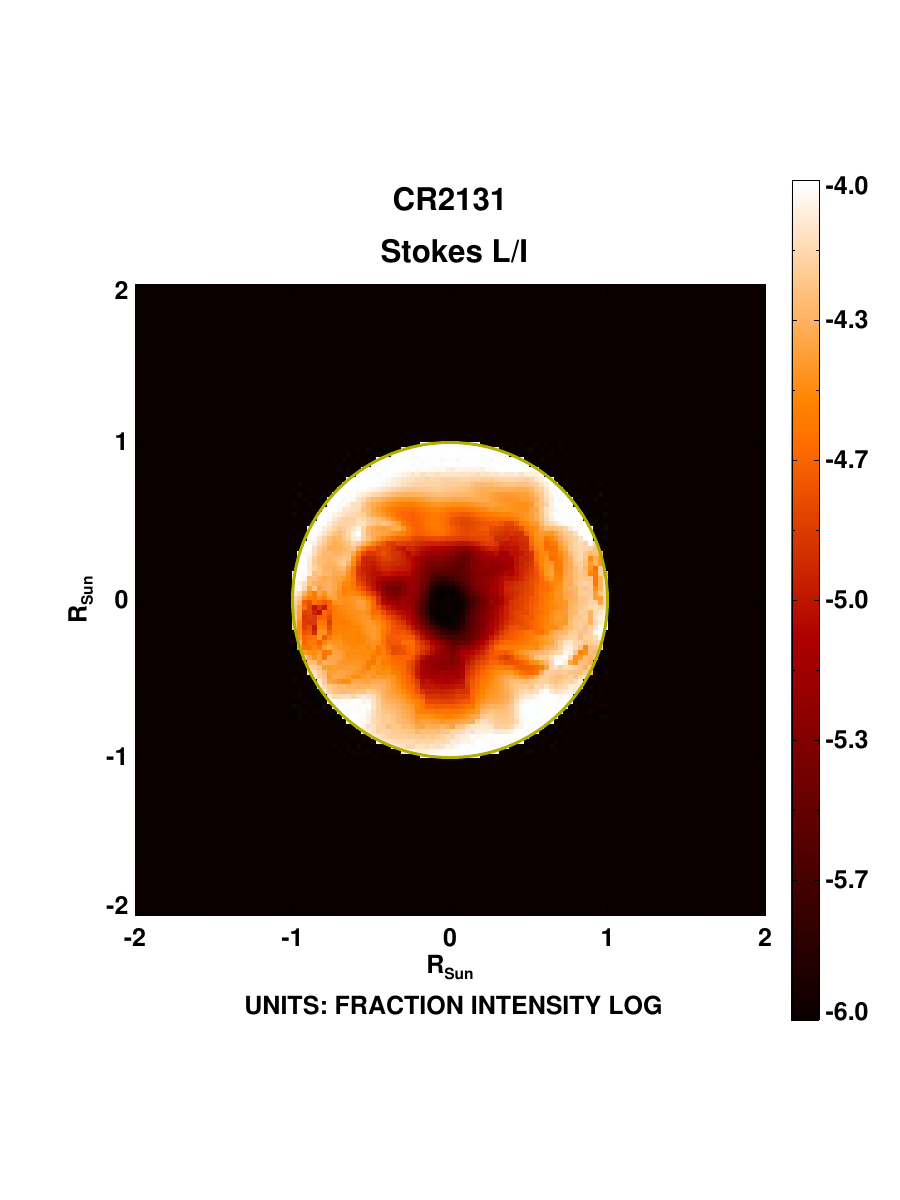}
    \includegraphics[width=0.11\textwidth,trim={1.1cm 2.7cm 0cm 2.9cm},clip]{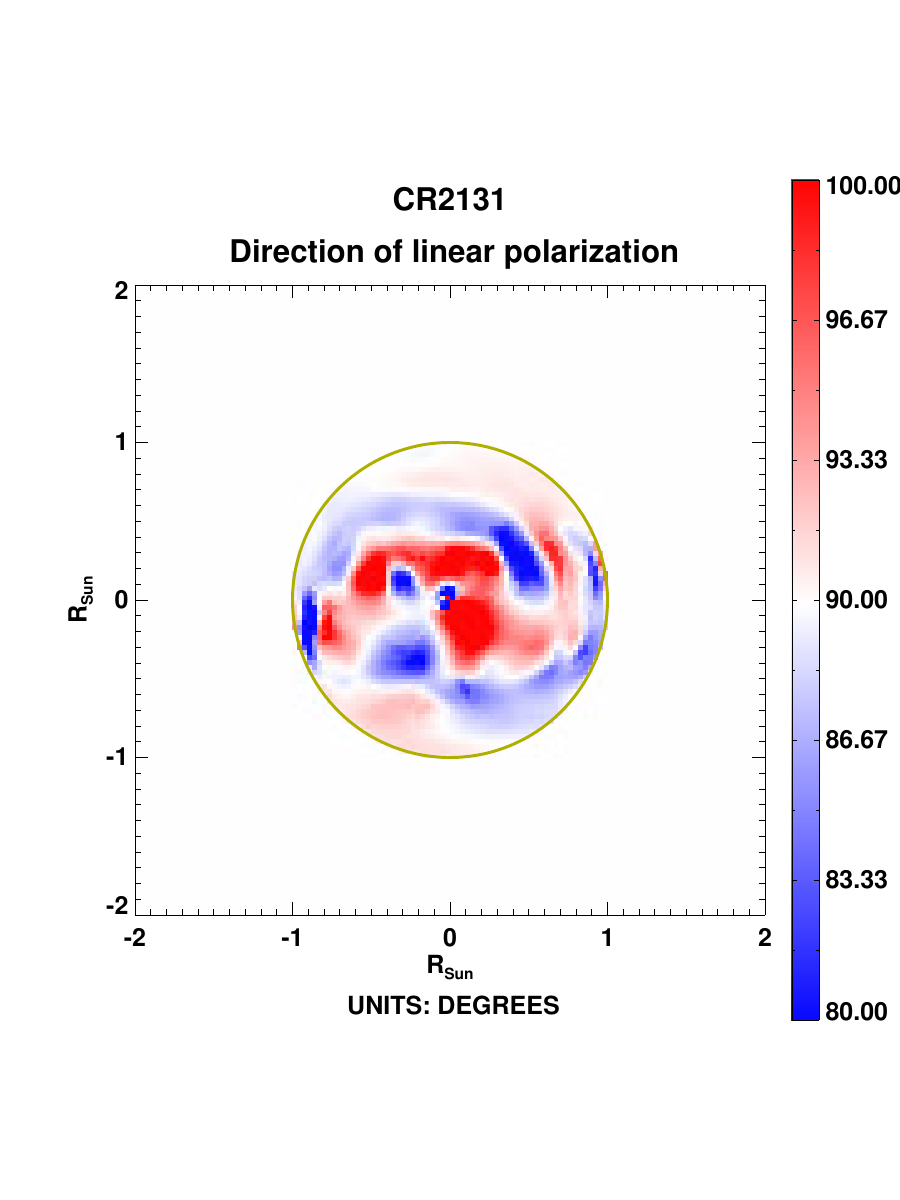}
    \includegraphics[width=0.115\textwidth,trim={0.3cm 1.2cm 0cm 0.5cm},clip]{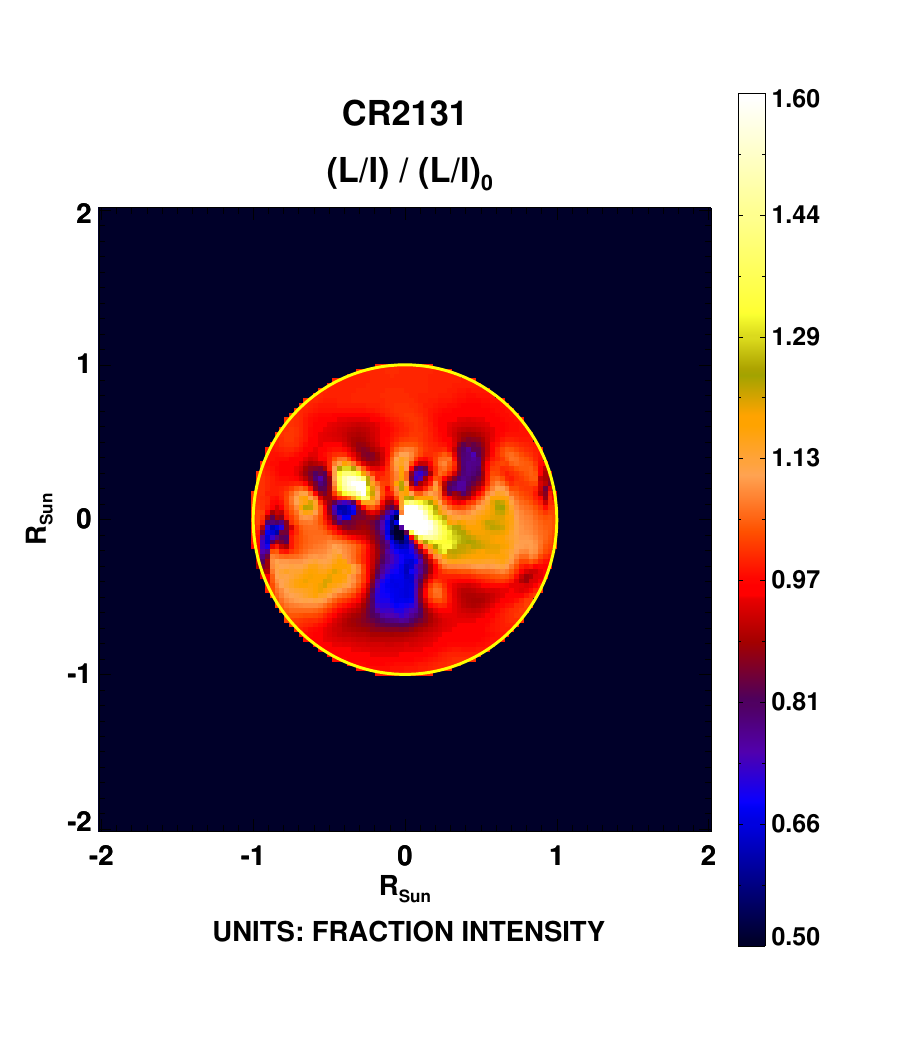}
    \includegraphics[width=0.11\textwidth,trim={1.1cm 2.5cm 0cm 2.9cm},clip]{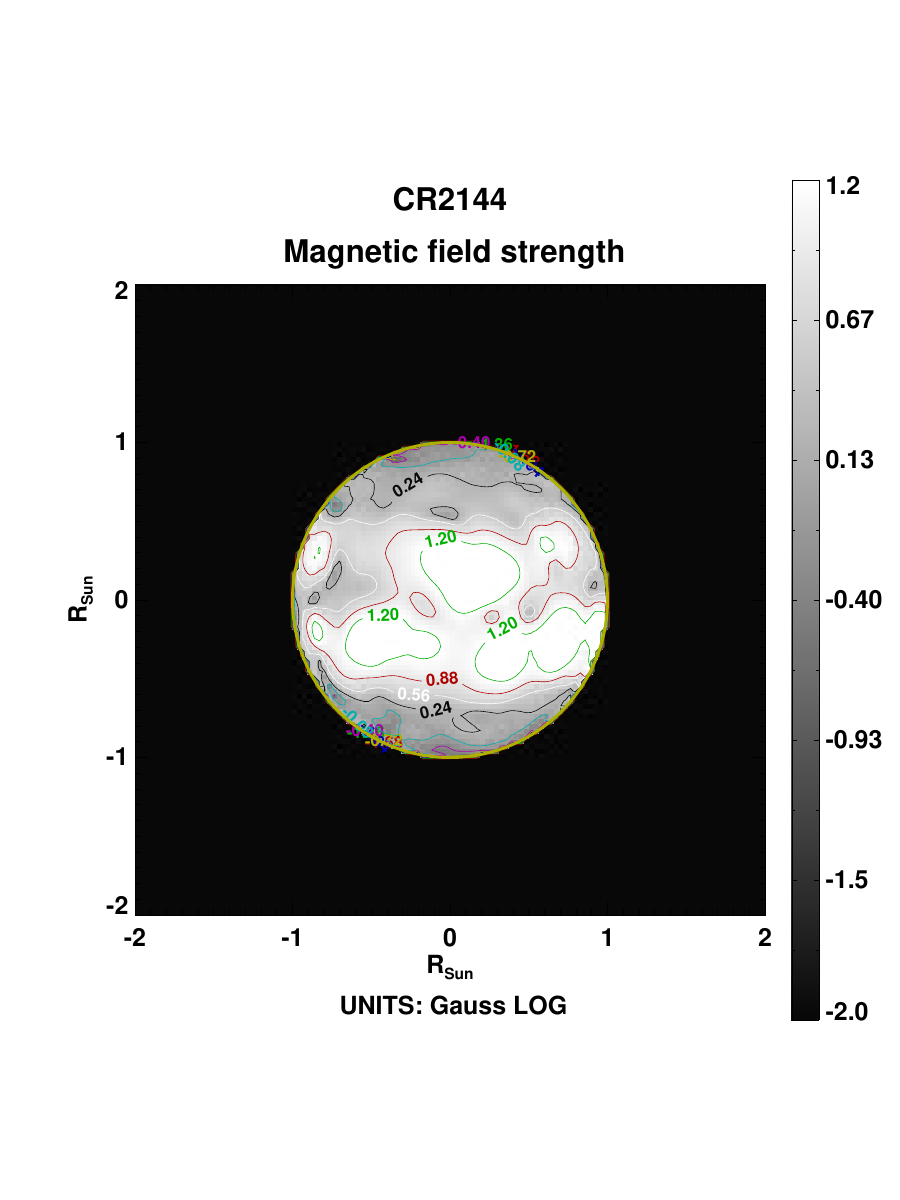}
    \includegraphics[width=0.11\textwidth,trim={1.1cm 2.5cm 0cm 2.9cm},clip]{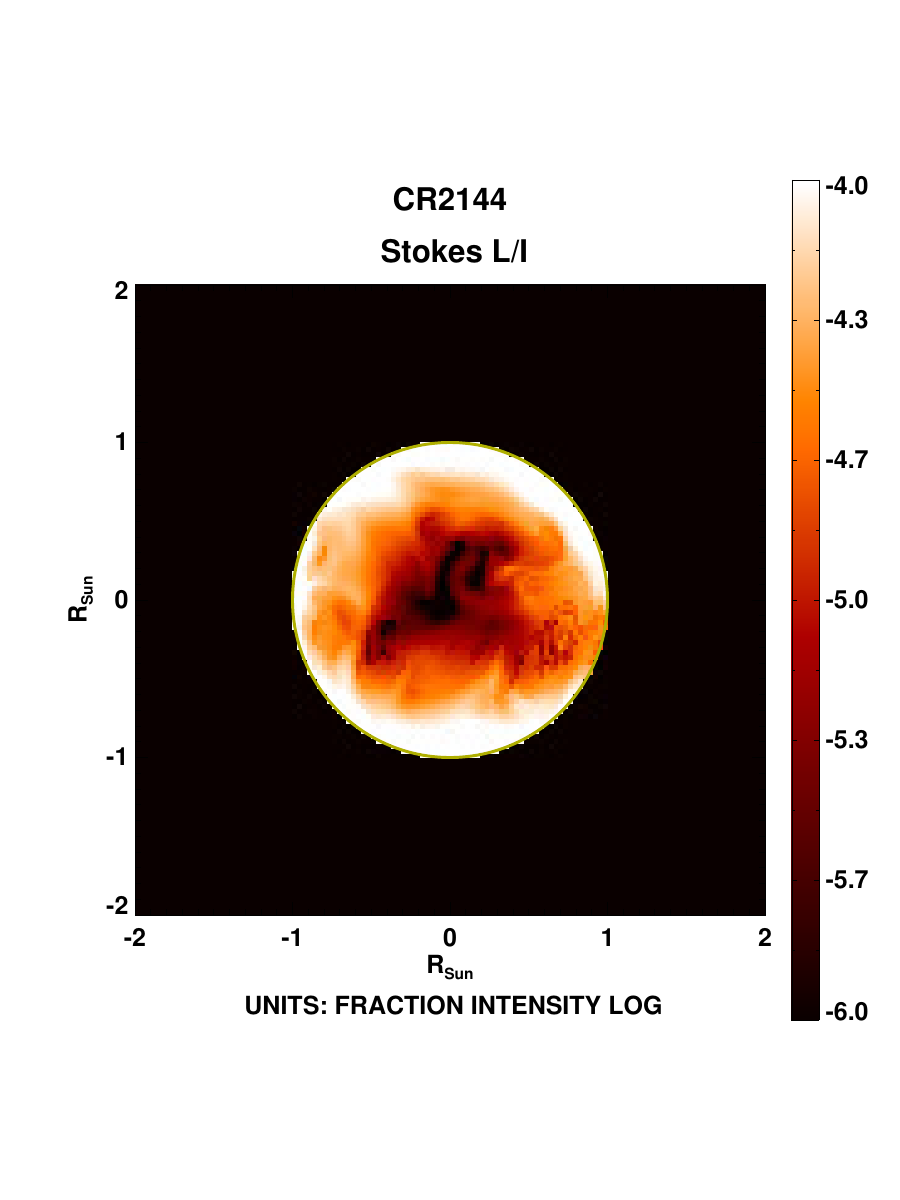}
    \includegraphics[width=0.11\textwidth,trim={1.1cm 2.7cm 0cm 2.9cm},clip]{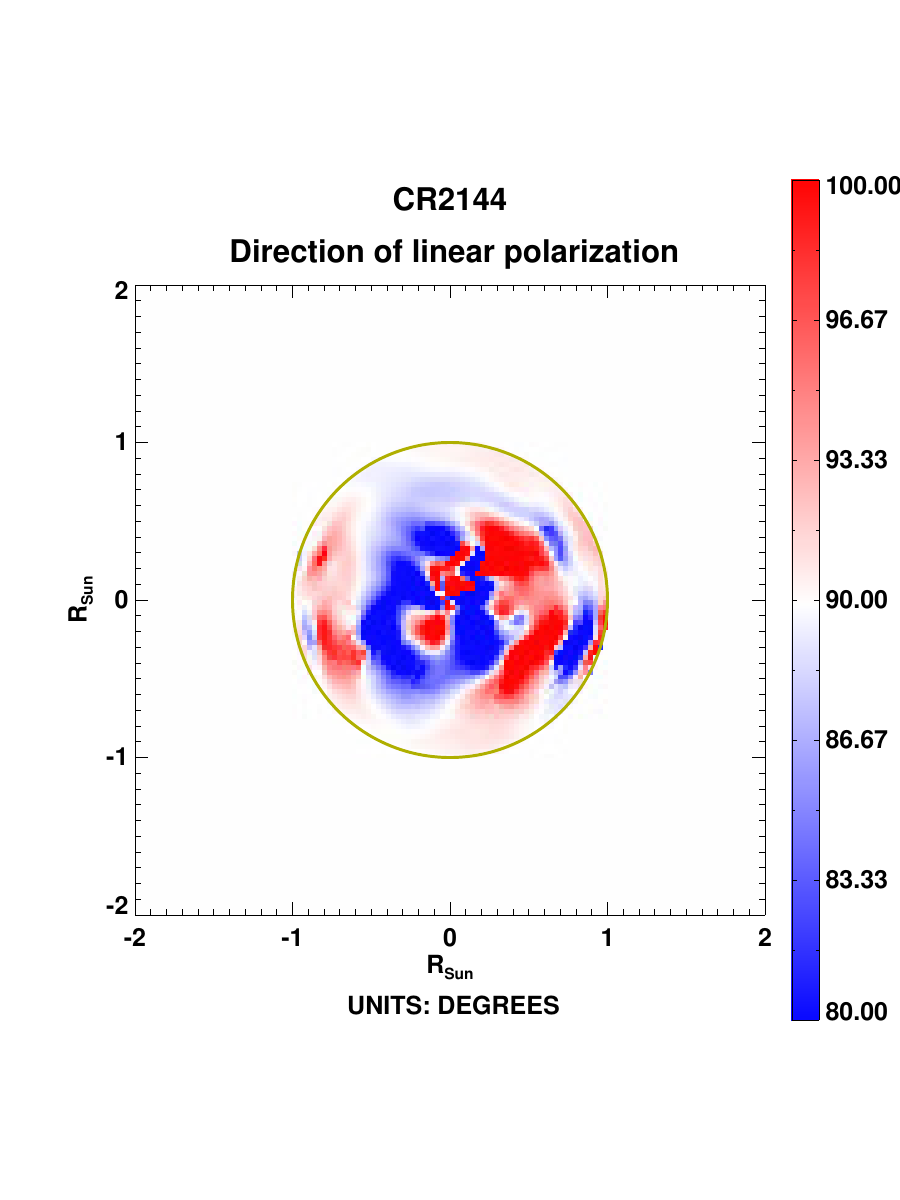}
    \includegraphics[width=0.115\textwidth,trim={0.3cm 1.2cm 0cm 0.5cm},clip]{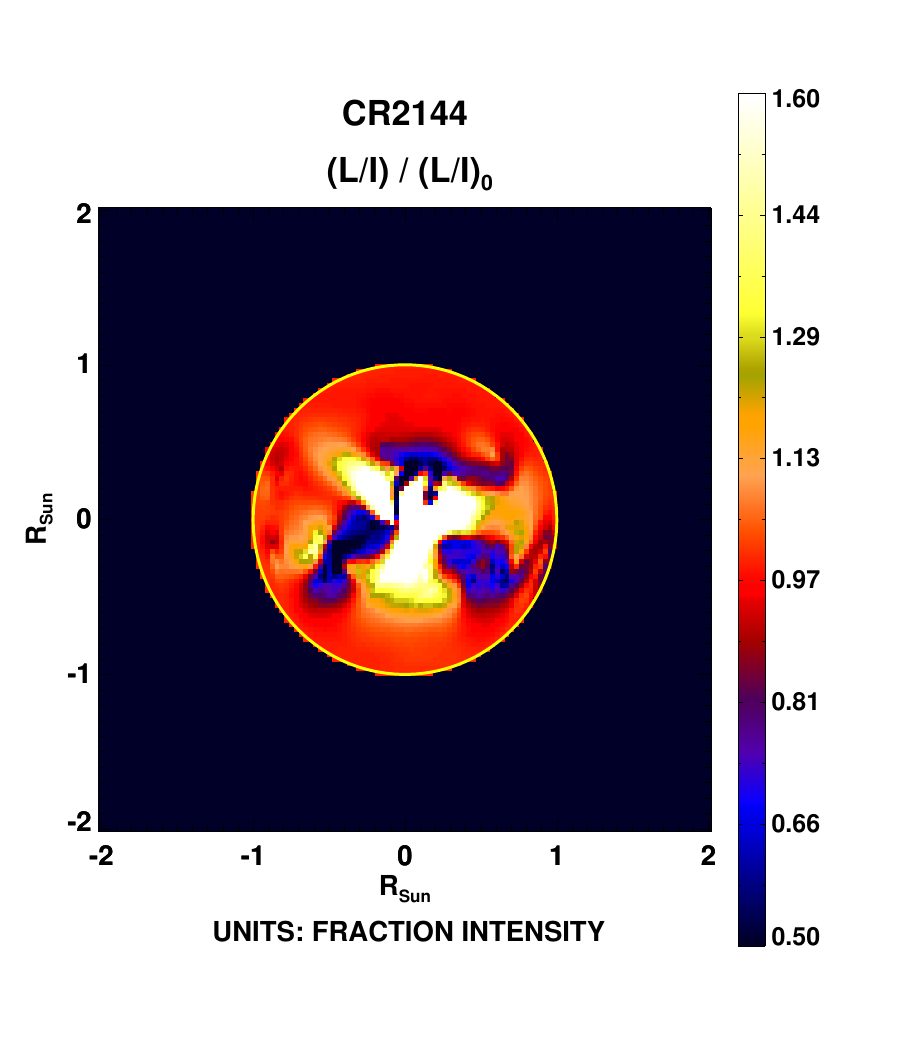}
    \includegraphics[width=0.11\textwidth,trim={1.1cm 2.5cm 0cm 2.9cm},clip]{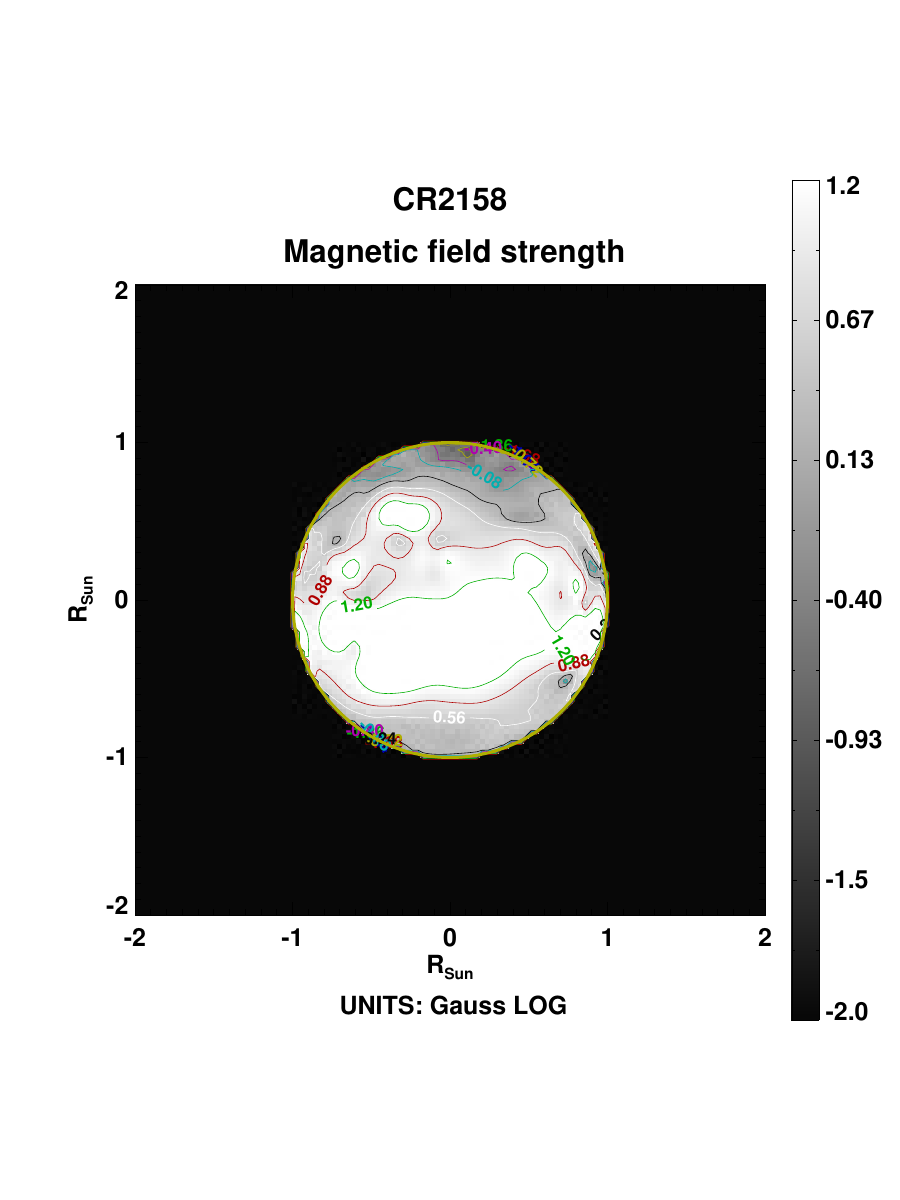}
    \includegraphics[width=0.11\textwidth,trim={1.1cm 2.5cm 0cm 2.9cm},clip]{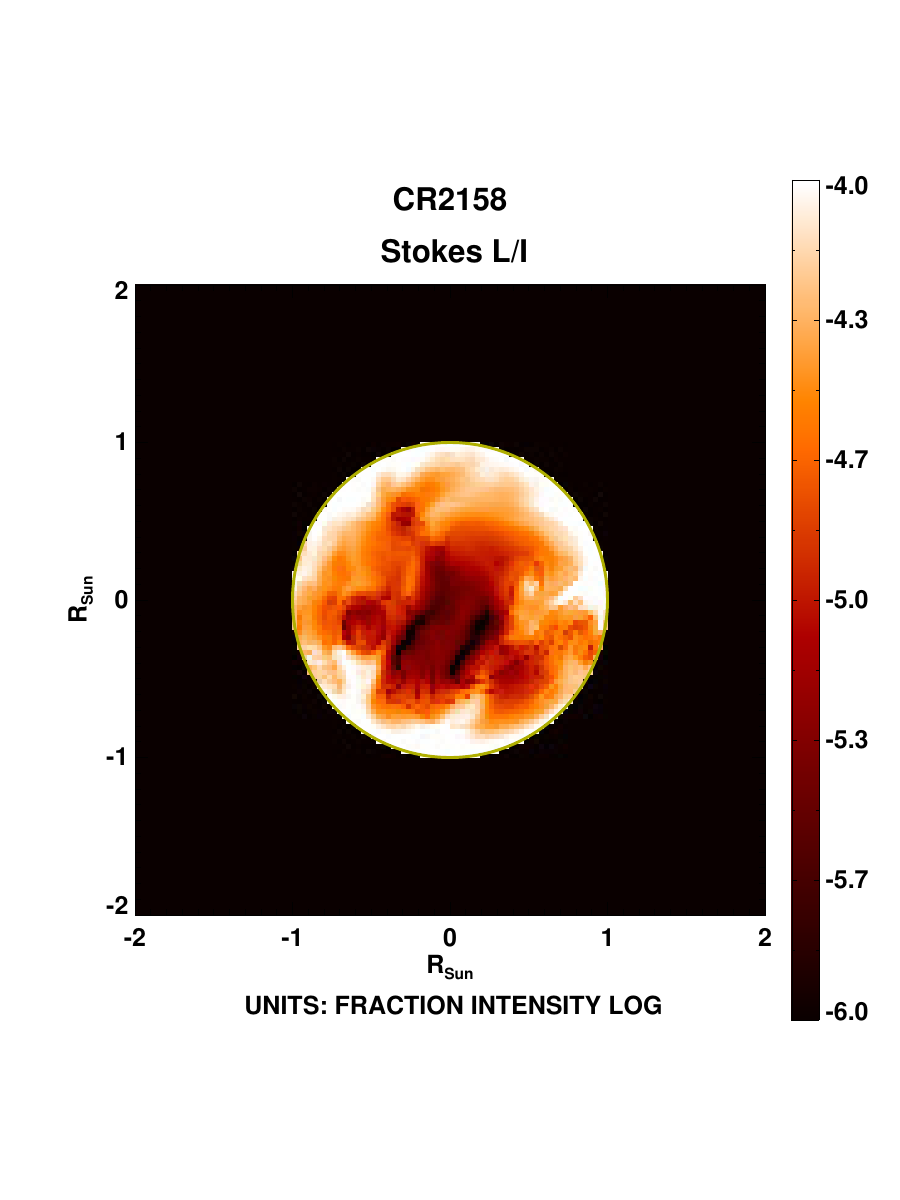}
    \includegraphics[width=0.11\textwidth,trim={1.1cm 2.7cm 0cm 2.9cm},clip]{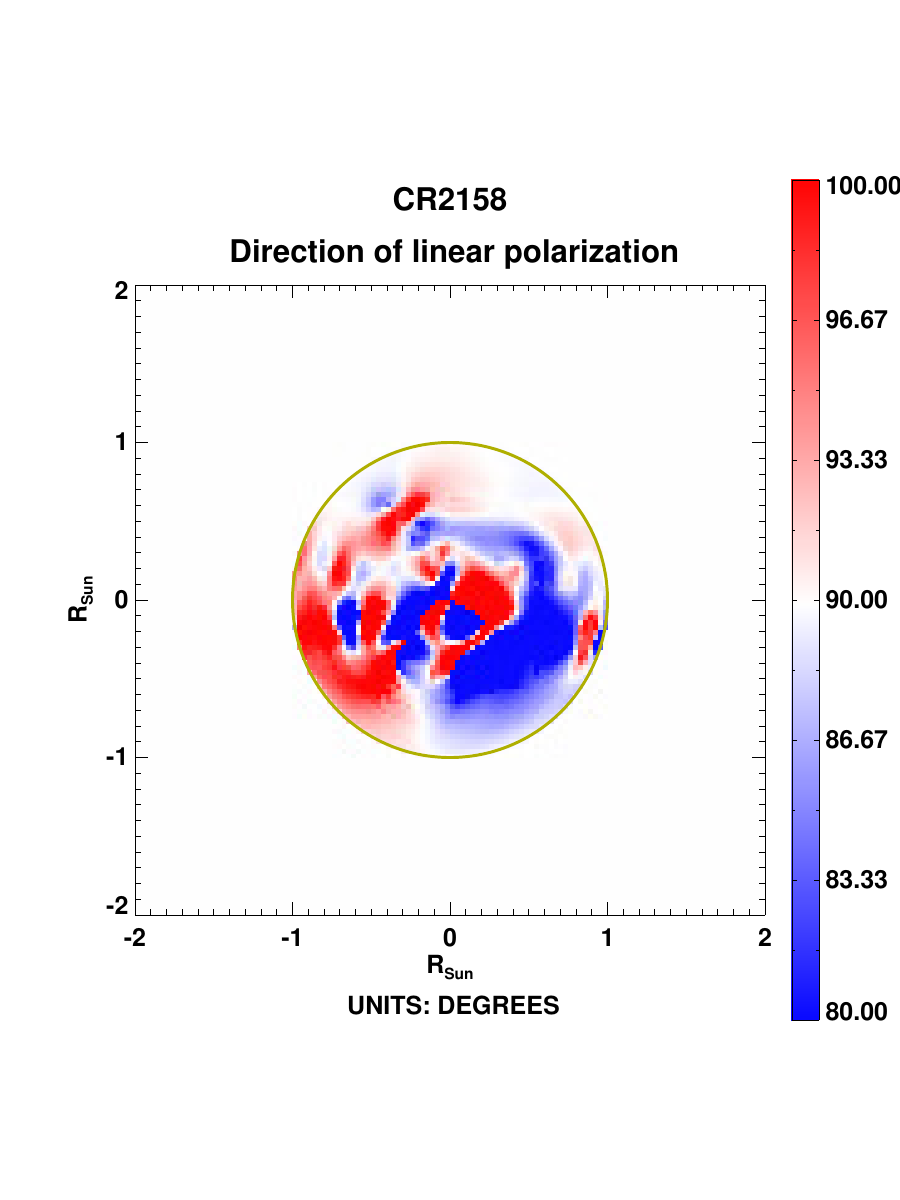}
    \includegraphics[width=0.115\textwidth,trim={0.3cm 1.2cm 0cm 0.5cm},clip]{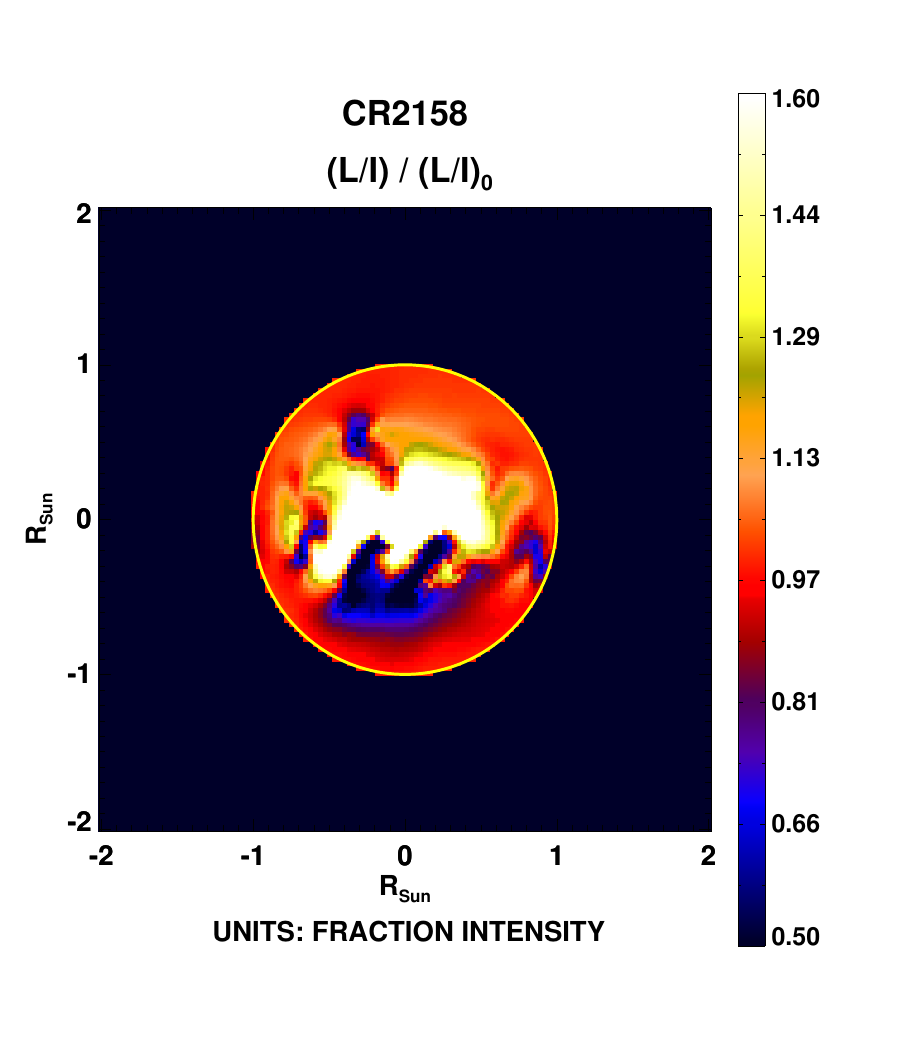}
    \includegraphics[width=0.11\textwidth,trim={1.1cm 2.5cm 0cm 2.9cm},clip]{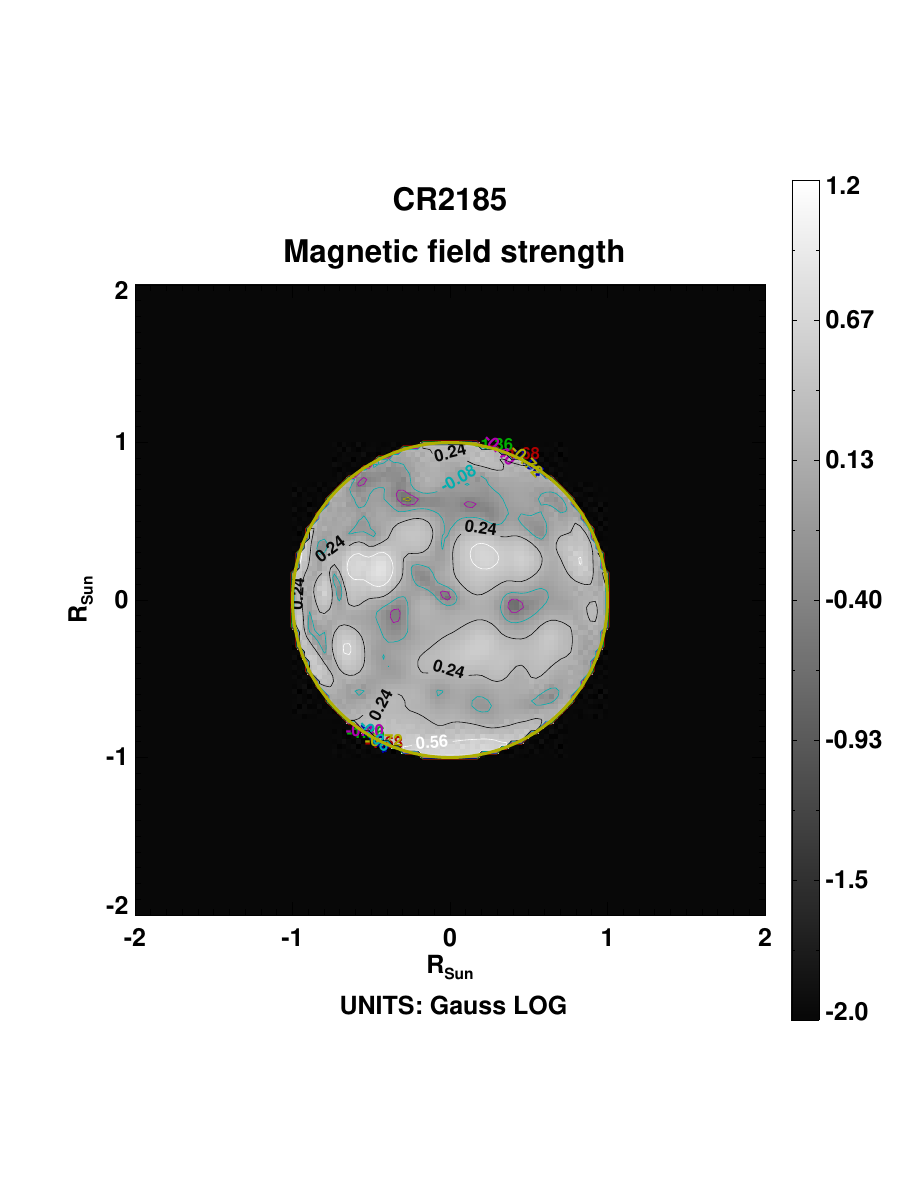}
    \includegraphics[width=0.11\textwidth,trim={1.1cm 2.5cm 0cm 2.9cm},clip]{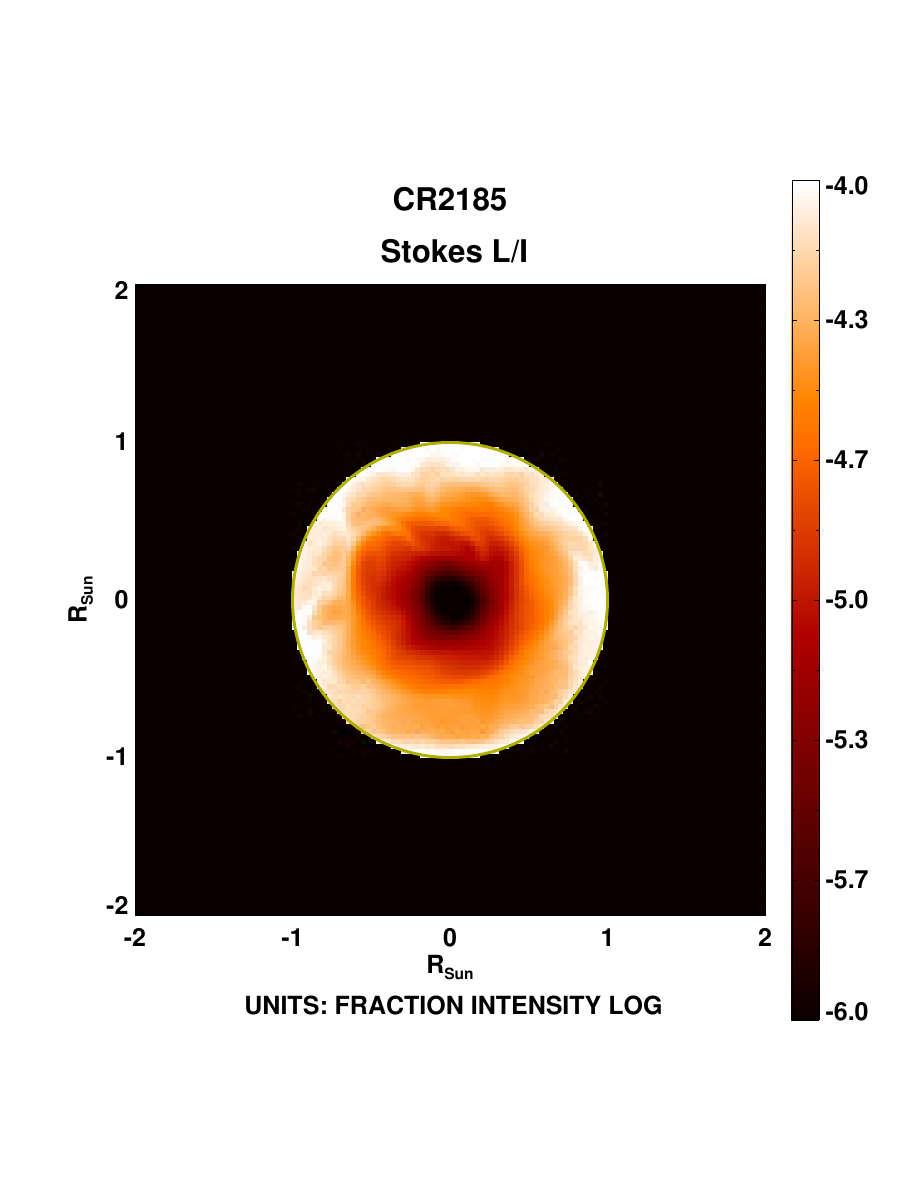}
    \includegraphics[width=0.11\textwidth,trim={1.1cm 2.7cm 0cm 2.9cm},clip]{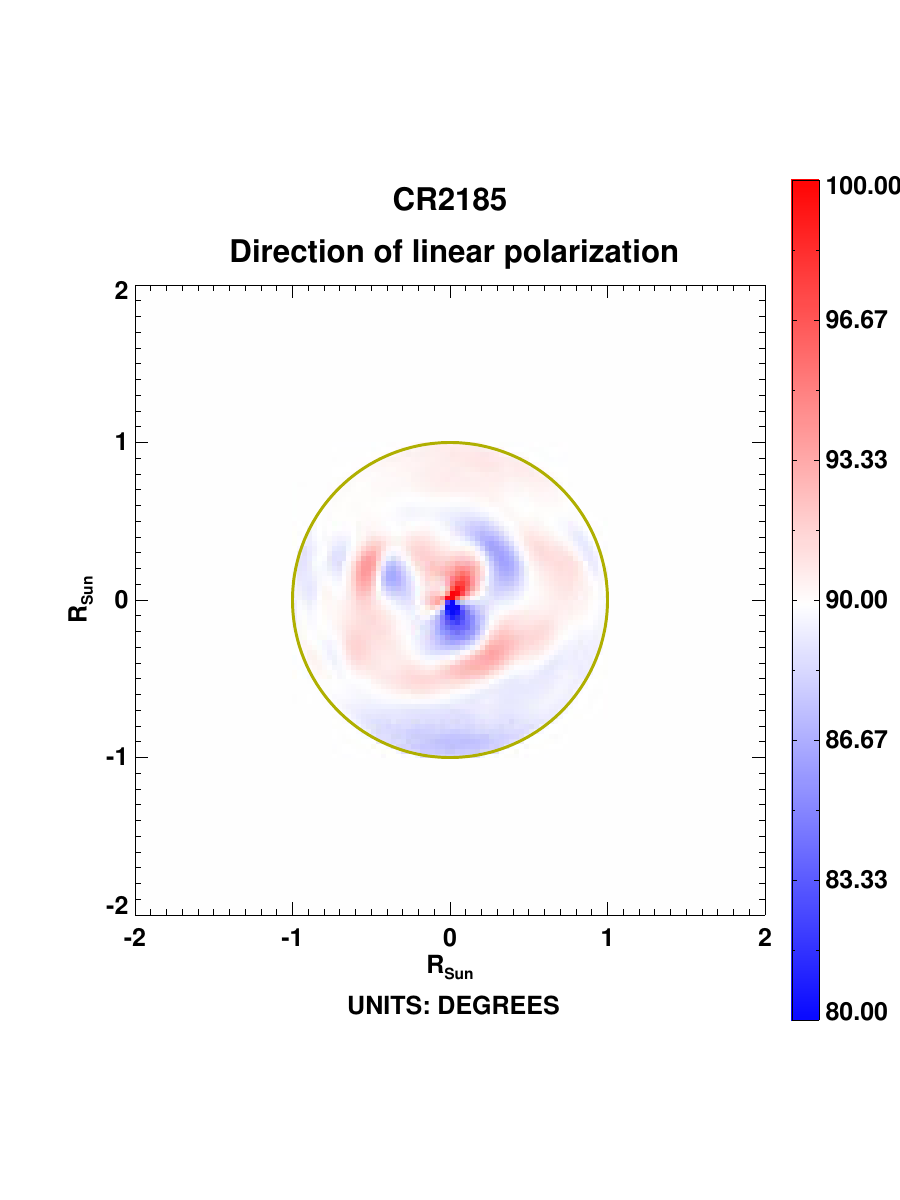}
    \includegraphics[width=0.115\textwidth,trim={0.3cm 1.2cm 0cm 0.5cm},clip]{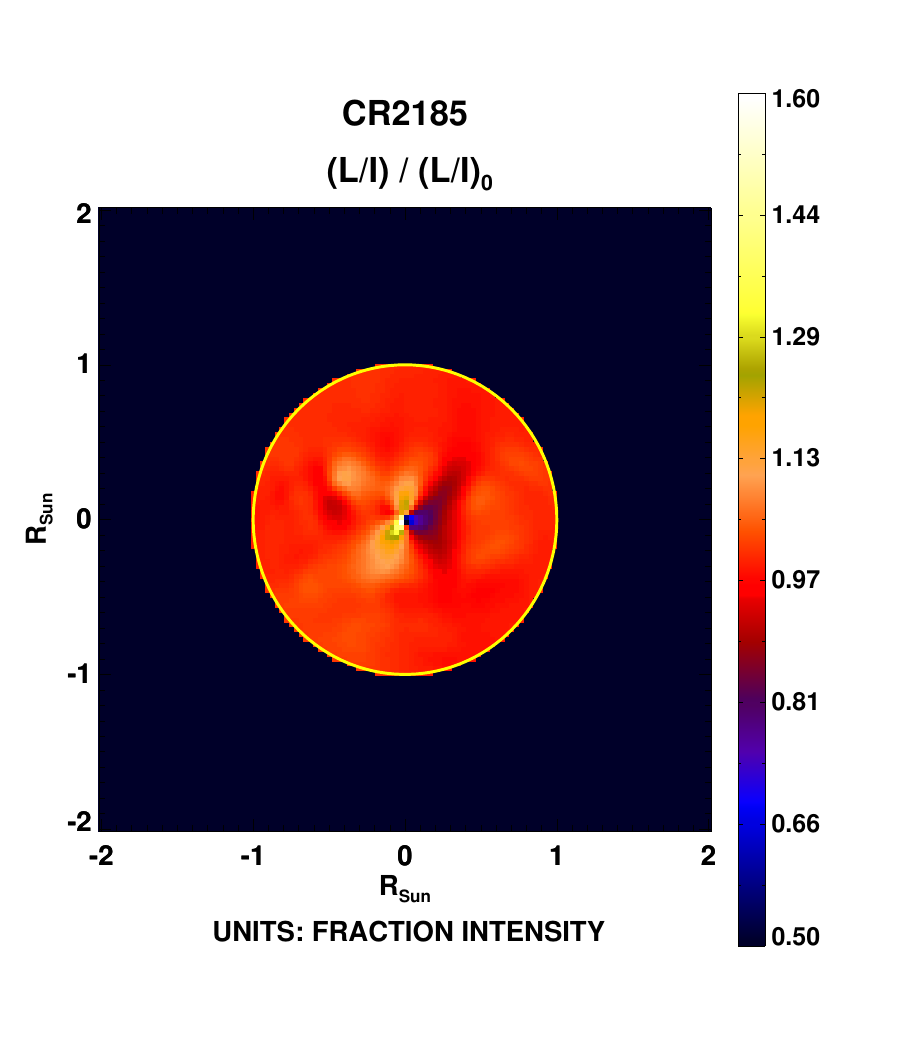}
    \includegraphics[width=0.11\textwidth,trim={1.1cm 2.5cm 0cm 2.9cm},clip]{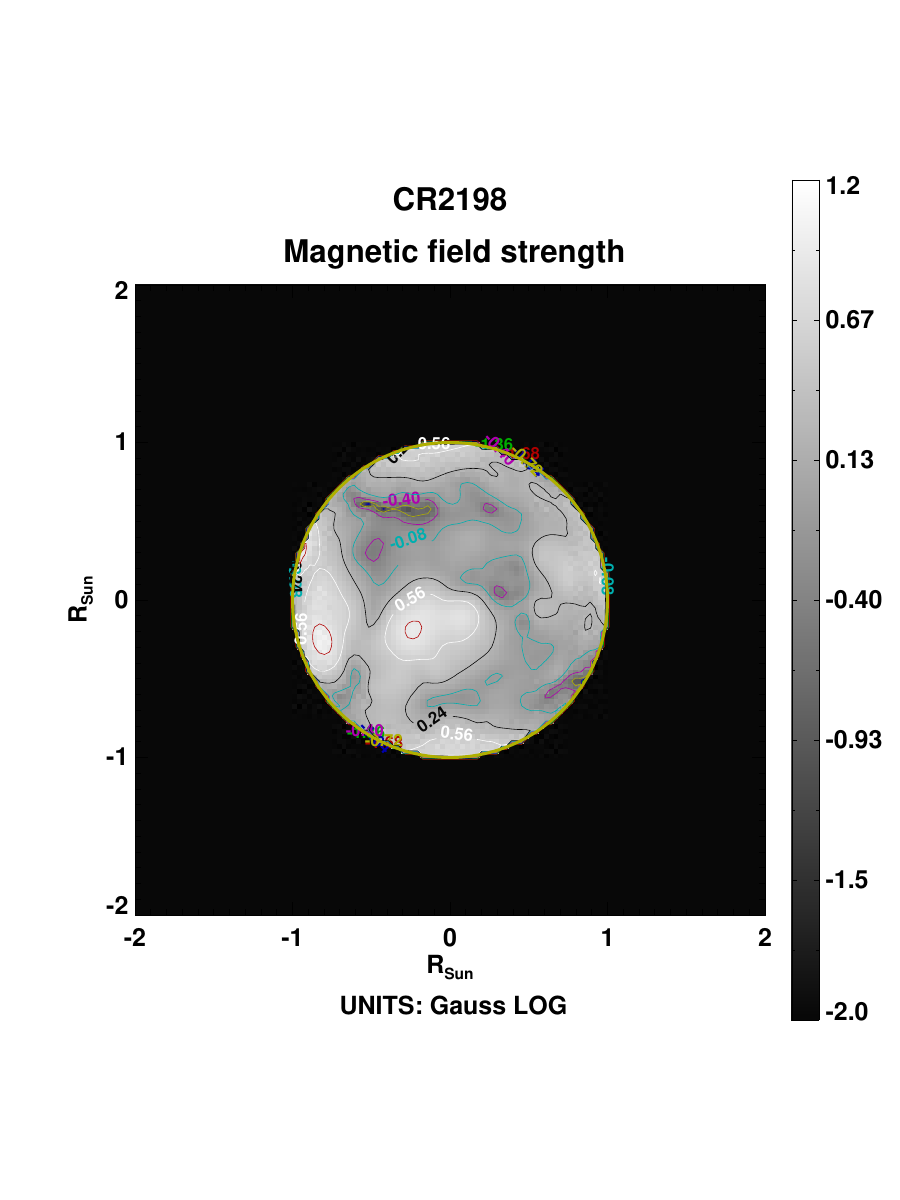}
    \includegraphics[width=0.11\textwidth,trim={1.1cm 2.5cm 0cm 2.9cm},clip]{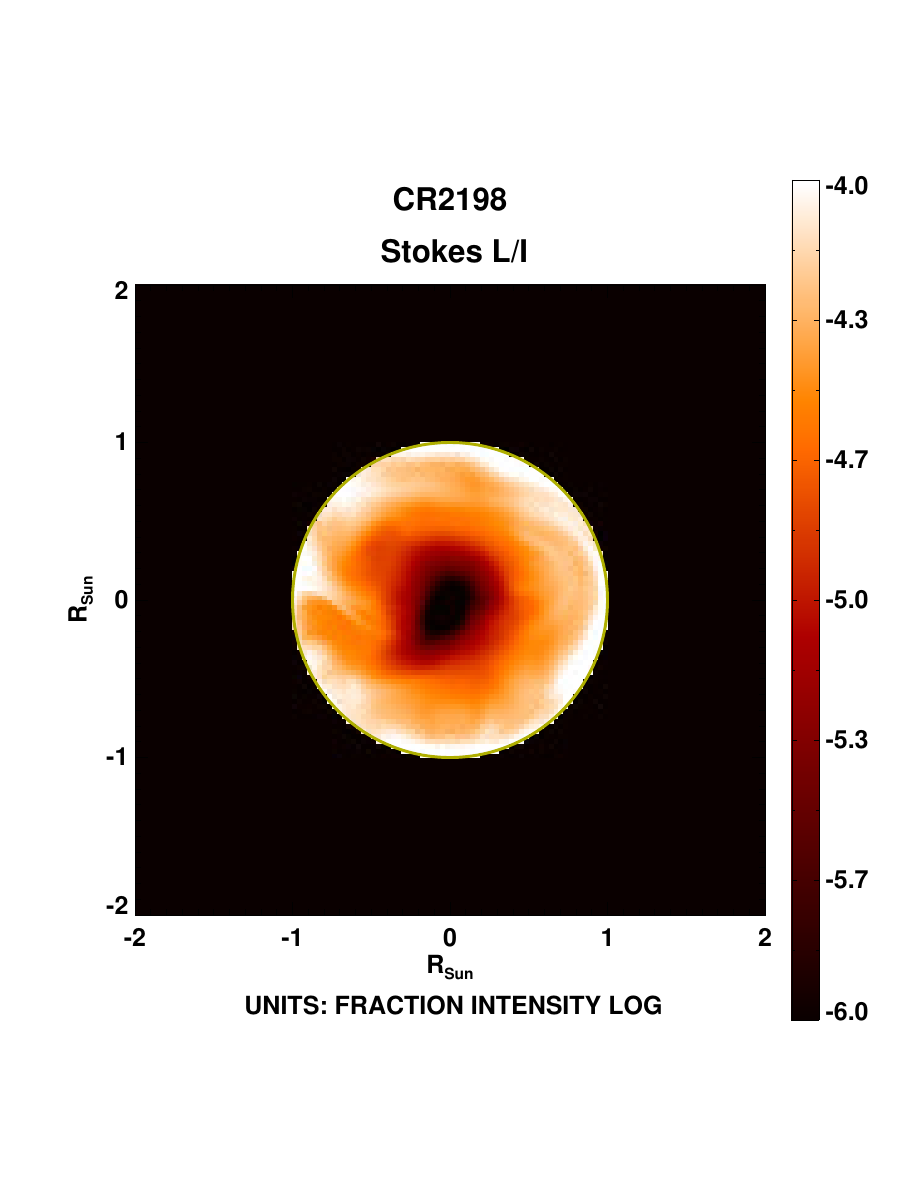}
    \includegraphics[width=0.11\textwidth,trim={1.1cm 2.7cm 0cm 2.9cm},clip]{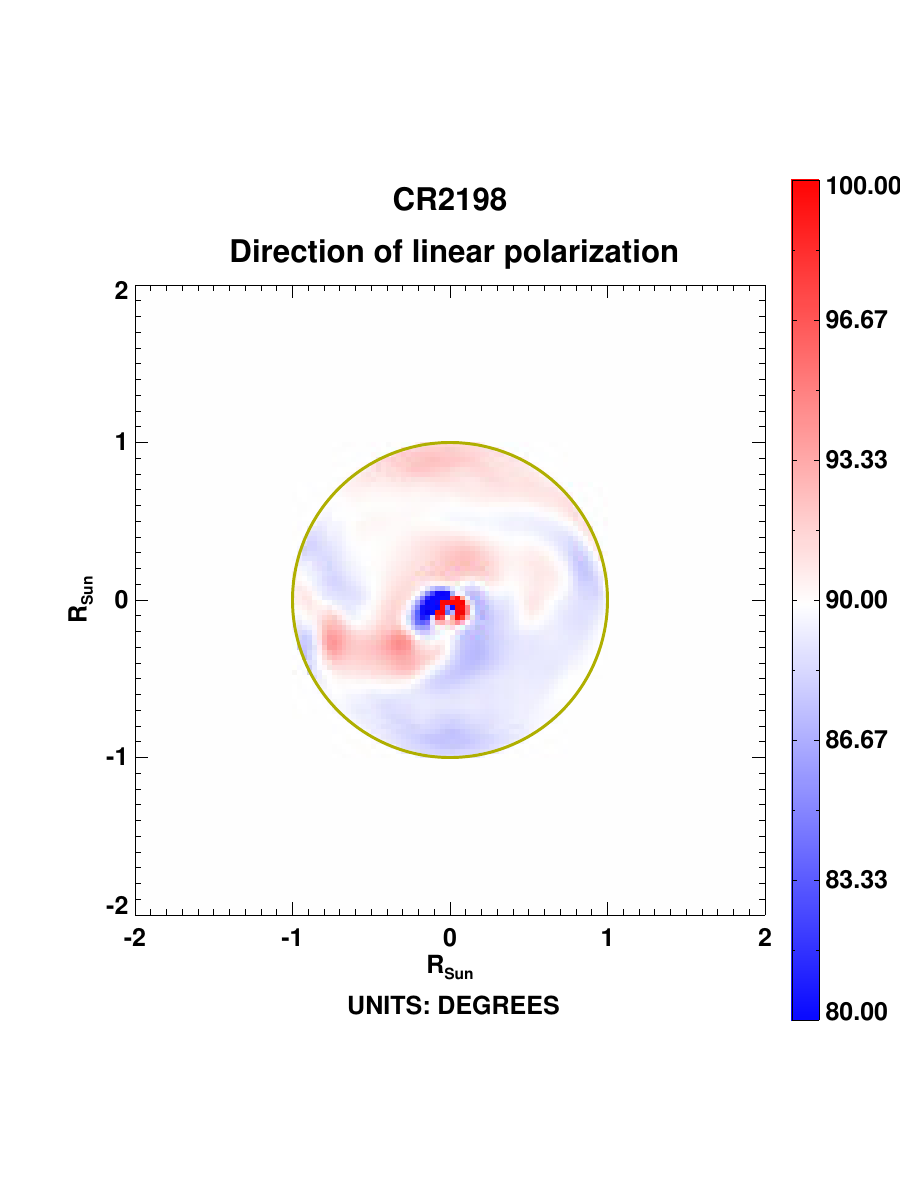}
    \includegraphics[width=0.115\textwidth,trim={0.3cm 1.2cm 0cm 0.5cm},clip]{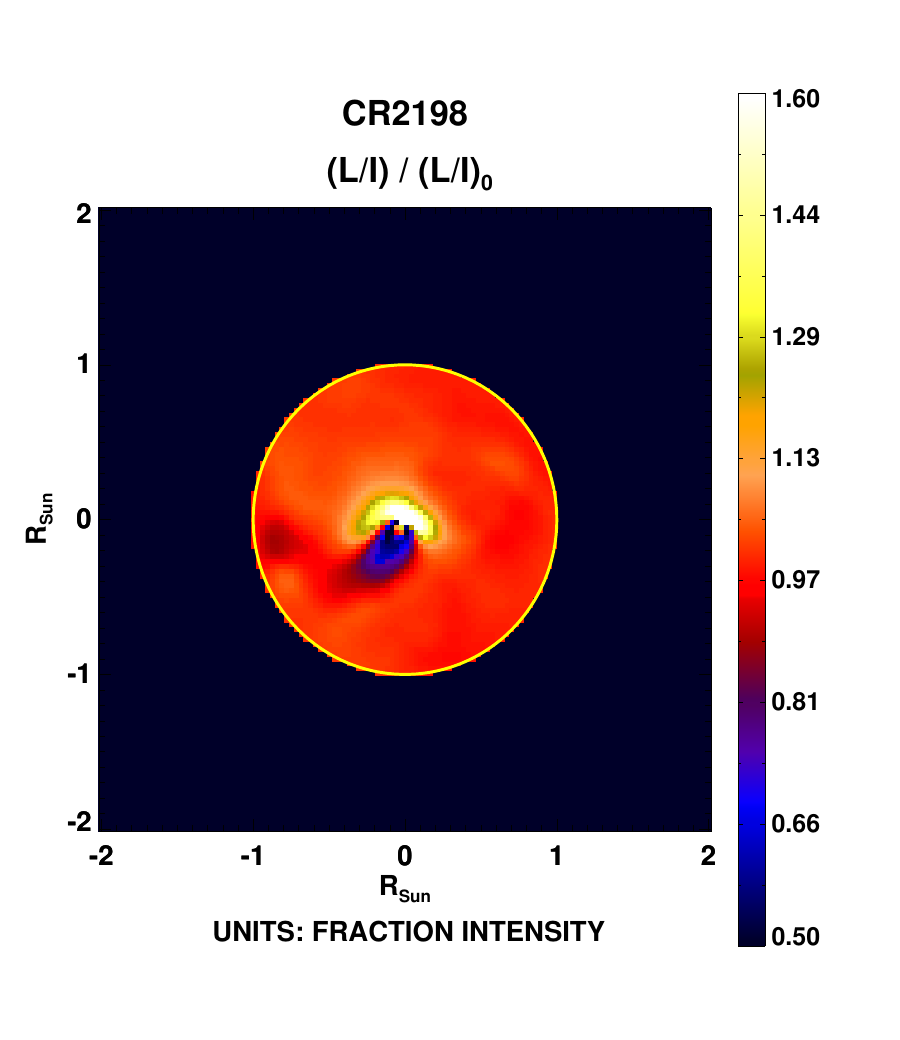}
    \includegraphics[width=0.11\textwidth,trim={1.1cm 2.5cm 0cm 2.9cm},clip]{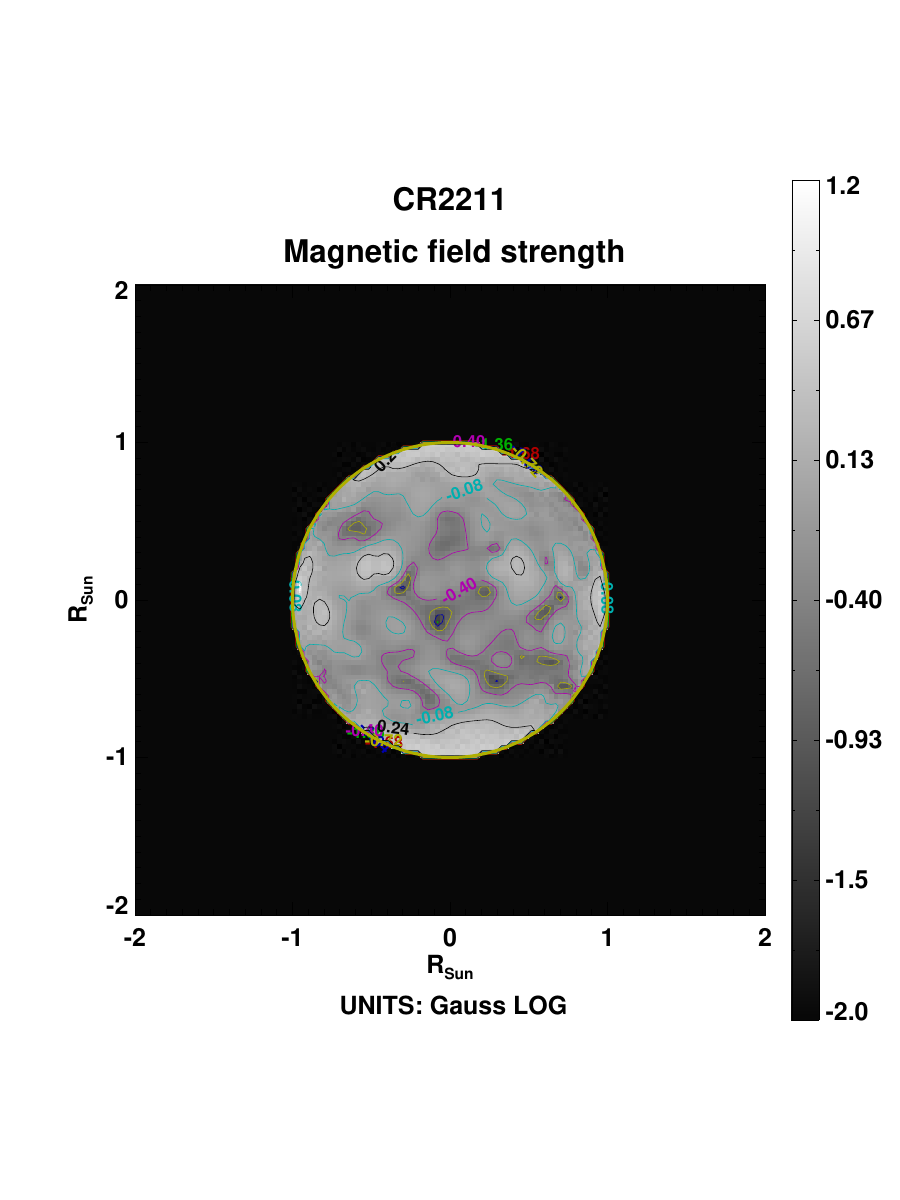}
    \includegraphics[width=0.11\textwidth,trim={1.1cm 2.5cm 0cm 2.9cm},clip]{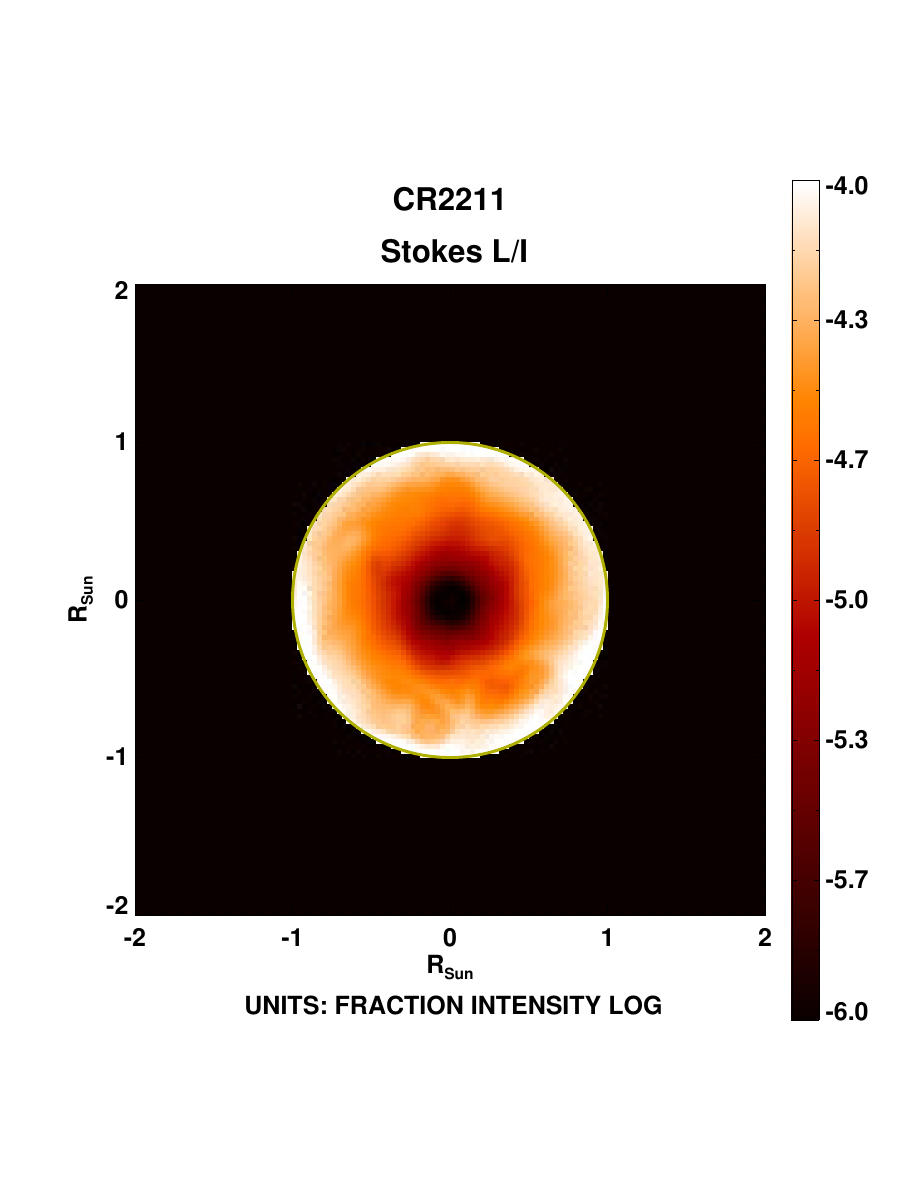}
    \includegraphics[width=0.11\textwidth,trim={1.1cm 2.7cm 0cm 2.9cm},clip]{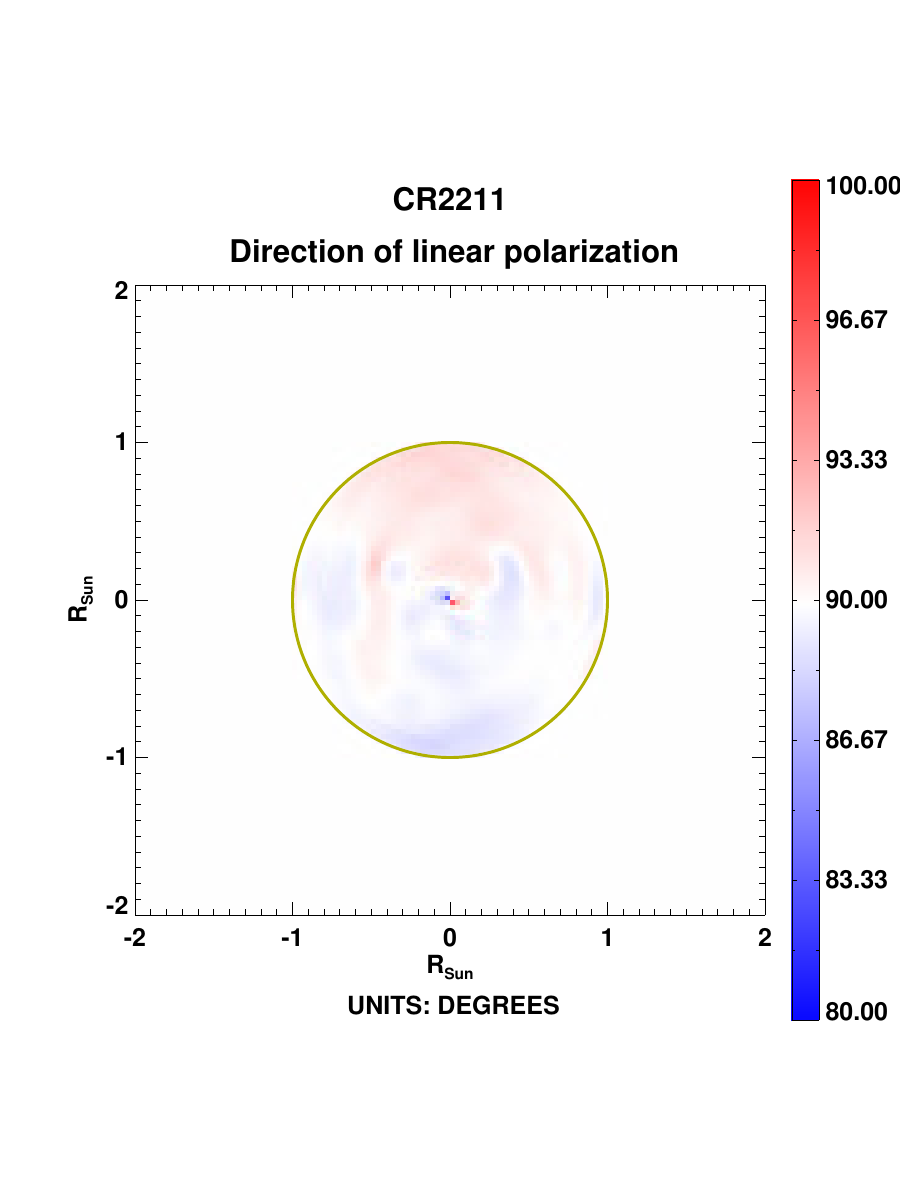}
    \includegraphics[width=0.115\textwidth,trim={0.3cm 1.2cm 0cm 0.5cm},clip]{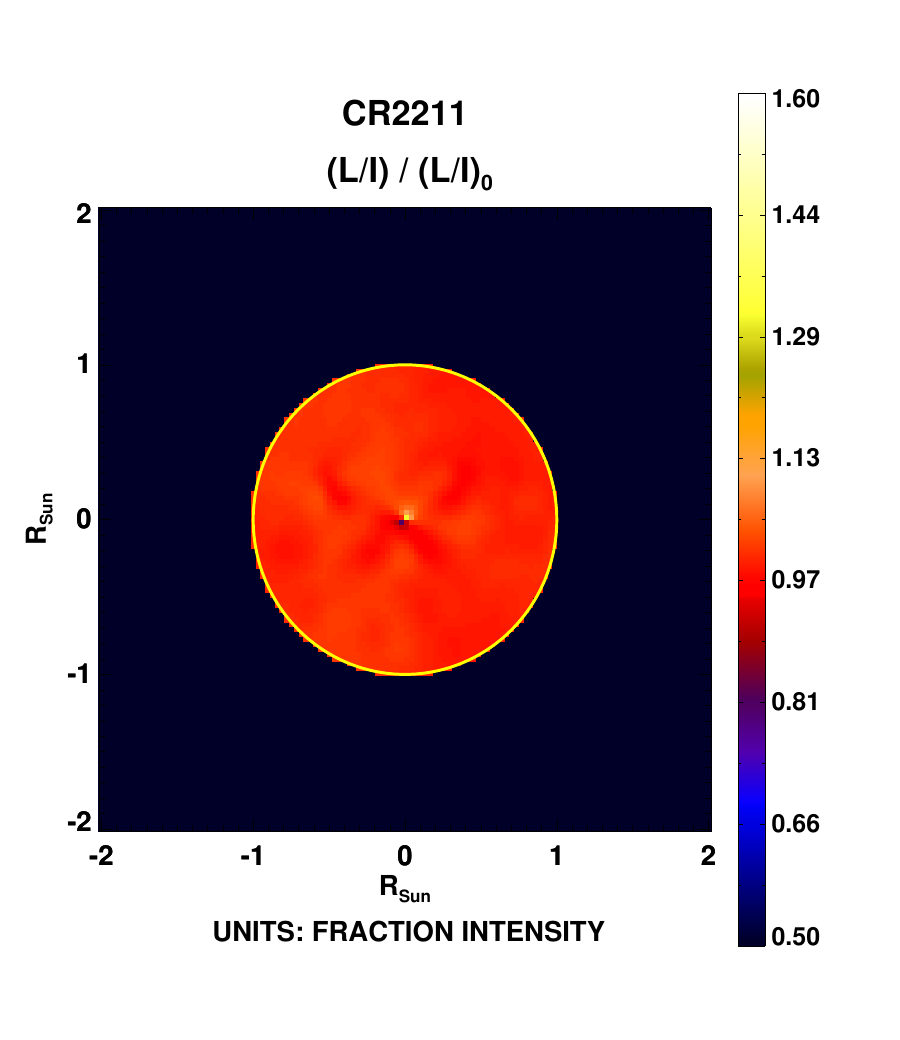}
    \includegraphics[width=0.11\textwidth,trim={1.1cm 2.5cm 0cm 2.9cm},clip]{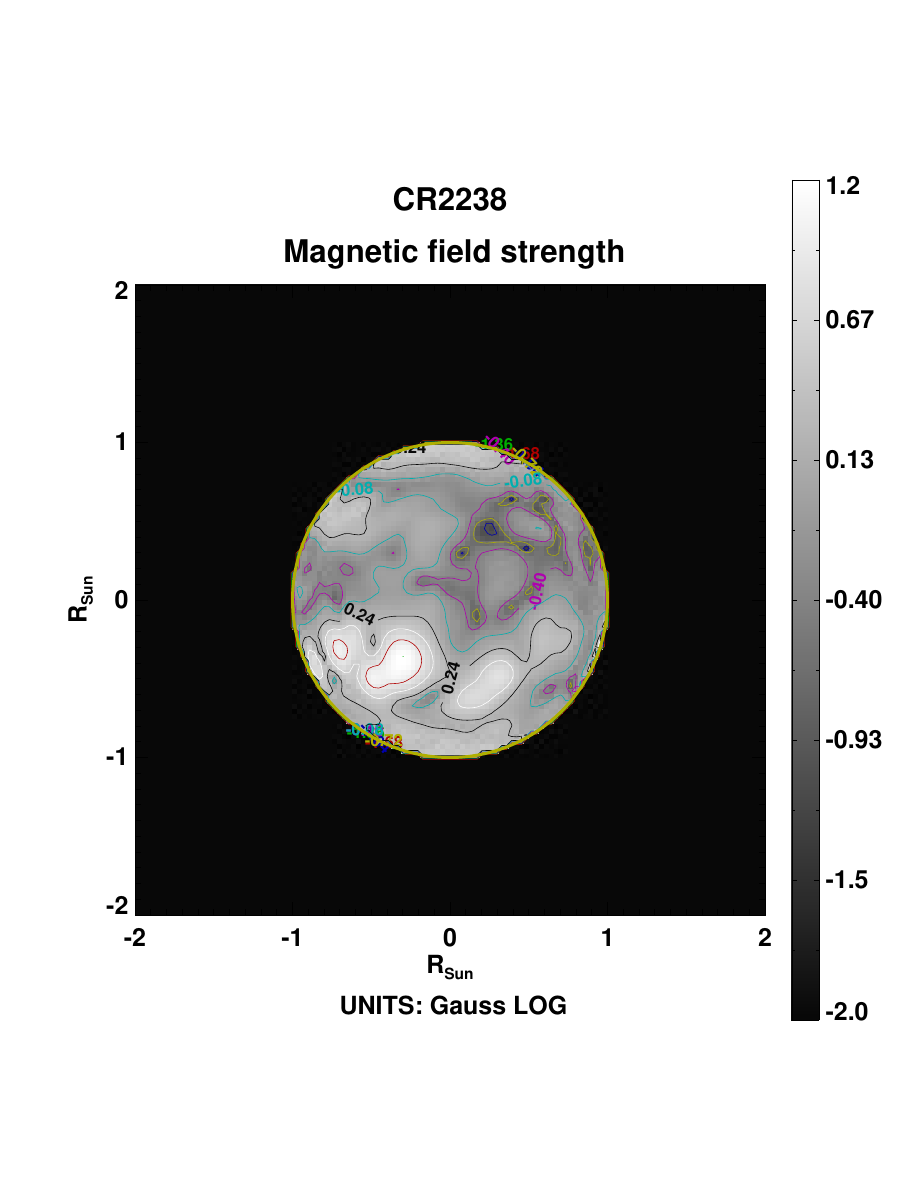}
    \includegraphics[width=0.11\textwidth,trim={1.1cm 2.5cm 0cm 2.9cm},clip]{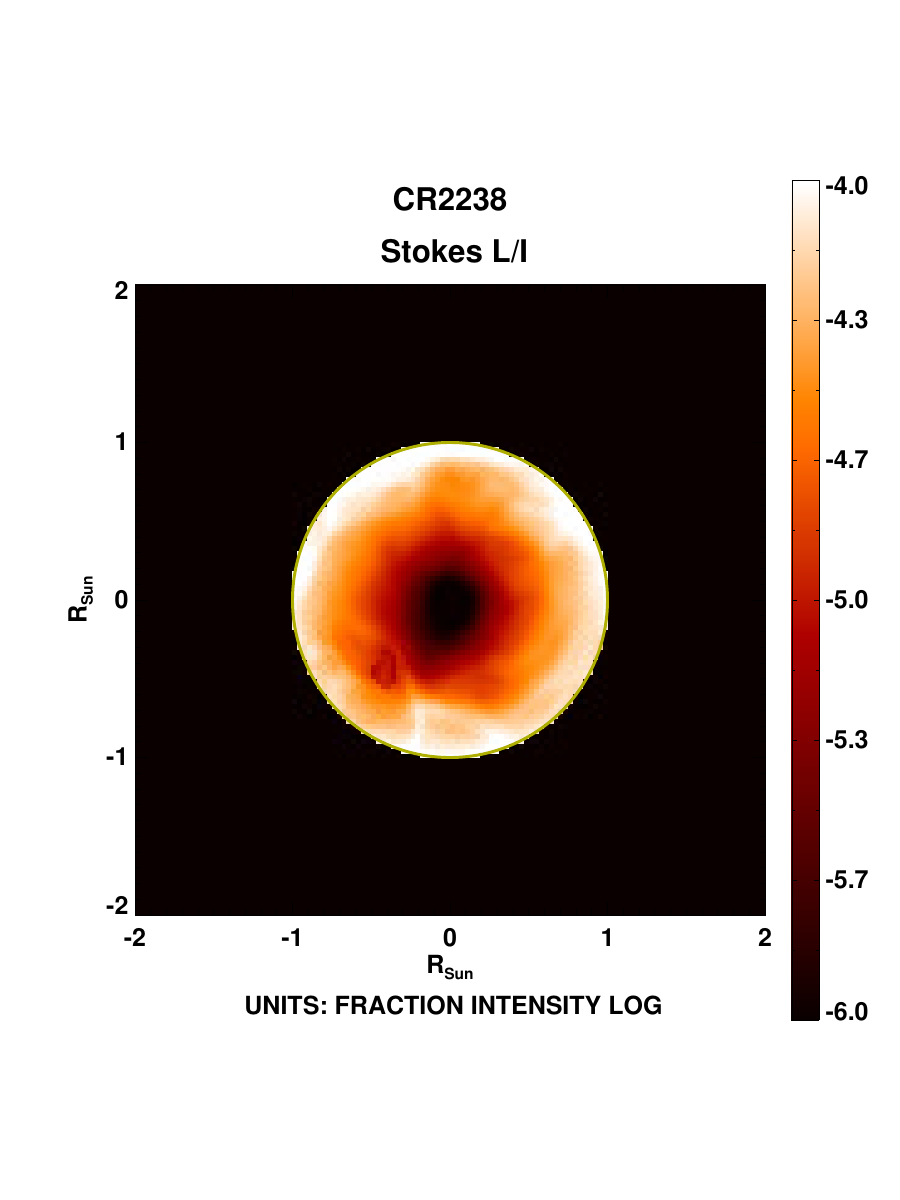}
    \includegraphics[width=0.11\textwidth,trim={1.1cm 2.7cm 0cm 2.9cm},clip]{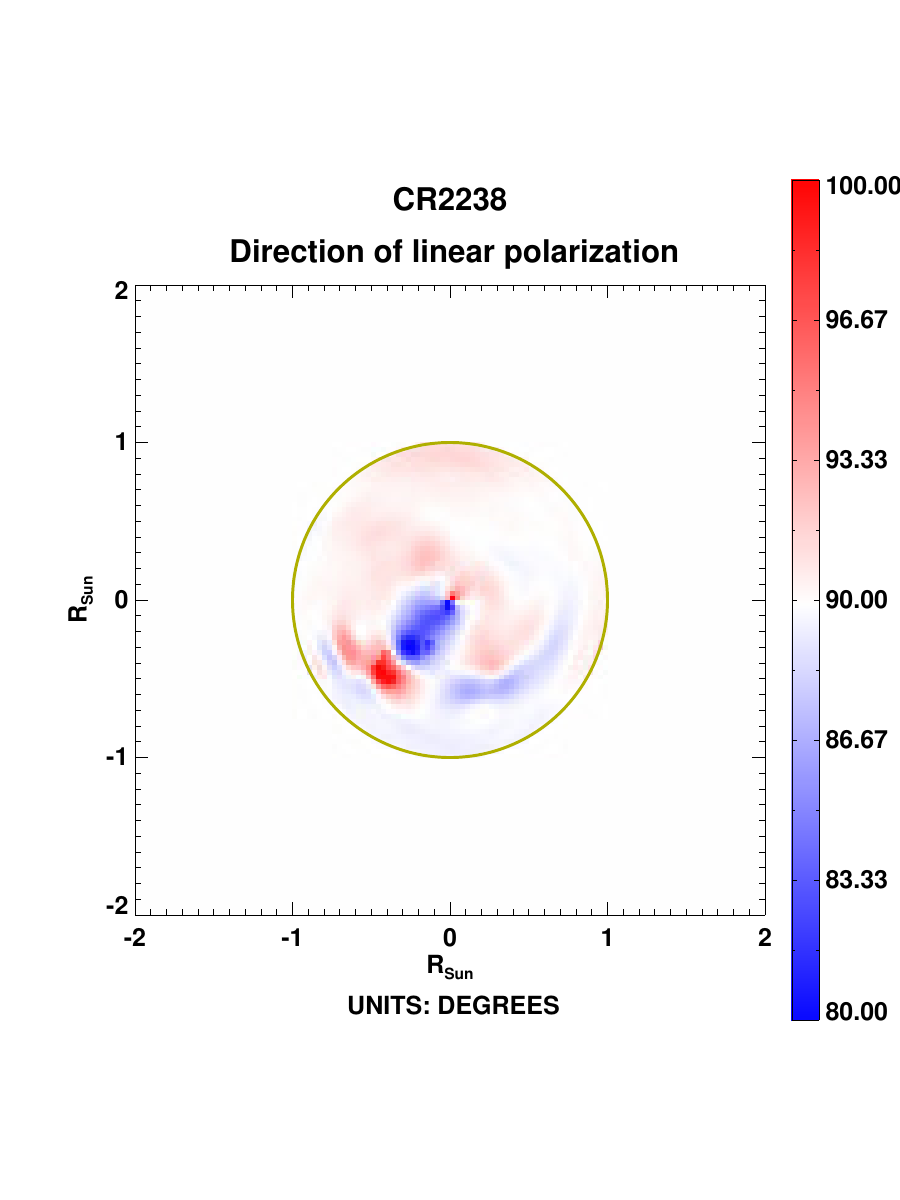}
    \includegraphics[width=0.115\textwidth,trim={0.3cm 1.2cm 0cm 0.5cm},clip]{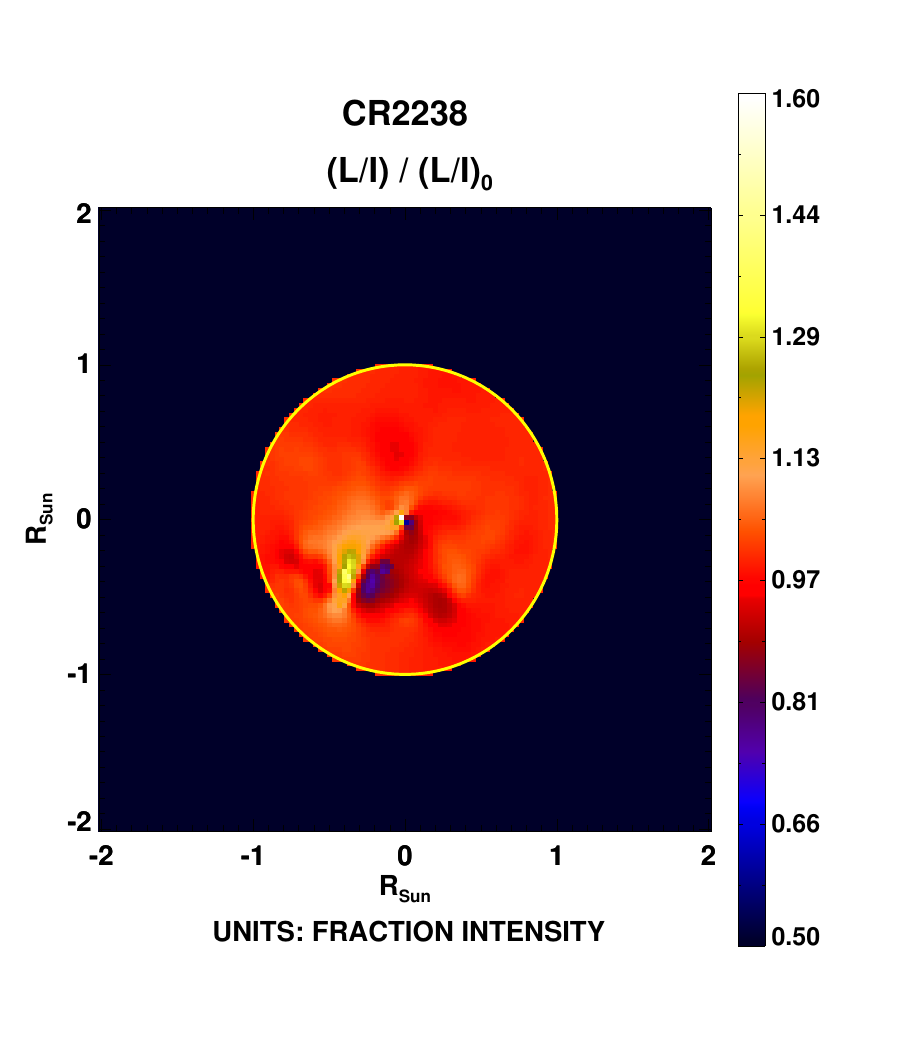}
    \caption{Same as Figure \ref{fig:bmag_stokesloi_az_ondisk} but for rest of the CR simulations.}
    \label{fig:supp_bmag_stokesloi_az_ondisk}
\end{figure}
\begin{figure}[htbp]
\centering
    \includegraphics[width=0.11\textwidth,trim={1.1cm 2.5cm 0cm 2.9cm},clip]{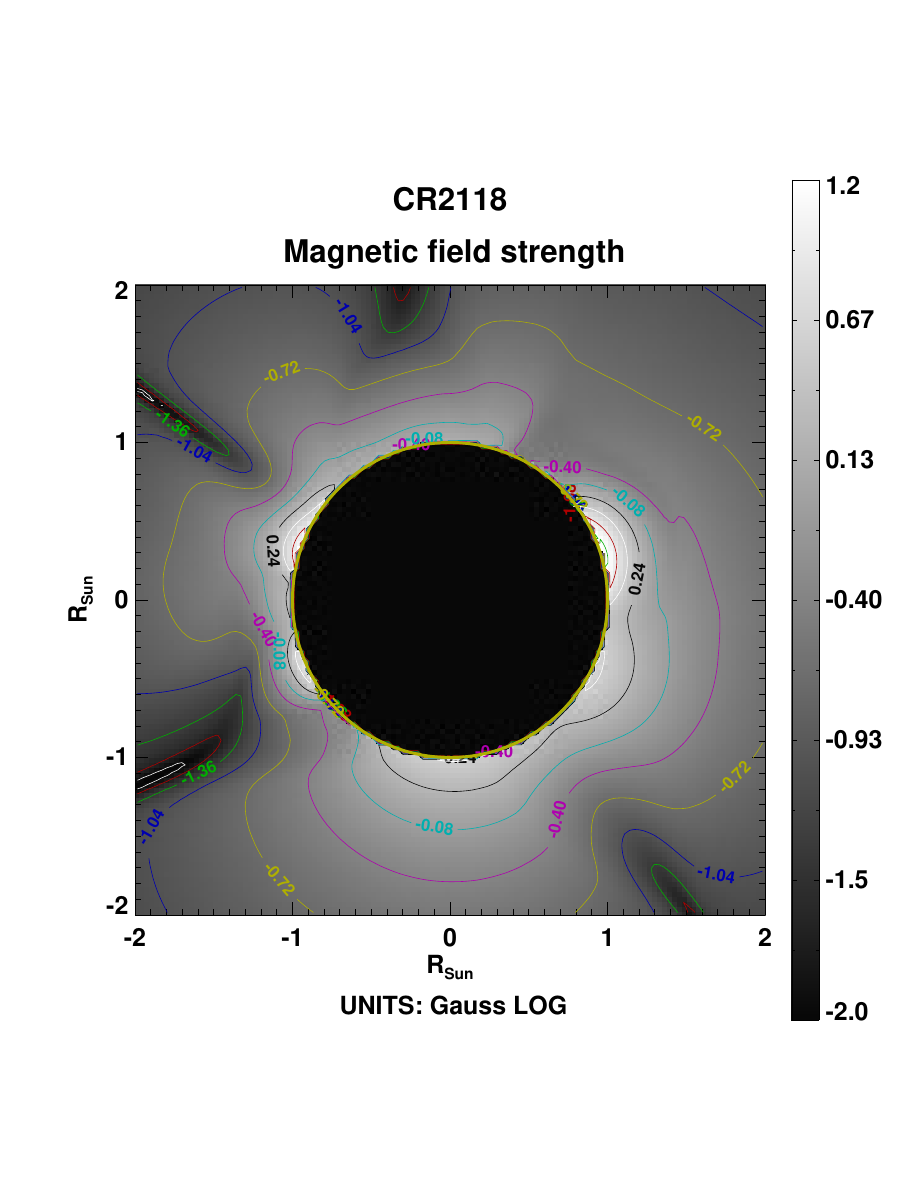}
    \includegraphics[width=0.11\textwidth,trim={1.1cm 2.5cm 0cm 2.9cm},clip]{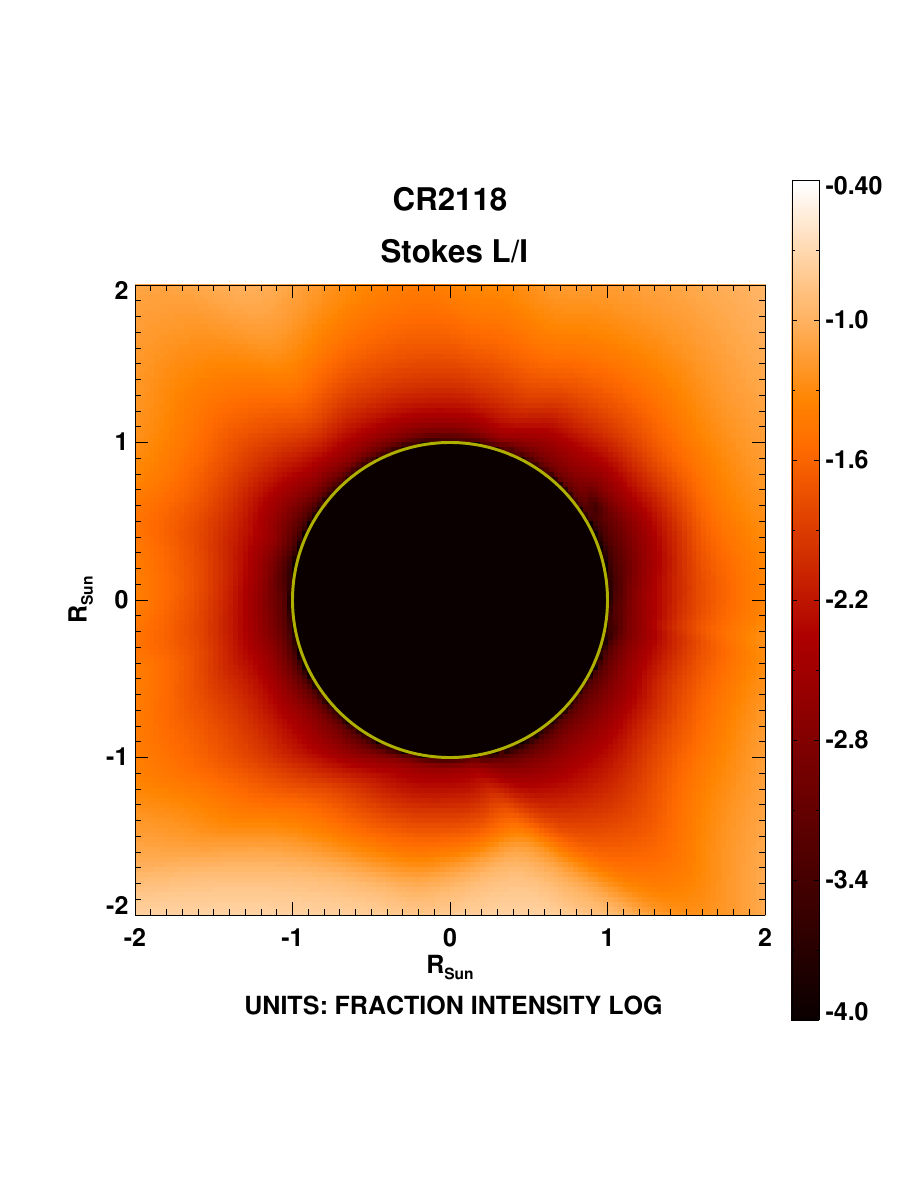}
    \includegraphics[width=0.11\textwidth,trim={1.1cm 2.7cm 0cm 2.9cm},clip]{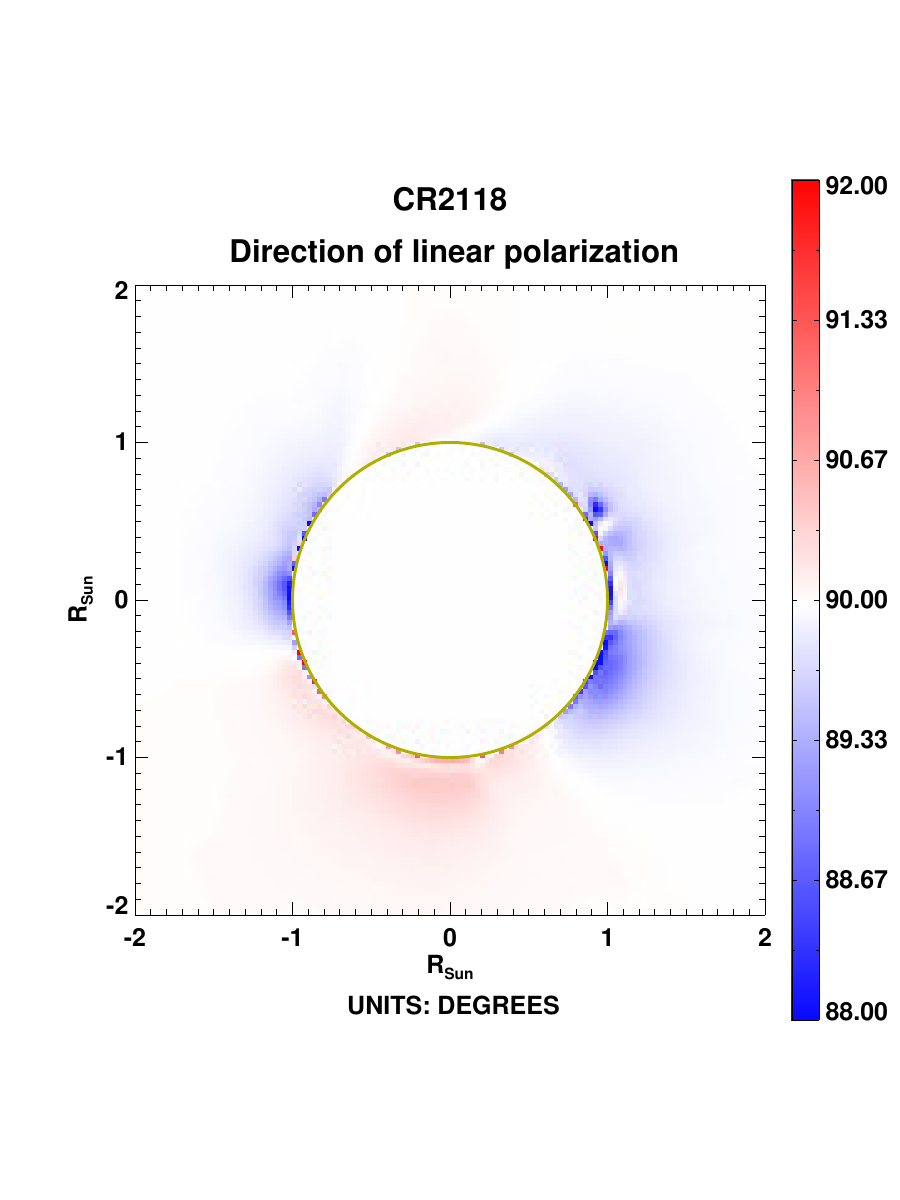}
    \includegraphics[width=0.115\textwidth,trim={0.3cm 1.2cm 0cm 0.5cm},clip]{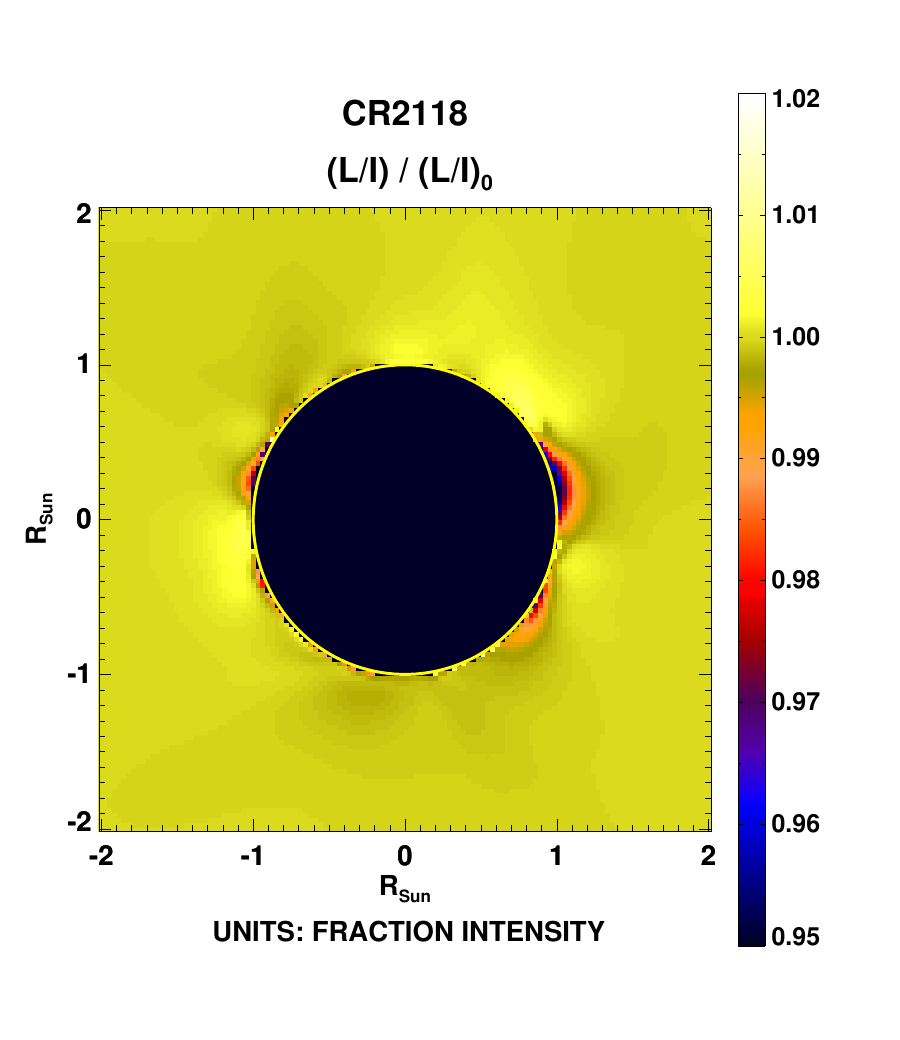}
    \includegraphics[width=0.11\textwidth,trim={1.1cm 2.5cm 0cm 2.9cm},clip]{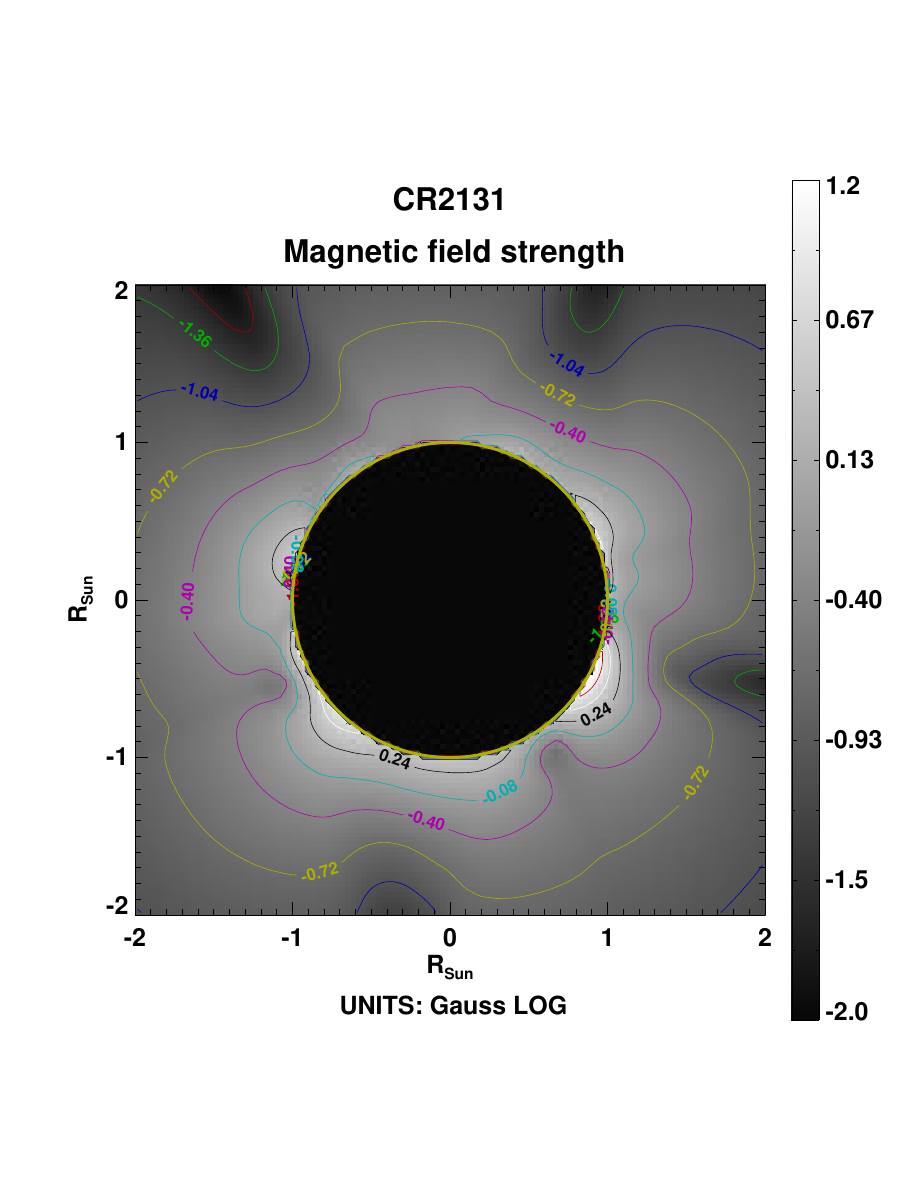}
    \includegraphics[width=0.11\textwidth,trim={1.1cm 2.5cm 0cm 2.9cm},clip]{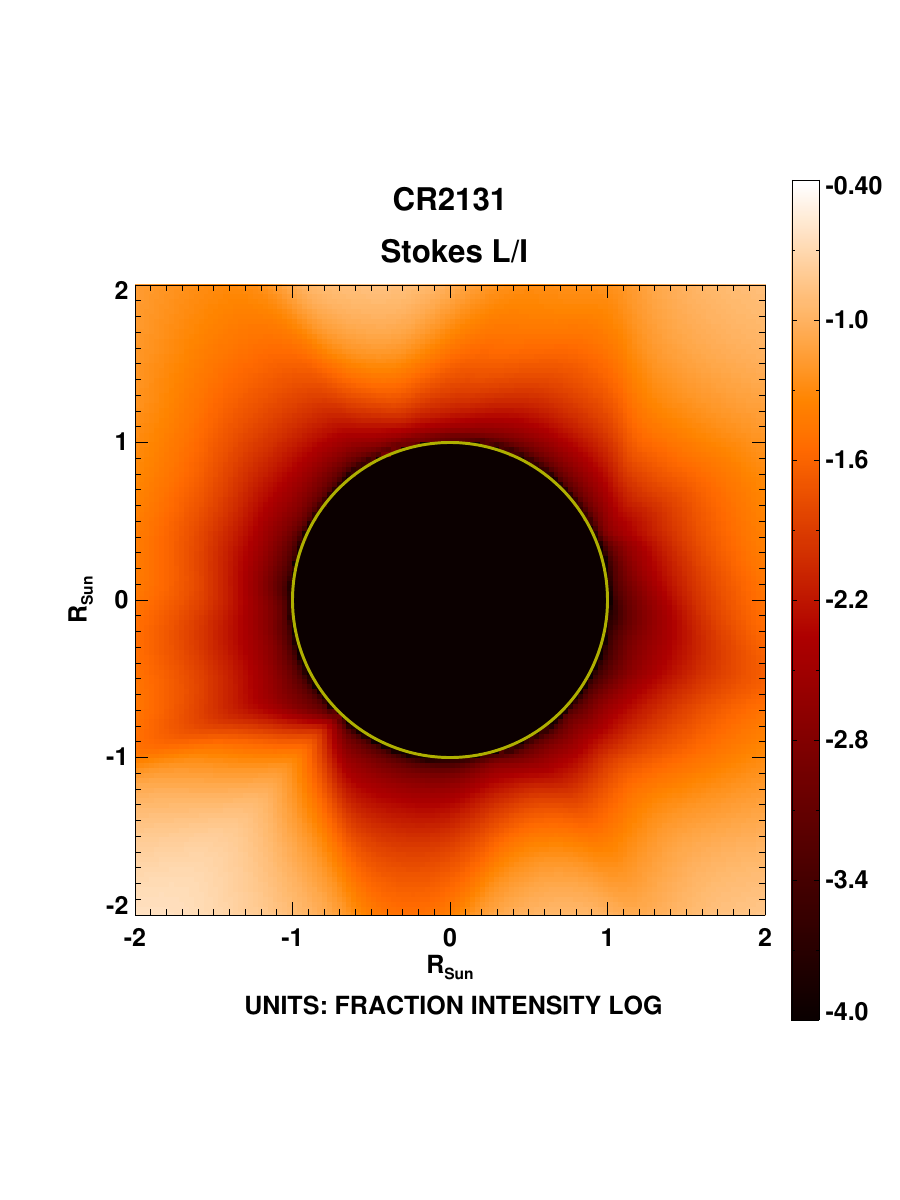}
    \includegraphics[width=0.11\textwidth,trim={1.1cm 2.7cm 0cm 2.9cm},clip]{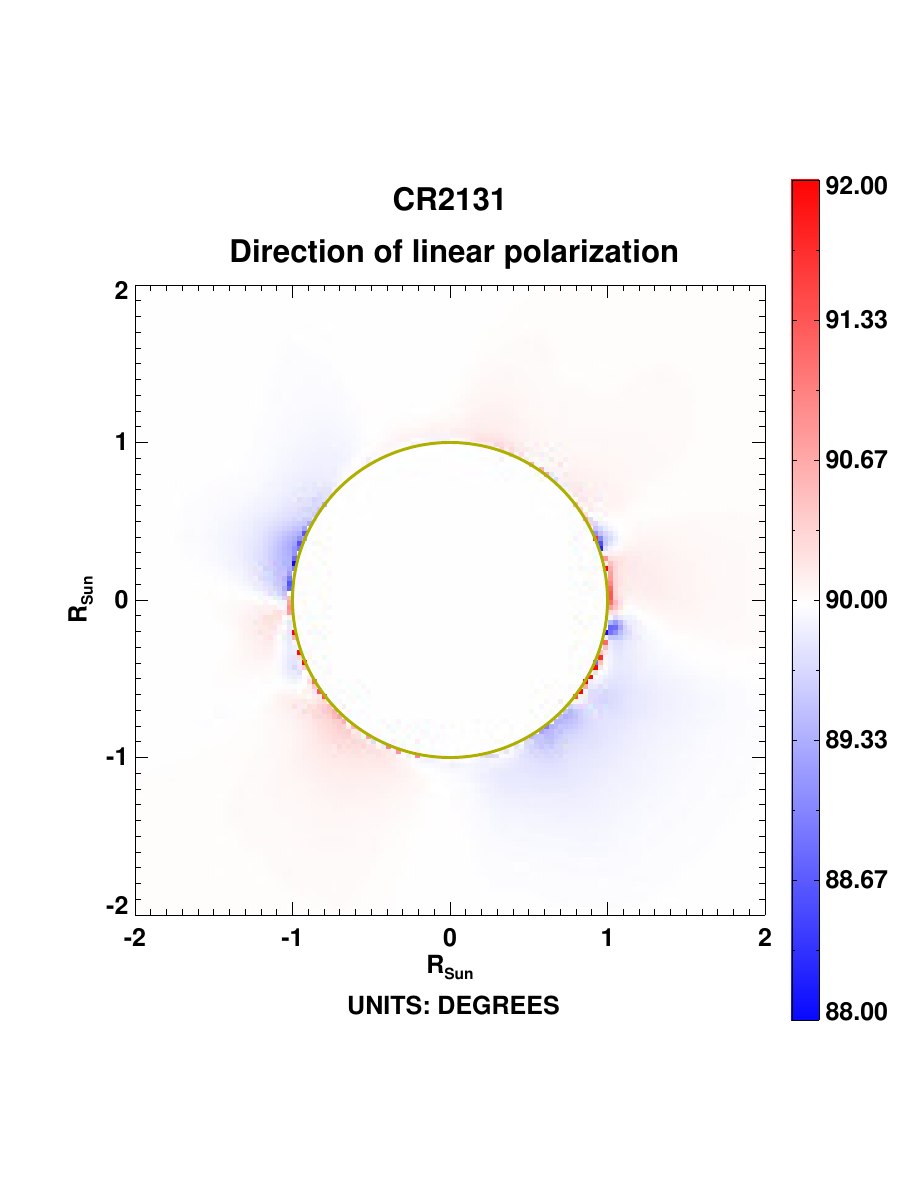}
    \includegraphics[width=0.115\textwidth,trim={0.3cm 1.2cm 0cm 0.5cm},clip]{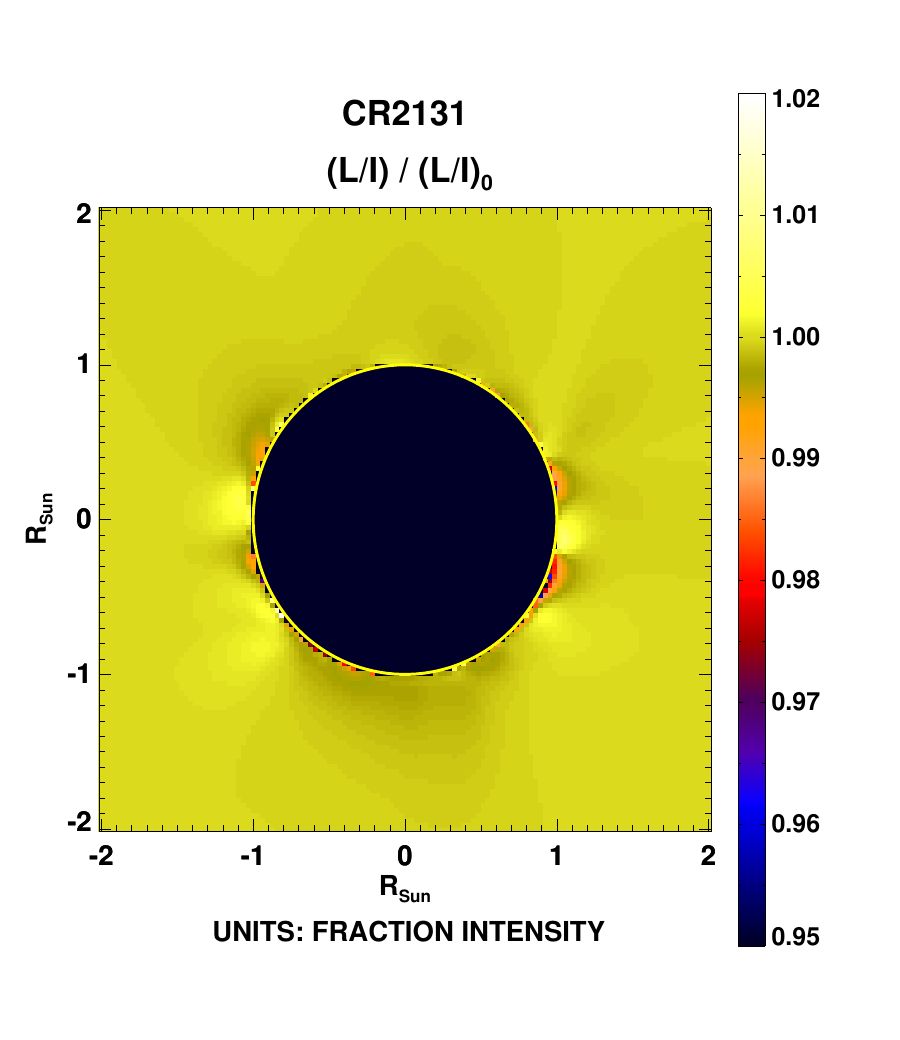}
    \includegraphics[width=0.11\textwidth,trim={1.1cm 2.5cm 0cm 2.9cm},clip]{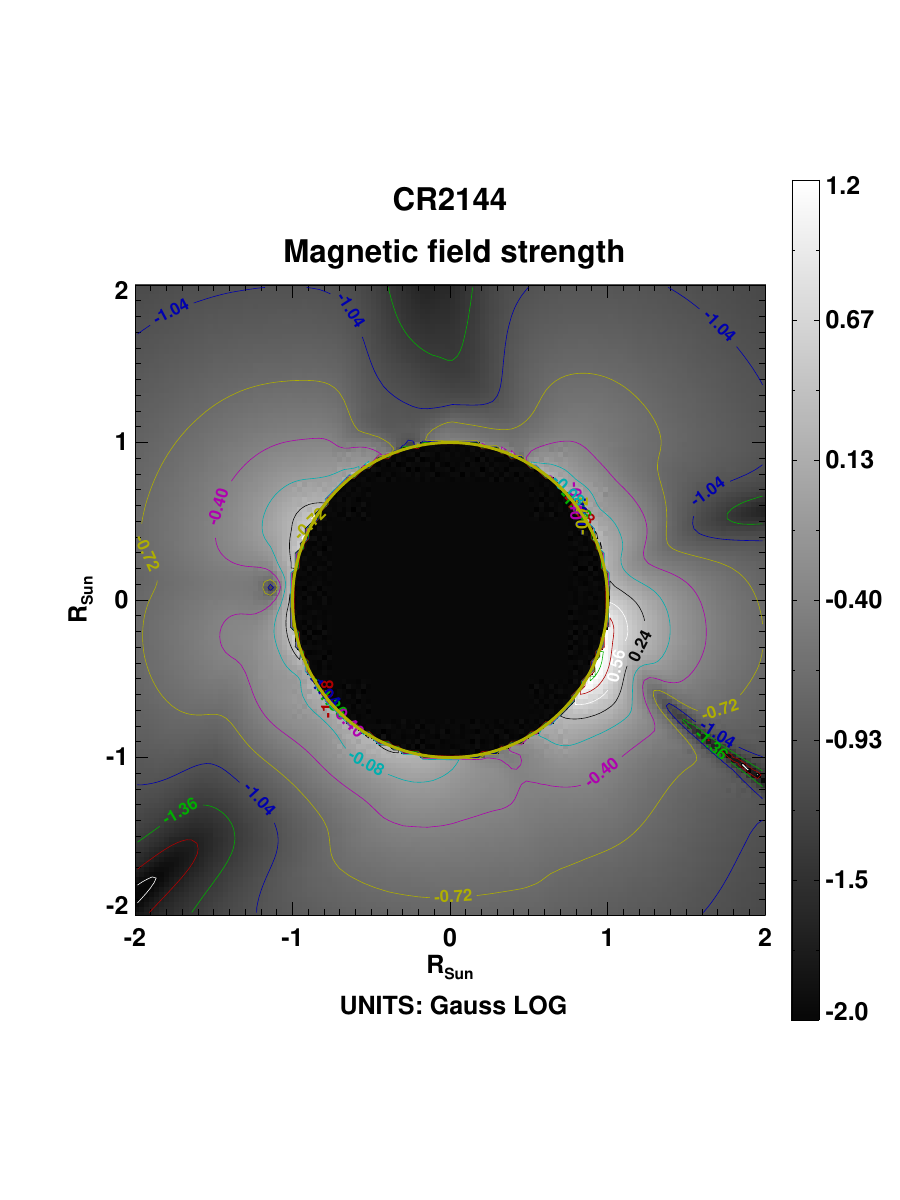}
    \includegraphics[width=0.11\textwidth,trim={1.1cm 2.5cm 0cm 2.9cm},clip]{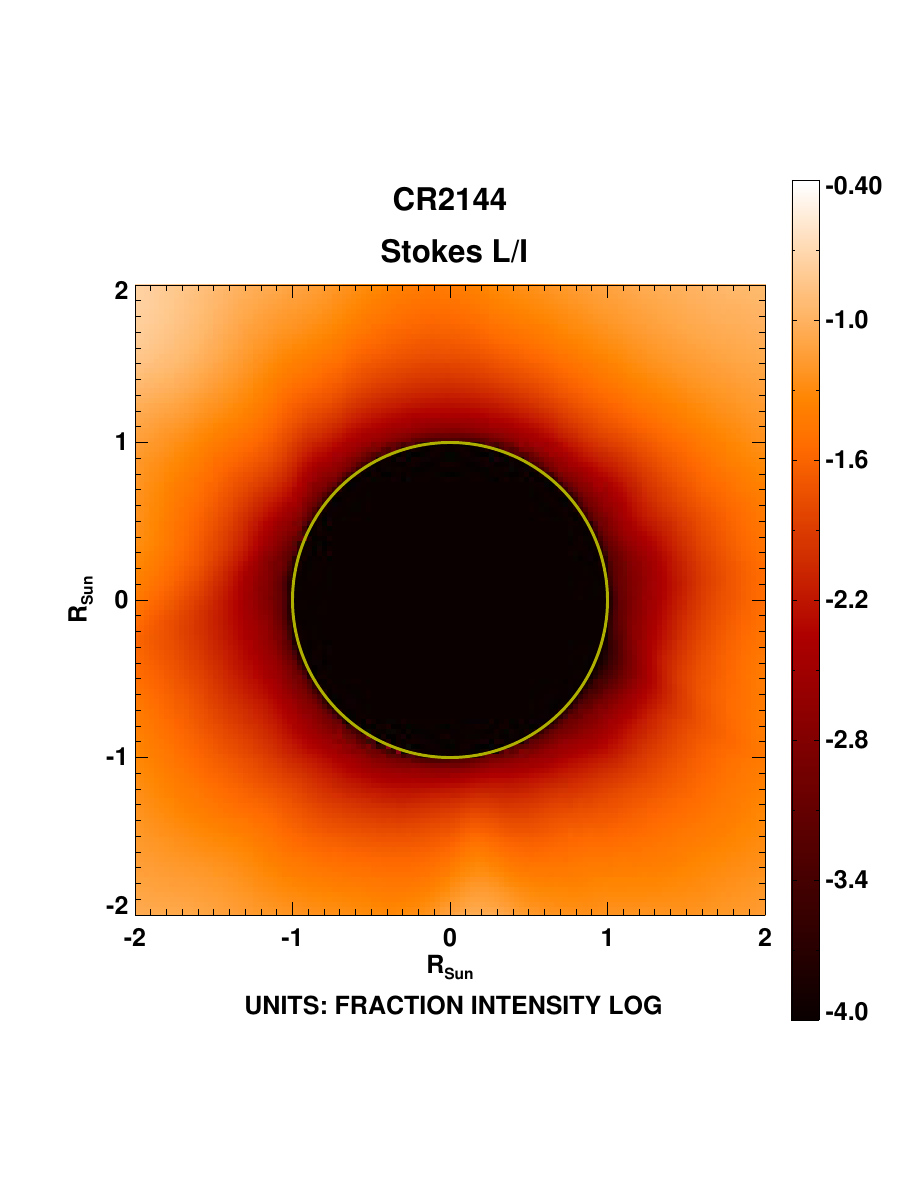}
    \includegraphics[width=0.11\textwidth,trim={1.1cm 2.7cm 0cm 2.9cm},clip]{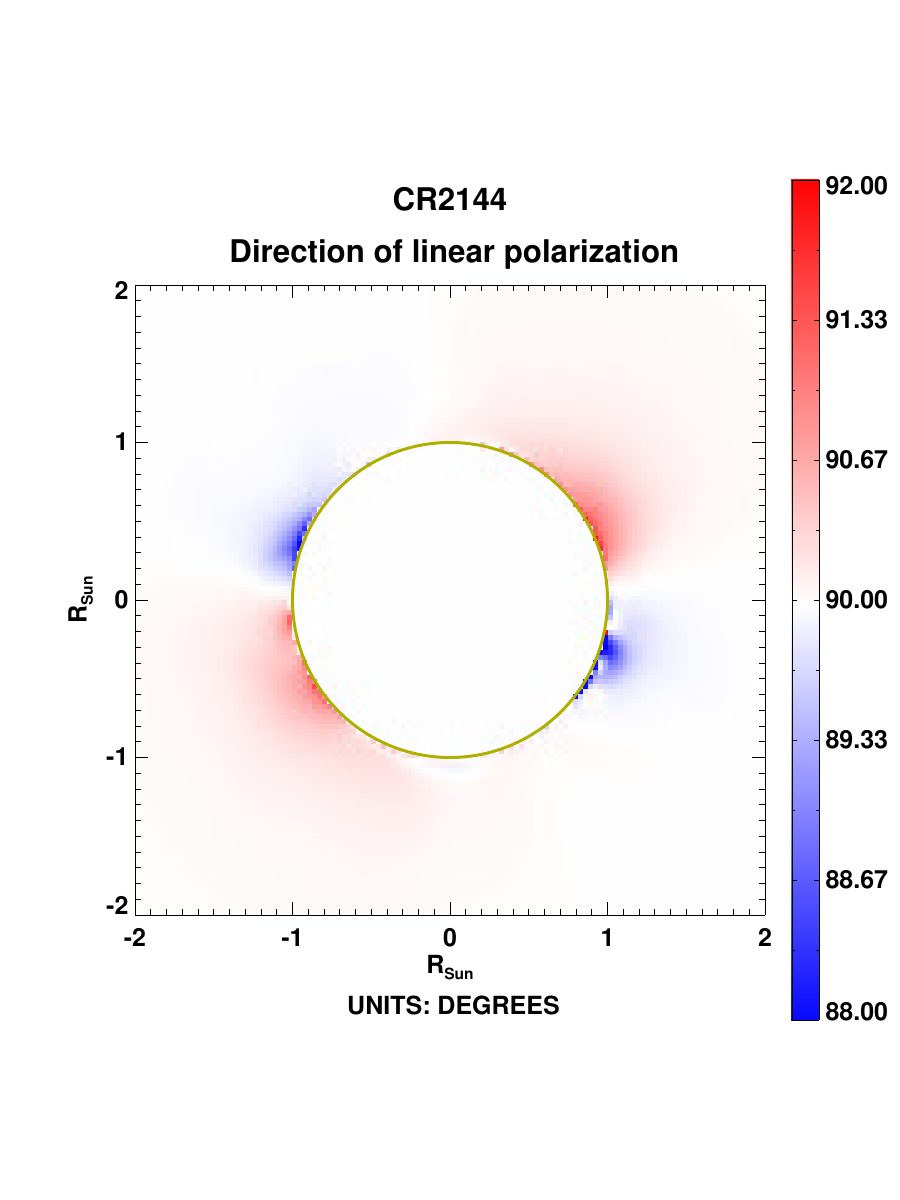}
    \includegraphics[width=0.115\textwidth,trim={0.3cm 1.2cm 0cm 0.5cm},clip]{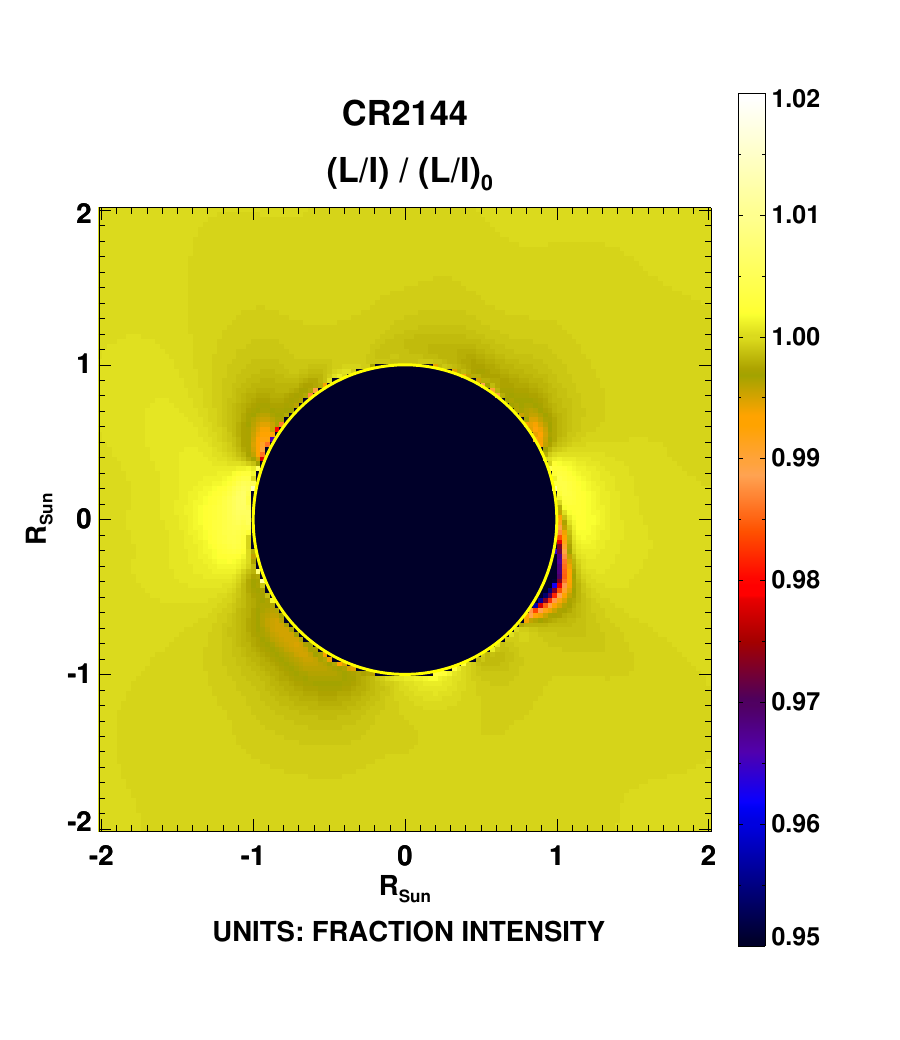}
    \includegraphics[width=0.11\textwidth,trim={1.1cm 2.5cm 0cm 2.9cm},clip]{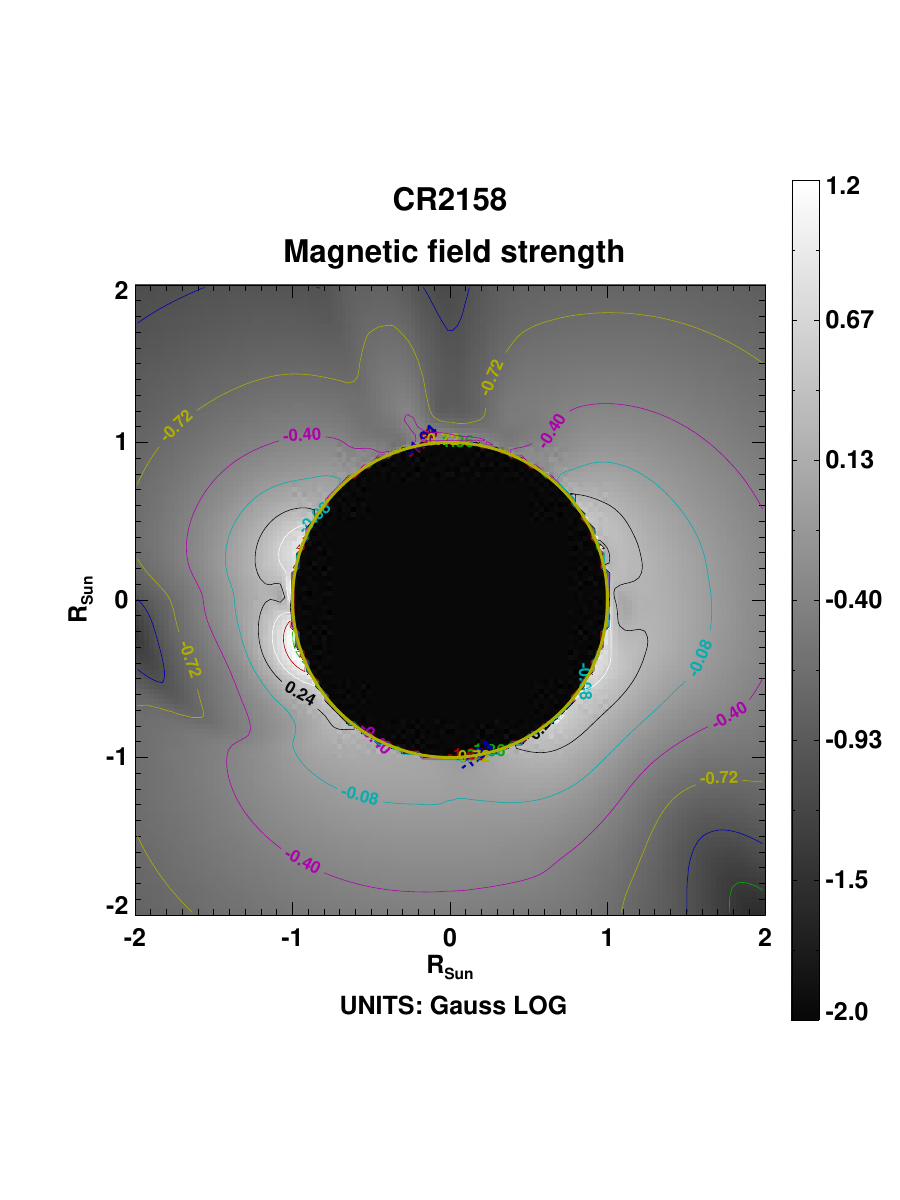}
    \includegraphics[width=0.11\textwidth,trim={1.1cm 2.5cm 0cm 2.9cm},clip]{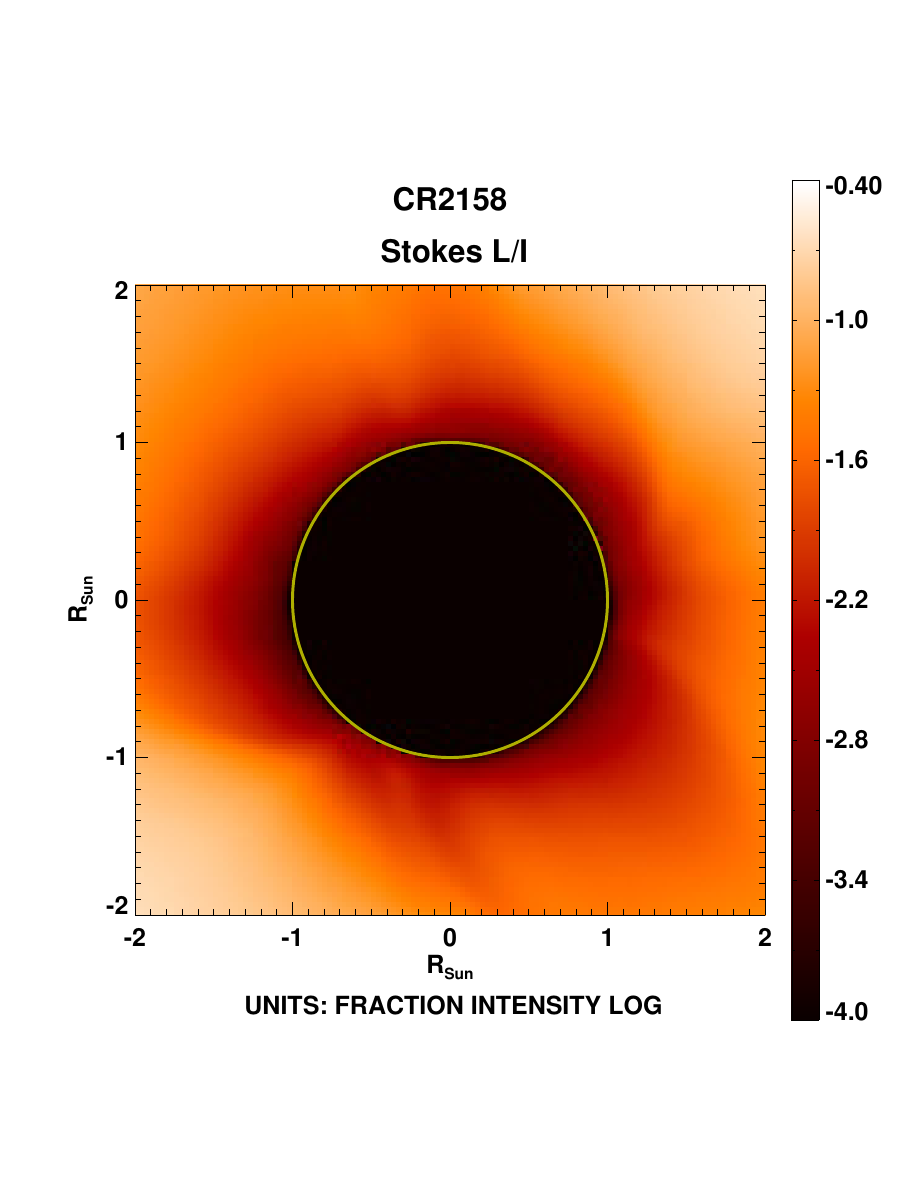}
    \includegraphics[width=0.11\textwidth,trim={1.1cm 2.7cm 0cm 2.9cm},clip]{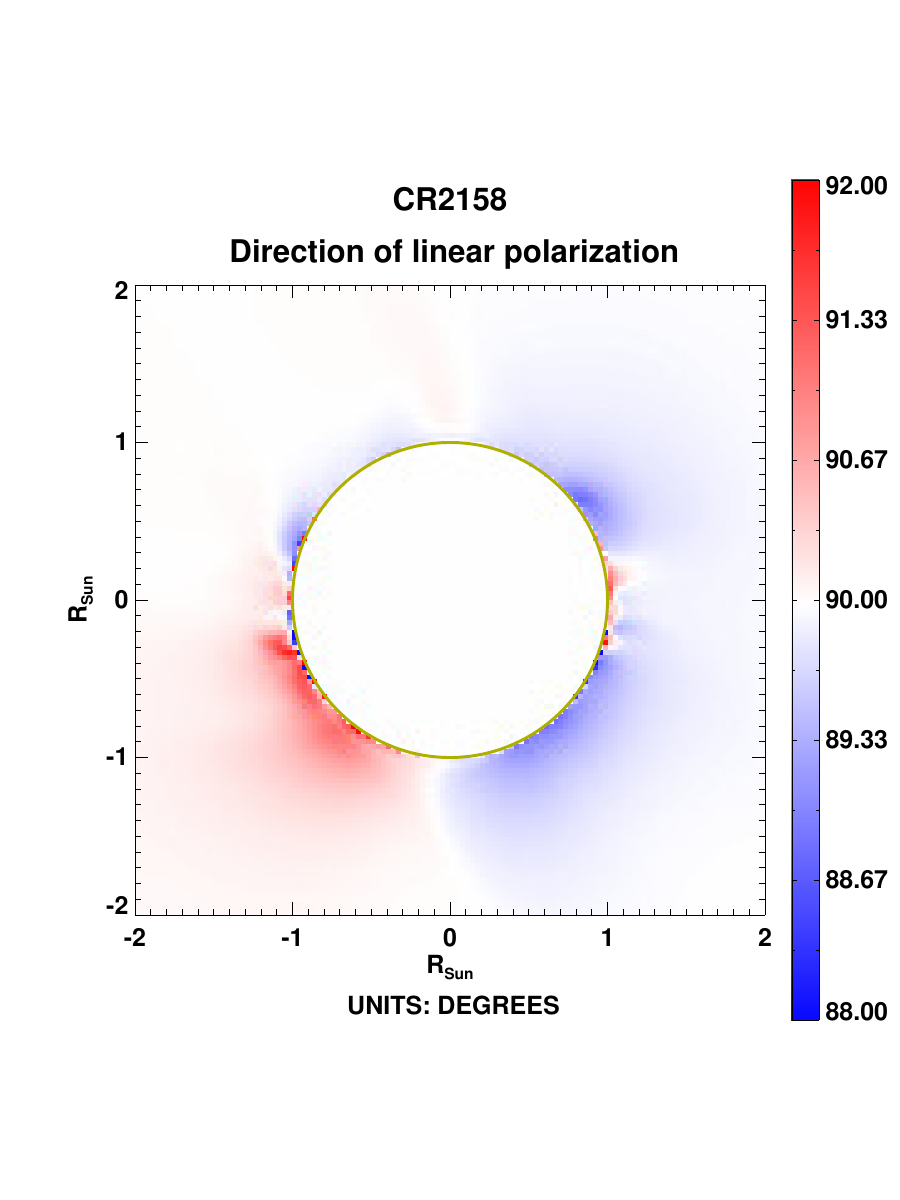}
    \includegraphics[width=0.115\textwidth,trim={0.3cm 1.2cm 0cm 0.5cm},clip]{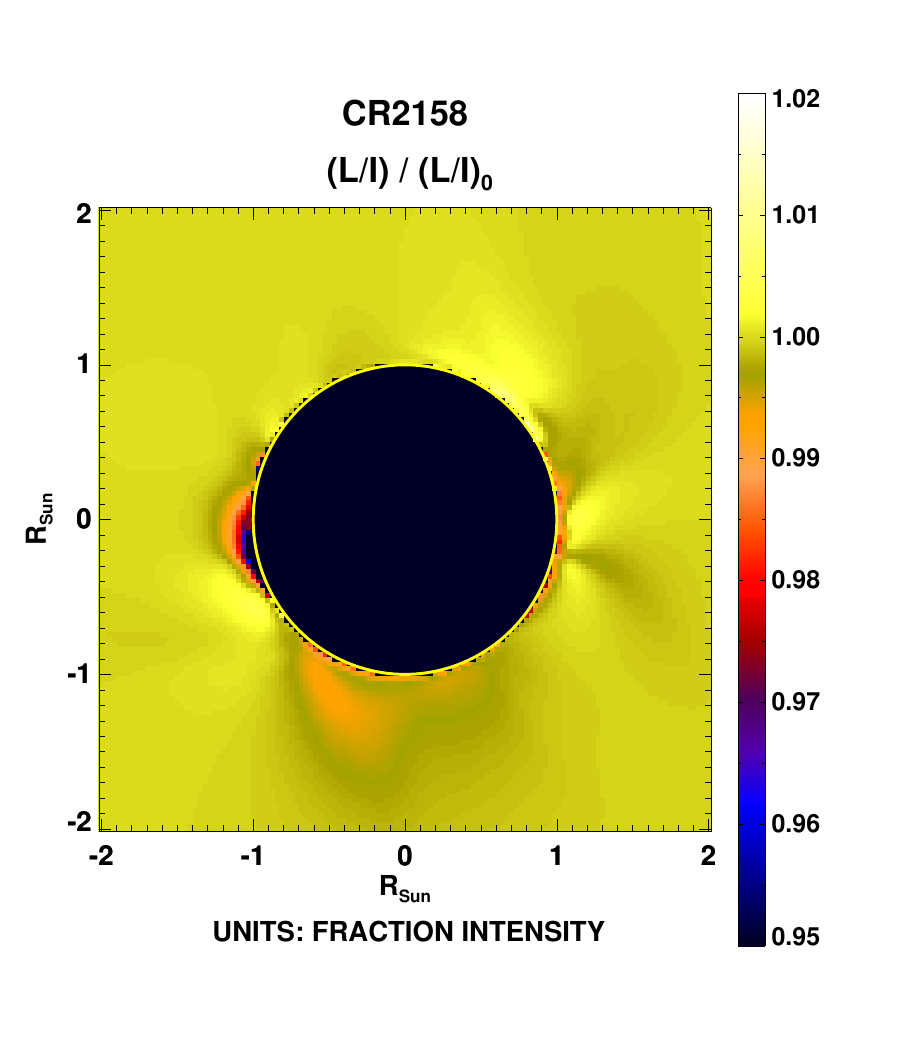}
    \includegraphics[width=0.11\textwidth,trim={1.1cm 2.5cm 0cm 2.9cm},clip]{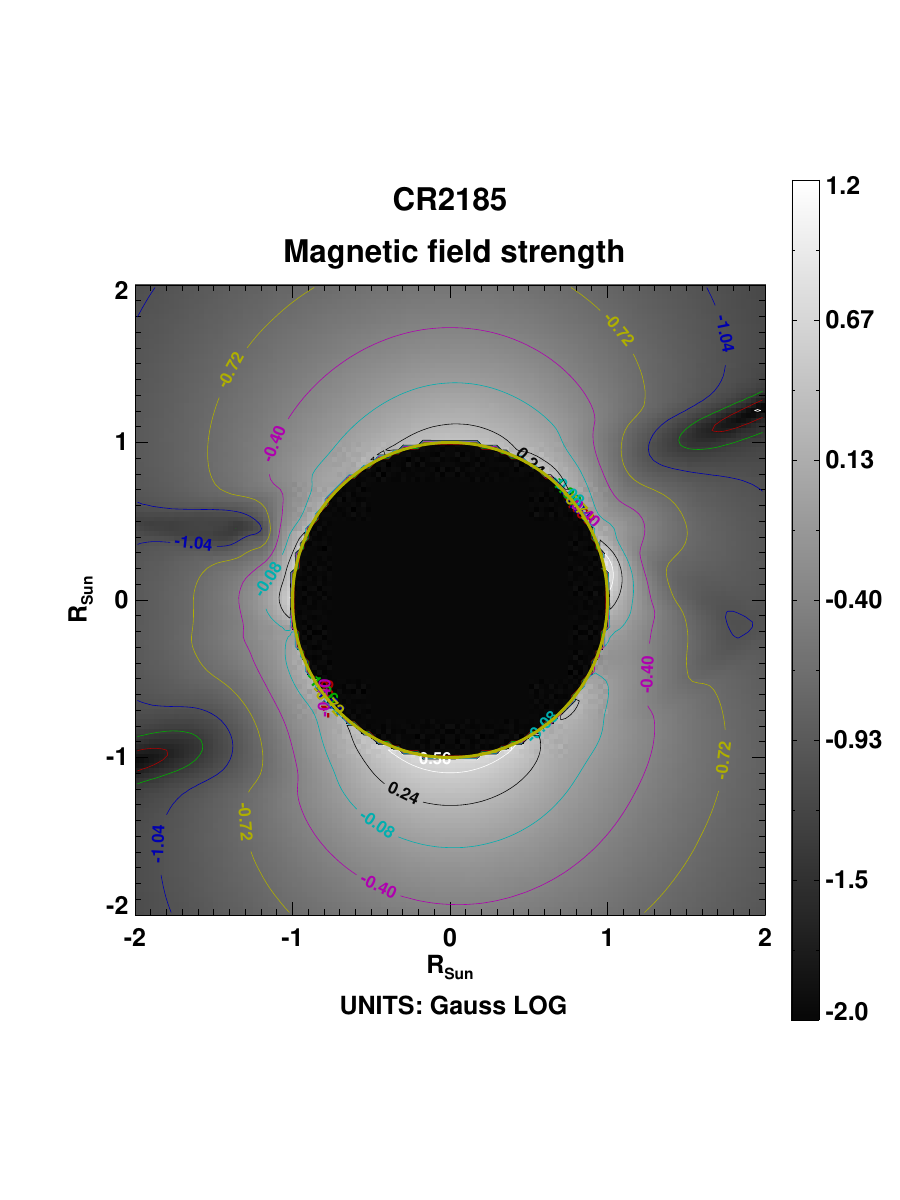}
    \includegraphics[width=0.11\textwidth,trim={1.1cm 2.5cm 0cm 2.9cm},clip]{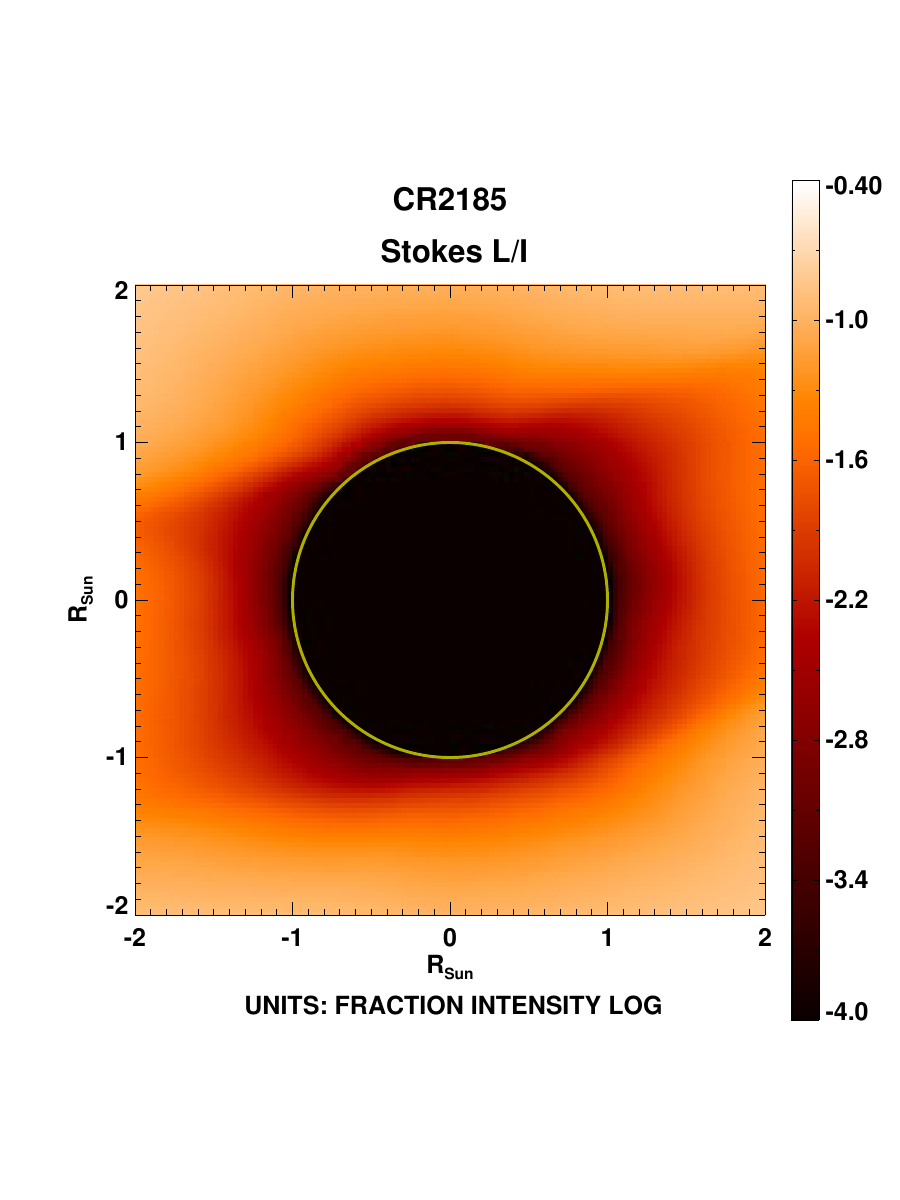}
    \includegraphics[width=0.11\textwidth,trim={1.1cm 2.7cm 0cm 2.9cm},clip]{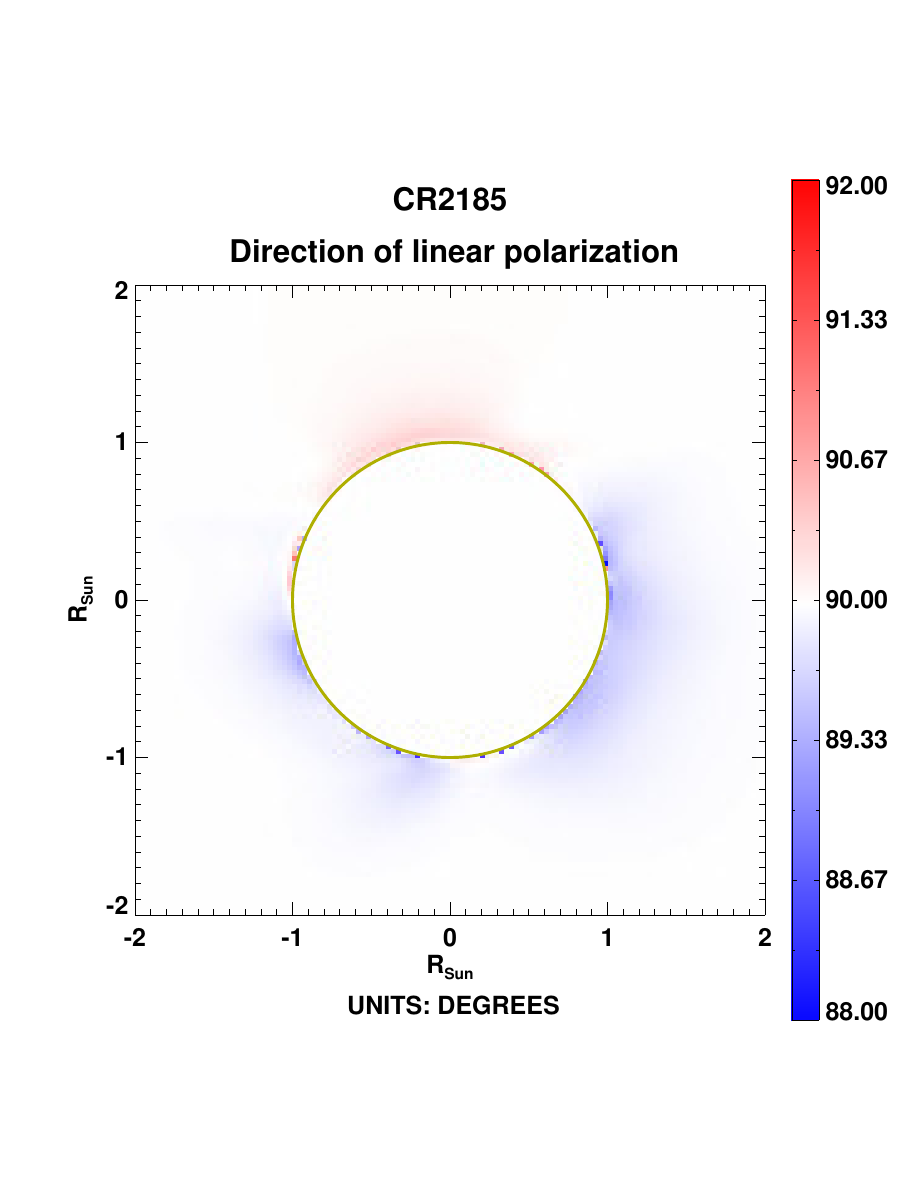}
    \includegraphics[width=0.115\textwidth,trim={0.3cm 1.2cm 0cm 0.5cm},clip]{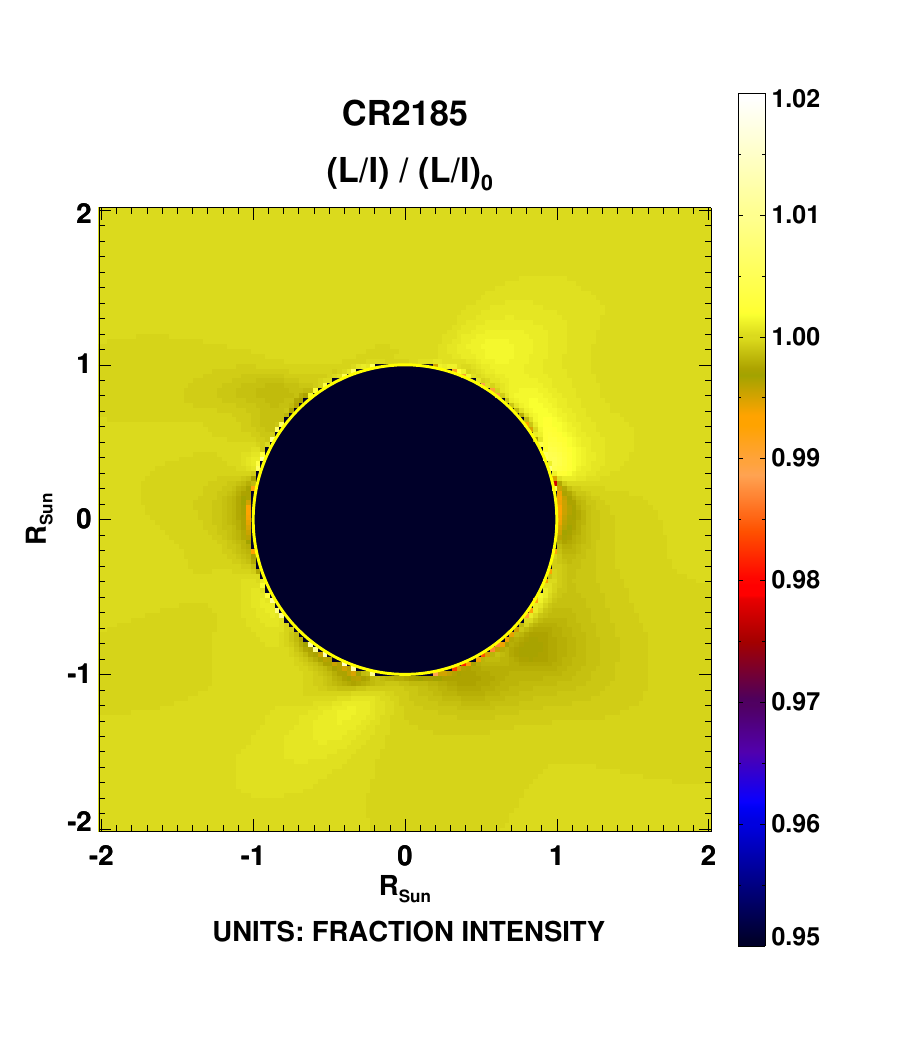}
    \includegraphics[width=0.11\textwidth,trim={1.1cm 2.5cm 0cm 2.9cm},clip]{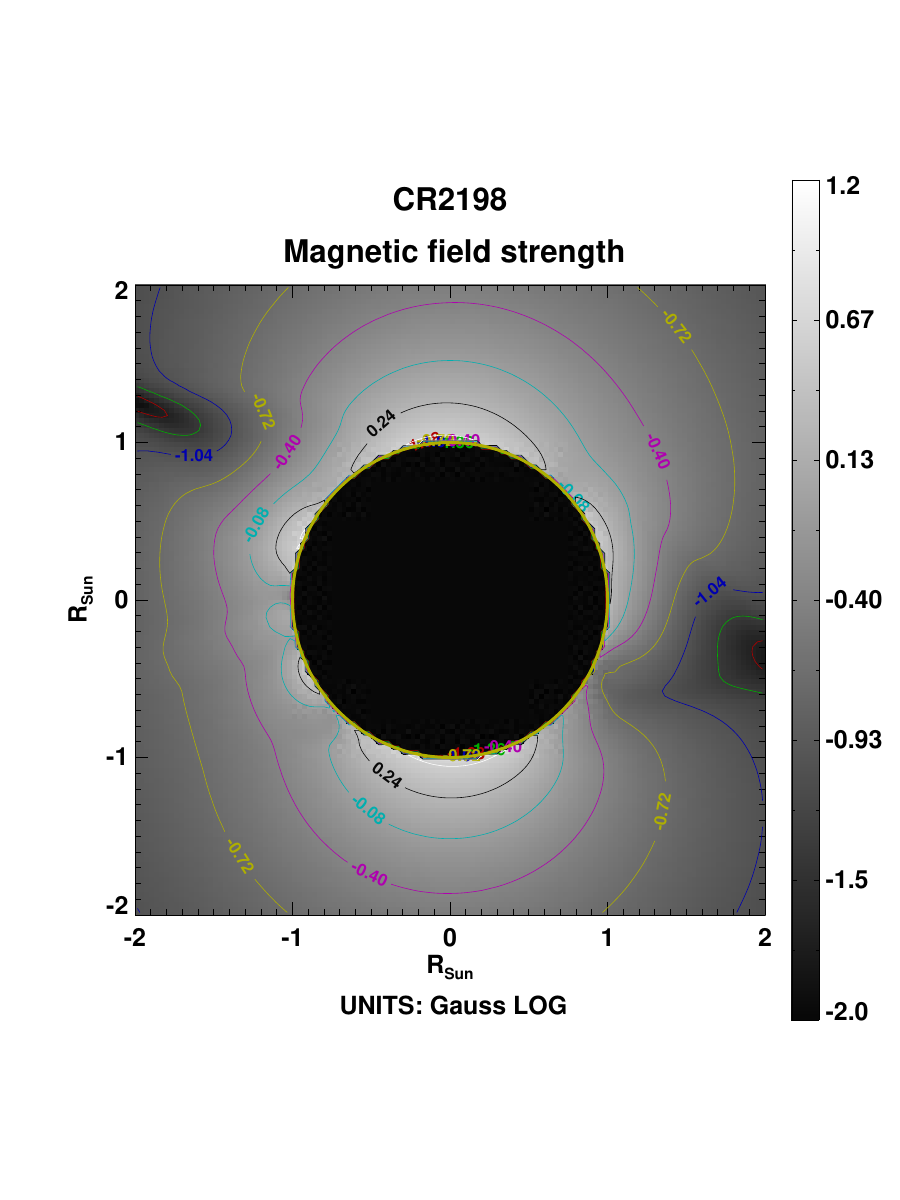}
    \includegraphics[width=0.11\textwidth,trim={1.1cm 2.5cm 0cm 2.9cm},clip]{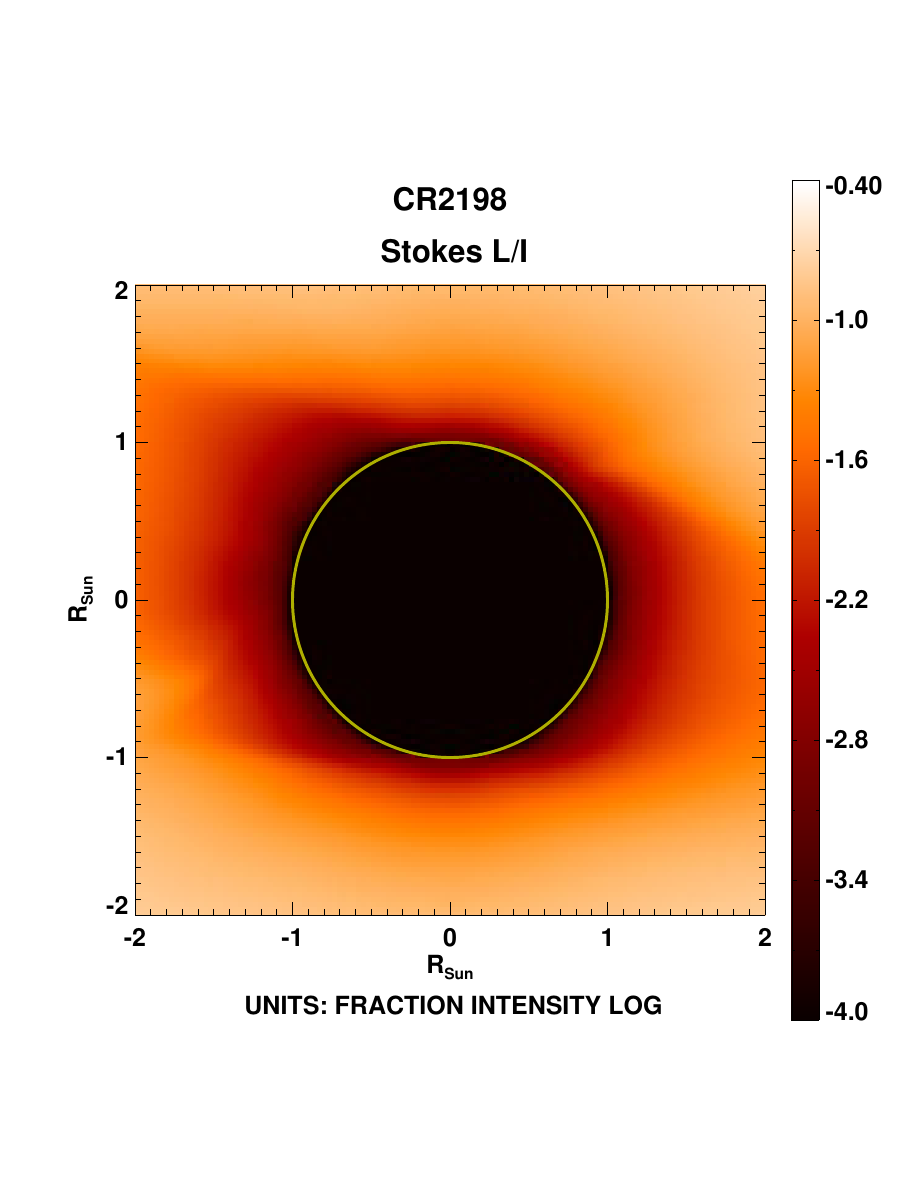}
    \includegraphics[width=0.11\textwidth,trim={1.1cm 2.7cm 0cm 2.9cm},clip]{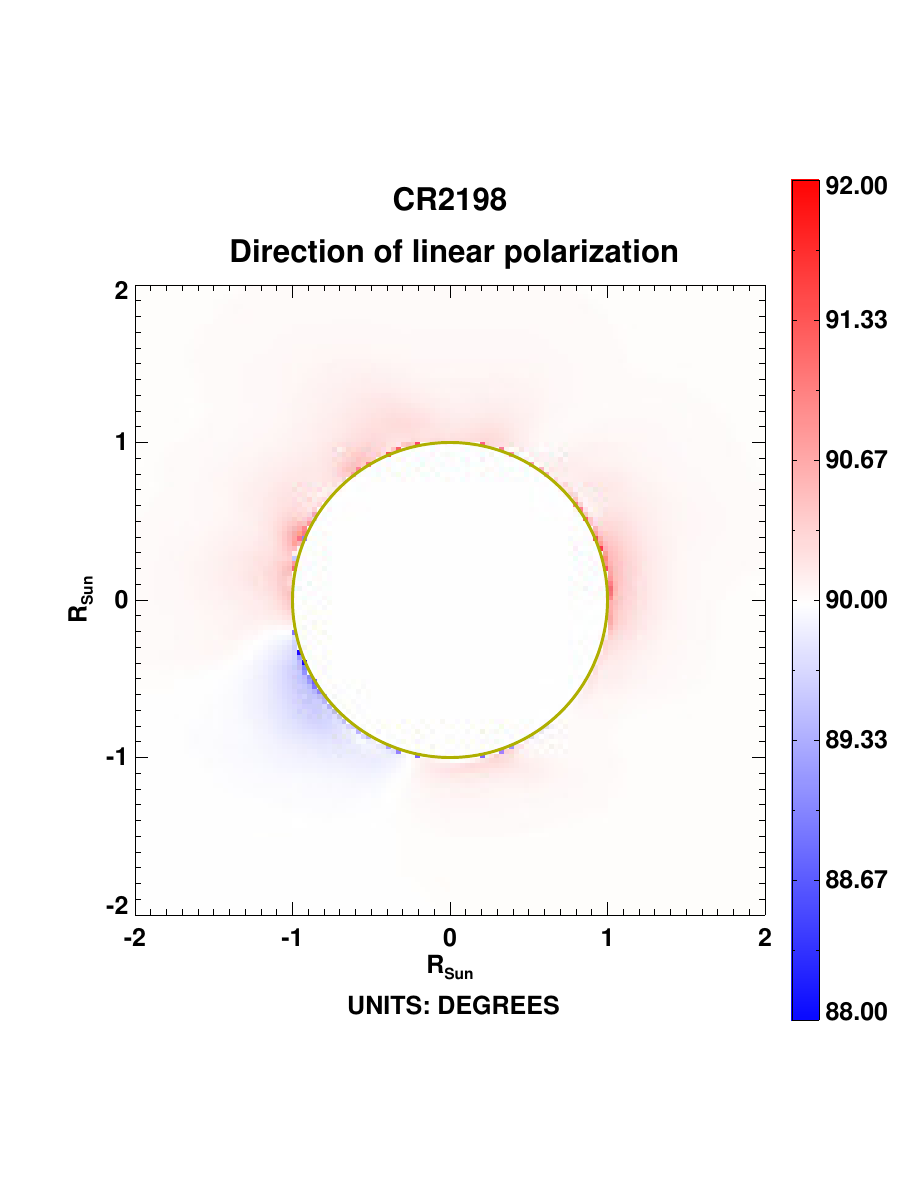}
    \includegraphics[width=0.115\textwidth,trim={0.3cm 1.2cm 0cm 0.5cm},clip]{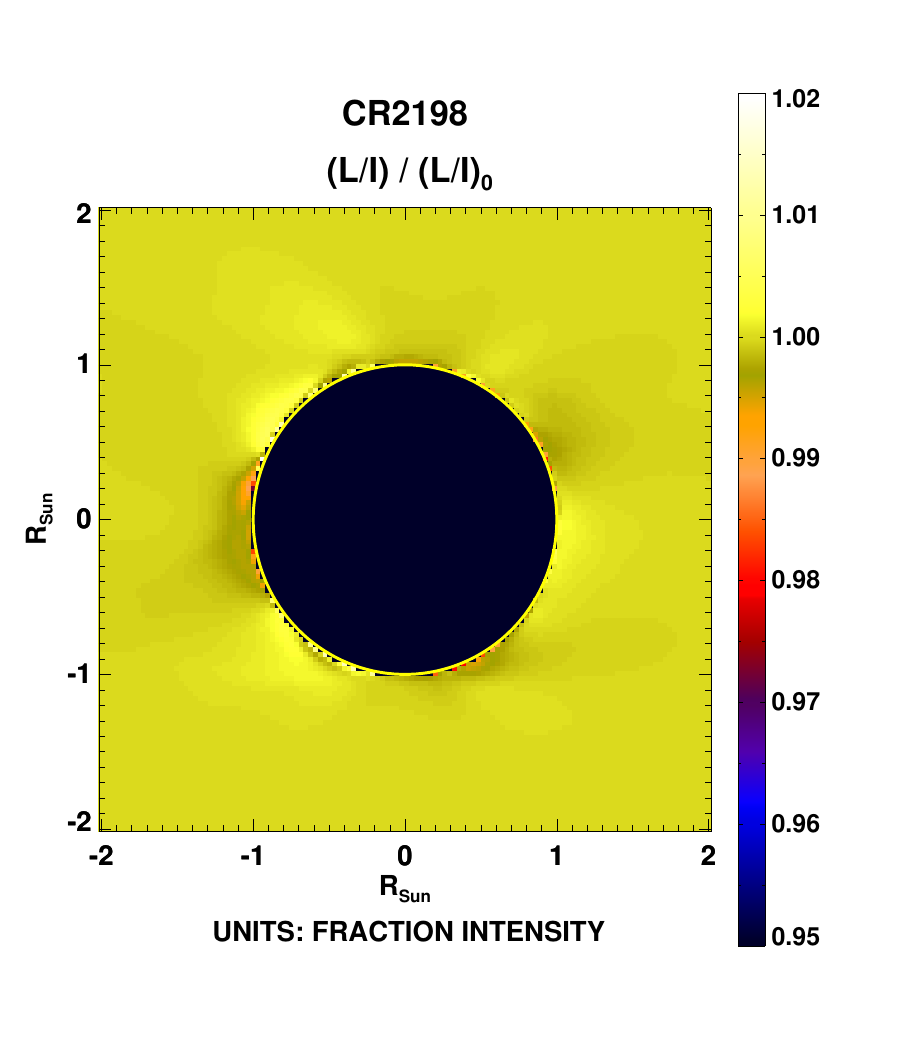}
    \includegraphics[width=0.11\textwidth,trim={1.1cm 2.5cm 0cm 2.9cm},clip]{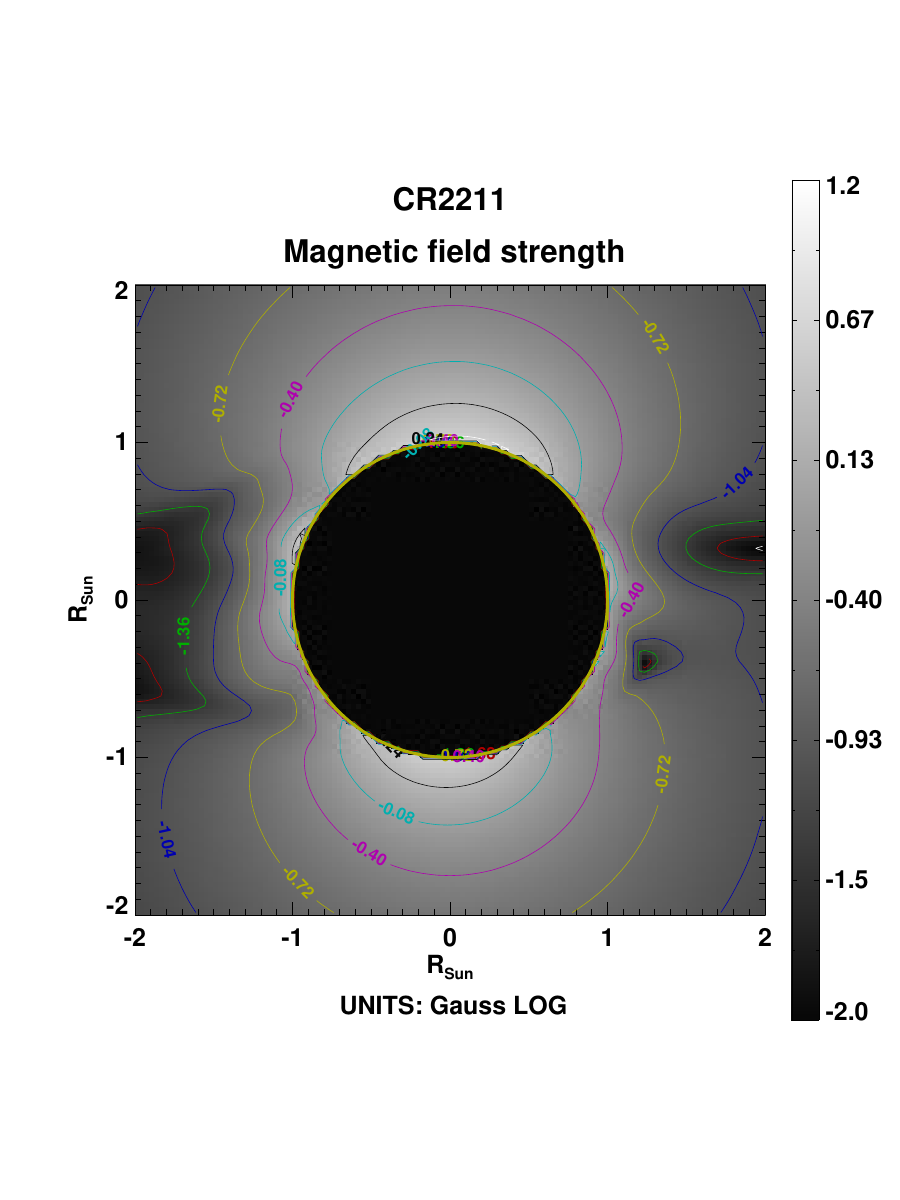}
    \includegraphics[width=0.11\textwidth,trim={1.1cm 2.5cm 0cm 2.9cm},clip]{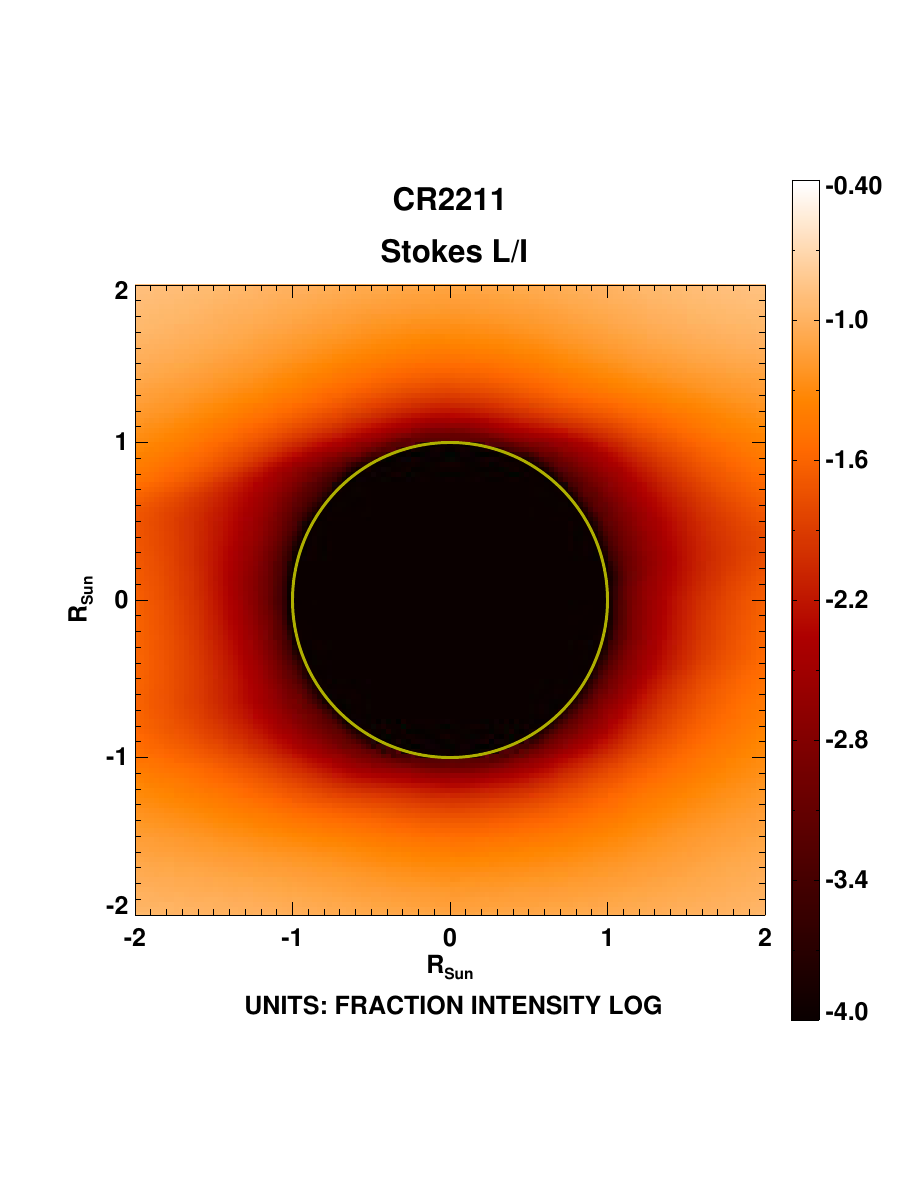}
    \includegraphics[width=0.11\textwidth,trim={1.1cm 2.7cm 0cm 2.9cm},clip]{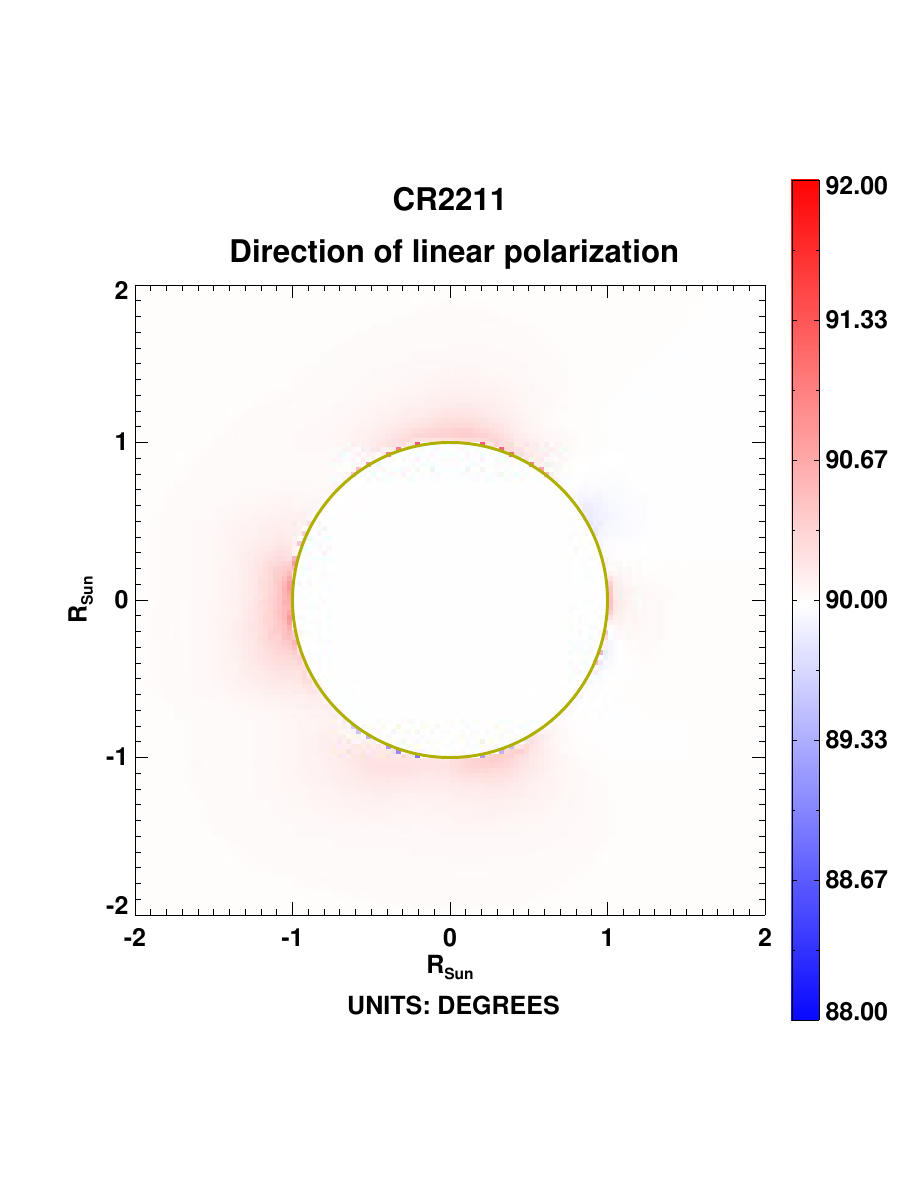}
    \includegraphics[width=0.115\textwidth,trim={0.3cm 1.2cm 0cm 0.5cm},clip]{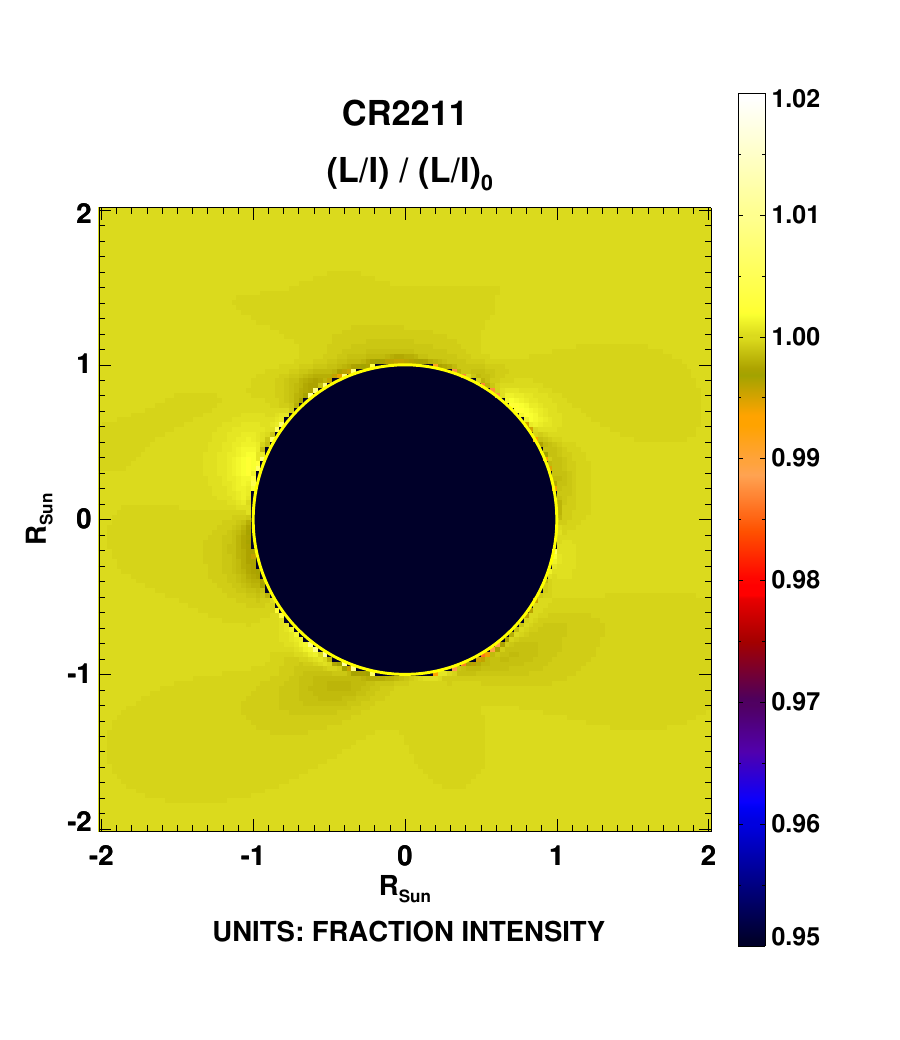}
    \includegraphics[width=0.11\textwidth,trim={1.1cm 2.5cm 0cm 2.9cm},clip]{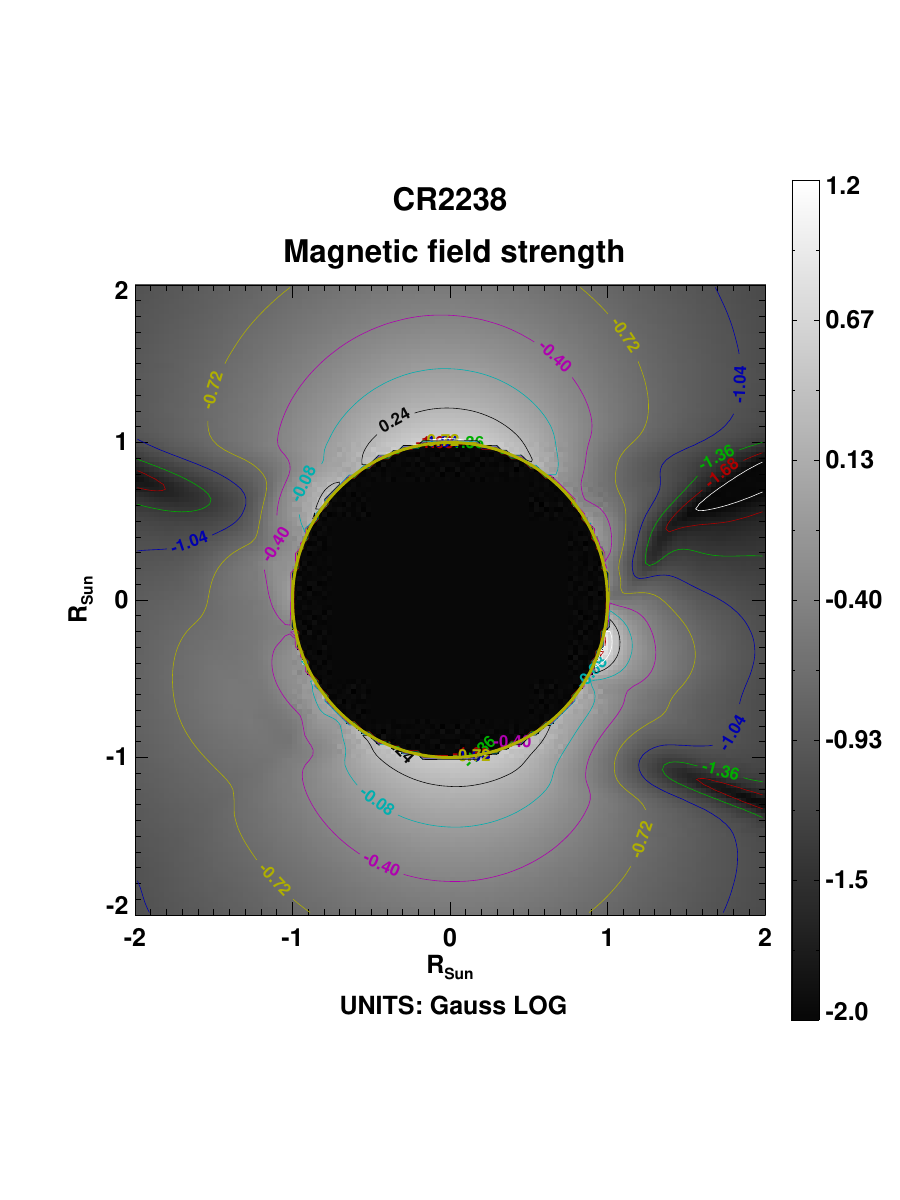}
    \includegraphics[width=0.11\textwidth,trim={1.1cm 2.5cm 0cm 2.9cm},clip]{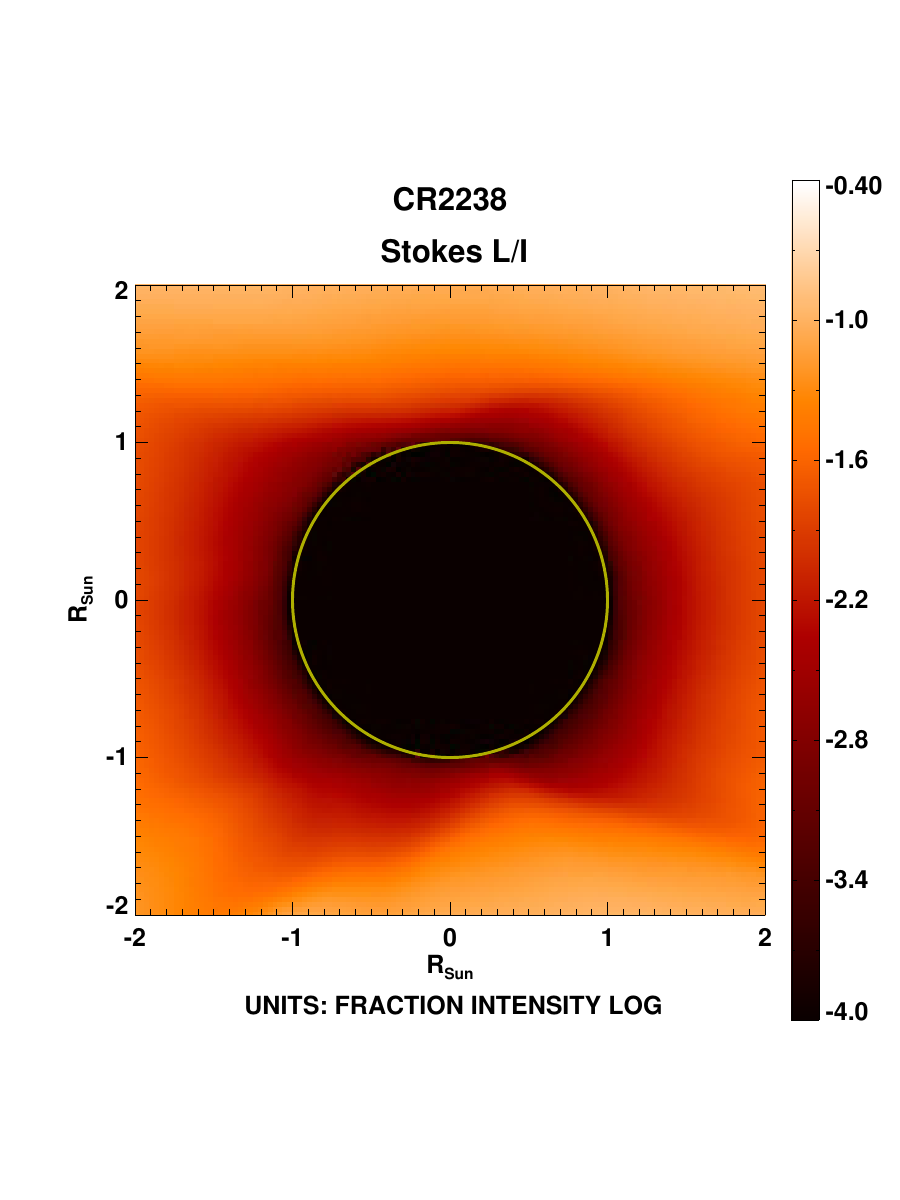}
    \includegraphics[width=0.11\textwidth,trim={1.1cm 2.7cm 0cm 2.9cm},clip]{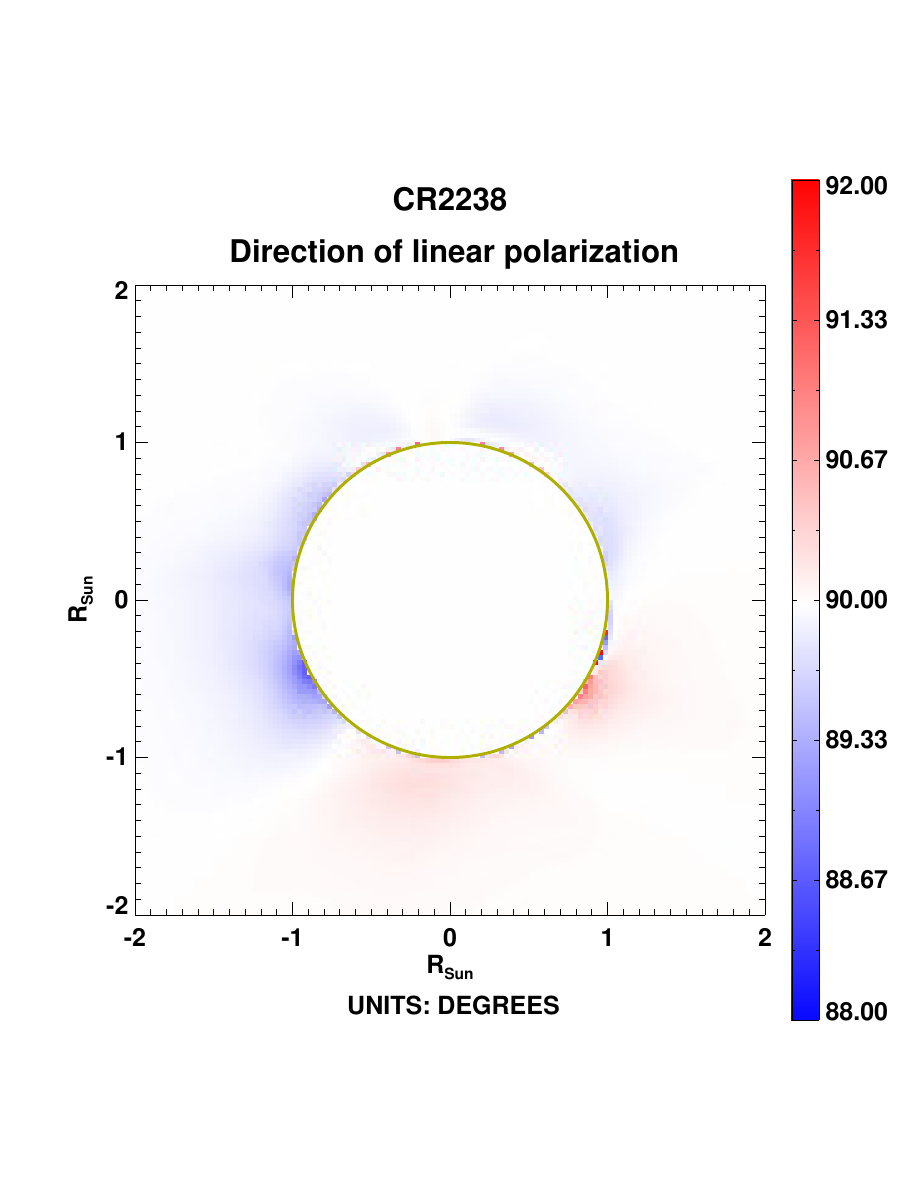}
    \includegraphics[width=0.115\textwidth,trim={0.3cm 1.2cm 0cm 0.5cm},clip]{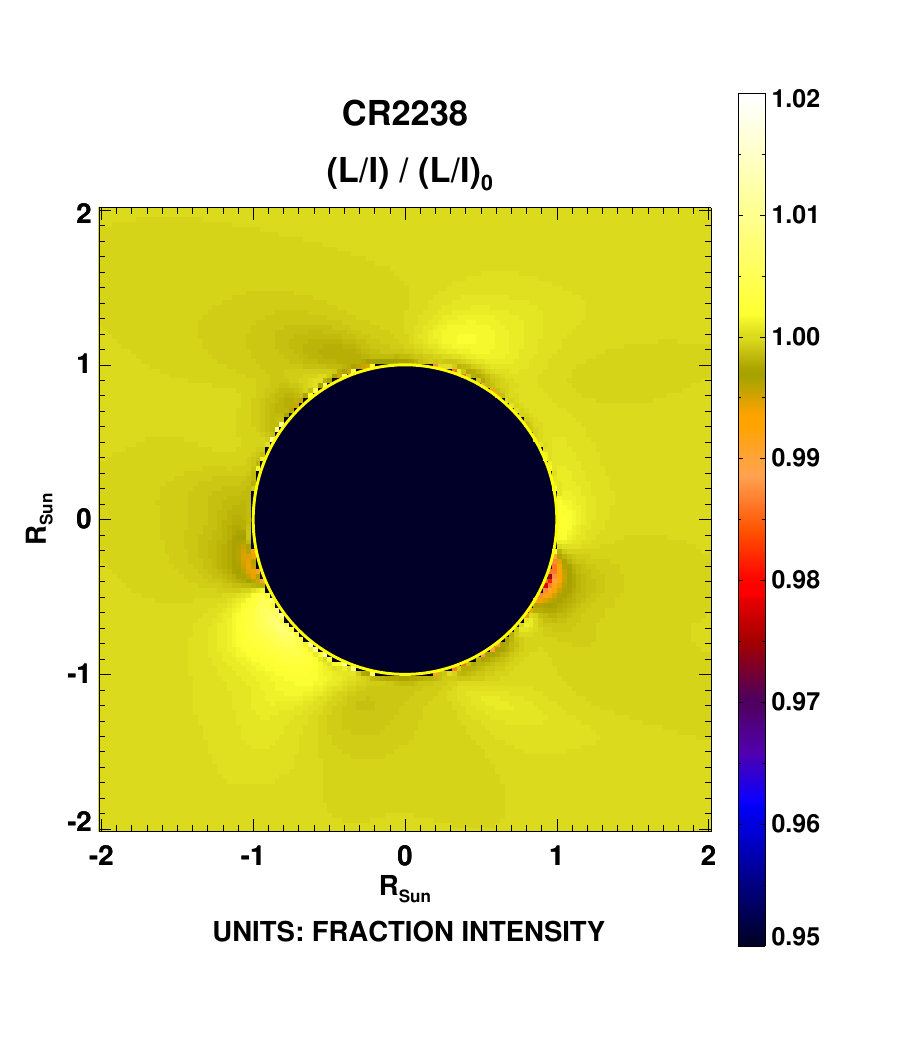}
    \caption{Same as Figure \ref{fig:bmag_stokesloi_az_offdisk} but for rest of the CR simulations.}
    \label{fig:supp_bmag_stokesloi_az_offdisk}
\end{figure}

\end{appendix}

\newpage
\bibliography{euv}{}
\bibliographystyle{aasjournal}

\end{document}